\documentclass[USenglish, a4paper, thm-restate,numberwithinsect, cleveref]{lipics-v2021}

\pdfoutput=1 
\hideLIPIcs  

\usepackage{graphicx}

\usepackage{amsmath,amsthm,amssymb,amsfonts}
\usepackage{nicefrac}
\usepackage{hyperref}
\usepackage{cleveref}
\crefname{appendix}{Appendix}{Appendices}
\Crefname{appendix}{Appendix}{Appendices}
\usepackage{xspace}

\usepackage{tabularx}
\usepackage{cite}
\usepackage{mathtools}
\usepackage{tikz}
\usetikzlibrary{arrows.meta,calc,positioning,fit,decorations.pathreplacing,decorations.pathmorphing,backgrounds,shapes.geometric}
\tikzset{
  vertex/.style={circle,draw=black!75,fill=white,inner sep=0pt,minimum size=5.2mm},
  smallv/.style={circle,draw=black!75,fill=white,inner sep=0pt,minimum size=3.2mm},
  root/.style={vertex,fill=black!18,draw=black,line width=.9pt},
  cycleedge/.style={black,line width=1pt},
  blueforest/.style={witnessblue,line width=1pt},
  orangeforest/.style={witnessorange,line width=1pt,dash pattern=on 3pt off 2pt},
  removed/.style={black!35,line width=.8pt},
  panel/.style={font=\small},
}

\definecolor{witnessblue}{HTML}{0072B2}
\definecolor{witnessorange}{HTML}{D55E00}
\definecolor{auxgreen}{HTML}{009E73}
\definecolor{auxpurple}{HTML}{7A3E9D}
\definecolor{complexitygreen}{HTML}{CFEAD6}
\definecolor{complexityred}{HTML}{F4CCCC}
\definecolor{complexitygray}{HTML}{DDDDDD}

\tikzstyle{knoten}=[circle,fill=white,draw=black,minimum size=7pt,inner sep=2pt]

\newcommand{\Oh}{\mathcal{O}}

\newcommand{\problemdef}[3]{
	\begin{center}\fbox{
	\begin{minipage}{0.95\textwidth}
        \vspace{3pt}
		\noindent
		\textsc{#1}
		\vspace{5pt}\\
		\setlength{\tabcolsep}{3pt}
		\begin{tabularx}{\textwidth}{@{}lX@{}}
			\text{Input:}     & #2 \\
			\text{Question:}  & #3
		\end{tabularx}
	\end{minipage}}
	\end{center}
}

\newtheorem*{rruleinner}{Tree Contraction Rule}
\newenvironment{rrule}{%
    \begin{rruleinner}%
    \phantomsection%
}{%
    \end{rruleinner}%
}
\newcommand{\TreeContractionRule}{%
    \hyperref[rule:contraction]{Tree Contraction Rule}\xspace}

\newtheorem*{cleanupruleinner}{Clean-Up Rule}
\newenvironment{cleanuprule}{%
    \begin{cleanupruleinner}%
    \phantomsection%
}{%
    \end{cleanupruleinner}%
}
\newcommand{\CleanupRule}{\hyperref[rule:cleanup]{Clean-Up Rule}\xspace}

\newcommand{\Gossip}{\textnormal{\textsc{Gossiping}}\xspace}

\newcommand{\tw}{\mathrm{tw}}
\newcommand{\treealpha}{\operatorname{tree\textnormal{-}\alpha}}
\newcommand{\cw}{\mathrm{cw}}

\usepackage[textsize=footnotesize,color=green!40]{todonotes}

\title{Recognizing Label-Connected Graphs}

\titlerunning{Recognizing Label-Connected Graphs}


\author{Ralf Klasing}{CNRS, LaBRI, Université de Bordeaux, Talence, France}{ralf.klasing@labri.fr}{https://orcid.org/0009-0004-9421-8525}{}

\author{George B. Mertzios}{Department of Computer Science, Durham University, UK}{george.mertzios@durham.ac.uk}{https://orcid.org/0000-0001-7182-585X}{}

\author{Hendrik Molter}{Chair for Algorithm Engineering, Hasso Plattner Institute, Potsdam, Germany \and Department of Computer Science, Ben-Gurion~University~of~the~Negev, 
	Beer-Sheva, 
	Israel}{hendrik.molter@hpi.de}{https://orcid.org/0000-0002-4590-798X}{}
	
\author{Nils Morawietz}{Institute of Computer Science, Friedrich Schiller University Jena,  Germany}{nils.morawietz@uni-jena.de}{https://orcid.org/0000-0002-7283-4982}{}

\author{Paul G. Spirakis}{School  of Computer Science and Informatics, University of Liverpool, UK}{p.spirakis@liverpool.ac.uk}{https://orcid.org/0000-0001-5396-3749}{}

\funding{Klasing is partially supported by the French ANR, project ANR-22-CE48-0001 (TEMPOGRAL). Mertzios is supported by the EPSRC grant EP/P020372/1. 
Molter is supported by the Israel Science Foundation, grant nr.~1470/24, by the European Union's Horizon Europe research and innovation programme under grant agreement 949707, by the European Research Council, grant nr.~101039913 (PARAPATH), and by the Deutsche Forschungsgemeinschaft (DFG, German Research Foundation) -- 565381415.
Spirakis is supported by the EPSRC grant EP/P02002X/1.
}

\acknowledgements{We used ChatGPT and Claude for mathematical
brainstorming, literature discovery, proof writing, consistency checks, and language
editing. For all results, we provided the LLMs with proof sketches and ideas, and subsequently checked and revised all mathematical
arguments, bibliographic claims, and formulations. We take full
responsibility for the manuscript.}

\authorrunning{R. Klasing, G. B. Mertzios, H. Molter, N. Morawietz, and P. G. Spirakis}

\Copyright{Ralf Klasing, George B. Mertzios, Hendrik Molter, Nils Morawietz, and Paul G. Spirakis}

\nolinenumbers 

\begin{document}
\renewcommand{\theHtheorem}{\thesection.\arabic{theorem}}
\providecommand{\theHlemma}{}
\renewcommand{\theHlemma}{\thesection.\arabic{theorem}}
\providecommand{\theHcorollary}{}
\renewcommand{\theHcorollary}{\thesection.\arabic{theorem}}
\maketitle
\begin{abstract}
In the classical gossiping setting, each participant initially knows a
distinct piece of information. Whenever two participants call each
other, they exchange everything they currently know. We study
\Gossip: given an undirected multigraph whose vertices represent
participants and whose edges represent available calls, decide whether
the calls can be ordered so that every participant learns all
information, using each edge at most once.
Equivalently, we ask whether assigning one time label to each edge
can make the graph temporally connected by strict temporal paths.
This recognition problem is known to be NP-complete; we study its parameterized
complexity.

After linear-time preprocessing, we give an exact
$\Oh(3^{n/3}n^3)$-time algorithm and a deterministic
$2^{\Oh(\tw)}\cdot n$-time algorithm, where $n$ is the number of vertices
and $\tw$ is the treewidth of the underlying simple graph. We introduce
a Tree Contraction Rule that extends the treewidth algorithm to single-exponential algorithms
for structural parameters that may remain small on dense graphs, such
as tree-independence number and clique-width, without requiring a
decomposition as input. For the excess $\delta=m-(2n-4)$ over the minimum
number of calls, where $m$ is the number of edges, we give a randomized
$2^{\Oh(\delta)}\cdot n^3$-time algorithm with
one-sided error. Its main subroutine solves \textsc{Rooted Independent
Cycle}: finding a cycle through four prescribed vertices whose edges
are independent in a represented linear matroid. We solve this problem
deterministically in $2^{\Oh(r)}m$ arithmetic operations over the finite
field representing the matroid, where $r$ is the matroid rank.
We also show that \Gossip is solvable in
polynomial time at maximum degree four, counting parallel edges, but
NP-complete on planar simple graphs of maximum degree five. As a direct consequence, the
treewidth algorithm gives a $2^{\Oh(\sqrt n)}$-time algorithm on planar
graphs.
\end{abstract}

\clearpage
\section{Introduction}

Distributing information among participants is a basic task in
communication networks. In a peer-to-peer exchange, two participants
can share both their own information and everything they have learned
from earlier exchanges. When the available pairwise exchanges are
fixed, their order determines how far information can spread. We ask
whether these exchanges can be scheduled so that every participant
receives all information, without introducing additional exchanges.

The classical gossiping problem captures this question in a particularly
clean form.  Each participant initially knows a distinct piece of
information, and whenever two participants call each other, they
exchange everything they currently know.  If every pair may communicate,
$2n-4$ calls are necessary and sufficient for $n\geq4$
participants~\cite{BakerS72,HajnalMS72}.  We instead prescribe the
available calls by an undirected multigraph: vertices represent
participants, and each edge represents one call between its endpoints.
The task is to decide whether these calls can be ordered so that all
information reaches every participant.  Thus the difficulty lies not in
selecting sufficiently many calls, but in coordinating a fixed and
possibly highly constrained set of calls.

This question is also a temporal graph realization problem.
A temporal graph $(G,\lambda)$ consists of an undirected multigraph
$G=(V,E)$ and a labeling $\lambda\colon E\to\mathbb{N}$ assigning a time
to each edge. A \emph{strict temporal path} is a path whose edge labels
strictly increase in traversal order.
The temporal graph is \emph{temporally connected} if such
a path exists for every ordered pair of vertices. 
Each edge receives exactly one label.\footnote{We use the temporal-graph terminology of edge labels and the gossiping terminology of calls interchangeably: an edge represents a call, and its label specifies when that call takes place.}
Different edges may receive equal labels. However, we can give all edges
distinct labels while preserving the order of edges whose labels were
already different. Every strict temporal path is preserved, so it suffices
to consider a total order of the calls (\Cref{sec:preliminaries}).
Loops are irrelevant, and parallel edges
represent distinct calls. A graph with at most one vertex is a
yes-instance. Disconnected graphs are immediate no-instances.
\problemdef{\Gossip}{An undirected multigraph $G=(V,E)$.}{Is there a
labeling $\lambda\colon E\to\mathbb{N}$ such that $(G,\lambda)$ is
temporally connected?}
Following the classical literature, we call a multigraph
\emph{label-connected} if it is a yes-instance of \Gossip.
Throughout, $n$ and $m$ denote the numbers of vertices and edges, and
the \emph{support} is the underlying simple graph (see
\Cref{sec:preliminaries} for the formal conventions).
G\"obel, Cerdeira, and Veldman~\cite{gobel1991label} proved that
recognizing label-connected graphs is NP-complete. A classical characterization
describes the yes-instances using either two spanning trees (a
\emph{$T$-witness}) or a cycle together with two spanning forests (an
\emph{$M$-witness})~\cite{BuroschLM88,BuroschL92,Krumme96}.
Both witnesses describe how information spreads. For two edge-disjoint
spanning trees, information is gathered at a root along one tree and
broadcast along the other. The root acts as a \emph{pivot vertex},
separating gathering from broadcasting~\cite{CasteigtsPS21,CasteigtsRRZ24}.
If the trees share an edge, information is gathered at its two
endpoints, exchanged by calling that edge, and then broadcast along
the second tree. The shared edge acts as a \emph{pivot edge}~\cite{CarnevaleCC25}.
We discuss these connections to temporal spanners in
\Cref{sec:witness-labelings}.
In the second witness, each
forest has four components, all meeting the cycle. Information is gathered
in four groups along the first forest, merged along the cycle via four
\emph{essential} cycle edges, and broadcast along the second forest;
\Cref{fig:labeling-example} in \Cref{sec:witnesses} illustrates such a schedule.
The characterization reduces recognition to finding one of these
witnesses. We determine which graph parameters and structural
restrictions make this task tractable and where this tractability ends.

\subsection{Our Contribution}
Our algorithms find structural witnesses and turn them into temporally
connecting labelings in $\Oh(n+m)$ additional time
(\Cref{sec:witnesses}). Thus, they construct a labeling whenever they
recognize a yes-instance. Since the classical witness characterization
is the structural foundation of our algorithms, we include a
self-contained proof in \Cref{app:characterization} using the notation
and multigraph conventions of this paper. Additionally, we provide a Lean formalization\footnote{Lean 4 is an automatic checkable proof system.} of this proof in~\cite{lax-630294}.
We summarize our main results below. \Cref{sec:technical-overview}
gives a technical overview of the paper, explaining the ideas behind
our algorithms and hardness reductions and how the structural results
connect them, before the detailed proofs in the subsequent sections.

\subparagraph*{Preprocessing and Exact Recognition (\Cref{sec:exact-preprocessing}).}
We first apply a simple \CleanupRule that removes loops and reduces the number of parallel edges
between any pair of vertices to at most three in linear time;
afterwards, $m\in\Oh(n^2)$.
We introduce the \TreeContractionRule, which contracts vertex sets that induce three
edge-disjoint spanning trees. It preserves label-connectivity and can
be applied exhaustively in $\Oh(n^2)$ time
(\Cref{sec:exact-preprocessing}). This reduction is central to our
algorithms for structural parameters: although the input may contain
arbitrarily dense regions, the reduced graph satisfies a local sparsity
bound. We use this bound to control its treewidth in terms of parameters
that need not bound the treewidth of the original graph. Independently, enumerating candidate cycles
and testing whether they extend to a witness gives an exact
$\Oh(3^{n/3}n^3)$-time algorithm (\Cref{thm:exact}).

\subparagraph*{Structural Parameters (\Cref{sec:structural-parameters}).}
We show how to solve \Gossip in $2^{\Oh(\tw)}\cdot n$ time, where
$\tw$ is the treewidth of the support (\Cref{thm:treewidth}).
Courcelle's theorem~\cite{courcelle1990monadic} already yields
fixed-parameter tractability, but in general with a non-elementary
parameter dependence.
As an immediate consequence of the treewidth bound for planar graphs,
we also obtain a $2^{\Oh(\sqrt n)}$-time algorithm on planar multigraphs
(\Cref{cor:planar-subexponential}).

The Tree Contraction Rule yields a kernel with $\Oh(\alpha)$ vertices
for the independence number~$\alpha$ (\Cref{cor:independence-kernel}).
Combined with our treewidth algorithm, it also gives
$\Oh(n^2)+2^{\Oh(p)}\cdot n$-time algorithms, where $p$ is the
tree-independence number, chordal vertex deletion number, or
clique-width\footnote{Formal definitions of these parameters are given in \Cref{sec:preliminaries}.}
(\Cref{cor:tree-independence,cor:clique-width}). No decomposition or deletion set
is needed as input.

\subparagraph*{The Excess over the Minimum Number of Calls (\Cref{sec:above-bound}).}
Since $T$-graphs can be recognized in polynomial time, we focus on
finding an $M$-witness whose cycle has length at most a given bound $k$.
We give a deterministic $n^{k+\Oh(1)}$-time XP algorithm and a
randomized $2^{\Oh(k)}m^3$-time FPT algorithm for this task
(\Cref{prop:above-xp,thm:short-m-witness}). The randomized algorithm
never accepts a no-instance and errs on yes-instances with probability
at most $1/n$.

Every connected yes-instance with $n\geq4$ has at least $2n-4$ edges.
For the \emph{excess} $\delta=m-(2n-4)$, every $M$-witness has cycle
length at most $\delta+4$. Our algorithms therefore yield a deterministic
$n^{\delta+\Oh(1)}$-time algorithm and a randomized
$2^{\Oh(\delta)}n^3$-time algorithm for \Gossip, with the same error
guarantee. A supplied bound $k$ can be smaller than $\delta+4$.

The XP algorithm enumerates short cycles and tests whether they extend to an
$M$-witness. For the randomized FPT algorithm, we encode the two
forests of an $M$-witness by a linear matroid union. Duality and random
projection reduce the remaining task to \textsc{Rooted Independent
Cycle} (RIC): given a graph, four prescribed vertices, and a linear
matroid on its edges represented by a matrix, find a cycle through all
four vertices whose edge set is independent in the matroid. We solve
RIC deterministically in $2^{\Oh(r)}m$ arithmetic operations over the finite
field representing the matroid, where $r$ is the matroid rank
(\Cref{thm:ric-fpt}), which our reduction bounds by $k$.

\subparagraph*{A Sharp Degree Threshold and Hardness (\Cref{sec:degree-complexity}).}
We show that \Gossip is solvable in $\Oh(n^3)$ time on multigraphs of maximum degree
four, where degrees count parallel edges, but NP-complete on planar
simple graphs of maximum degree five
(\Cref{cor:degree-four,thm:deg5-hard}). Moreover, we prove that \Gossip remains
NP-complete on multigraphs with at most four edges of multiplicity one
(\Cref{lem:multiplicity-one}). As standard consequences of the hardness
reductions, ETH\footnote{A claimed refutation of ETH was announced in
October 2026 and is undergoing community examination~\cite{ImpagliazzoPaturiWilliams2026ETH}.
Our ETH-based statements express conditional consequences of the
reductions and should be read subject to the outcome of that examination.} rules out
$2^{o(n+m)}$ time on simple graphs of maximum degree five
(\Cref{cor:eth}) and $2^{o(\sqrt n)}n^{\Oh(1)}$ time on planar ones
(\Cref{cor:planar-eth}). These bounds match the treewidth and planar
dependences up to constants in the exponent and exclude a
subexponential exact algorithm, but leave the best exponential base open.

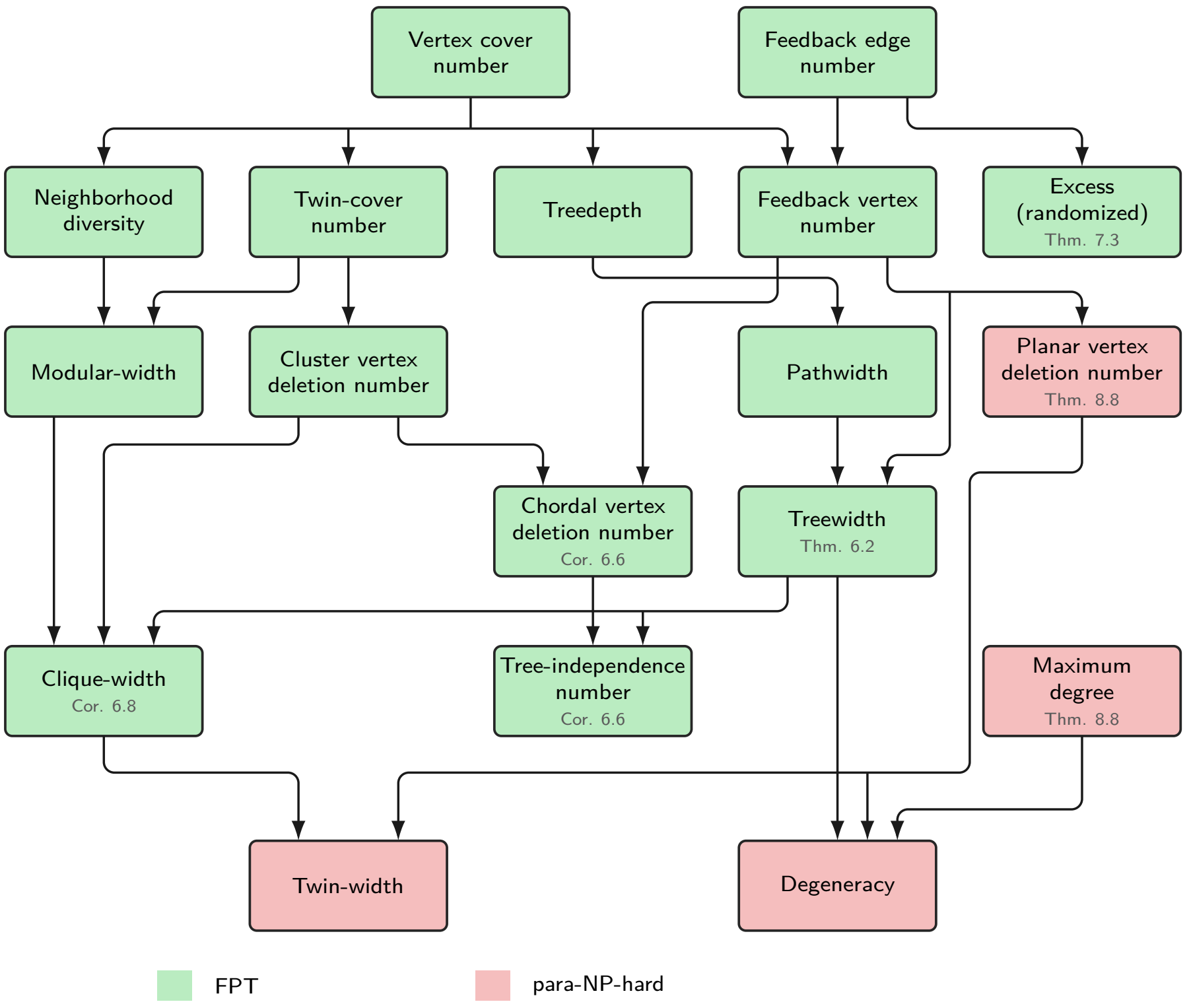
\begin{figure}[t]
\centering
\providecommand{\refnote}[1]{{\tiny\color{black!65}#1}}
\begin{tikzpicture}[
  every node/.style={font=\fontsize{6.3}{7.4}\selectfont\sffamily,align=center},
  parameter/.style={draw=black!85,line width=.7pt,rounded corners=2pt,
                    text width=18.7mm,minimum height=9mm,inner sep=1.5pt},
  fpt/.style={parameter,fill=complexitygreen!90!green},
  hard/.style={parameter,fill=complexityred!93!red},
  unknown/.style={parameter,fill=complexitygray},
  line/.style={draw=black!90,line width=.65pt,rounded corners=3pt},
  relation/.style={line,-{Latex[length=2mm,width=1.4mm]}}
]
  \node[fpt] (vc) at (-1.225,8.000) {Vertex cover\\number};
  \node[fpt] (fen) at (2.450,8.000) {Feedback edge\\number};
  \node[fpt] (nd) at (-4.900,6.400) {Neighborhood\\diversity};
  \node[fpt] (tc) at (-2.450,6.400) {Twin-cover\\number};
  \node[fpt] (td) at (0.000,6.400) {Treedepth};
  \node[fpt] (fvs) at (2.450,6.400) {Feedback vertex\\number};
  \node[fpt] (ex) at (4.900,6.400) {Excess\\(randomized)\\\refnote{Thm.~\ref{thm:short-m-witness}}};
  \node[fpt] (mw) at (-4.900,4.800) {Modular-width};
  \node[fpt] (cvd) at (-2.450,4.800) {Cluster vertex\\deletion number};
  \node[fpt] (pw) at (2.450,4.800) {Pathwidth};
  \node[hard] (dp) at (4.900,4.800) {Planar vertex\\deletion number\\\refnote{Thm.~\ref{thm:deg5-hard}}};
  \node[fpt] (chd) at (0.000,3.200) {Chordal vertex\\deletion number\\\refnote{Cor.~\ref{cor:tree-independence}}};
  \node[fpt] (tw) at (2.450,3.200) {Treewidth\\\refnote{Thm.~\ref{thm:treewidth}}};
  \node[fpt] (cw) at (-4.900,1.600) {Clique-width\\\refnote{Cor.~\ref{cor:clique-width}}};
  \node[fpt] (ta) at (0.000,1.600) {Tree-independence\\number\\\refnote{Cor.~\ref{cor:tree-independence}}};
  \node[hard] (tww) at (-2.450,-0.350) {Twin-width};
  \node[hard] (deg) at (2.450,-0.350) {Degeneracy};
  \node[hard] (delta) at (4.900,1.600) {Maximum\\degree\\\refnote{Thm.~\ref{thm:deg5-hard}}};
  \draw[line] (-1.225,7.550) -- (-1.225,7.235);
  \draw[relation] (-1.225,7.235) -- (-4.900,7.235) -- (-4.900,6.850);
  \draw[relation] (-1.225,7.235) -- (1.950,7.235) -- (1.950,6.850);
  \draw[relation] (-2.450,7.235) -- (-2.450,6.850);
  \draw[relation] (0.000,7.235) -- (0.000,6.850);
  \draw[relation] (2.450,7.550) -- (2.450,6.850);
  \draw[relation] (3.150,7.550) -- (3.150,7.235) -- (4.900,7.235) -- (4.900,6.850);
  \draw[relation] (-4.900,5.950) -- (-4.900,5.250);
  \draw[relation] (-2.950,5.950) -- (-2.950,5.600) -- (-4.400,5.600) -- (-4.400,5.250);
  \draw[relation] (-2.450,5.950) -- (-2.450,5.250);
  \draw[relation] (0.000,5.950) -- (0.000,5.740) -- (2.450,5.740) -- (2.450,5.250);
  \draw[relation] (1.850,5.950) -- (1.850,5.495) -- (0.500,5.495) -- (0.500,3.650);
  \draw[relation] (2.950,5.950) -- (2.950,5.600) -- (4.900,5.600) -- (4.900,5.250);
  \draw[relation] (3.575,5.600) -- (3.575,3.965) -- (2.950,3.965) -- (2.950,3.650);
  \draw[relation] (-1.950,4.350) -- (-1.950,4.070) -- (-0.500,4.070) -- (-0.500,3.650);
  \draw[relation] (2.450,4.350) -- (2.450,3.650);
  \draw[relation] (-2.950,4.350) -- (-2.950,4.070) -- (-4.900,4.070) -- (-4.900,2.050);
  \draw[relation] (-5.400,4.350) -- (-5.400,2.050);
  \draw[relation] (0.000,2.750) -- (0.000,2.050);
  \draw[relation] (1.950,2.750) -- (1.950,2.400) -- (-4.400,2.400) -- (-4.400,2.050);
  \draw[relation] (0.500,2.400) -- (0.500,2.050);
  \draw[relation] (4.900,4.350) -- (4.900,3.790) -- (3.775,3.790) -- (3.775,0.783) -- (-1.950,0.783) -- (-1.950,0.100);
  \draw[relation] (2.750,0.783) -- (2.750,0.100);
  \draw[relation] (2.450,2.750) -- (2.450,0.100);
  \draw[relation] (4.900,1.150) -- (4.900,0.415) -- (3.050,0.415) -- (3.050,0.100);
  \draw[relation] (-4.900,1.150) -- (-4.900,0.783) -- (-2.950,0.783) -- (-2.950,0.100);
  \fill[complexitygreen!90!green] (-4.365,-1.500) rectangle ++(0.35,0.3);
  \node[anchor=west] at (-3.915,-1.350) {FPT};
  \fill[complexityred!93!red] (-1.180,-1.500) rectangle ++(0.35,0.3);
  \node[anchor=west] at (-0.730,-1.350) {para-NP-hard};
\end{tikzpicture}
\caption{The parameterized complexity landscape established by our
results and standard parameter bounds.  Each box names one
functional-boundedness equivalence class.  An arrow from $p$ to $q$
means that bounded $p$ implies bounded $q$.  Green boxes are FPT, red
boxes are para-NP-hard. The excess,
the feedback edge number $m-n+1$, and the maximum degree count parallel
edges; all other parameters are those of the support. For the excess
$\delta=m-(2n-4)$, our FPT algorithm is randomized; bounded feedback edge
number even bounds $n$ on yes-instances.}
\label{fig:parameter-landscape}
\end{figure}

\subparagraph*{Parameter Landscape.}
\Cref{fig:parameter-landscape} brings together the tractability and hardness
results from the preceding paragraphs and their corresponding sections.
An arrow from $p$ to $q$ means that $q$ is bounded by a function of
$p$; thus tractability propagates against the arrows and
para-NP-hardness along them. It is easy to see that, on yes-instances
with more than four vertices, the feedback edge number upper-bounds the excess.
Para-NP-hardness for degeneracy and twin-width follows
from our hardness result for planar graphs of maximum degree five, since planar graphs are $5$-degenerate and have bounded
twin-width~\cite{HlinenyJ25,KralPekarkovaStorgel2024}.

\subparagraph*{Organization.}
In \Cref{sec:preliminaries}, we introduce the terminology, notation, and
computational model used throughout the paper.
\Cref{sec:technical-overview} then gives an overview of the main
algorithmic ideas. We next recall the classical witness characterization
and show how to construct a temporal labeling from either witness. We
then develop the preprocessing rules and the exact algorithm, followed
by the algorithms for structural parameters and for the excess.
Finally, we establish the degree and edge-multiplicity hardness
results. Proof details for the classical characterization and the
required matroid-representation tools are deferred to the appendices.

\subsection{Related Work}

The classical $2n-4$ call bound was proved independently
in~\cite{BakerS72,HajnalMS72}. Later work studied prescribed
communication networks and minimum-length schemes~\cite{HararyS74,Bumby81,KleitmanS80};
see the survey~\cite{HedetniemiHL88}. G\"obel, Cerdeira, and
Veldman~\cite{gobel1991label} introduced label-connected graph
recognition, proved NP-completeness, and connected it to spanning-tree
packings. Fault-tolerant variants based on ascending paths are studied
in~\cite{BermanH86} and surveyed in~\cite{Pelc96}.

Our structural starting point is the spanning-witness
characterization~\cite{BuroschLM88,BuroschL92,Krumme96}, including how
forest components meet the cycle and support a temporally connecting
labeling. Related order-theoretic aspects appear in~\cite{Labahn95}.
Carnevale, Casteigts, and Schindl~\cite{CarnevaleCS26} reformulate
classical results in temporal-graph terminology without considering
recognition; we use this structure for recognition algorithms and
complexity classifications.

Parallel edges correspond to multiple labels on a static edge;
see~\cite{CasteigtsCS22} for model comparisons. Temporal-labeling
problems address connectivity~\cite{MertziosMS19,akrida2017complexity,KlobasMMS22,CarnevaleDO25},
prescribed degree sequences~\cite{CDM25}, reachability relations~\cite{EMM25},
and ordered connectivity requirements~\cite{BellittoEtAl2026}.
Periodic variants include realization with multiple labels per
edge~\cite{EMW24}, prescribed or bounded fastest-path
durations~\cite{KlobasMMS24,MMMS24}, and bounded stretch relative to
static distances~\cite{Realization-Stretch-Arxiv-2025}.
Cauvi, Morawietz, and Viennot~\cite{CauviMV26} study several path metrics
in periodic and non-periodic settings. In our problem, the calls and
their endpoints are fixed; only their ordering is variable.

Our algorithms use rank-based treewidth
methods~\cite{BodlaenderEtAl2015}, pebble games for spanning-tree
packing~\cite{LeeStreinu2008,StreinuT09}, and representative
families~\cite{Marx09,FominLPS16}. As in deterministic
$k$-\textsc{Path}~\cite{FominLPS16}, a uniform vertex matroid and
representative families preserve continuations without repeated
vertices, avoiding color-coding~\cite{AYZ95}.

Our \textsc{Rooted Independent Cycle} (RIC) algorithm is most closely
related to algorithms for cycles through prescribed elements and
matroid-constrained paths. A shortest
cycle through $t$ prescribed vertices or edges can be found in
randomized $2^t n^{\Oh(1)}$ time~\cite{BjorklundHT2012}.
\textsc{Cycle Through Terminals} requires all prescribed terminal
edges and has a deterministic single-exponential algorithm parameterized
by the permitted weight of the remaining edges~\cite{FominGLS2019}.
Agrawal et al.~\cite{AgrawalJKS20} find bounded-length paths with
matroid-independent edge sets; Casteigts et al.~\cite{casteigts2021finding}
find dipaths with vertex sets independent in a represented linear
matroid. RIC instead requires a cycle through four prescribed vertices
with an edge set independent in a represented linear matroid.
We adapt representative families by tracking prescribed vertices,
combining the edge matroid with a uniform vertex matroid, and
preserving the information needed to close the path into a cycle.

\section{Preliminaries}
\label{sec:preliminaries}
In this section, we fix our graph and temporal-labeling conventions,
define the two spanning witnesses, and recall the parameters, the
computational model, and the matroid notions used in the algorithms.

\subparagraph*{Graphs.}
We use standard graph-theoretic terminology; see, for example,
Diestel~\cite{Die16}. Unless stated otherwise, graphs are finite
undirected multigraphs, with parallel edges treated as distinct. We
write $n=|V(G)|$ and $m=|E(G)|$. Degrees count multiplicities; the
maximum degree of the support is explicitly identified and denoted
by $\Delta$. For $X\subseteq V(G)$, let $G[X]$ be the induced subgraph
and $E[X]$ its edge set. For $F\subseteq E(G)$, the graph $(V(G),F)$
retains isolated vertices, and $V(F)$ is the set of endpoints of $F$.
We write $[a]=\{1,\ldots,a\}$ for a positive integer $a$, use $G-S$
and $G-F$ for vertex and edge deletion, and $A\setminus B$ for set difference.

The \emph{support} of a multigraph is the simple graph obtained by
deleting loops and replacing each nonempty parallel class by one
edge. A path has no repeated vertices; a path consisting of one vertex
is allowed. A \emph{cycle} of length $\ell\geq1$ is a closed walk
$v_0e_1v_1\cdots e_\ell v_\ell$ with $v_\ell=v_0$, pairwise distinct
edges $e_1,\ldots,e_\ell$, and pairwise distinct vertices
$v_0,\ldots,v_{\ell-1}$; we identify it with the subgraph formed by
these vertices and edges. Thus a cycle of length one is a loop, and a
cycle of length two consists of two parallel edges. The length of a
path is its number of edges.
A forest contains no cycle; in particular, it contains no loop and no
pair of parallel edges.
A spanning forest includes every vertex, and isolated vertices count
as components. A spanning forest on $n$ vertices with $c$ components
has $n-c$ edges. A \emph{matching} is a set of pairwise vertex-disjoint edges.
Edge-disjoint subgraphs may share vertices.

We discard loops in instances of \Gossip, since they do not
affect communication. Contraction of a connected vertex set identifies
its vertices, retains external parallel edges, and deletes the resulting
loops. We call the graph obtained by contracting one or more pairwise
disjoint connected vertex sets a \emph{quotient} of the original graph.
In the matroid constructions, we also identify vertex sets that
need not be connected. There we explicitly retain all edges, including
resulting loops, to preserve the matroid ground set.

\subparagraph*{Temporal Connectivity.}
A general introduction to temporal graphs is given by Casteigts et
al.~\cite{casteigts2012time}; for the distinctions between simple,
strict, and proper temporal graphs, we follow Casteigts, Corsini, and
Sarkar~\cite{CasteigtsCS22}.
A \emph{temporal labeling} of a multigraph $G=(V,E)$ is a function
$\lambda\colon E\to\mathbb{N}$ assigning one positive integer to each
edge. A \emph{strict temporal path} traverses edges
$e_1,\ldots,e_\ell$ with
$\lambda(e_1)<\cdots<\lambda(e_\ell)$. The temporal graph
$(G,\lambda)$ is \emph{temporally connected} if a strict temporal path
exists from $u$ to $v$ for every ordered pair $u,v\in V$.
Paths of length zero are allowed, so a graph with at most one vertex
is temporally connected. 
Order the edges by their labels, placing edges with equal labels in
an arbitrary order, and assign distinct labels $1,\ldots,m$ in that
order. This preserves every strict comparison between original labels
and hence every strict temporal path. Thus every temporally connectable
graph admits a bijective labeling $\lambda\colon E\to[m]$.

A bijective temporal labeling defines a \emph{gossip scheme}, that is,
a total order of the calls. Initially each vertex knows its own distinct
secret; when an edge is called, its endpoints
exchange all secrets they currently know. After any prefix of the
order, a vertex $v$ knows the secret of a vertex $u$ if and only if
some strict temporal path from $u$ to $v$ uses only calls of that prefix. 
A gossip scheme is \emph{complete} if every vertex knows every
secret after all calls. Thus $(G,\lambda)$ is temporally connected if
and only if the corresponding gossip scheme is complete.

\subparagraph*{Spanning Witnesses.}
The classical characterization uses the following two types of
subgraphs. We follow the terminology of Burosch and Laborde and of
Krumme~\cite{BuroschL92,Krumme96}.
\begin{definition}[$T$-graph and $T$-witness]
A multigraph is a \emph{$T$-graph} if it has two spanning trees sharing
at most one edge.
We call such a pair a \emph{$T$-witness}.
\end{definition}

The second witness type replaces the two almost edge-disjoint trees by
two forests connected through a common cycle.
\begin{definition}[$M$-witness and $M$-graph]\label{def:m-graph}
An \emph{$M$-witness} of a multigraph $G$ consists of pairwise
edge-disjoint subgraphs $C,F_1,F_2$, where $C$ is a cycle of length at
least four and each $F_i$ is a spanning forest with exactly four
components, each containing a vertex of $C$. A graph is an \emph{$M$-graph} if it
admits an $M$-witness and is not a $T$-graph. An $M$-witness is
\emph{minimum} if $|E(C)|$ is minimum among all $M$-witnesses of $G$.
\end{definition}

In this terminology, $T$-graphs and $M$-graphs form disjoint classes.
The gossiping literature calls an $M$-witness a \emph{realizing set}. We use ``witness'' to emphasize its
algorithmic role. For statements that also
apply to $T$-graphs containing such a cycle-and-forests structure, we
say explicitly that the graph \emph{admits an $M$-witness}.

In intuitive descriptions and figures, we call $F_1$ the
\emph{\textcolor{witnessblue}{blue forest}} and $F_2$ the
\emph{\textcolor{witnessorange}{orange forest}}. In figures showing
witnesses, we consistently use blue for $F_1$ and orange for $F_2$,
and the same colors for the corresponding witness trees. We retain
the notation $F_1,F_2$ in formal statements.

\begin{figure}[t]
\centering
\begin{tikzpicture}[every node/.style={font=\footnotesize}]
\begin{scope}[x=0.78cm,y=0.9cm]
  \foreach \i/\x/\y in {1/0/0,2/1/1.2,3/2/0,4/3/1.2,5/4/0,6/5/1.2}
    \coordinate (v\i) at (\x,\y);
  \draw[blueforest] (v1)--(v2)--(v3) (v4)--(v5)--(v6);
  \draw[orangeforest] (v1)--(v3)--(v5) (v2)--(v4)--(v6);
  \draw[witnessblue,line width=1pt,transform canvas={xshift=-1.2pt,yshift=0.8pt}] (v3)--(v4);
  \draw[orangeforest,transform canvas={xshift=1.2pt,yshift=-0.8pt}] (v3)--(v4);
  \foreach \i in {1,...,6} \node[vertex] at (v\i) {$v_\i$};
  \node[panel] at (2.5,-1.05) {(a) $T$-witness};
\end{scope}
\begin{scope}[xshift=7.3cm,yshift=-0.2cm,scale=0.85,every node/.append style={transform shape=false}]
  \coordinate (c1) at (-0.95,0.95);
  \coordinate (c2) at (0.95,0.95);
  \coordinate (c3) at (1.45,-0.55);
  \coordinate (c4) at (0,-1.45);
  \coordinate (c5) at (-1.45,-0.55);
  \coordinate (x3) at (0,1.55);
  \coordinate (x1) at (0,2.45);
  \coordinate (x4) at (0,3.35);
  \coordinate (x2) at (0.1,-0.4);
  \draw[cycleedge] (c1)--(c2)--(c3)--(c4)--(c5)--(c1);
  \draw[blueforest] (x3)--(c1) (x3)--(c2) (x4) to[out=190,in=100] (c1) (x4)--(x1) (x2)--(c3);
  \draw[orangeforest] (x3)--(x1) (x1)--(c1) (x1)--(c2) (x4) to[out=-10,in=80] (c2) (x2)--(c5);
  \foreach \v/\l in {c1/c_1,c2/c_2,c3/c_3,c4/c_4,c5/c_5,x1/x_1,x2/x_2,x3/x_3,x4/x_4}
    \node[vertex] at (\v) {$\l$};
  \node[panel] at (0,-2.25) {(b) $M$-witness};
\end{scope}
\begin{scope}[xshift=0cm,yshift=2.9cm]
  \draw[cycleedge] (0,0)--(0.6,0) node[right,text=black] {cycle $C$};
  \draw[blueforest] (0,-0.45)--(0.6,-0.45) node[right,text=black] {tree $T_1$ / forest $F_1$};
  \draw[orangeforest] (0,-0.9)--(0.6,-0.9) node[right,text=black] {tree $T_2$ / forest $F_2$};
\end{scope}
\end{tikzpicture}
\caption{The two witness types.
(a) A $T$-witness: two spanning trees $T_1$ (solid) and $T_2$ (dashed)
that share only the edge $v_3v_4$. (b) An $M$-witness on nine vertices:
the black edges form the cycle $C$, and the blue and orange edges form
spanning forests with four components each, all meeting $C$. The graph
in (b) has $15=2n-3$ edges, but it is not a $T$-graph (see the text). We use it as a running example
in \Cref{fig:labeling-example,fig:excess-quotient-example}.}
\label{fig:m-graph}
\end{figure}

\Cref{fig:m-graph} shows both witness types. The graph in
\Cref{fig:m-graph}(b) has nine vertices and $15=2n-3$ edges, which is
exactly the number of edges of two spanning trees sharing one edge.
Nevertheless, it is not a $T$-graph: the five vertices
$c_1,c_2,x_1,x_3,x_4$ induce nine edges, but at most
$2(5-1)=8$ of them lie in a union of two forests. Hence, every union of
two forests has at most $2n-4$ edges. We use this graph as a running
example.

\subparagraph*{Parameters.}
Graph parameters are measured on the support unless stated otherwise.
The excess, maximum degree, and feedback edge number $m-n+1$ of a
connected multigraph count parallel edges. The \emph{excess} is
$\delta=m-(2n-4)$ for instances with $n\geq4$ and $m\geq2n-4$.
For $n\geq4$, instances below this edge bound are no-instances by the classical gossiping
bound~\cite{BakerS72,HajnalMS72}. Instances with fewer than four vertices
can be handled directly.

A \emph{tree decomposition} of a simple graph assigns a vertex set,
called a bag, to each node of a tree such that every vertex occurs
in a nonempty connected set of bags
and every edge has both endpoints in a bag. Its width is the maximum
bag size minus one; \emph{treewidth} $\tw(G)$ is the minimum width for
the support of $G$.

A simple graph is
\emph{chordal} if it has no induced cycle of length at least four;
its \emph{chordal vertex deletion number} is the minimum number of
vertices whose deletion makes it chordal.

The \emph{independence
number} $\alpha(Q)$ is the maximum size of an independent set in $Q$.
The \emph{tree-independence number} $\treealpha(Q)$ is the minimum,
over tree decompositions of $Q$, of the maximum $\alpha(Q[B])$ over
their bags $B$~\cite{DallardMS24}.

The \emph{clique-width} $\cw(G)$ is the
minimum number of auxiliary labels needed to build $G$ by creating
labeled vertices, taking disjoint unions, relabeling every occurrence
of one label, and adding all edges between two distinct labels.
These construction labels are discarded afterward.

\subparagraph*{Complexity and Computational Model.}
We follow Cygan et al.~\cite{C+15} for parameterized complexity.
For input length $L$ and a computable function $f$, a problem is
\emph{fixed-parameter tractable} for $p$ if it admits an
$f(p)L^{\Oh(1)}$-time algorithm, and \emph{XP} if it admits an
$L^{f(p)}$-time algorithm.
A problem is \emph{para-NP-hard} for $p$ if it is NP-hard for a constant
value of $p$. Our randomized algorithm has one-sided error: it never accepts a no-instance
and accepts every yes-instance with probability at least $1-1/n$.
The Exponential Time Hypothesis (ETH) asserts that 3SAT on $p$ variables
cannot be solved in $2^{o(p)}$ time; its sparse formulation follows
from the sparsification lemma~\cite{ImpagliazzoPaturiZane2001}.

Running times use the word RAM model with $\Theta(\log(n+m))$-bit words.
In the randomized reduction, we use a prime field $\mathbb{F}$ of size
$\Oh(n^2)$; its elements fit in constantly many words, so field
operations take constant time. For the standalone RIC algorithm, we
count arithmetic operations over the field of the supplied
representation. We denote the matrix multiplication exponent by
$\omega<2.372$~\cite{AlmanDVXXZ25}.

\subparagraph*{Matroids.}
We use standard matroid terminology; see Oxley~\cite{Oxley2011}.
A \emph{matroid} $M=(E,\mathcal{I})$ has a finite ground set $E$ and
a family $\mathcal{I}\subseteq2^E$ of \emph{independent sets} containing
$\varnothing$, closed under subsets, and satisfying the exchange
axiom: for $A,B\in\mathcal{I}$ with $|A|<|B|$, some $b\in B\setminus A$
has $A\cup\{b\}\in\mathcal{I}$. We also write $\mathcal{I}(M)$ for
this family. A \emph{basis} is an inclusion-maximal independent set;
all bases have the same size, the \emph{rank} $r(M)$. A \emph{loop}
belongs to no independent set.

The \emph{graphic matroid} $M(G)$ of a multigraph $G=(V,E)$ has ground
set $E$ and independent sets $F$ for which $(V,F)$ is a forest.
Its loops are exactly those of $G$, and parallel edges remain distinct
elements. A matrix $A$ over $\mathbb{F}$ with columns indexed by $E$
is a \emph{representation} of the \emph{linear matroid} $M[A]$ on $E$: a set is
independent exactly when its columns are linearly independent over
$\mathbb{F}$, which can be tested by their rank.

Intuitively, we encode the two forests of an $M$-witness by a matroid
union and express that the cycle avoids them by duality. Formally, the
\emph{union} of two matroids $M_1$ and $M_2$ on the same ground set $E$
is the matroid $M_1\vee M_2$ on $E$ whose independent sets are the
sets $X_1\cup X_2$ with $X_i\in\mathcal{I}(M_i)$ for $i\in[2]$.
The sets $X_1,X_2$ may be chosen disjoint. Hence
$r(M_1\vee M_2)\leq r(M_1)+r(M_2)$.
The \emph{dual} of a matroid $M$ on $E$ is the matroid $M^*$ whose
independent sets are disjoint from some basis of $M$. Its bases are
exactly the complements of bases of $M$, so $r(M^*)=|E|-r(M)$.
Equivalently, a set $X\subseteq E$ is independent in $M^*$ if and only
if $E\setminus X$ contains a basis of $M$.

\section{Technical Overview of Our Results}
\label{sec:technical-overview}
This section explains the main ideas behind our algorithms and hardness
results before their detailed treatment in the subsequent sections.
Our algorithms build on a classical characterization: a graph admits
a temporally connecting labeling if and only if it contains a
$T$-witness or an $M$-witness. These witnesses describe subgraphs that
support a successful call schedule. Since $T$-witnesses can be found in
polynomial time, our parameterized algorithms focus on finding
$M$-witnesses: a cycle and two spanning forests satisfying the required
component and edge-disjointness conditions. We explain how cycle
enumeration, connectivity dynamic programming, and matroid independence
address this task, and how structural reductions extend the resulting
algorithms to further parameters. All running times below assume the
linear-time clean-up described in \Cref{sec:exact-preprocessing}, which
removes loops and retains at most three edges between each pair of
vertices, ensuring $m\in\Oh(n^2)$.

\subsection{Classical Characterization and Witness Structure}

In \Cref{sec:witnesses}, we recall the classical characterization of
temporally connectable graphs by $T$- and $M$-witnesses~\cite{BuroschLM88,BuroschL92,Krumme96} and the common
structure of the two forests along the cycle of an $M$-witness. We
provide a self-contained proof of the characterization in
\Cref{app:characterization}. Additionally, we provide a Lean formalization of this proof in~\cite{lax-630294}.
Our main new structural result shows that
every minimum $M$-witness in an $M$-graph has an induced cycle,
allowing our exact algorithm to restrict its cycle search. We also
implement the conversion from either type of witness to a temporally
connecting labeling in $\Oh(n+m)$ time.

A $T$-witness
supports a gather-and-broadcast schedule along its two trees. An
$M$-witness consists of a cycle and two spanning forests, each with
four components. Information is first gathered within the components
of the first forest, exchanged through the cycle, and then distributed
within the components of the second forest.

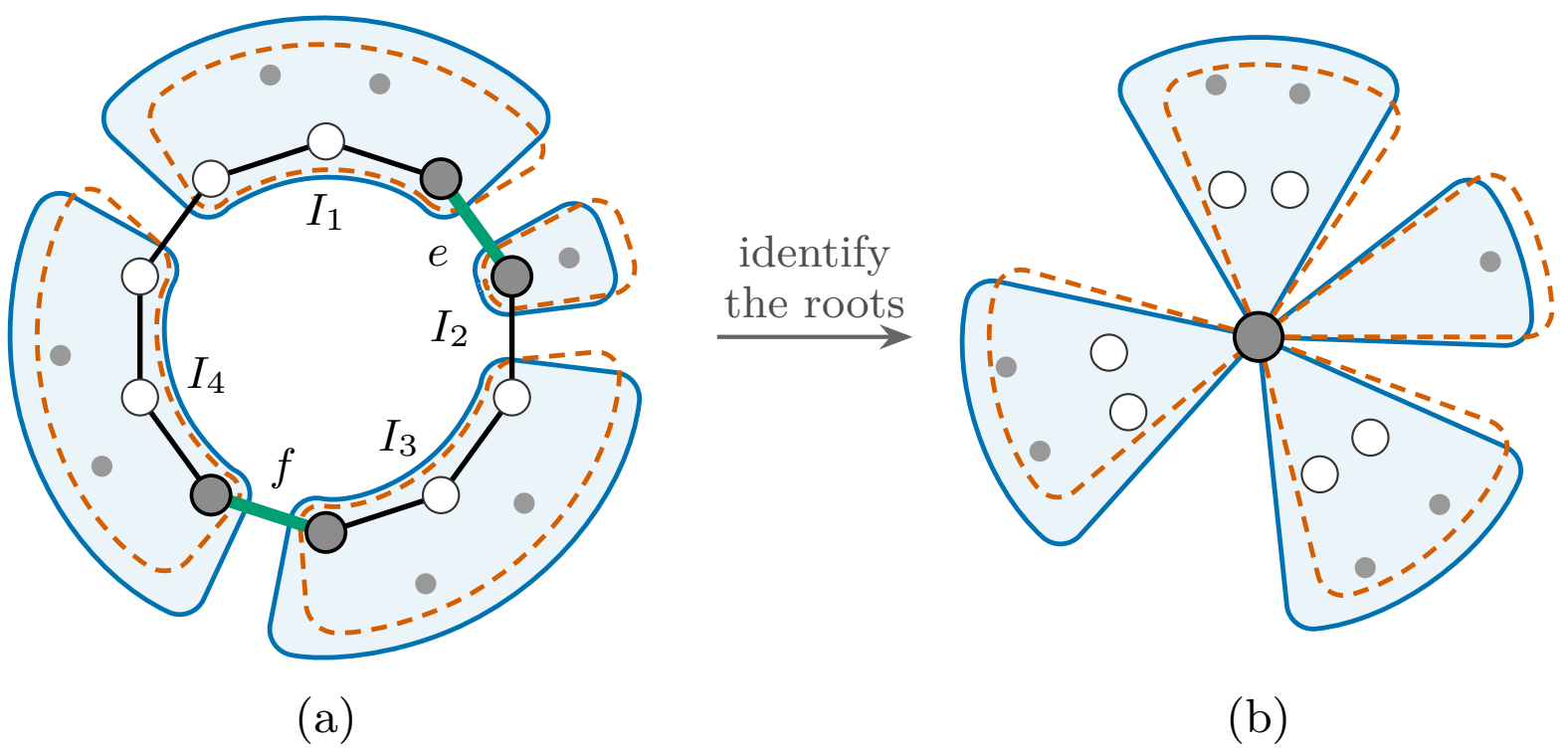
\begin{figure}[t]
\centering
\begin{tikzpicture}[
  cyv/.style={circle,draw=black!80,fill=white,inner sep=0pt,minimum size=2.4mm},
  rootv/.style={circle,draw=black,fill=black!45,line width=.7pt,inner sep=0pt,minimum size=2.6mm},
  odot/.style={circle,fill=black!40,inner sep=0pt,minimum size=1.4mm},
  matchedge/.style={auxgreen,line width=2.4pt},
  f1blob/.style={draw=witnessblue,line width=.9pt,fill=witnessblue!8,rounded corners=5pt},
  f2blob/.style={draw=witnessorange,line width=.9pt,dash pattern=on 3pt off 2pt,rounded corners=5pt},
]
\begin{scope}
\coordinate (c0) at (126:1.3);
\coordinate (c1) at (90:1.3);
\coordinate (c2) at (54:1.3);
\coordinate (c3) at (18:1.3);
\coordinate (c4) at (-18:1.3);
\coordinate (c5) at (-54:1.3);
\coordinate (c6) at (-90:1.3);
\coordinate (c7) at (-126:1.3);
\coordinate (c8) at (-162:1.3);
\coordinate (c9) at (-198:1.3);
\path[f1blob] (43.0:1.060) arc[start angle=43.0,end angle=137.0,radius=1.060cm] -- (137.0:2.120) arc[start angle=137.0,end angle=43.0,radius=2.120cm] -- cycle;
\path[f1blob] (7.0:1.060) arc[start angle=7.0,end angle=29.0,radius=1.060cm] -- (29.0:1.960) arc[start angle=29.0,end angle=7.0,radius=1.960cm] -- cycle;
\path[f1blob] (-101.0:1.060) arc[start angle=-101.0,end angle=-7.0,radius=1.060cm] -- (-7.0:2.120) arc[start angle=-7.0,end angle=-101.0,radius=2.120cm] -- cycle;
\path[f1blob] (-209.0:1.060) arc[start angle=-209.0,end angle=-115.0,radius=1.060cm] -- (-115.0:2.100) arc[start angle=-115.0,end angle=-209.0,radius=2.100cm] -- cycle;
\path[f2blob] (44.5:1.110) arc[start angle=44.5,end angle=135.5,radius=1.110cm] -- (129.5:1.940) arc[start angle=129.5,end angle=38.5,radius=1.940cm] -- cycle;
\path[f2blob] (8.5:1.110) arc[start angle=8.5,end angle=27.5,radius=1.110cm] -- (27.5:2.100) arc[start angle=27.5,end angle=8.5,radius=2.100cm] -- cycle;
\path[f2blob] (-99.5:1.110) arc[start angle=-99.5,end angle=-8.5,radius=1.110cm] -- (-2.5:1.960) arc[start angle=-2.5,end angle=-93.5,radius=1.960cm] -- cycle;
\path[f2blob] (-207.5:1.110) arc[start angle=-207.5,end angle=-116.5,radius=1.110cm] -- (-122.5:1.920) arc[start angle=-122.5,end angle=-213.5,radius=1.920cm] -- cycle;
\draw[cycleedge] (c0) -- (c1) -- (c2) -- (c3) -- (c4) -- (c5) -- (c6) -- (c7) -- (c8) -- (c9) -- cycle;
\draw[matchedge] (c2)--(c3) (c6)--(c7);
\node[font=\footnotesize] at (36:0.92) {$e$};
\node[font=\footnotesize] at (-108:0.92) {$f$};
\node[odot] at (78.0:1.720) {};
\node[odot] at (102.0:1.780) {};
\node[odot] at (18.0:1.700) {};
\node[odot] at (-40.0:1.720) {};
\node[odot] at (-68.0:1.770) {};
\node[odot] at (-150.0:1.720) {};
\node[odot] at (-176.0:1.770) {};
\node[cyv] at (c0) {};
\node[cyv] at (c1) {};
\node[rootv] at (c2) {};
\node[rootv] at (c3) {};
\node[cyv] at (c4) {};
\node[cyv] at (c5) {};
\node[rootv] at (c6) {};
\node[rootv] at (c7) {};
\node[cyv] at (c8) {};
\node[cyv] at (c9) {};
\node[font=\footnotesize] at (90.0:0.83) {$I_1$};
\node[font=\footnotesize] at (4.0:0.83) {$I_2$};
\node[font=\footnotesize] at (-54.0:0.83) {$I_3$};
\node[font=\footnotesize] at (-162.0:0.83) {$I_4$};
\node[panel] at (0,-2.55) {(a)};
\end{scope}
\begin{scope}[xshift=6.2cm]
\path[f1blob] (0,0) -- (60.0:1.970) arc[start angle=60.0,end angle=120.0,radius=1.970cm] -- cycle;
\path[f1blob] (0,0) -- (-2.0:1.810) arc[start angle=-2.0,end angle=38.0,radius=1.810cm] -- cycle;
\path[f1blob] (0,0) -- (-84.0:1.970) arc[start angle=-84.0,end angle=-24.0,radius=1.970cm] -- cycle;
\path[f1blob] (0,0) -- (-192.0:1.950) arc[start angle=-192.0,end angle=-132.0,radius=1.950cm] -- cycle;
\path[f2blob] (0,0) -- (56.0:1.790) arc[start angle=56.0,end angle=112.0,radius=1.790cm] -- cycle;
\path[f2blob] (0,0) -- (0.0:1.950) arc[start angle=0.0,end angle=36.0,radius=1.950cm] -- cycle;
\path[f2blob] (0,0) -- (-76.0:1.810) arc[start angle=-76.0,end angle=-20.0,radius=1.810cm] -- cycle;
\path[f2blob] (0,0) -- (-196.0:1.770) arc[start angle=-196.0,end angle=-140.0,radius=1.770cm] -- cycle;
\node[cyv] at (78.0:1.000) {};
\node[cyv] at (102.0:1.000) {};
\node[odot] at (80.4:1.640) {};
\node[odot] at (99.6:1.700) {};
\node[odot] at (18.0:1.620) {};
\node[cyv] at (-66.0:1.000) {};
\node[cyv] at (-42.0:1.000) {};
\node[odot] at (-42.8:1.640) {};
\node[odot] at (-65.2:1.690) {};
\node[cyv] at (-174.0:1.000) {};
\node[cyv] at (-150.0:1.000) {};
\node[odot] at (-152.4:1.640) {};
\node[odot] at (-173.2:1.690) {};
\node[rootv,minimum size=3.2mm] at (0,0) {};
\node[panel] at (0,-2.55) {(b)};
\end{scope}
\draw[-{Stealth[length=2.2mm]},black!60,line width=.8pt] (2.6,0) -- node[above,font=\footnotesize,align=center,text=black!70]{identify\\the roots} (3.9,0);
\end{tikzpicture}
\caption{The common roots of an $M$-witness. (a) The cycle (black)
splits into four intervals of consecutive vertices $I_1,\ldots,I_4$. Each component of the first
forest (blue, solid) and of the second forest (orange, dashed) meets the
cycle in exactly the vertices of one interval; away from the cycle, the components of
the two forests can differ. The endpoints of the two disjoint edges $e$
and $f$ (green) give one root (gray) in each interval. (b) Removing the
cycle edges and identifying the four roots turns each forest into a
spanning tree.}
\label{fig:overview-roots}
\end{figure}

A known structural result describes how these components meet the
cycle when the graph is an $M$-graph~\cite[Corollary~1 and proof of Theorem~3]{Krumme96}. The vertices of the cycle split into
four nonempty consecutive sets, called \emph{intervals}; each interval
contains the vertices of a path along the cycle, possibly consisting of a single vertex.
Each of the four components of the first forest meets the cycle in exactly the vertices of one of these four intervals, one per component. The same holds for the second
forest, with exactly the same intervals, although its components may
differ outside the cycle. Exactly four cycle edges connect the
intervals, one between each consecutive pair. Taking every other one
of these four edges in cyclic order gives two pairs. The two edges in
each pair have no common endpoint, so each pair is a matching of size two.
Choose either matching. Its four endpoints include exactly one vertex
in each interval, and hence exactly one vertex in each component of
either forest. We can therefore use the same four vertices as roots
for both forests, choosing one root per component.
\Cref{fig:overview-roots} illustrates the common intervals, the selected
roots, and their identification.

This common choice of roots is useful in two different ways. For a
fixed cycle, it reduces the root choices to the endpoints of two
disjoint cycle edges. For the treewidth algorithm, it allows both
forests to use the same auxiliary connections.

To establish the induced-cycle result, consider an $M$-witness whose
cycle has minimum length. A chord between different intervals would
yield a $T$-witness. A chord
inside one interval shortcuts the cycle; if the chord belongs to a
forest, an exchange with an edge of the replaced cycle segment
preserves that forest's components.

Finally, we implement the classical conversion of Burosch and
Laborde~\cite[Lemmas~7--8]{BuroschL92} and
Krumme~\cite[Theorem~3]{Krumme96} from either witness to a temporally
connecting labeling in $\Oh(n+m)$ time. Consequently,
finding witnesses also solves the constructive version of recognition.

\subsection{Preprocessing and Exact Recognition}

In \Cref{sec:exact-preprocessing}, we develop two answer-preserving
reduction rules and an $\Oh(3^{n/3}n^3)$-time exact recognition
algorithm. The reduction rules remove unnecessary edges and contract
sufficiently dense vertex sets, preparing the input for our later
algorithms. The exact algorithm enumerates induced cycles and tests
whether any of them can be completed to an $M$-witness, after first
checking for a $T$-witness.

The \CleanupRule deletes loops and retains at most three edges from
each parallel class. This suffices because each of the three parts of
an $M$-witness uses at most one edge of any parallel class. It gives
$m\in\Oh(n^2)$ without changing the support or losing a witness with
a prescribed bound on its cycle length.

The \TreeContractionRule contracts a vertex set $S$ whenever $G[S]$
contains three edge-disjoint spanning trees. The reason for using
three trees is that an $M$-witness has three edge-disjoint parts. Two
internal trees extend the two forests through $S$, while a path in
the third tree joins the original endpoints of the cycle edges at the
contracted vertex. In the other direction, contraction preserves
temporal connectivity. Using pebble games~\cite{LeeStreinu2008,StreinuT09}, described in
\Cref{sec:pebble-games}, we find all maximal sets eligible for contraction
in one computation, giving exhaustive application in
$\Oh(n^2)$ time. The resulting reduced graph $H$ satisfies
\[
 |E(H[X])|\leq3|X|-4\qquad\text{for every }X\subseteq V(H),\ |X|\geq2.
\]
This local bound, rather than only a bound on the total number of
edges, is what makes the rule useful for structural parameters.

For exact recognition, we first test for a $T$-witness in
$\Oh(n^2)$ time. Otherwise, fix a candidate cycle $C$ of length $\ell$.
For each matching of size two in $C$, let $R$ be its four endpoints.
Identifying the four vertices of $R$ in $G-E(C)$ turns two spanning
forests, each with exactly one vertex of $R$ per component, into two
edge-disjoint spanning trees, as illustrated in
\Cref{fig:overview-roots}(b). Conversely, undoing the identification
turns these trees back into the required forests. We refer to
transferring a structure back to the original graph in this way as
\emph{lifting}. We test whether these two edge-disjoint spanning trees exist
by adding two copies of a path through $R$ to $G-E(C)$.
We preprocess $G-E(C)$ once in $\Oh(n^2)$ time. For each choice of $R$,
the pebble-game algorithm then tests the six added edges in $\Oh(n)$
time. Since there are $\Oh(\ell^2)$ choices of $R$, testing whether $C$
extends to an $M$-witness takes $\Oh(n^2+\ell^2n)$ time in total.
Enumerating induced cycles of the support with the algorithm of
Uno and Satoh~\cite{UnoS14} and using one fixed
representative of each parallel class therefore suffices. The bound
of $3^{n/3}+\Oh(n)$ induced cycles due to Morrison and Scott~\cite[Corollary~1.3]{MorrisonS17} yields the
$\Oh(3^{n/3}n^3)$-time exact algorithm.

\subsection{Algorithms for Structural Parameters}
In \Cref{sec:structural-parameters}, we first give a deterministic
$2^{\Oh(\tw)}\cdot n$-time algorithm, where $\tw$ is the treewidth of
the support. We then use the \TreeContractionRule to extend
tractability to structural parameters that permit dense graphs of
unbounded treewidth. The key is to bound the treewidth of the reduced
graph in terms of a parameter of the original input. For some
parameters, preprocessing alone already produces a small kernel.

The treewidth algorithm first tests for a $T$-witness using a simpler
connectivity dynamic program. If this test fails, it remains to find
an $M$-witness. We focus here on this more involved part of the algorithm.
A direct dynamic program for
$M$-witnesses would have to account for the four components of each
forest and their intersections with the cycle. We avoid guessing roots
or maintaining large edge counters by relaxing the forests to
edge sets $D_1,D_2$ with at most four components each, all meeting the
cycle. Spanning forests of these components can be split further
until they have exactly four components, each still meeting the cycle.
Hence this relaxation preserves the existence of an $M$-witness.

We express these component conditions as connectivity requirements by
adding an auxiliary vertex $a$, adjacent to every original vertex.
A solution selects at most four edges incident with $a$, all ending
at vertices of the chosen cycle, and adds the same selected edges to
both spanning subgraphs. We require each augmented subgraph to be
connected. Every component of either original subgraph must therefore
contain an endpoint of a selected edge. Consequently, there are at
most four components, and each meets the cycle. Conversely, if an
$M$-witness exists after the $T$-witness test fails, its two forests
have four common roots on the cycle. Joining $a$ to these roots makes
both forests connected.
At each bag of the tree decomposition, the dynamic program summarizes
the choices made in the processed part of the graph:
\begin{itemize}
\item \emph{Cycle degrees:} for each vertex in the bag, we record
whether it belongs to the cycle and how many incident cycle edges
have already been selected. This ensures that every cycle vertex
ultimately has degree two.
\item \emph{Counters and a closure flag:} we track the number of
selected auxiliary edges, whether the cycle has at least four
vertices, and whether it has already been closed. Only counts up to
four are needed.
\item \emph{Connectivity:} three partitions record which vertices in
the bag are already connected through the partial cycle and through
each of the two augmented spanning subgraphs. These partitions let
us determine how later choices join the existing components.
\end{itemize}
The number of possible connectivity partitions is too large for the
desired single-exponential running time. We reduce it using the
rank-based approach of Bodlaender et al.~\cite{BodlaenderEtAl2015}.
For partial solutions with the same cycle degrees, counters, and
closure flag, the connectivity tests admit a matrix representation
of single-exponential rank. Retaining partial solutions corresponding
to a basis of its rows preserves the following guarantee: whenever
a discarded partial solution can be completed, some retained partial
solution can be completed using the same remaining choices. Thus only
$2^{\Oh(\tw)}$ combinations of partitions need to be stored. This gives a
single-exponential dependence on treewidth, including witness
reconstruction and computation of a suitable decomposition.

Applying the \TreeContractionRule first gives a linear vertex kernel
for independence number. Combining the rule with the treewidth
algorithm gives $\Oh(n^2)+2^{\Oh(p)}\cdot n$-time algorithms, where $p$
is the tree-independence number, chordal vertex deletion number, or
clique-width. Neither a
parameter decomposition nor a deletion set is required as input. We
explain the kernel first, followed by the treewidth bounds underlying
the other algorithms.

Recall that the reduced graph $H$ has at most $3|X|-4$ edges within
any vertex set $X$ of size at least two. Every induced subgraph of
its support therefore has average degree less than six and contains
a vertex of degree at most five. Thus the support is 5-degenerate
and hence 6-colorable. Each color class is an independent set, so
$|V(H)|\leq6\alpha(H)$. Since independence number does not increase
under induced minors, this gives a linear vertex kernel for the
independence number of the input.

Recall that tree-independence number is the smallest $a$ for which a
graph has a tree decomposition whose bags each contain no independent
set larger than $a$. The support of $H$ is an induced minor of the
input support, and tree-independence number does not increase under
induced minors~\cite{DallardMS24}. Thus, if the input has tree-independence number $a$,
the support of $H$ admits a decomposition with this same bound on its
bags. Since the support of $H$ is 6-colorable, each bag has at most
$6a$ vertices: each of its six color classes has size at most $a$.
Its treewidth is therefore at most $6a-1$, allowing us to apply the
treewidth algorithm.

Chordal graphs admit tree decompositions whose bags are cliques, so
they have tree-independence number at most one. By the preceding
argument, reduced graphs with chordal support therefore have constant
treewidth. A chordal deletion set transfers through contraction
without increasing its size; adding the transferred set to every bag
bounds the treewidth by the original deletion-set size plus a constant.
This gives the algorithm parameterized by chordal vertex deletion number.

For clique-width, we work with a construction expression whose operations
create labeled vertices, take disjoint unions, relabel vertices, and add
all edges between two labels. At a cut separating a subexpression from
the remaining vertices, vertices with the same construction label have
the same neighbors on the other side. Excluding
$K_6$ bounds the number of contracted sets that cross the cut.
Among sets wholly on either side, each label defines a complete
bipartite graph of adjacencies across the cut. Excluding
$K_{6,5}$ lets us cover its edges with at most five vertices. Over all
labels, these covers and the crossing contracted sets bound the
treewidth of the reduced graph linearly in the input clique-width.

These applications use the same algorithm:
contract first and run the treewidth algorithm on the result. The
parameter decomposition or deletion set is needed only for the
analysis, not as input to the algorithm.

\subsection{Parameterization by the Excess}

In \Cref{sec:above-bound}, we find an $M$-witness whose cycle has length
at most a supplied bound $k$, using a deterministic
$n^{k+\Oh(1)}$-time XP algorithm or a randomized
$2^{\Oh(k)}m^3$-time FPT algorithm. The randomized
algorithm never accepts a no-instance and fails on a yes-instance
with probability at most $1/n$. These algorithms also yield algorithms
parameterized by the excess $\delta=m-(2n-4)$ over the minimum number
of calls.

We first test for a $T$-witness in polynomial time. If this test fails,
it remains to find an $M$-witness. Let $C$ denote its cycle. Each of
its two forests has $n-4$ edges, so
\[
 |E(C)|+2n-8\leq m=2n-4+\delta,
 \qquad\text{and hence}\qquad |E(C)|\leq\delta+4.
\]
Taking $k=\delta+4$ therefore gives a deterministic
$n^{\delta+\Oh(1)}$-time algorithm and a randomized
$2^{\Oh(\delta)}n^3$-time algorithm for \Gossip, with the same error
guarantee. A supplied bound $k$ can be smaller than $\delta+4$.
Enumerating short cycles and applying the extension test
gives the deterministic XP algorithm.

The FPT algorithm avoids enumerating all short cycles by using
matroids to express whether the edges outside a candidate cycle can
support the two required forests. For four proposed common roots
$\rho$, let $G/\rho$ be obtained by identifying these roots into one
vertex while retaining all edges. Recall that $M(Q)$ denotes the
graphic matroid of a graph $Q$, whose independent sets are the edge
sets of forests. We consider the union of two copies of this matroid:
\[
 \mathcal M_\rho=M(G/\rho)\vee M(G/\rho).
\]
A basis of size $2n-8$ partitions into two spanning forests, each
with four components rooted at $\rho$. A cycle through $\rho$ can
avoid such a basis exactly when its edge set is independent in
$\mathcal M_\rho^*$, provided $r(\mathcal M_\rho)=2n-8$.
Thus duality expresses the existence of the two forests as an
independence condition on the cycle alone.

We construct a randomized linear representation of the union and
project its dual representation to rank at most $k$. A fixed solution
survives with high probability, while the construction cannot create
a false witness. The remaining
problem is \textsc{Rooted Independent Cycle}: find a cycle through
four prescribed vertices whose edges are independent in a represented
linear matroid of rank $r\leq k$.

We solve this problem deterministically in $2^{\Oh(r)}m$ field
operations using representative families of paths, following the
approach of Fomin et al.~\cite{FominLPS16}. A state records
the current endpoint, the prescribed vertices already visited, and
the path size. Combining the edge matroid with a uniform matroid on
vertices lets representative families preserve continuations that
share only their endpoints with the retained path and whose edges,
together with those of the retained path, are independent in the matroid.
The extension budget explicitly includes the final edge closing the
cycle. Solving at most $m^2$ RIC instances gives the claimed
$2^{\Oh(k)}m^3$ running time.


\subsection{The Degree Threshold and Edges of Multiplicity One}
In \Cref{sec:degree-complexity}, we establish a sharp degree threshold:
recognition takes $\Oh(n^3)$ time on multigraphs of maximum degree
four, but is NP-complete on planar simple graphs of maximum degree
five. We also derive ETH lower bounds matching the single-exponential
dependence on treewidth and the subexponential dependence on the
number of vertices for planar inputs. Finally, we prove NP-completeness
even with at most four edges of multiplicity one, whereas connected
instances without such edges always admit a temporally connecting labeling.

The degree-four algorithm follows from the excess parameterization.
Degrees count parallel edges, so maximum degree four implies $m\leq2n$.
We reject instances with fewer than $2n-4$ edges, the minimum required
for temporal connectivity. The remaining instances have excess
$0\leq\delta=m-(2n-4)\leq4$, so every $M$-witness has a cycle of
length at most eight. Bounded degree allows us to enumerate all such
cycles in $\Oh(n)$ time. Testing each cycle for extension to an
$M$-witness takes $\Oh(n^2)$ time, giving $\Oh(n^3)$ time overall,
including the preliminary $T$-witness test. At maximum degree five,
the excess can grow linearly with $n$, so this argument no longer
bounds the cycle length by a constant.

For hardness at maximum degree five, we build on the approach of
G\"obel, Cerdeira, and Veldman~\cite[Theorem~16 and proof of Theorem~17]{gobel1991label}.
Their intermediate problem, \textsc{PathTreeTree}, asks for an $s$-$t$
path and two spanning trees that are pairwise edge-disjoint. Their
reduction to recognition adds an $s$-$t$ path of length four with
three new internal vertices. These degree-two vertices exclude a
$T$-witness and force an $M$-witness to use the added path as part of
its cycle. Removing that path recovers precisely a
\textsc{PathTreeTree} solution.

\begin{figure}[t]
\centering
\begin{tikzpicture}[
  x=0.95cm,y=0.95cm,
  d/.style={circle,draw=black!80,fill=white,inner sep=0pt,minimum size=2.3mm},
  term/.style={circle,draw=black,fill=white,inner sep=0pt,minimum size=4.2mm,font=\footnotesize},
  zv/.style={circle,draw=black!80,fill=white,inner sep=0pt,minimum size=2.3mm},
  attach/.style={black!45,line width=.6pt},
  hl/.style={auxgreen!45,line width=5pt,line cap=round,line join=round},
  gadget/.style={rounded corners=3pt,fill=black!7,draw=black!50},
  lbl/.style={font=\footnotesize},
]
  \foreach \i/\o in {1/0,2/3,3/6}{
    \coordinate (s\i) at (\o,0);      \coordinate (t\i) at (\o+2.2,0);
    \coordinate (u\i) at (\o+0.75,0.75); \coordinate (v\i) at (\o+1.45,0.75);
    \coordinate (p\i) at (\o+0.75,-0.75); \coordinate (q\i) at (\o+1.45,-0.75);
  }
  \coordinate (z1) at (-0.7,-1.4); \coordinate (z2) at (4.1,-3.0); \coordinate (z3) at (8.9,-1.4);
  \draw[hl] (s1)--(p1)--(q1)--(t1)--(s2)--(u2)--(v2)--(t2)--(s3)--(u3)--(v3)--(t3);
  \draw[hl,rounded corners=8pt] (t3)--(8.9,0)--(z3)--(8.9,-3.0)--(z2)--(-0.7,-3.0)--(z1)--(-0.7,0)--(s1);
  \draw[gadget] (0.4,1.75) rectangle (7.8,2.35);
  \node[lbl] at (4.1,2.05) {clause $x_1\lor x_2\lor x_3$};
  \draw[gadget] (0.45,-1.75) rectangle (4.0,-2.35);
  \node[lbl] at (2.2,-2.05) {clause $\lnot x_1\lor\lnot x_2$};
  \draw[gadget] (4.2,-1.75) rectangle (7.75,-2.35);
  \node[lbl] at (6.0,-2.05) {clause $\lnot x_2\lor\lnot x_3$};
  \foreach \u in {u1,u2,u3} \draw[attach] (\u)--++(0,1.0);
  \foreach \u in {p1,p2} \draw[attach] (\u)--++(0,-1.0);
  \foreach \u in {q2,q3} \draw[attach] (\u)--++(0,-1.0);
  \foreach \i in {1,2,3}{
    \draw[gadget] ($(s\i)+(0.5,-0.32)$) rectangle ($(t\i)+(-0.5,0.32)$);
    \foreach \a/\b in {s\i/0.5,t\i/-0.5} \draw[attach] (\a)--++(\b,0);
    \foreach \a/\b in {u\i/-0.43,v\i/-0.43,p\i/0.43,q\i/0.43} \draw[attach] (\a)--++(0,\b);
    \draw[line width=.7pt] (s\i)--(u\i)--(v\i)--(t\i)--(q\i)--(p\i)--(s\i);
    \node[lbl,anchor=south east,inner sep=1pt] at ($(s\i)!0.5!(u\i)$) {$x_\i$};
    \node[lbl,anchor=north east,inner sep=1pt] at ($(s\i)!0.5!(p\i)$) {$\lnot x_\i$};
  }
  \draw[line width=.7pt] (t1)--(s2) (t2)--(s3);
  \draw[line width=.7pt,rounded corners=8pt] (t3)--(8.9,0)--(z3)--(8.9,-3.0)--(z2)--(-0.7,-3.0)--(z1)--(-0.7,0)--(s1);
  \foreach \i in {1,2,3} \foreach \u in {u\i,v\i,p\i,q\i,s\i,t\i} \node[d] at (\u) {};
  \foreach \u in {z1,z2,z3} \node[zv,fill=black!25] at (\u) {};
  \node[lbl,anchor=south] at ($(s1)+(-0.35,0.06)$) {$s$};
  \node[lbl,anchor=south] at ($(t3)+(0.35,0.06)$) {$t$};
\end{tikzpicture}
\caption{A sketch of the hardness construction for the formula
$(x_1\lor x_2\lor x_3)\land(\lnot x_1\lor\lnot x_2)\land(\lnot x_2\lor\lnot x_3)$.
Each variable is a cycle with one branch per literal. The gray boxes
are gadgets that contain two edge-disjoint spanning trees. The
highlighted $s$-$t$ path encodes the assignment $x_1=1$, $x_2=x_3=0$:
for each variable, it traverses the branch of the false literal. Each
clause remains attached to the two spanning trees through a branch
avoided by the path, that is, through a true literal. The added path
with three new vertices (gray) closes the $s$-$t$ path into the cycle
of an $M$-witness.}
\label{fig:overview-hardness}
\end{figure}
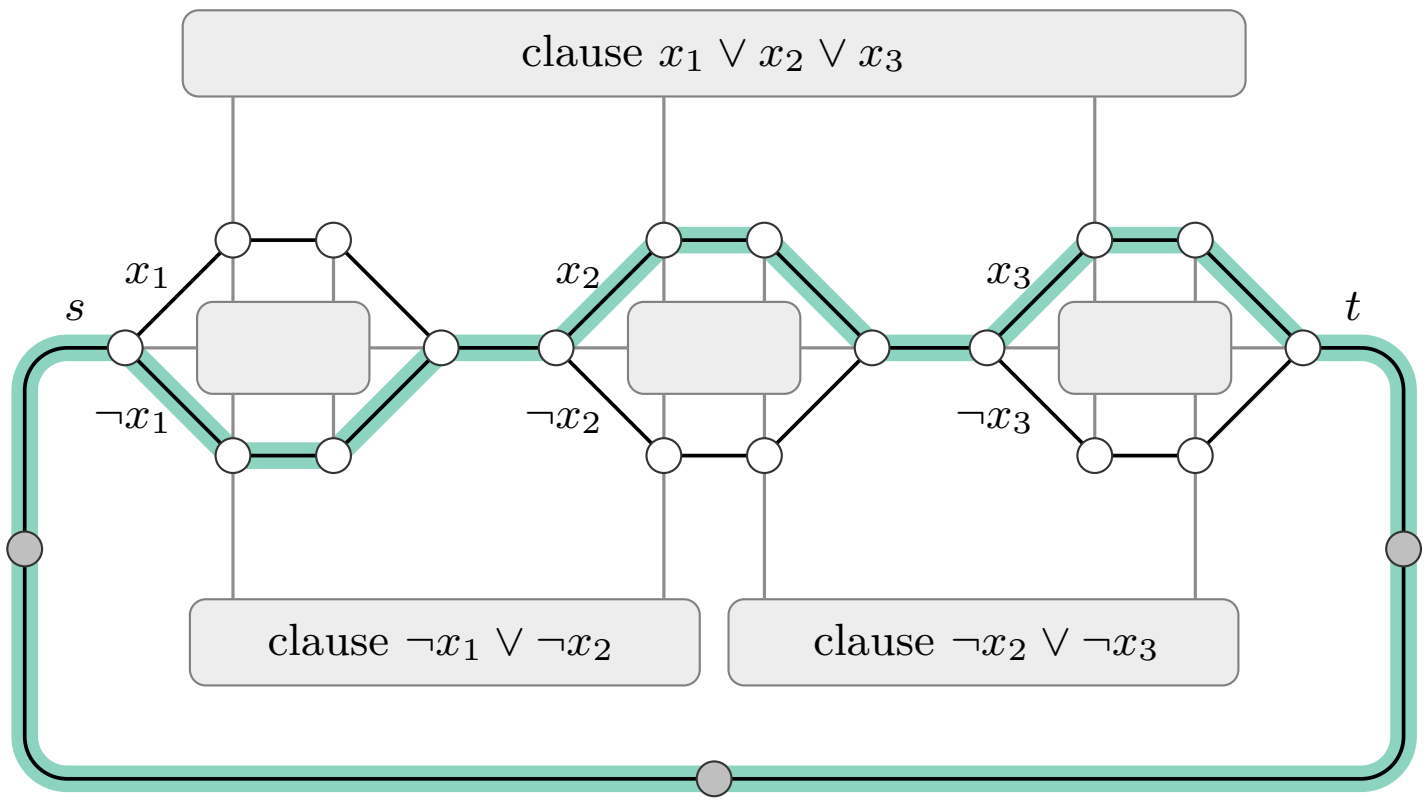

Our main reduction establishes hardness of this intermediate problem
on planar simple graphs while controlling degrees and the positions
of the terminals. Variable cycles offer two branches for the solution
path, encoding a truth assignment. \Cref{fig:overview-hardness} sketches the construction. Clause attachments force a
satisfied literal because, otherwise, deleting the path disconnects
the graph needed by the two spanning trees. Planar docking gadgets
provide two internal edge-disjoint spanning trees with attachment
ports on the outer face. Their used ports have degree three, so a
solution path cannot enter them: two incident path edges would leave
too few edges for the two spanning trees. This separates the choices
made by the path from the internal tree packings and keeps the final
maximum degree at five.

The ETH bounds exclude $2^{o(n+m)}$ time in general and
$2^{o(\sqrt n)}n^{\Oh(1)}$ time on planar inputs. The latter also
excludes $2^{o(\tw)}n^{\Oh(1)}$ time. For the multiplicity result,
we apply the classical path addition to hard \textsc{PathTreeTree}
instances with doubled edges. The added path supplies the only four
edges of multiplicity one. In contrast, connected instances with no
edges of multiplicity one always contain two edge-disjoint spanning trees.

\subsection{Consequences and Open Questions}

We conclude in \Cref{sec:conclusion} by discussing the limits of our
techniques and directions for further research. For the excess
$\delta=m-(2n-4)$, the deterministic
independent-cycle algorithm isolates the remaining use of randomness
in the matroid representation: the dual already has rank
$m-(2n-8)=\delta+4$, so no projection is needed for this parameter.
Derandomizing the representation of the union of the two graphic
matroids would therefore suffice for a deterministic FPT algorithm.
Other questions concern a polynomial kernel for the excess and
parameterization by the number of vertices of degree at least five.

\section{Classical Characterization and Witness Structure}
\label{sec:witnesses}

We first recall the classical characterization and the four-interval
structure of its witnesses. We then show that the cycle of every minimum
$M$-witness in an $M$-graph is induced in the underlying simple graph
(\Cref{lem:minimum-induced-witness}); this allows our exact algorithm in
\Cref{sec:exact-preprocessing} to
restrict its search to induced cycles. Finally, we give a linear-time
implementation of the classical construction of a temporal labeling
from either type of witness.

\begin{theorem}[{\cite{BuroschLM88,BuroschL92,Krumme96}}]\label{thm:characterization}
Let $G$ be a multigraph. Then $G$ is a yes-instance of \Gossip if and only
if at least one of the following conditions holds:
\begin{enumerate}
\item\label{Gtrivial} $G$ has at most one vertex;
\item\label{GisT} $G$ is a $T$-graph;
\item\label{GisM} $G$ is an $M$-graph.
\end{enumerate}
\end{theorem}

For completeness, \Cref{app:characterization} gives a self-contained
proof using our notation and multigraph conventions. Additionally, we provide a Lean formalization of this proof in~\cite{lax-630294}.

\subsection{Classical Witness Structure}

We first describe how the components of the two forests meet the
cycle of an $M$-witness in an $M$-graph. An \emph{interval} of the
cycle is a nonempty set of consecutive cycle vertices, forming a
path; a single vertex is allowed. The following lemma states that
the four components of each forest meet the cycle in the same four
intervals. The four edges between consecutive intervals can be paired
by taking every other edge in cyclic order. Each pair is a matching
whose endpoints provide one root in each component of either forest.
The proof is given in \Cref{app:construct-labeling}.

\begin{lemma}[{\cite[Corollary~1 and proof of Theorem~3]{Krumme96}}]\label{lem:four-intervals}
Let $G$ be an $M$-graph, and let $(C,F_1,F_2)$ be an $M$-witness of $G$.
There is a partition of $V(C)$ into four nonempty consecutive intervals
$X_1,X_2,X_3,X_4$, indexed in cyclic order, with the following properties:
\label{lem:structured-m-witness}%
\begin{itemize}
\item \emph{Common intervals.} For each forest $F_j$, the intersections
of its four components with $V(C)$ are exactly $X_1,X_2,X_3,X_4$.
\item \emph{Alternating matchings.} Let $b_i$ be the cycle edge joining
$X_i$ to $X_{i+1}$, with indices modulo four. Then
$B_1=\{b_1,b_3\}$ and $B_2=\{b_2,b_4\}$ are edge-disjoint matchings.
Every component of each forest contains exactly one vertex of $V(B_1)$
and exactly one vertex of $V(B_2)$.
\end{itemize}
\end{lemma}

Krumme calls $b_1,\ldots,b_4$ the \emph{essential} edges.
If an interval consists of a single vertex, then $B_1$ and $B_2$
share that vertex. Thus the two matchings need not have disjoint
vertex sets.
In particular, both forests of an $M$-witness of an $M$-graph can be
rooted\footnote{Rooting a forest means choosing one vertex as the root of each tree component.} at the same four vertices, namely at the endpoints of~$B_1$.
We use this repeatedly to reduce the number of guessed roots. The
labeling construction of \Cref{lem:construct-labeling} instead roots
$F_1$ at the endpoints of $B_1$ and $F_2$ at those of $B_2$.

We use the following additional lemma to restrict the cycles
enumerated by our exact algorithm.
\begin{figure}[t]
\centering
\begin{tikzpicture}[every node/.style={font=\footnotesize},
  qv/.style={circle,draw=black!75,fill=white,inner sep=0pt,minimum size=6.2mm,font=\scriptsize},
  treeA/.style={witnessblue,line width=1.4pt},
  treeB/.style={witnessorange,line width=1.4pt,dash pattern=on 3pt off 2pt},
  chord/.style={auxpurple,line width=1.3pt}]
\begin{scope}
  \coordinate (X1) at (0,1.6); \coordinate (X2) at (1.6,1.6);
  \coordinate (X3) at (1.6,0); \coordinate (X4) at (0,0);
  \draw[treeA] (X1)--(X2) node[midway,above] {$b_1$};
  \draw[treeB] (X2)--(X3) node[midway,right] {$b_2$};
  \draw[treeA] (X3)--(X4) node[midway,below] {$b_3$};
  \draw[treeB] (X4)--(X1) node[midway,left] {$b_4$};
  \draw[treeA,transform canvas={xshift=1.1pt,yshift=1.1pt}] (X1)--(X3);
  \draw[treeB,transform canvas={xshift=-1.1pt,yshift=-1.1pt}] (X1)--(X3);
  \node at (1.05,0.98) {$e$};
  \foreach \v/\l in {X1/X_1,X2/X_2,X3/X_3,X4/X_4} \node[qv] at (\v) {$\l$};
  \node[panel,text width=48mm,align=center] at (0.8,-1.2)
    {(a) chord between $X_1$ and $X_3$};
\end{scope}
\begin{scope}[xshift=5.2cm,yshift=0.2cm]
  \foreach \i in {0,...,4} \coordinate (p\i) at (0.75*\i,0);
  \coordinate (l) at (-0.75,0); \coordinate (r) at (3.75,0);
  \draw[black!40,line width=1pt] (l)--(p0) (p4)--(r);
  \draw[line width=1pt] (p0)--(p1)--(p2)--(p3)--(p4);
  \draw[black!45,line width=4pt,opacity=.35] (p0)--(p1)--(p2)--(p3)--(p4);
  \node at (1.5,-0.45) {$Q$};
  \draw[chord] (p0) to[bend left=50] node[above] {$e$} (p4);
  \foreach \i in {0,...,4} \node[smallv] at (p\i) {};
  \draw[decorate,decoration={brace,amplitude=4pt,mirror}] (-0.1,-0.8)--(3.1,-0.8)
    node[midway,below=4pt] {inside interval $X_i$};
  \node[align=left,text width=62mm,font=\scriptsize] at (1.45,-2.25)
    {Replace $Q$ by $e$ in the cycle. If $e\in E(F_1)$, replace $e$ in
     $F_1$ by an edge $e'$ of $Q$ joining the two components of $F_1-e$.};
  \node[panel] at (1.5,-3.2) {(b) chord inside one interval};
\end{scope}
\end{tikzpicture}
\caption{The two cases in the proof of
\Cref{lem:minimum-induced-witness}. (a) A chord $e$ whose endpoints lie
in different intervals, shown in the multigraph obtained from $C+e$ by
identifying each interval. The edge sets $\{e,b_1,b_3\}$ (solid) and
$\{e,b_2,b_4\}$ (dashed) are spanning trees sharing only $e$; completing
$F_1$ and $F_2$ with them gives a $T$-witness. The case
of a chord between consecutive intervals is analogous. Hence, no
$M$-witness of an $M$-graph has such a chord. (b) A chord
inside one interval shortcuts the cycle, which contradicts minimality;
gray stubs indicate the rest of~$C$.}
\label{fig:chord-cases}
\end{figure}
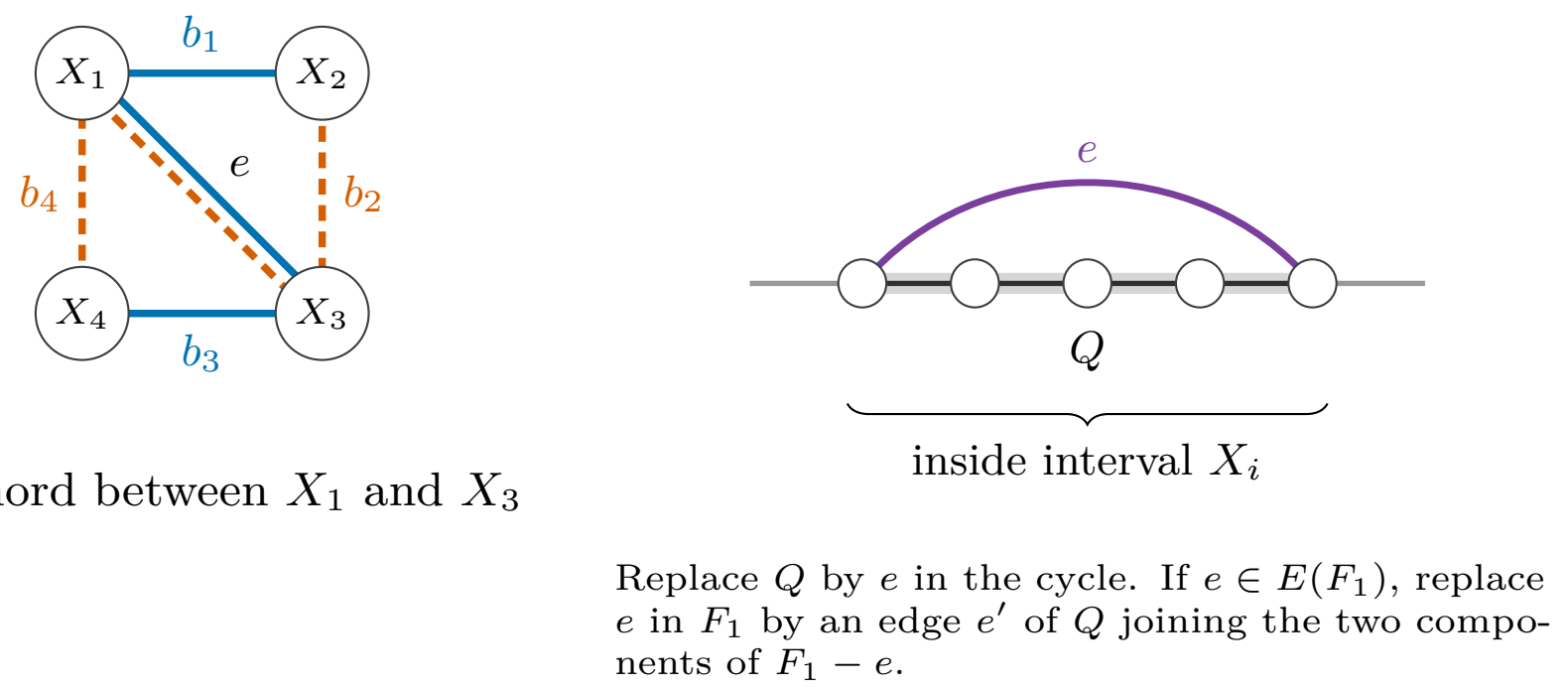

\begin{lemma}\label{lem:minimum-induced-witness}
Let $G$ be an $M$-graph.
The cycle of every minimum $M$-witness is induced in the support of $G$.
\end{lemma}
\begin{proof}
Use the intervals $X_1,\ldots,X_4$ of \Cref{lem:four-intervals} and the
essential edges $b_1,\ldots,b_4$.
Suppose that a minimum witness has a chord $e$,
that is, an edge joining two nonconsecutive vertices of $C$.
First suppose its endpoints lie in distinct intervals. Then $e$ belongs
to neither forest. Consider the four-vertex multigraph obtained from
$C+e$ by identifying each interval into one vertex
(\Cref{fig:chord-cases}(a)). If $e$ joins
$X_1$ to $X_2$, the edge sets $\{e,b_2,b_3\}$ and $\{b_1,b_3,b_4\}$
are spanning trees sharing only $b_3$. If it joins $X_1$ to $X_3$,
the sets $\{e,b_1,b_3\}$ and $\{e,b_2,b_4\}$ share only $e$.
Rotation and reversal cover all cases. Completing $F_1,F_2$ with these
sets contradicts that $G$ is not a $T$-graph. This case does not use
minimality: no $M$-witness of an $M$-graph has a chord between distinct
intervals.

Thus both endpoints of $e$ lie in one interval $X_i$. Let $Q$ be the
path between them contained in that interval (\Cref{fig:chord-cases}(b)). Replacing $Q$ by $e$
gives a shorter cycle $C'$ that still meets all four intervals.
If $e$ belongs to neither forest, this immediately gives a smaller
witness. Otherwise, say $e\in E(F_1)$, deleting $e$ splits its tree
into two components. 
The path $Q$ lies entirely in that tree's vertex set 
and joins the two endpoints of $e$, so some edge $e'$ of $Q$ joins these two components. 
Replace $e$ by $e'$ in $F_1$.
This preserves the component partition, and the new forest is
edge-disjoint from $C'$ and $F_2$. Again we obtain a smaller witness.
\end{proof}

\subsection{From Witnesses to Labelings}
\label{sec:witness-labelings}

The classical constructions of Burosch and Laborde~\cite[Lemmas~7--8]{BuroschL92}
and Krumme~\cite[Theorem~3]{Krumme96} turn either witness into a call
schedule.

The $T$-witness construction also connects to pivotability in temporal
spanners. A \emph{pivot vertex} is a vertex that all vertices can reach
by a common time and from which all vertices can be reached afterward
\cite{CasteigtsPS21,CasteigtsRRZ24}. Gathering and broadcasting along
two edge-disjoint spanning trees produces such a pivot at their common
root. If the trees share an edge, the construction instead makes that
edge a \emph{pivot edge}: every vertex admits a strict temporal path
ending with it and a strict temporal path starting with it
\cite{CarnevaleCC25}.

We use the following explicit running-time guarantee.

\begin{lemma}[Linear-time implementation of the classical construction]
\label{lem:construct-labeling}
Given a $T$-witness or an $M$-witness of $G$, a temporal labeling under
which $G$ is temporally connected can be constructed in $\Oh(n+m)$ time.
\end{lemma}

For a $T$-witness, the construction gathers information along one tree
and distributes it along the other, as described above. For an
$M$-witness with the common four-interval structure, let $p_i$ and $q_i$
be the endpoints of interval $X_i$ incident with the matchings
$B_1=\{b_1,b_3\}$ and $B_2=\{b_2,b_4\}$, respectively. First gather
information along each component of $F_1$ at its root $p_i$. Call
$b_1,b_3$, then traverse each interval from $p_i$ to $q_i$, and finally
call $b_2,b_4$. Each $q_i$ now knows all information, which is
distributed along $F_2$ rooted at these vertices. Thus, the labeling
roots $F_1$ at the endpoints of $B_1$ and $F_2$ at the endpoints of
$B_2$, whereas recognition can root both forests at the endpoints of $B_1$.
The two roots of an
interval coincide when it consists of a single vertex.
\Cref{fig:labeling-example} illustrates these phases on the running
example of \Cref{fig:m-graph}: calls 1--5 gather information along
$F_1$, calls 6--10 exchange it through the cycle, and calls 11--15
distribute it along $F_2$.

If the supplied witness does not
have the common four-interval structure, a linear-time procedure
extracts a $T$-witness instead. Thus the construction also handles
$M$-witnesses in graphs that are $T$-graphs. The implementation and
its running-time analysis are given in
\Cref{app:construct-labeling}.

\begin{figure}[t]
\centering
\begin{tikzpicture}[every node/.style={font=\footnotesize},
  lab/.style={fill=white,inner sep=1pt,font=\scriptsize\bfseries,rounded corners=1pt,draw=black!25,solid,text=black}]
\begin{scope}[x=1.35cm,y=1.2cm]
  \coordinate (c1) at (-0.95,0.95);
  \coordinate (c2) at (0.95,0.95);
  \coordinate (c3) at (1.45,-0.55);
  \coordinate (c4) at (0,-1.45);
  \coordinate (c5) at (-1.45,-0.55);
  \coordinate (x3) at (0,1.55);
  \coordinate (x1) at (0,2.45);
  \coordinate (x4) at (0,3.35);
  \coordinate (x2) at (0.1,-0.4);
  \draw[cycleedge] (c1)--(c2) node[lab,pos=.5] {8};
  \draw[cycleedge] (c2)--(c3) node[lab,pos=.5] {6};
  \draw[cycleedge] (c3)--(c4) node[lab,pos=.5] {9};
  \draw[cycleedge] (c4)--(c5) node[lab,pos=.5] {7};
  \draw[cycleedge] (c5)--(c1) node[lab,pos=.5] {10};
  \draw[blueforest] (x4)--(x1) node[lab,pos=.5] {1};
  \draw[blueforest] (x4) to[out=190,in=100] node[lab,pos=.5] {2} (c1);
  \draw[blueforest] (c1)--(x3) node[lab,pos=.4] {3};
  \draw[blueforest] (x3)--(c2) node[lab,pos=.6] {4};
  \draw[blueforest] (x2)--(c3) node[lab,pos=.5] {5};
  \draw[orangeforest] (c1)--(x1) node[lab,pos=.42] {11};
  \draw[orangeforest] (x1)--(x3) node[lab,pos=.5] {12};
  \draw[orangeforest] (x1)--(c2) node[lab,pos=.58] {13};
  \draw[orangeforest] (c2) to[out=80,in=-10] node[lab,pos=.5] {14} (x4);
  \draw[orangeforest] (c5)--(x2) node[lab,pos=.5] {15};
  \foreach \v/\l in {c1/c_1,c2/c_2,c3/c_3,c4/c_4,c5/c_5,x1/x_1,x2/x_2,x3/x_3,x4/x_4}
    \node[vertex] at (\v) {$\l$};
\end{scope}
\begin{scope}[xshift=5.1cm,yshift=2.6cm,
  desc/.style={anchor=north west,text width=60mm,align=left,inner sep=0pt,execute at begin node={\hyphenpenalty=10000\exhyphenpenalty=10000}},
  rng/.style={anchor=north east,inner sep=0pt}]
  \node[desc] (d1) at (0,0) {gather information along $F_1$ at $c_2,c_3,c_4,c_5$};
  \node[rng] (r1) at ([xshift=-9pt]d1.north west) {\textbf{1--5}:};
  \draw[blueforest] ([xshift=-1.85cm,yshift=-0.6ex]d1.north west) -- ++(0.32,0);
  \node[desc] (d2) at ([yshift=-5pt]d1.south west) {exchange gathered information within the pairs $c_2,c_3$ and $c_4,c_5$};
  \node[rng] (r2) at ([xshift=-9pt]d2.north west) {\textbf{6--7}:};
  \draw[cycleedge] ([xshift=-1.85cm,yshift=-0.6ex]d2.north west) -- ++(0.32,0);
  \node[desc] (d3) at ([yshift=-5pt]d2.south west) {transfer gathered information from $c_2$ to $c_1$};
  \node[rng] (r3) at ([xshift=-9pt]d3.north west) {\textbf{8}:};
  \draw[cycleedge] ([xshift=-1.85cm,yshift=-0.6ex]d3.north west) -- ++(0.32,0);
  \node[desc] (d4) at ([yshift=-5pt]d3.south west) {exchange information between $c_3,c_4$ and between $c_5,c_1$; all four distribution roots now know everything};
  \node[rng] (r4) at ([xshift=-9pt]d4.north west) {\textbf{9--10}:};
  \draw[cycleedge] ([xshift=-1.85cm,yshift=-0.6ex]d4.north west) -- ++(0.32,0);
  \node[desc] (d5) at ([yshift=-5pt]d4.south west) {distribute information along $F_2$ from $c_1,c_3,c_4,c_5$};
  \node[rng] (r5) at ([xshift=-9pt]d5.north west) {\textbf{11--15}:};
  \draw[orangeforest] ([xshift=-1.85cm,yshift=-0.6ex]d5.north west) -- ++(0.32,0);
\end{scope}
\end{tikzpicture}
\caption{A temporally connecting labeling of the running example from
\Cref{fig:m-graph}, obtained by the construction of
\Cref{lem:construct-labeling}. Each edge is labeled by its position in
the gossip scheme. Blue solid edges gather information at
$c_2,c_3,c_4,c_5$; black cycle edges exchange it so that the distribution
roots $c_1,c_3,c_4,c_5$ know everything after call 10; orange dashed
edges distribute it along $F_2$. Call 8 transfers information within
the only interval with more than one vertex, from $c_2$ to $c_1$.
Every edge is used exactly once.}
\label{fig:labeling-example}
\end{figure}

\section{Preprocessing and Exact Recognition}
\label{sec:exact-preprocessing}
The Clean-Up Rule provides universal preprocessing and bounds the number
of edges in terms of the number of vertices. We then recall the
pebble-game algorithms used for the packing tests, introduce the Tree Contraction Rule, and finally derive the exact recognition
algorithm.

\subsection{The Clean-Up Rule}

\Cref{thm:characterization} allows us to bound the number of edges by a
function of the number of vertices. We call a multigraph \emph{clean}
if it has no loops and every parallel class has at most three edges.

\begin{cleanuprule}\label{rule:cleanup}
Delete a loop, or delete an arbitrary edge from a parallel class that
contains at least four edges.
\end{cleanuprule}

\begin{lemma}\label{lem:cleanup}
The \CleanupRule preserves whether the input is a yes-instance and can be applied exhaustively in
$\Oh(n+m)$ time. The resulting graph $G'$ is a clean spanning subgraph
of $G$ with at most $3\binom n2\in\Oh(n^2)$ edges. For every $k\geq4$,
the graph $G$ admits an $M$-witness whose cycle has length at most $k$
if and only if $G'$ does. Moreover, $G'$ has the
same support as $G$, and its excess $|E(G')|-(2n-4)$ and its maximum degree are
at most those of $G$. Every temporally connecting labeling of $G'$ can
be extended to one of $G$ in $\Oh(n+m)$ time, where $m=|E(G)|$.
\end{lemma}
\begin{proof}
To apply the rule exhaustively, number the vertices $1,\ldots,n$ and
write every non-loop edge as the ordered pair $(\min,\max)$ of its
endpoints. Sort these pairs lexicographically with two rounds of bucket
sort, and keep the first three edges of each resulting run. This takes
$\Oh(n+m)$ time and produces a clean graph~$G'$.

If $G'$ is a yes-instance, then so is $G$: label the edges of
$E(G)\setminus E(G')$ with pairwise distinct labels larger than all
labels of $G'$. Additional calls cannot destroy temporal connectivity,
and this extends the labeling in $\Oh(n+m)$ time. 
Conversely, suppose that $G$ is a
yes-instance. If $n\leq1$, then so is $G'$. Otherwise,
\Cref{thm:characterization} gives a $T$-witness or an $M$-witness of $G$.
Such a witness contains no loop. Each spanning tree, each forest, and
the cycle of length at least four contain at most one edge of any
parallel class. Hence, a $T$-witness uses at most two edges of each
parallel class, and an $M$-witness uses at most three. Since parallel
edges are interchangeable, we can replace the used edges of each class
by retained edges of that class. This gives a witness of the same type
in $G'$, so $G'$ is a yes-instance by \Cref{thm:characterization}.
The same replacements apply to any supplied $M$-witness and preserve
its cycle length; conversely, every $M$-witness of $G'$ is one of $G$.
This proves the bounded-witness claim. The remaining statements are
immediate, since $G'$ is obtained by deleting
loops and parallel edges.
\end{proof}

Henceforth, we assume that the input has been cleaned up, so
$m\in\Oh(n^2)$. All running times are stated under this convention;
for an arbitrary input, add $\Oh(n+m)$ time to read and clean up its edge
list. Whenever an algorithm modifies the
multigraph, for example by contracting vertex sets, we say explicitly
whether it restores this assumption.

\subsection{Sparsity Matroids and Pebble Games}
\label{sec:pebble-games}

A pebble game is played on a graph using tokens, called \emph{pebbles},
placed at its vertices. We use the following version of the game of
Lee and Streinu~\cite{LeeStreinu2008}. It starts with $c$ pebbles at each
vertex, where $c\geq1$ is an integer. Initially, no edges are included
in the game; during play, we add edges from the input graph. A pebble
is \emph{free} if it is not covering an edge. The goal is to add as many
edges as possible while obeying two rules:
\begin{itemize}
\item \emph{Add an edge.} If its endpoints together hold at least $c+1$
free pebbles, take one pebble from an endpoint and use it to cover the
edge. Orient the edge away from that endpoint.
\item \emph{Move a pebble.} If an added edge is directed from $u$ to $v$
and $v$ has a free pebble, use that pebble to cover the edge, release the
pebble previously covering it at $u$, and reverse the edge. This
transfers a free pebble from $v$ to $u$.
\end{itemize}
To add an edge, the algorithm uses pebble moves to try to gather enough
free pebbles at its endpoints. We call the orientation of the added
edges, together with the free-pebble counts at the vertices, a
\emph{pebble orientation}. Intuitively, requiring $c+1$ free pebbles
before adding an edge prevents too many edges from accumulating within
any set of vertices.

Formally, an edge set $A\subseteq E(G)$ is
\emph{$(c,c)$-sparse}\footnote{The two entries specify the coefficient of
$|X|$ and the subtracted constant in the sparsity bound.} if
\[
 |A[X]|\leq c|X|-c
 \qquad\text{for every nonempty }X\subseteq V(G),
\]
where $A[X]$ denotes the edges of $A$ with both endpoints in $X$.
By Nash-Williams's theorem~\cite{Nash-Williams64}, these are exactly the
edge sets that can be partitioned into $c$ forests, or equivalently,
the independent sets of the union of $c$ copies of the graphic matroid.
In particular, $G$ contains $c$ edge-disjoint spanning trees if and only
if a maximum $(c,c)$-sparse edge set has $c(n-1)$ edges. Here,
\emph{maximum} means having the largest possible number of edges.

A vertex set $X$ with $|X|\geq2$ is \emph{tight} for a $(c,c)$-sparse
edge set $Y$ if $|Y[X]|=c|X|-c$. Adding a new edge $uv$, with $u\neq v$,
preserves sparsity precisely when no tight set contains both endpoints:
only vertex sets containing both $u$ and $v$ gain an internal edge.
The following theorem summarizes the pebble-game guarantees used by
our algorithms.

\begin{theorem}[{\cite{LeeStreinu2008,StreinuT09}}]\label{thm:pebble}
Let $c\geq1$ be a constant and let $G$ be a multigraph with $n$ vertices
and $m\in\Oh(n^2)$ edges.
\begin{enumerate}
\item\label{it:pebble-basis} In $\Oh(n^2)$ time, one can compute a
maximum $(c,c)$-sparse set $Y\subseteq E(G)$ together with its
inclusion-maximal tight sets. These maximal tight sets are pairwise
vertex-disjoint.
\item\label{it:pebble-insert} The computation maintains a pebble
orientation of $Y$, of size $\Oh(n)$. Given a pebble orientation of a
$(c,c)$-sparse set $Y$ and a new edge $e$, one can decide in $\Oh(n)$
worst-case time whether $Y\cup\{e\}$ is $(c,c)$-sparse, and if so,
update the orientation.
\item\label{it:pebble-decomp} Given a $(c,c)$-sparse set $Y$ with
$|Y|=c(n-1)$, a partition of $Y$ into $c$ spanning trees can be
computed in $\Oh(n^2)$ time.
\end{enumerate}
\end{theorem}

Lee and Streinu~\cite{LeeStreinu2008} provide a basic pebble game for
testing edge insertions and an extension that also identifies all
maximal tight sets. These give Statements~\ref{it:pebble-insert}
and~\ref{it:pebble-basis}, respectively. The maximal tight sets are
pairwise vertex-disjoint: if two tight sets $V_1,V_2$ intersect,
subtracting their common edges and applying sparsity to $V_1\cap V_2$
gives
\[
 |Y[V_1\cup V_2]|\geq c|V_1\cup V_2|-c.
\]
The reverse inequality follows from sparsity, so their union is tight.
Thus two distinct maximal tight sets cannot intersect.

For the insertion bound in Statement~\ref{it:pebble-insert}, each search
for a free pebble is a depth-first search in the current orientation.
This orientation has $n$ vertices and at most $c(n-1)$ edges, so a
search takes $\Oh(n)$ time. An insertion requires at most $c+1$
searches and stops at the first unsuccessful search. Since $c$ is
constant, the insertion test takes $\Oh(n)$ time.

The decomposition in Statement~\ref{it:pebble-decomp} follows from the
colored pebble game of Streinu and Theran~\cite{StreinuT09}.
Alternatively, the spanning-tree packing algorithm of Roskind and
Tarjan~\cite{RoskindT85} gives the same bound because the sparse edge
set has $\Oh(n)$ edges.

We will also use the pebble orientation to process additional edges
without restarting the computation. Suppose $Y$ is a maximum sparse
subset of an edge set $E'$. Starting from its pebble orientation,
process the edges of another set $E''$, retaining each edge whose
insertion succeeds. The resulting set is a maximum sparse subset of
$E'\cup E''$. This follows from the matroid property: $Y$ is a basis of
$E'$, and the insertion tests extend it greedily to a basis of
$E'\cup E''$.

\subsection{The Tree Contraction Rule}

We next introduce a rule that replaces a vertex set by a single vertex.
The set's internal structure ensures that a witness in the contracted
graph can be transformed into a witness in the original graph. We call
this transformation \emph{lifting} the witness.

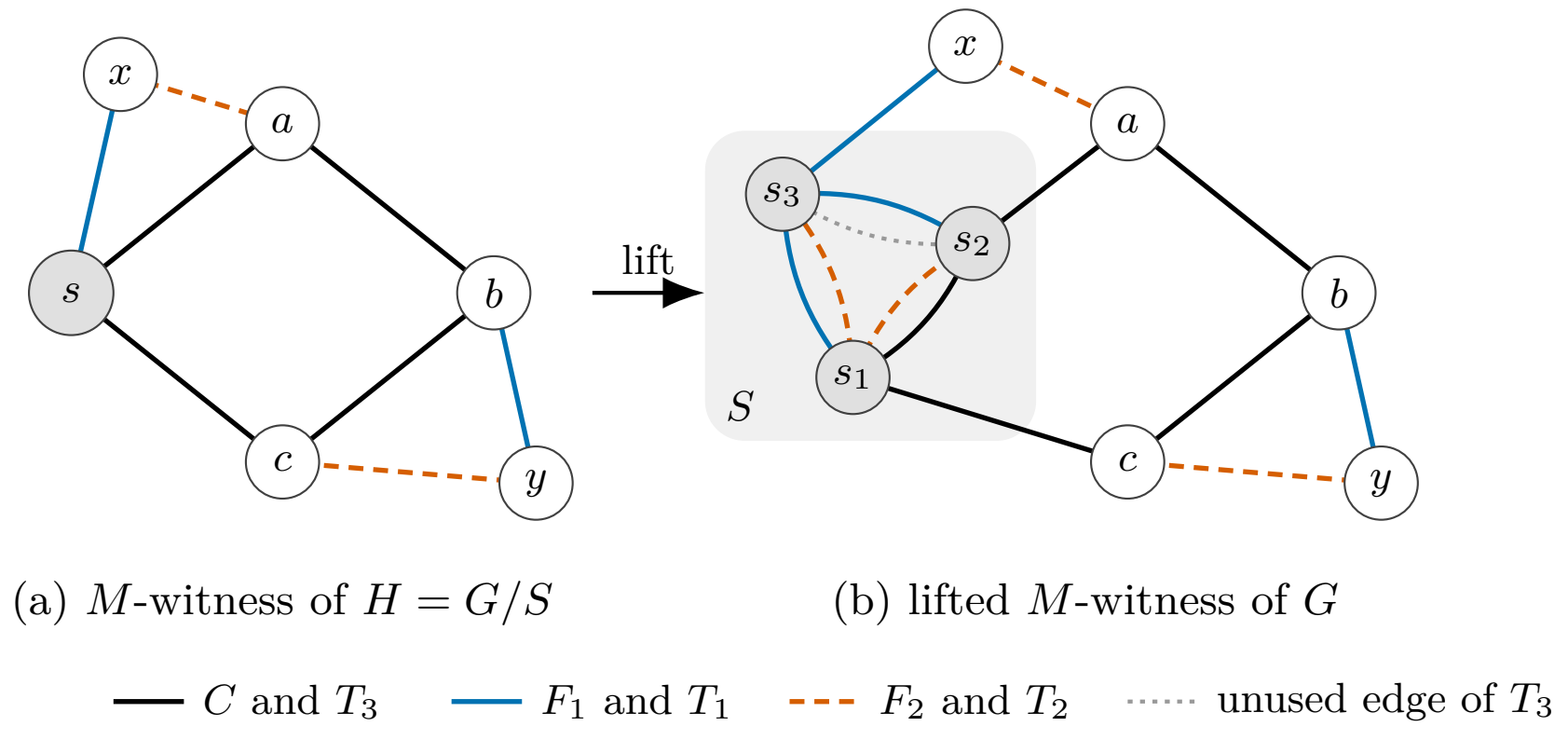
\begin{figure}[t]
\centering
\begin{tikzpicture}[every node/.style={font=\footnotesize},
  tthree/.style={black,line width=1pt},
  tthreeunused/.style={black!40,line width=.8pt,dash pattern=on 1pt off 1.5pt}]
\begin{scope}
  \coordinate (s) at (0,0);
  \coordinate (a) at (1.5,1.2);
  \coordinate (b) at (3,0);
  \coordinate (c) at (1.5,-1.2);
  \coordinate (x) at (0.35,1.55);
  \coordinate (y) at (3.3,-1.35);
  \draw[cycleedge] (s)--(a)--(b)--(c)--(s);
  \draw[blueforest] (x)--(s) (y)--(b);
  \draw[orangeforest] (x)--(a) (y)--(c);
  \foreach \v in {a,b,c,x,y} \node[vertex] at (\v) {$\v$};
  \node[vertex,fill=black!12,minimum size=6mm] at (s) {$s$};
  \node[panel] at (1.5,-2.2) {(a) $M$-witness of $H=G/S$};
\end{scope}
\draw[-{Latex[length=3mm]},thick] (3.7,0)--(4.5,0) node[midway,above] {lift};
\begin{scope}[xshift=6.0cm]
  \coordinate (s1) at (-0.45,-0.6);
  \coordinate (s2) at (0.4,0.35);
  \coordinate (s3) at (-0.95,0.7);
  \coordinate (a) at (1.5,1.2);
  \coordinate (b) at (3,0);
  \coordinate (c) at (1.5,-1.2);
  \coordinate (x) at (0.35,1.75);
  \coordinate (y) at (3.3,-1.35);
  \fill[black!6,rounded corners=8pt] (-1.5,-1.05) rectangle (0.85,1.15);
  \node at (-1.25,-0.8) {$S$};
  \draw[witnessblue,line width=1pt] (s1) to[bend left=18] (s3) (s2) to[bend right=18] (s3);
  \draw[witnessorange,line width=1pt,dash pattern=on 3pt off 2pt] (s1) to[bend left=18] (s2) (s1) to[bend right=18] (s3);
  \draw[tthree] (s1) to[bend right=18] (s2);
  \draw[tthreeunused] (s2) to[bend left=18] (s3);
  \draw[cycleedge] (s2)--(a)--(b)--(c)--(s1);
  \draw[blueforest] (x)--(s3) (y)--(b);
  \draw[orangeforest] (x)--(a) (y)--(c);
  \foreach \v in {a,b,c,x,y} \node[vertex] at (\v) {$\v$};
  \foreach \v/\l in {s1/s_1,s2/s_2,s3/s_3} \node[vertex,fill=black!12] at (\v) {$\l$};
  \node[panel] at (1.2,-2.2) {(b) lifted $M$-witness of $G$};
\end{scope}
\begin{scope}[xshift=0.3cm,yshift=-2.9cm]
  \draw[cycleedge] (0,0)--(0.5,0) node[right,text=black] {$C$ and $T_3$};
  \draw[blueforest] (2.4,0)--(2.9,0) node[right,text=black] {$F_1$ and $T_1$};
  \draw[orangeforest] (4.8,0)--(5.3,0) node[right,text=black] {$F_2$ and $T_2$};
  \draw[tthreeunused] (7.2,0)--(7.7,0) node[right,text=black] {unused edge of $T_3$};
\end{scope}
\end{tikzpicture}
\caption{Lifting an $M$-witness through the \TreeContractionRule
(\Cref{lem:rule-safe}). Here $G[S]$ contains three edge-disjoint
spanning trees $T_1,T_2,T_3$; every pair of vertices of $S$ is joined
by two parallel edges. (a) An $M$-witness of the quotient $H=G/S$ whose
cycle passes through the contracted vertex~$s$. (b) Lifting it to $G$:
the forests are completed by $T_1$ and $T_2$, respectively, and the
cycle is rerouted inside $S$ along the path in $T_3$ between the
endpoints $s_1$ and $s_2$ of its two edges at~$s$.}
\label{fig:contraction-rule}
\end{figure}

\begin{rrule}\label{rule:contraction}
If $|S|\geq2$ and $G[S]$ contains three edge-disjoint spanning trees,
contract $S$. That is, replace $S$ by a single new vertex $s$, delete
the edges with both endpoints in $S$, and replace the endpoint in $S$ of
every other edge by $s$, keeping parallel edges. We denote the result
by $G/S$.
\end{rrule}

In particular, the \TreeContractionRule creates no loops. We first prove
that it preserves label-connectivity. The forward direction is a special case
of the known closure of label-connectivity under contraction; for the
reverse direction, the three internal trees suffice to lift the
pairwise edge-disjoint parts of either witness.
\begin{lemma}\label{lem:rule-safe}
An application of the \TreeContractionRule preserves whether the instance is a
yes-instance of \Gossip.
\end{lemma}
\begin{proof}
Write $H=G/S$ and let $s$ be the contracted vertex.
If $G$ is a yes-instance, retain the labels of all edges surviving
in $H$. Every strict temporal path in $G$ maps, after deleting edges
inside $S$, to a strict temporal walk in $H$. Removing portions between
repeated vertices gives a strict temporal path. Hence $H$ is temporally
connected. This is the contraction property established by G\"obel,
Cerdeira, and Veldman~\cite[Lemma~8]{gobel1991label}.

For the converse, use \Cref{thm:characterization}. Fix three edge-disjoint
spanning trees $T_1,T_2,T_3$ of $G[S]$. If $H$ has one vertex, then
$G$ is already a $T$-graph. If $H$ is a $T$-graph, lift its
two witnessing trees to their original edges and add $T_1$ and $T_2$,
respectively. This gives two spanning trees of $G$ with the same
intersection as before.

Finally suppose $H$ has an $M$-witness $(C,F_1,F_2)$.
\Cref{fig:contraction-rule} illustrates how this witness lifts to $G$.
Lift the original
edges of $F_i$ and add $T_i$. This replaces $s$ in its forest component
by a tree and preserves the four components.
If $s\notin V(C)$, retain $C$. Otherwise let $u,v\in S$ be the original
endpoints of the two cycle edges incident with $s$, and replace $s$
along $C$ by the $u$-$v$ path in $T_3$ (trivial if $u=v$).
The lifted cycle is simple, has length at least four, and is edge-disjoint
from both lifted forests. All their components still meet the cycle.
Hence $G$ has an $M$-witness and is a yes-instance by
\Cref{thm:characterization}; a labeling can be constructed with
\Cref{lem:construct-labeling}.
\end{proof}

Call a multigraph \emph{reduced} if it has no loops and
the \TreeContractionRule is inapplicable. Since the input is clean and
contraction creates no loops, the graph remains loopless throughout.
We next show that all applicable sets can
be found in one pebble-game computation; there is no need to restart
the search after each contraction.
\begin{lemma}\label{lem:rule-time}
The \TreeContractionRule can be applied exhaustively in $\Oh(n^2)$ time. Moreover,
the resulting multigraph can be assumed to be clean.
Furthermore, a temporally connecting labeling of $G$ can be computed from a
witness of the reduced multigraph in $\Oh(n^2)$ time.
\end{lemma}
\begin{proof}
It suffices to find all inclusion-maximal vertex sets supporting
three edge-disjoint spanning trees in one computation and to show that
contracting them leaves no further eligible set. We obtain these sets
as the maximal tight sets of a maximum $(3,3)$-sparse set.

Compute a maximum $(3,3)$-sparse set $Y\subseteq E(G)$ and its
maximal tight sets in $\Oh(n^2)$ time with \Cref{thm:pebble}. Each
maximal tight set $X$ satisfies $|Y[X]|=3(|X|-1)$. Since $Y[X]$ is a
union of three forests on $X$, each of these forests is a spanning tree
of $G[X]$. Hence, every maximal tight set is eligible for the rule.

\emph{Every eligible set is contained in a maximal tight set.}
Suppose that $G[S]$ contains
three edge-disjoint spanning trees, and let $u,v\in S$ be distinct.
Consider the union of three graphic matroids on $E(G)$ together with a
fresh auxiliary edge $uv$. The union of the three trees is independent
of size $3(|S|-1)$, so adding $uv$ creates a dependent set. Hence, $uv$
lies in the closure of $E(G)$. Since $Y$ is a maximal independent subset
of $E(G)$, it has the same closure as $E(G)$, and therefore
$Y\cup\{uv\}$ is not $(3,3)$-sparse. Hence, some tight set of $Y$
contains $u$ and $v$. Each tight set lies in a maximal tight set, and
the maximal tight sets are pairwise vertex-disjoint. Consequently, all
of $S$ lies in one maximal tight set.

\emph{The quotient is reduced.}
Contract all maximal tight sets. Contracting them one after another
is a sequence of applications of the \TreeContractionRule, since
contracting one set does not change the edges inside the others. Every
quotient vertex has a \emph{branch set}: a contracted maximal tight set,
or a single vertex outside all maximal tight sets. Let $W$ be a set of
at least two quotient vertices inducing three edge-disjoint spanning
trees, and let $S'$ be the union of their branch sets. Lifting the
trees and adding three internal spanning trees in every contracted
branch set, one to each lifted tree, shows that $S'$ is eligible
in~$G$. By the previous paragraph, $S'$ lies in one maximal tight set.
But $S'$ meets two distinct branch sets, and each maximal tight set is
a single branch set, a contradiction. The quotient is therefore
reduced after a single global computation.

\emph{Running time and reconstruction.}
The contraction takes $\Oh(n+m)=\Oh(n^2)$ time. Contracting may create
parallel classes with more than three edges. We therefore apply
the \CleanupRule to the quotient in $\Oh(n^2)$ time. This preserves
the answer, and it does not create eligible sets, since it only
deletes edges. Hence, the clean quotient is still reduced.

Finally, a witness of the reduced multigraph lifts to $G$ as in the proof of
\Cref{lem:rule-safe}, one contracted set at a time. The required three
edge-disjoint spanning trees of each contracted maximal tight set $X$
come from the pebble game: $Y[X]$ is a $(3,3)$-sparse set of $G[X]$ with
$3(|X|-1)$ edges, so Statement~\ref{it:pebble-decomp} of
\Cref{thm:pebble} partitions it into three spanning trees in
$\Oh(|X|^2)$ time. Since the maximal tight sets are disjoint, this takes
$\Oh(\sum_X|X|^2)=\Oh(n^2)$ time in total. The lifting itself takes
$\Oh(n^2)$ time, and \Cref{lem:construct-labeling} then yields a
temporally connecting labeling of $G$ in $\Oh(n+m)=\Oh(n^2)$ time.
This proves the reconstruction bound.
\end{proof}

Reducedness imposes a sparsity bound on every induced subgraph.

\begin{lemma}\label{lem:reduced-edges}
Let $G=(V,E)$ be reduced. For every $X\subseteq V$ with $|X|\geq2$,
\[
 |E[X]|\leq3|X|-4.
\]
\end{lemma}
\begin{proof}
Otherwise choose an inclusion-minimal set $S$ of size at least two
with $|E[S]|\geq3(|S|-1)$.
Every nonempty proper subset $Z\subset S$ satisfies
$|E[Z]|\leq3(|Z|-1)$; singletons satisfy this because there are no loops.
For any partition of $S$ into $q\geq2$ nonempty parts, the number
of edges crossing between parts is at least
\[
 3(|S|-1)-\sum_{Z\text{ a part}}3(|Z|-1)=3(q-1).
\]
The Tutte--Nash-Williams spanning-tree packing
theorem~\cite{Nash-Williams61,Tutte61} yields three edge-disjoint
spanning trees in $G[S]$, contradicting reducedness.
\end{proof}
In particular, a reduced graph with $n\geq2$ has at most $3n-4$ edges.

\subsection{Exact Recognition}

Having established the preprocessing rules, we turn to exact
recognition. We first test for a $T$-witness and then show how to decide whether a
prescribed cycle extends to an $M$-witness. Enumerating induced cycles
and applying the extension test yields the exact algorithm.

\begin{lemma}\label{lem:t-graph-test}
One can decide in $\Oh(n^2)$ time whether a multigraph is
a $T$-graph and, if so, find a $T$-witness.
\end{lemma}
\begin{proof}
Since the input is clean, $m\in\Oh(n^2)$ and \Cref{thm:pebble}
applies. Compute a maximum
$(2,2)$-sparse set $Y\subseteq E(G)$ and its maximal tight sets with
\Cref{thm:pebble}. We claim that $G$ is a $T$-graph if
and only if $|Y|\geq2n-3$. If two spanning trees share at most one
edge, their union is a union of two forests with at least $2n-3$ edges.
Conversely, suppose $|Y|\geq2n-3$. If $|Y|=2n-2$, then
Statement~\ref{it:pebble-decomp} of \Cref{thm:pebble} partitions $Y$
into two edge-disjoint spanning trees. Otherwise $|Y|=2n-3$. Then $V(G)$
is not tight for $Y$. Since the maximal tight sets are vertex-disjoint
and $V(G)$ is not one of them, some two vertices $u,v$ lie in no common
maximal tight set, and hence in no common tight set. Thus, $Y\cup\{uv\}$ is $(2,2)$-sparse for a new edge $uv$, and it
partitions into two spanning trees $T_1,T_2$.
Say $uv\in E(T_2)$. Then $T_2-uv$ has two components, and the spanning
tree $T_1$ contains an edge $e$ joining them. Now $T_1$ and
$(T_2-uv)+e$ are spanning trees of $G$ sharing only $e$.

All steps take $\Oh(n^2)$ time.
\end{proof}

We next test whether a prescribed cycle can serve as the cycle of an
$M$-witness. Intuitively, identifying the four roots of the forests is
equivalent to adding two copies of a path through the roots. A path
through four roots has three edges, so two copies contribute six edges.
For each candidate root set, we test these six edges using a copy of a
pebble orientation computed once for the graph with the prescribed cycle
edges removed.

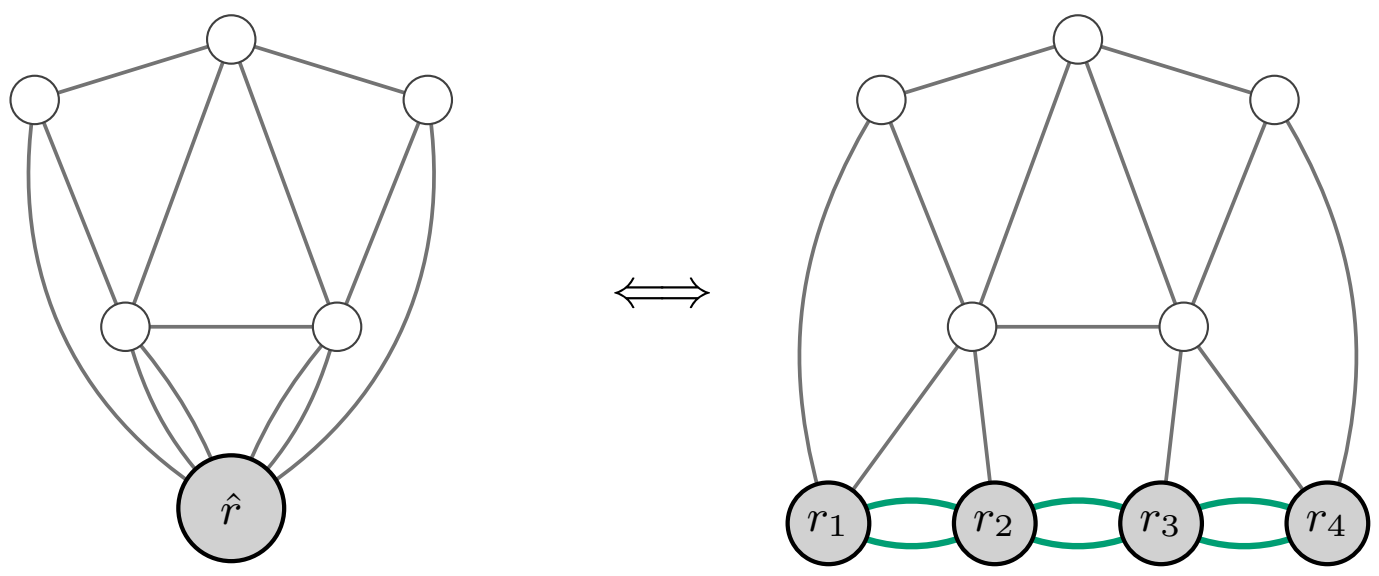
\begin{figure}[t]
\centering
\begin{tikzpicture}[every node/.style={font=\footnotesize},
  xv/.style={circle,draw=black,fill=black!18,line width=.8pt,inner sep=0pt,minimum size=5.4mm},
  gedge/.style={black!55,line width=.7pt},
  added/.style={auxgreen,line width=1.2pt}]
\newcommand{\background}{%
  \coordinate (u1) at (-0.4,1.3); \coordinate (u2) at (0.9,1.7); \coordinate (u3) at (2.2,1.3);
  \coordinate (u4) at (0.2,-0.2); \coordinate (u5) at (1.6,-0.2);
}
\begin{scope}
  \background
  \coordinate (x) at (0.9,-1.4);
  \draw[gedge] (u1)--(u2)--(u3) (u1)--(u4) (u2)--(u4) (u2)--(u5) (u3)--(u5) (u4)--(u5);
  \draw[gedge] (u4) to[bend right=15] (x) (u4) to[bend left=10] (x)
               (u5) to[bend left=15] (x) (u5) to[bend right=10] (x)
               (u1) to[bend right=35] (x) (u3) to[bend left=35] (x);
  \foreach \p in {u1,u2,u3,u4,u5} \node[smallv] at (\p) {};
  \node[xv,minimum size=7mm] at (x) {$\hat r$};
  \node[panel,text width=60mm,align=center] at (0.9,-2.45)
    {(a) $H_R$: identify $R$ in $G-E(C)$};
\end{scope}
\node[font=\large] at (3.75,0) {$\Longleftrightarrow$};
\begin{scope}[xshift=5.6cm]
  \background
  \coordinate (x1) at (-0.75,-1.5); \coordinate (x2) at (0.35,-1.5);
  \coordinate (x3) at (1.45,-1.5);  \coordinate (x4) at (2.55,-1.5);
  \draw[gedge] (u1)--(u2)--(u3) (u1)--(u4) (u2)--(u4) (u2)--(u5) (u3)--(u5) (u4)--(u5);
  \draw[gedge] (u4)--(x1) (u4)--(x2) (u5)--(x3) (u5)--(x4) (u1) to[bend right=25] (x1) (u3) to[bend left=25] (x4);
  \foreach \a/\b in {x1/x2,x2/x3,x3/x4}{
    \draw[added] (\a) to[bend left=28] (\b);
    \draw[added] (\a) to[bend right=28] (\b);
  }
  \foreach \p in {u1,u2,u3,u4,u5} \node[smallv] at (\p) {};
  \foreach \p/\l in {x1/r_1,x2/r_2,x3/r_3,x4/r_4} \node[xv] at (\p) {$\l$};
  \node[panel,text width=62mm,align=center] at (0.9,-2.45)
    {(b) $G_R^+$: add two copies of $r_1r_2r_3r_4$};
\end{scope}
\end{tikzpicture}
\caption{The equivalence used in \Cref{lem:cycle-extension}. The graph
$H_R$ has two edge-disjoint spanning trees if and only if $G_R^+$ does.
Since $G_R^+$ differs from $G-E(C)$ only by six added edges (green),
each candidate root set $R$ is tested by six insertions into a
precomputed pebble orientation, in $\Oh(n)$ time.}
\label{fig:cycle-extension}
\end{figure}

\begin{lemma}\label{lem:cycle-extension}
Let $G$ be a multigraph that is not a $T$-graph, and let $C$ be
a cycle of length $\ell\geq4$. In $\Oh(n^2+\ell^2n)$ time, one can
decide whether $C$ extends to an $M$-witness and, if so, construct its
forests.
\end{lemma}
\begin{proof}
For a set $R=\{r_1,r_2,r_3,r_4\}$ of four vertices, let $H_R$ be obtained
from $G-E(C)$ by identifying $R$ into one vertex. Moreover, let $G_R^+$
be obtained from $G-E(C)$ by adding two copies of each of the three
fresh edges $r_1r_2$, $r_2r_3$, and $r_3r_4$.

\emph{Auxiliary equivalence.}
We first show that $H_R$ has two edge-disjoint spanning trees if and
only if $G_R^+$ has. \Cref{fig:cycle-extension} illustrates this
equivalence. Given a spanning tree of $H_R$, lift it to its
original edges. Each lifted tree is a forest with four components, one
containing each vertex of $R$; adding one copy of the path
$r_1r_2r_3r_4$ gives a spanning tree of $G_R^+$. Conversely,
identifying $R$ in a spanning tree of $G_R^+$ turns the added edges into
loops and leaves a connected spanning subgraph of $H_R$, which contains a
spanning tree. Both transformations preserve edge-disjointness.

\emph{Correctness.}
The algorithm enumerates the $\ell(\ell-3)/2$ matchings $B$ of size two in
$C$, and for each, it tests whether $G_R^+$ has two edge-disjoint
spanning trees, where $R=V(B)$. It accepts if some test succeeds.
Suppose that $(C,F_1,F_2)$ is an $M$-witness. Then $G$ is an $M$-graph,
and by \Cref{lem:four-intervals}, the enumerated matching $B=B_1$
has exactly one endpoint in each component of each forest. Identifying
$R=V(B)$ therefore turns both forests into two edge-disjoint spanning
trees of $H_R$, and the algorithm accepts. Conversely, lift two
edge-disjoint spanning trees of $H_R$ to their original edges in $G$.
Each lifts to a spanning forest with four components, one containing
each vertex of $R\subseteq V(C)$. The forests are edge-disjoint and
avoid $E(C)$, so they form an $M$-witness with $C$.

\emph{Running time.}
Since the input is clean, \Cref{thm:pebble} applies. Compute a maximum
$(2,2)$-sparse set $Y_0$ of
$E(G)\setminus E(C)$ and its pebble orientation once in $\Oh(n^2)$ time.
For each matching, copy this orientation in $\Oh(n)$ time; the maximal
tight sets are not needed here. Then insert the six added edges with
Statement~\ref{it:pebble-insert} of \Cref{thm:pebble}, in $\Oh(n)$
worst-case time each. By the matroid property, the
resulting sparse set is maximum in $E(G_R^+)$. Hence, $G_R^+$ has two
edge-disjoint spanning trees if and only if at least $2n-2-|Y_0|$ of the
six insertions succeed. This takes $\Oh(n)$ time per matching. For the
successful matching, partition the final sparse set into two spanning
trees in $\Oh(n^2)$ time, identify $R$, choose spanning trees of the
resulting connected subgraphs of $H_R$, and lift them in $\Oh(n)$ time.
The total running time is $\Oh(n^2+\ell^2n)$.
\end{proof}

The induced-cycle property gives an exact algorithm whose exponential
factor depends only on the number of vertices.

\begin{lemma}\label{lem:representative-cycle}
Fix one representative edge of every parallel class of a multigraph
$G$. Let $C$ be the lift of a cycle of the support using these
representatives. If $G$ admits an $M$-witness whose cycle has the same
support as $C$, then $C$ extends to an $M$-witness of $G$.
\end{lemma}
\begin{proof}
Let $(C',F_1,F_2)$ be such a witness, and let $f\in E(C')$ be a
non-representative edge with representative $r$. The cycle uses at most
one edge of each parallel class, so $r$ is unused or belongs to one
forest. Replace $f$ by $r$ in the cycle and, in the latter case, replace
$r$ by $f$ in that forest. Because $f$ and $r$ have the same endpoints,
the result is again an $M$-witness. Repeating this exchange gives a
witness with cycle $C$.
\end{proof}

\begin{theorem}\label{thm:exact}
\Gossip can be solved in $\Oh(3^{n/3}n^3)$ time.
\end{theorem}
\begin{proof}
Accept if $G$ has at most one vertex. Otherwise, test whether $G$ is a
$T$-graph in $\Oh(n^2)$ time using \Cref{lem:t-graph-test}, and accept
if so. If this test fails, \Cref{thm:characterization} reduces the task
to finding an $M$-witness. By
\Cref{lem:minimum-induced-witness}, it suffices to enumerate induced cycles
of length at least four in the support of $G$ and test each with
\Cref{lem:cycle-extension}.

Fix one representative edge of every parallel class and lift each
enumerated support cycle using these representatives.
\Cref{lem:representative-cycle} shows that this loses no witness.

Morrison and Scott~\cite[Corollary~1.3]{MorrisonS17} determined the
maximum number of induced cycles in an $n$-vertex graph for all
$n\geq n_0$, where $n_0$ is a constant. Their formula implies that
there are at most $3^{n/3}+12n+51$ induced cycles for $n\geq n_0$, and
hence $3^{n/3}+\Oh(n)$ induced cycles for all $n$. The algorithm of
Uno and Satoh~\cite{UnoS14} enumerates the induced cycles of a simple
graph with $n'$ vertices and $m'$ edges in $\Oh(n'+m')$ time per cycle.
Write $h$ for the number of induced cycles. The enumeration also outputs
triangles, which we skip. By
\Cref{lem:cycle-extension}, each remaining extension test takes
$\Oh(n^2+\ell^2n)=\Oh(n^3)$ time. Thus, using the per-cycle bound of
Uno and Satoh, the total running time is $\Oh(n^2+hn^3)$. Substituting
$h\leq3^{n/3}+\Oh(n)$ gives $\Oh(3^{n/3}n^3)$.
\end{proof}
More generally, the proof gives an $\Oh(n^2+hn^3)$-time algorithm,
where $h$ is the number of induced cycles in the support. Thus,
recognition is polynomial-time solvable on graph classes with
polynomially many induced cycles. If the support is chordal, then
\Cref{lem:minimum-induced-witness} excludes the $M$-graph case, so the
$T$-graph test of \Cref{lem:t-graph-test} solves \Gossip in $\Oh(n^2)$ time.

\section{Algorithms for Structural Parameters}
\label{sec:structural-parameters}

We first give a deterministic single-exponential algorithm parameterized
by treewidth. We then combine it with the \TreeContractionRule to obtain
algorithms for several parameters that permit dense inputs of large
treewidth.

\subsection{Recognition Parameterized by Treewidth}

We search for the witnesses by connectivity dynamic programs over a
tree decomposition. Intuitively, the exact number of forest edges and
the exact root sets of an $M$-witness are only needed to certify that
the forests have four components. We first show that a relaxed
condition suffices; it can be checked with a constant amount of
counting information, which removes both the guessing of roots and all
counters of size $\Theta(n)$ from the dynamic program.

\begin{lemma}\label{lem:relaxed-witness}
A multigraph $G$ admits an $M$-witness if and only if there are pairwise
disjoint edge sets $C,D_1,D_2\subseteq E(G)$ such that $C$ is the edge
set of a cycle of length at least four and, for each $i\in\{1,2\}$, the
graph $(V(G),D_i)$ has at most four components, each of which meets
$V(C)$. Moreover, given such sets, an $M$-witness can be computed in
$\Oh(n+m)$ time.
\end{lemma}
\begin{proof}
Every $M$-witness satisfies the conditions, with $C$ being the edge set
of its cycle. Conversely, replace each
$D_i$ by a spanning forest $F_i$ of $(V(G),D_i)$ with the same
components. While $F_i$ has fewer than four components, some component
contains two vertices of $C$, since $C$ has at least four vertices.
Delete an edge of the path in $F_i$ between two such vertices. This
splits one component into two components, each containing a vertex of
$C$. After at most three deletions, the cycle with edge set $C$
together with $F_1,F_2$ forms an $M$-witness.
Computing the forests and performing the deletions takes $\Oh(n+m)$
time.
\end{proof}

The relaxed formulation lets us count components through a constant
number of auxiliary edges, rather than through the sizes of the forests.
We use it in the following dynamic program.

\begin{theorem}\label{thm:treewidth}
\Gossip can be solved deterministically in
$2^{\Oh(\tw)}\cdot n$ time, where $\tw$ is the treewidth of
the support of the input multigraph. If the input is a yes-instance, a
temporally connecting labeling can be computed within the same time.
\end{theorem}
\begin{proof}

Handle $n\leq1$ directly and reject disconnected inputs. Thus we may
assume that $G$ is connected and $n\geq2$. Let $t=\tw\geq1$.
Since the input is clean and its support has at most $tn$ edges,
we have $m\leq3tn$. Using Korhonen's algorithm~\cite{Korhonen21}, we compute a
tree decomposition of the support of width at most $2t+1$ in
$2^{\Oh(t)}\cdot n$ time.
By \Cref{thm:characterization}, $G$ is a
yes-instance if and only if it is a $T$-graph or admits an $M$-witness.
We test these alternatives in this order. The first test uses a short
connectivity dynamic program. Only if it fails do we run the more
involved $M$-witness program. We use the rank-based approach of
Bodlaender et al.~\cite{BodlaenderEtAl2015} to keep both tables
single-exponential in the bag size.

Each program augments the graph and the decomposition as described
below. After this augmentation, we use a nice tree decomposition with
introduce-edge nodes, $\Oh(tn)$ nodes, and every relevant edge introduced
exactly once. This conversion takes $\Oh(t^2n)$
time~\cite[Chapter~7]{C+15}. Let $b=\Oh(t)$ be the maximum bag size.
For a node $x$, let $G_x$ consist of
the vertices in bags at or below $x$ and the edges introduced there. In
particular, every edge incident with a vertex $v$ is introduced below
the node forgetting $v$. A \emph{partial solution} at $x$ consists of
the choices in $G_x$ accepted by all transitions at or below $x$.

\subparagraph*{The $T$-Graph Pretest.}
Fix an arbitrary vertex $v_0$ and add it to every bag of the
decomposition, increasing its width by at most one. We may assume that
the root and leaf bags contain only $v_0$.
The graph $G$ is a $T$-graph if and only if there are edge sets
$E_1,E_2$ with $|E_1\cap E_2|\leq1$ such that $(V,E_1)$ and $(V,E_2)$
are connected: spanning trees of these two graphs then share at most
one edge.
At a bag $B$, a partial solution assigns each introduced edge to
$E_1$, $E_2$, both, or neither. Its state consists of a counter
$\sigma\in\{0,1\}$ of edges assigned to both sets, and one partition of $B$
for each $E_i$ describing the connectivity of the partial graph on the
bag. Introducing a vertex adds a singleton block. Introducing an edge
merges the blocks of its endpoints in the chosen partitions. Forgetting
a vertex whose block is a singleton in some partition is rejected,
since its component could never reach the permanent vertex $v_0$.
Otherwise, the vertex is removed from both partitions. Join nodes add
the counters $\sigma$ of the two children and join the partitions.
Whenever $\sigma$ would exceed one, the partial solution is discarded. The root accepts if both
partitions consist of a single block.
Thus every component of each partial graph $(V(G_x),E_i\cap E(G_x))$
contains a vertex of the bag. Conversely, the restriction of a solution
to $G_x$ is never rejected, since both graphs $(V,E_i)$ are connected.

For each value of $\sigma$, the standard rank-based reduction for two
connectivity partitions retains at most $4^{|B|}$ states and preserves
all extensions~\cite{BodlaenderEtAl2015}. The transitions and reductions
therefore take $2^{\Oh(b)}$ time per bag. Predecessor pointers recover
$E_1,E_2$, from which we extract the required spanning trees. We run
this pretest first and stop if it accepts. Hence, in the remainder of
the proof, $G$ is not a $T$-graph.

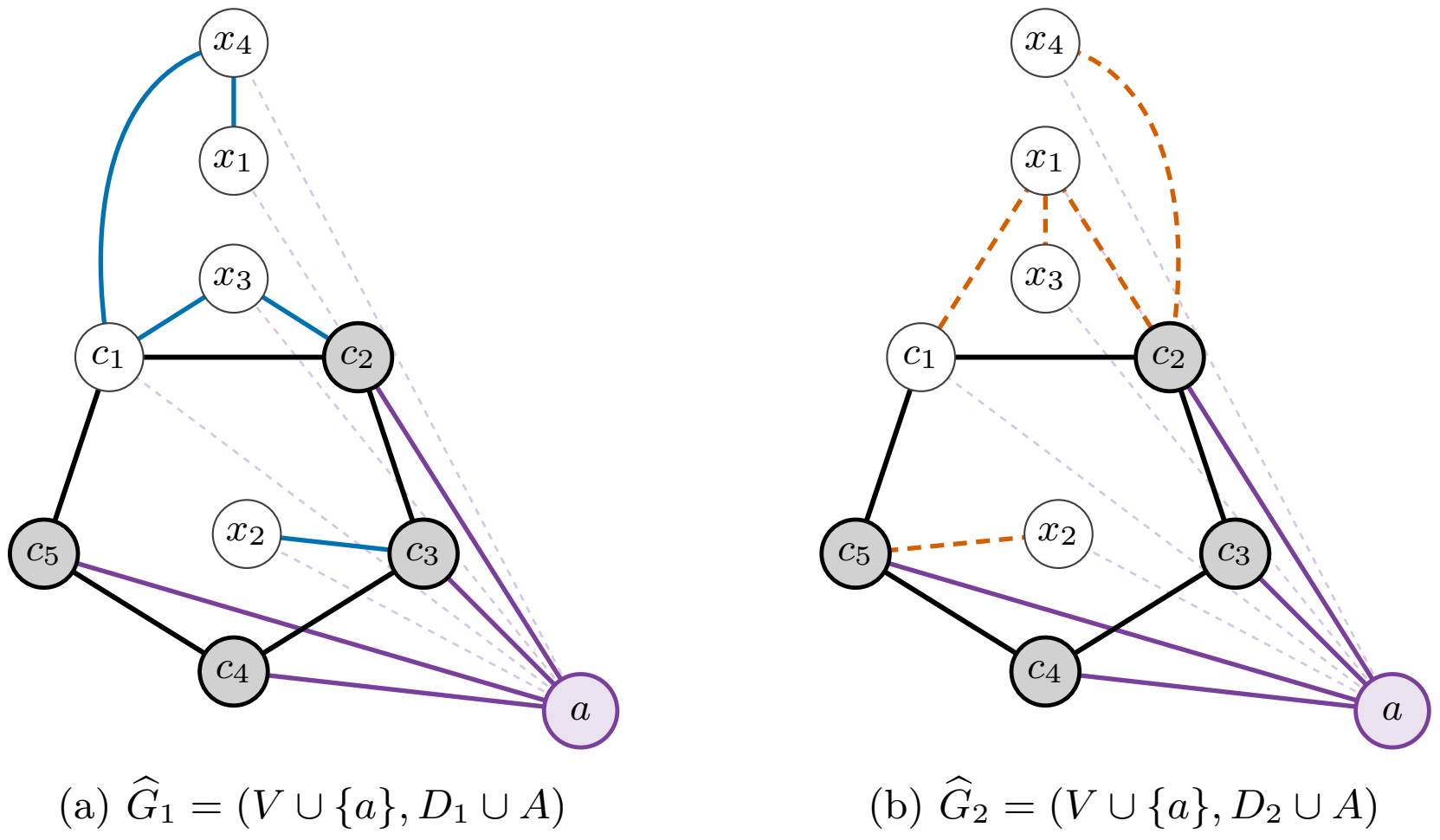
\begin{figure}[t]
\centering
\begin{tikzpicture}[
    every node/.style={font=\footnotesize},
    vertex/.style={circle,draw=black!75,fill=white,inner sep=0pt,minimum size=5.2mm},
    root/.style={vertex,fill=black!18,draw=black,line width=.9pt},
    apexv/.style={circle,draw=auxpurple,fill=auxpurple!15,line width=.9pt,inner sep=0pt,minimum size=5.6mm},
    cycleedge/.style={black,line width=1.1pt},
    apexedge/.style={auxpurple,line width=1pt},
    apexoff/.style={auxpurple!30,line width=.5pt,dash pattern=on 1.5pt off 1.5pt},
    panel/.style={font=\small}
]
\newcommand{\apexexample}[1]{%
  \coordinate (c1) at (-0.95,0.95);
  \coordinate (c2) at (0.95,0.95);
  \coordinate (c3) at (1.45,-0.55);
  \coordinate (c4) at (0,-1.45);
  \coordinate (c5) at (-1.45,-0.55);
  \coordinate (x3) at (0,1.55);
  \coordinate (x1) at (0,2.45);
  \coordinate (x4) at (0,3.35);
  \coordinate (x2) at (0.1,-0.4);
  \coordinate (a) at (2.65,-1.75);
  \foreach \v in {c1,x1,x2,x3,x4} \draw[apexoff] (a)--(\v);
  \foreach \v in {c2,c3,c4,c5} \draw[apexedge] (a)--(\v);
  \draw[cycleedge] (c1)--(c2)--(c3)--(c4)--(c5)--(c1);
  #1
  \foreach \v/\l in {c1/c_1,x1/x_1,x3/x_3,x4/x_4,x2/x_2}
    \node[vertex] at (\v) {$\l$};
  \foreach \v/\l in {c2/c_2,c3/c_3,c4/c_4,c5/c_5}
    \node[root] at (\v) {$\l$};
  \node[apexv] at (a) {$a$};}
\begin{scope}
  \apexexample{\draw[blueforest] (x3)--(c1) (x3)--(c2) (x4) to[out=190,in=100] (c1) (x4)--(x1) (x2)--(c3);}
  \node[panel,align=center] at (0.6,-2.45) {(a) $\widehat G_1=(V\cup\{a\},D_1\cup A)$};
\end{scope}
\begin{scope}[xshift=6.2cm]
  \apexexample{\draw[orangeforest] (x3)--(x1) (x1)--(c1) (x1)--(c2) (x4) to[out=-10,in=80] (c2) (x2)--(c5);}
  \node[panel,align=center] at (0.6,-2.45) {(b) $\widehat G_2=(V\cup\{a\},D_2\cup A)$};
\end{scope}
\end{tikzpicture}
\caption{The apex formulation in the proof of \Cref{thm:treewidth} on the
running example of \Cref{fig:m-graph}(b). A new vertex $a$ is added to
every bag and joined to every vertex by an apex edge (a few unselected
ones are shown dotted). A solution selects the cycle $C$ (black), the
forest edges $D_1$ and $D_2$, and the four apex edges $A$ to the common
roots $c_2,\ldots,c_5$ (purple). Each forest has four components, one
per root, so $D_i\cup A$ is connected. Conversely, if $D_i\cup A$ is
connected with $|A|\leq4$ and all apex edges ending on the cycle, then
$D_i$ has at most four components, all meeting the cycle. Hence, for
the two forests, the dynamic program only needs to track connectivity;
the cycle is handled separately through vertex degrees, connectivity,
and a closing step.}
\label{fig:treewidth-apex}
\end{figure}

\subparagraph*{The $M$-Witness Formulation.}
Add a new vertex $a$ to the graph and to every bag of the decomposition.
For every $v\in V$, introduce an auxiliary edge $av$, called an
\emph{apex edge}. This increases the width by at most one. We may assume
that the root and leaf bags contain only $a$. We seek pairwise disjoint
sets $C,D_1,D_2$ of original edges and a common set $A$ of apex edges such that
\begin{enumerate}
\item $C$ is the edge set of a connected 2-regular graph with at least
      four vertices;
\item every apex edge in $A$ ends at a vertex of $C$, and $|A|\leq4$;
\item $\widehat G_i=(V\cup\{a\},D_i\cup A)$ is connected for each
      $i\in\{1,2\}$.
\end{enumerate}
A connected 2-regular multigraph with at least four vertices is a cycle
of length at least four. If the conditions hold, every component of
$(V,D_i)$ meets $V(C)$ and there are at most $|A|\leq4$ components.
Conversely, since the pretest rejected, $G$ is an $M$-graph whenever it
admits an $M$-witness. By \Cref{lem:four-intervals}, its two witness
forests have four common roots on the cycle. Choose $D_i=E(F_i)$ and
let $A$ consist of the four apex edges to these roots. Both augmented
graphs are connected. \Cref{fig:treewidth-apex} illustrates this common
apex construction on the running example. They share the edges of $A$; only the original
edge sets $C,D_1,D_2$ are required to be pairwise disjoint. By
\Cref{lem:relaxed-witness}, these conditions are equivalent to the
existence of an $M$-witness. We call a tuple
$(C,D_1,D_2,A)$ satisfying them a \emph{solution}.

\subparagraph*{The $M$-Witness Program.}
At a bag $B$, let $B^\circ=B\setminus\{a\}$. A partial solution at $x$ determines which
original vertices of $G_x$ are \emph{cycle vertices}, assigns each
original edge of $G_x$ to at most one of $C,D_1,D_2$, and selects apex
edges of $G_x$ for $A$. Its \emph{partial cycle graph} consists of
the cycle vertices and the edges assigned to $C$; its \emph{partial
augmented graphs} are $\widehat G_{x,i}=(V(G_x),(D_i\cup A)\cap E(G_x))$.
The state of a partial solution consists of its \emph{signature}

\[
 (Z,\mathsf{cycdeg},\mathsf{apex},\mathsf{forgot},\mathsf{closed})
\]
and three partitions. Here $Z\subseteq B^\circ$ is the set of
\emph{active} vertices, that is, the cycle vertices in the bag;
$\mathsf{cycdeg}\colon Z\to\{0,1,2\}$ records their degrees in the partial cycle
graph; $\mathsf{apex}\in\{0,\ldots,4\}$ is the number of selected apex edges
at $a$; $\mathsf{forgot}\in\{0,\ldots,4\}$ is the number of forgotten cycle
vertices, truncated at four; and $\mathsf{closed}\in\{0,1\}$ indicates that the
cycle has been closed. The partitions are a partition $\pi_C$ of $Z$
and, for each $i$, a partition $\pi_i$ of $B$, describing the
connectivity of the partial cycle graph and of $\widehat G_{x,i}$ on
the bag. Each vertex of $B^\circ$ is either inactive or active with one
of three degrees, so there are at most $4^{|B^\circ|}\cdot5^2\cdot2$
signatures per bag. The transitions are as follows.
\begin{itemize}
\item \emph{Leaf}: all counters are zero, $Z=\varnothing$, $\mathsf{closed}=0$,
      and $\pi_1,\pi_2$ are discrete.
\item \emph{Introduce an original vertex $v$}: insert $\{v\}$ into
      $\pi_1$ and $\pi_2$. If $\mathsf{closed}=0$, optionally make $v$ a cycle
      vertex: add it to $Z$ with $\mathsf{cycdeg}(v)=0$ and insert $\{v\}$ into
      $\pi_C$.
\item \emph{Introduce an original edge $uv$}: assign it to at most one
      of $C,D_1,D_2$. Assigning it to $C$ requires $u,v\in Z$ and
      $\mathsf{cycdeg}(u),\mathsf{cycdeg}(v)\leq1$; it increments both degrees and merges
      the blocks of $u,v$ in $\pi_C$. Assigning it to $D_i$ merges the
      blocks of $u,v$ in $\pi_i$.
\item \emph{Introduce an apex edge $av$}: optionally select it. This
      requires $v\in Z$ and $\mathsf{apex}\leq3$; it increments $\mathsf{apex}$ and
      merges the blocks of $v$ and $a$ in both $\pi_1$ and $\pi_2$.
\item \emph{Forget $v$}: reject if $\{v\}$ is a block of $\pi_1$ or
      $\pi_2$, and otherwise remove $v$ from both. If $v\in Z$, require
      $\mathsf{cycdeg}(v)=2$ and set $\mathsf{forgot}:=\min\{4,\mathsf{forgot}+1\}$. If moreover $\{v\}$ is
      not a block of $\pi_C$, remove $v$ from $Z$ and $\pi_C$.
      Otherwise, the component of $v$ in the partial cycle graph is
      complete; require $Z=\{v\}$ and $\mathsf{closed}=0$, and \emph{close the
      cycle} by setting $\mathsf{closed}:=1$ and $Z:=\varnothing$.
\item \emph{Join}: both children must have the same set $Z$. Add the
      degrees $\mathsf{cycdeg}$ and the counters $\mathsf{apex}$, rejecting values
      above two and four, respectively; add the counters $\mathsf{forgot}$,
      truncating at four; set
      $\mathsf{closed}=\mathsf{closed}_{\mathrm{left}}\lor\mathsf{closed}_{\mathrm{right}}$,
      where the subscripts indicate the left and right child; and join the
      partitions. Reject if both child flags are one, or if either child
      has $\mathsf{closed}=1$ and the other has $\mathsf{forgot}>0$,
      since the cycle would then have two components.
\item \emph{Root} (bag $\{a\}$): accept if $\mathsf{closed}=1$ and $\mathsf{forgot}=4$.
\end{itemize}

These transitions ensure the following invariant for every partial
solution at a node $x$:
\begin{itemize}
\item[(I)] every component of $\widehat G_{x,i}$ contains a vertex of
      $B$, for $i\in\{1,2\}$; if $\mathsf{closed}=0$, every component of the
      partial cycle graph contains a vertex of $Z$; and if $\mathsf{closed}=1$,
      then $Z=\varnothing$ and the partial cycle graph is a cycle.
\end{itemize}
The first part is exactly the rejection rule for singleton blocks of
$\pi_i$. For the second part, a component of the partial cycle graph
loses its last active vertex only when the cycle is closed; at this
point $Z=\{v\}$, so it is the only component, and all its vertices were
forgotten with degree two. Afterwards no vertex becomes active, and by
the join rules no other branch contains cycle vertices.
Hence every accepted run yields a solution: $C$ is connected, every
cycle vertex has degree two, $\mathsf{forgot}=4$ certifies at least four cycle
vertices, and at the root every component of $\widehat G_i$ contains
$a$.

Conversely, fix a solution $(C,D_1,D_2,A)$ with cycle vertices
$V(C)$. We show that its restriction to $G_x$ is never rejected.
Degrees and the apex counter only grow towards their final values, since
every edge is introduced exactly once. When a vertex $v$ is
forgotten, all its edges are already introduced, so a cycle vertex has
degree two, and a singleton block $\{v\}$ of $\pi_i$ would make the
component of $v$ in $\widehat G_{x,i}$ a component of $\widehat G_i$
avoiding $a$. For the cycle, observe that the nodes whose bags meet
$V(C)$ form a subtree, since $C$ is connected. Let $v$ be the cycle
vertex whose forget node $x_C$ is the parent of the top node of this
subtree; every other cycle vertex is forgotten strictly below $x_C$.
If at the forget node of some cycle vertex $u$ the block $\{u\}$ of
$\pi_C$ is a singleton, then all other vertices of the component $K$ of
$u$ in the partial cycle graph are already forgotten. Hence all
$C$-edges at vertices of $K$ are introduced, and $K=C$ since $C$ is
connected. As $v\in K$ is not forgotten below $x_C$, this happens only
for $u=v$. At $x_C$, it does happen: $K=C$, and $Z=\{v\}$ because all
other cycle vertices are forgotten. Thus the cycle is closed exactly
once, namely at $x_C$, with $\mathsf{closed}=0$ before. No cycle vertex is
introduced above $x_C$, since all of them are forgotten there.
At a join node, both children have $Z=V(C)\cap B$. If the subtree of
one child contains $x_C$, then all cycle vertices are forgotten in that
subtree, so the other child has $\mathsf{forgot}=0$ and $\mathsf{closed}=0$. Finally, the root
has $\mathsf{closed}=1$ and $\mathsf{forgot}=\min\{4,|V(C)|\}=4$.

\subparagraph*{Compressing the Partitions.}
It remains to bound the partition tables. They admit single-exponential
compression despite the three simultaneous connectivity conditions.
For partitions $\pi,\pi'$ of a set $U$, let $\pi\vee\pi'$ denote their
finest common coarsening: two elements are in the same block if a
sequence of identifications from $\pi$ or $\pi'$ connects them.
For a nonempty set $U$, let $\Phi_U$ be the matrix over $\mathbb{F}_2$
with rows and columns indexed by the partitions of $U$, where
$\Phi_U[\pi,\pi']=1$ if and only if $\pi\vee\pi'=\{U\}$.
Fix an element $u_0\in U$, and call a set $\kappa\subseteq U$ with
$u_0\in\kappa$ a \emph{cut}. Let $\Psi_U$ be the matrix with rows
indexed by partitions and columns indexed by cuts, where
$\Psi_U[\pi,\kappa]=1$ if and only if every block of $\pi$ is contained
in $\kappa$ or in $U\setminus\kappa$. Then $\Phi_U=\Psi_U\Psi_U^{\mathsf T}$
over $\mathbb{F}_2$: the number of cuts respected by both $\pi$ and
$\pi'$ is $2^{h-1}$, where $h$ is the number of blocks of $\pi\vee\pi'$,
and this number is odd precisely when $h=1$~\cite{BodlaenderEtAl2015}.
For $U=\varnothing$, set $\Phi_U=\Psi_U=(1)$, so the same
factorization holds.

An \emph{extension} $\Gamma$ of a node $x$ fixes the choices for all
vertices and edges outside $G_x$. It induces a partition $\pi_i^\Gamma$ of
$B$ by the connectivity of its part of $\widehat G_i$, and a
partition $\pi_C^\Gamma$ of $Z$ by the connectivity of its part of the cycle
graph. Let $P$ be a partial solution at $x$ with partitions
$\pi_C,\pi_1,\pi_2$. By~(I), the components of $\widehat G_i$ meeting
$B$ correspond to the blocks of $\pi_i\vee\pi_i^\Gamma$, and all other
components lie entirely in $\Gamma$. If $\mathsf{closed}=0$, the same holds for the
cycle graph with $Z$ in place of $B$; if $\mathsf{closed}=1$, then $Z=\varnothing$
and the partial cycle graph of $P$ is already a cycle. All other
requirements, namely the degrees, the apex counter, the number of
cycle vertices, and the admissibility of the parts of $\Gamma$ that avoid
the bag, depend only on the signature of $P$ and on $\Gamma$. Hence $P$
together with $\Gamma$ is a solution if and only if a condition on the
signature of $P$ and on $\Gamma$ holds and
\begin{equation}\label{eq:rank-test}
 \Phi_Z[\pi_C,\pi_C^\Gamma]\cdot\Phi_B[\pi_1,\pi_1^\Gamma]\cdot\Phi_B[\pi_2,\pi_2^\Gamma]=1.
\end{equation}
The left-hand side is an entry of the Kronecker product
$\Phi_Z\otimes\Phi_B\otimes\Phi_B$. It factors as
$\Psi\Psi^{\mathsf T}$ with $\Psi=\Psi_Z\otimes\Psi_B\otimes\Psi_B$,
which has at most
$2^{\max\{0,|Z|-1\}}\cdot2^{|B^\circ|}\cdot2^{|B^\circ|}
\leq2^{3b}$ columns; the factor $\Psi_Z$ has one column when
$Z=\varnothing$.

After computing the candidate states of a node from the tables of its
children, we compress: for each fixed signature, retain a row basis of
the matrix $\Psi$ restricted to the rows of the candidate partition
triples, keeping one witnessing partial solution for each retained row.

\begin{claim}
For every node $x$ and every extension $\Gamma$ of $x$, if some partial
solution $P$ at $x$ together with $\Gamma$ forms a solution, then so does
some partial solution retained at $x$ with the same signature as $P$, in
particular with the same set $Z$.
\end{claim}
\begin{claimproof}
We first show that the claim holds for the candidates at $x$: if a
candidate $P$ together with $\Gamma$ forms a solution, then the row of $P$
is a sum of retained rows with the same signature. Multiplying by the
row of $\Psi$ indexed by the partitions of $\Gamma$ shows that some retained
row satisfies~\eqref{eq:rank-test}. The corresponding partial solution
has the signature of $P$, so it forms a solution together with $\Gamma$.

We now proceed by induction from the leaves upwards. At a leaf there is
only one partial solution. Let $x$ be an introduce or forget node with
child $y$, and let $P$ together with $\Gamma$ form a solution $W$. Let
$P_y$ be the restriction of $P$ to $G_y$, and let $\Gamma_y$ consist of
$\Gamma$ and the choice made at $x$. By the induction hypothesis, some
partial solution $Q_y$ retained at $y$ with the signature of $P_y$ forms
a solution $W'$ together with $\Gamma_y$. Extending $Q_y$ by the choice
made at $x$ gives the restriction $Q$ of $W'$ to $G_x$. It is not
rejected, as shown above, so it is a candidate at $x$. Moreover, $Q$ has
the signature of $P$, since for a transition that does not reject, the
new signature depends only on the old signature and on the choice made
at $x$. For introduce nodes, this is immediate from the rules; in
particular, $Z$ changes only if the introduced vertex is made a cycle
vertex. At the node forgetting $v$, the set $Z$ becomes $Z\setminus\{v\}$
in either case: if $v\in Z$ and $\{v\}$ is a block of $\pi_C$, then
$Z=\{v\}$ is required, and closing the cycle sets $Z:=\varnothing$.
Whether the cycle is closed also depends only on the signature: if
$Z=\{v\}$, then $\{v\}$ is the only block of $\pi_C$; otherwise, a
singleton block $\{v\}$ of $\pi_C$ leads to rejection. Hence the first
paragraph applies to $Q$.

Finally, let $x$ be a join node with children
$y_1,y_2$, and let $P$ together with $\Gamma$ form a solution, where $P$
combines partial solutions $P_1$ at $y_1$ and $P_2$ at $y_2$. Apply the
induction hypothesis to $y_1$, treating $\Gamma$ and the part of $P_2$
outside $G_{y_1}$ as the extension; this yields a retained $Q_1$ with
the signature of $P_1$. Then apply it to $y_2$ with $\Gamma$ and the part
of $Q_1$ outside $G_{y_2}$ as the extension; this yields a retained
$Q_2$ with the signature of $P_2$. Since $Z_{Q_1}=Z_P=Z_{Q_2}$, where the
subscript indicates the partial solution, $Q_1$ and $Q_2$ agree on the
cycle vertices of the bag, and the two states are joined. The
combination of $Q_1$ and $Q_2$ together with $\Gamma$ forms a solution,
so it is not rejected and is a candidate at $x$. Its signature is that
of $P$, since the signature computed at a join node depends only on the
signatures of the children. Again, the first paragraph applies.
\end{claimproof}

Applying the claim at the root with the empty extension shows that the
compressed program accepts whenever a solution exists. Compression only
discards partial solutions, so every acceptance still yields a
solution.
For the $T$-graph pretest, the same argument uses only the two forest
partitions, giving the bound $4^{|B|}$ stated above.

\subparagraph*{Running Time.}
Recall that $b$ denotes the maximum bag size. There are at most
$4^{b}\cdot5^2\cdot2$ signatures per bag, and each
retains at most $2^{3b}$ partition triples after compression. Each
transition processes at most $2^{\Oh(b)}$ candidate states, each in
$b^{\Oh(1)}$ time; at a join node, this includes all pairs of retained
states. Gaussian elimination over $\mathbb{F}_2$ on a matrix with
$2^{\Oh(b)}$ rows and at most $2^{3b}$ columns takes $2^{\Oh(b)}$ time.
Hence, each node takes $2^{\Oh(t)}$ time, and both programs take
$2^{\Oh(t)}\cdot tn=2^{\Oh(t)}\cdot n$ time. Predecessor pointers
recover $C,D_1,D_2$. Together with \Cref{lem:relaxed-witness} and
\Cref{lem:construct-labeling}, they give a temporally connecting labeling
in $\Oh(n+m)=\Oh(tn)$ time. The same bound for a $T$-graph follows from
the witness returned by the preliminary test.
\end{proof}

The single-exponential dependence on treewidth is optimal under ETH:
by \Cref{cor:planar-eth} (\Cref{sec:degree-complexity}), no
$2^{o(\tw)}n^{\Oh(1)}$-time algorithm exists, even on planar simple
graphs of maximum degree five.

For planar inputs, the treewidth dependence immediately gives a
subexponential algorithm in the number of vertices.
\begin{corollary}\label{cor:planar-subexponential}
\Gossip can be solved deterministically in
$2^{\Oh(\sqrt n)}$ time on planar multigraphs.
\end{corollary}
\begin{proof}
The support of a planar multigraph on $n$ vertices has treewidth
$\Oh(\sqrt n)$, which follows from the planar separator
theorem~\cite{LiptonTarjan1979}. The claim follows from
\Cref{thm:treewidth}, since $2^{\Oh(\sqrt n)}\cdot n=2^{\Oh(\sqrt n)}$.
\end{proof}

The dependence on $n$ is optimal under ETH:
\Cref{cor:planar-eth} (\Cref{sec:degree-complexity}) rules out a
$2^{o(\sqrt n)}n^{\Oh(1)}$-time algorithm, even on planar simple graphs
of maximum degree five.

\subsection{Applications of the Tree Contraction Rule}
\label{sec:structural-applications}

We now apply the \TreeContractionRule to parameters that permit dense
inputs of large treewidth. We begin with independence number, where
the colorability of the reduced graph bounds its total number of
vertices and thus gives a small kernel. For tree-independence number
and deletion distance to chordal graphs, the same colorability
argument bounds bag sizes instead, allowing us to apply the treewidth
algorithm. Finally, we handle clique-width
by analyzing cuts of a construction expression.

Apply the \TreeContractionRule exhaustively, obtaining a
clean, reduced multigraph $H$ in $\Oh(n^2)$ time. This preserves the answer
by \Cref{lem:rule-safe,lem:rule-time}. For the treewidth-based applications,
apply \Cref{thm:treewidth}
to $H$. If $H$ is a yes-instance, lift the witness of $H$ found by the
dynamic program through all contracted maximal tight sets as in the
proof of \Cref{lem:rule-safe}, using the three edge-disjoint spanning
trees of each set provided by the pebble game. By
\Cref{lem:rule-time}, this gives a temporally connecting labeling of
$G$ in $\Oh(n^2)$ additional time. The algorithm does not need a
deletion set or a decomposition witnessing any parameter. Hence each
treewidth-based application only has to bound $\tw(H)$ in terms of a parameter of the
original input $G$.

We use three consequences of \Cref{lem:reduced-edges}. Every induced
subgraph of the support of $H$ on $\ell\geq2$ vertices has at most $3\ell-4$
edges and hence a vertex of degree at most five. Thus the support is
5-degenerate and therefore 6-colorable. Moreover,
it contains neither $K_6$ nor $K_{6,5}$ as a subgraph, since these
would have $15>3\cdot6-4$ and $30>3\cdot11-4$ edges, respectively.

We also use the following observation to transfer deletion sets.
Each vertex of $H$ represents a connected \emph{branch set} in $G$.
For $S\subseteq V(G)$, let $S_H$ consist of the vertices whose branch
sets meet $S$. Then $|S_H|\leq|S|$, and the support of $H-S_H$ is an
\emph{induced minor} of the support of $G-S$: it can be obtained by
vertex deletions and edge contractions, without deleting edges.
Indeed, delete all branch sets meeting $S$, then contract the remaining
ones. Adding $S_H$ to every bag of a decomposition of $H-S_H$ increases
its width by at most $|S|$. If $H-S_H$ is empty, use the single bag $S_H$.

\subparagraph*{Independence Number.}
For the independence number itself, preprocessing already yields a
small instance.

\begin{corollary}\label{cor:independence-kernel}
Parameterized by the independence number $\alpha(G)$, \Gossip admits a
kernel with at most $6\alpha(G)$ vertices and at most
$\max\{0,18\alpha(G)-4\}$ edges, computable in $\Oh(n^2)$ time.
\end{corollary}
\begin{proof}
The clean, reduced multigraph $H$ is an equivalent instance. Since its
support is 6-colorable, $|V(H)|\leq6\alpha(H)$. Moreover,
$\alpha(H)\leq\alpha(G)$: an independent set of $H$ lifts to one of $G$
by choosing one vertex from each of its branch sets. If $|V(H)|\geq2$,
\Cref{lem:reduced-edges} gives
$|E(H)|\leq3|V(H)|-4\leq18\alpha(G)-4$; otherwise $H$ has no edges.
\end{proof}

\subparagraph*{Tree-Independence Number and Deletion to Chordal Graphs.}
Tree-independence number does not increase under induced
minors~\cite{DallardMS24}. In a reduced graph, a bound on the independence
number of each bag also bounds its size, because the graph is
6-colorable. The chordal graphs are exactly the graphs of
tree-independence number at most one: they are the graphs admitting a
tree decomposition whose bags are cliques~\cite{DallardMS24}.

\begin{corollary}\label{cor:tree-independence}\label{cor:chordal-deletion}
Let $S\subseteq V(G)$ and $a\geq1$ satisfy $\treealpha(G-S)\leq a$.
Every reduced graph $H$ obtained from $G$ by the \TreeContractionRule satisfies
\[
 \tw(H)\leq |S|+6a-1,
\]
and $\tw(H)\leq|S|+4$ if $a=1$. Consequently, \Gossip can be solved
deterministically in $\Oh(n^2)+2^{\Oh(|S|+a)}\cdot n$ time.
\end{corollary}
\begin{proof}
The support of $H-S_H$ is an induced minor of the support of $G-S$ and
hence has tree-independence number at most $a$.
Consider a witnessing tree decomposition and a proper 6-coloring
of this support. Each color occurs at most $a$ times in any bag,
since each color class is independent. Hence every bag has at most
$6a$ vertices. If $a=1$, every bag is a clique and thus has at most
five vertices, since the support contains no $K_6$. Adding $S_H$ gives
the claimed bounds. For the algorithm, apply \Cref{thm:treewidth}.
Handle the empty graph directly.
\end{proof}

Taking $S=\varnothing$ gives the algorithm parameterized by
tree-independence number. Taking $S$ to be a chordal vertex deletion
set and $a=1$ gives running time $\Oh(n^2)+2^{\Oh(d)}\cdot n$, where
$d$ is the chordal vertex deletion number. Since cluster, interval,
split, and block graphs are chordal, this also applies to deletion
distance to these classes.

For example, a \emph{circular-arc graph} is the intersection graph of
arcs on a circle and has tree-independence number at most two.
To see this, choose a point of the circle. The arcs containing that
point form a clique; the other arcs form an interval graph. Add the
clique to every bag of a clique-bag decomposition of the interval
graph. Each resulting bag is the union of two cliques.
Consequently, if deleting $d$ vertices leaves a circular-arc support,
then $\tw(H)\leq d+11$, giving an $\Oh(n^2)+2^{\Oh(d)}\cdot n$-time
algorithm. More generally, deletion to any fixed class of bounded
tree-independence number has the same single-exponential dependence.

\subparagraph*{Clique-Width.}
We now bound the treewidth of the reduced graph in terms of the
clique-width of the input. The exclusion of $K_6$ bounds the number
of branch sets crossing a cut of a construction expression. For the
remaining branch sets, each construction label gives a complete
bipartite graph across the cut, whose edges can be covered by a
constant number of vertices because $K_{6,5}$ is excluded.

We use maximum matching width to turn these cut bounds into a
treewidth bound. A \emph{vertex branch decomposition} of a graph $J$
is a tree of maximum degree three whose leaves are in bijection with
$V(J)$. Each tree edge partitions the leaves and hence $V(J)$ into
two sides. Its width is the maximum, over these partitions, of the
size of a maximum matching consisting of edges of $J$ between the two sides.
The \emph{maximum matching width} $\operatorname{mmw}(J)$ is the minimum
width of such a decomposition. For graphs with at least one edge,
$\tw(J)+1\leq3\operatorname{mmw}(J)$~\cite{JeongST18}.

\begin{figure}[t]
\centering
\begin{tikzpicture}[
  x=1cm,y=1cm,
  v/.style={circle,draw=black!80,fill=white,inner sep=0pt,minimum size=2.4mm},
  li/.style={v,fill=auxpurple!75,draw=auxpurple!90!black},
  bs/.style={draw=black!55,fill=black!5,line width=.6pt,rounded corners=7pt},
  sel/.style={draw=auxgreen!80!black,fill=auxgreen!18,line width=1pt,rounded corners=7pt},
  cutline/.style={black!55,line width=.8pt,dash pattern=on 4pt off 2.5pt},
  e/.style={black,line width=.8pt},
  cross/.style={witnessorange,line width=1pt},
  lbl/.style={font=\footnotesize},
]
\begin{scope}
  \draw[cutline] (0,-1.75) -- (0,1.75);
  \node[lbl,anchor=south east] at (-0.1,1.75) {$X$};
  \node[lbl,anchor=south west] at (0.1,1.75) {$V(Q)\setminus X$};
  \draw[bs] (-1.75,0.35) rectangle (1.15,1.25);
  \draw[bs] (-1.15,-1.25) rectangle (1.75,-0.35);
  \coordinate (w1) at (-1.35,0.8); \coordinate (x1) at (-0.55,0.8); \coordinate (y1) at (0.6,0.8);
  \coordinate (x2) at (-0.6,-0.8); \coordinate (y2) at (0.55,-0.8); \coordinate (w2) at (1.35,-0.8);
  \draw[e] (w1)--(x1)--(y1) (x2)--(y2)--(w2);
  \draw[cross] (x1)--(y2) (x2)--(y1);
  \node[li] at (x1) {}; \node[li] at (x2) {};
  \foreach \u in {w1,y1,y2,w2} \node[v] at (\u) {};
  \draw[cutline] (0,-1.75) -- (0,1.75);
  \node[lbl,anchor=east] at (-1.8,0.8) {$U_1$};
  \node[lbl,anchor=west] at (1.8,-0.8) {$U_2$};
  \node[panel] at (0,-2.3) {(a)};
\end{scope}
\begin{scope}[xshift=6.3cm]
  \draw[cutline] (0,-1.75) -- (0,1.75);
  \node[lbl,anchor=south east] at (-0.1,1.75) {$X$};
  \node[lbl,anchor=south west] at (0.1,1.75) {$V(Q)\setminus X$};
  \foreach \k in {1,...,6}{
    \pgfmathsetmacro{\yy}{1.375-0.55*(\k-1)}
    \draw[bs,rounded corners=4pt] (-1.6,\yy-0.21) rectangle (-0.6,\yy+0.21);
    \coordinate (a\k) at (-0.82,\yy);
    \coordinate (b\k) at (-1.35,\yy);
  }
  \draw[sel] (0.6,0.35) rectangle (1.6,1.05);
  \draw[sel] (0.6,-1.05) rectangle (1.6,-0.35);
  \coordinate (p1) at (0.85,0.7); \coordinate (q1) at (1.35,0.7);
  \coordinate (p2) at (0.85,-0.7); \coordinate (q2) at (1.35,-0.7);
  \foreach \k in {1,...,6} \foreach \p in {p1,p2} \draw[black!45,line width=.5pt] (a\k)--(\p);
  \draw[e] (p1)--(q1) (p2)--(q2);
  \foreach \k in {1,...,6}{ \draw[e] (b\k)--(a\k); \node[v] at (b\k) {}; \node[li] at (a\k) {}; }
  \foreach \u in {p1,q1,p2,q2} \node[v] at (\u) {};
  \draw[cutline] (0,-1.75) -- (0,1.75);
  \draw[decorate,decoration={brace,amplitude=4pt,mirror},black!60] (-1.75,1.6) -- (-1.75,-1.6) node[midway,left=5pt,lbl,text=black] {$A_i$};
  \draw[decorate,decoration={brace,amplitude=4pt},black!60] (1.75,1.1) -- (1.75,-1.1) node[midway,right=5pt,lbl,text=black] {$B_i$};
  \node[panel] at (0,-2.3) {(b)};
\end{scope}
\begin{scope}[xshift=2.75cm,yshift=-2.3cm]
  \node[li] (L) at (0,0) {}; \node[lbl,anchor=west] at (0.15,0) {label $i$};
\end{scope}
\end{tikzpicture}
\caption{The cut argument in the proof of
\Cref{lem:clique-induced-minor-treewidth}. Boxes are branch sets in $Q$
representing vertices of an induced minor $J$. The dashed line separates
the vertex set $X$ of a subexpression from its complement. Vertices
with label $i$ (purple) at this subexpression have the same neighbors
outside $X$. (a) Two mixed branch sets $U_1,U_2$ meet both sides.
Their chosen crossing edges (black) have endpoints with label $i$ in
$X$. The orange edges therefore exist, making $U_1$ and $U_2$
adjacent in $J$. (b) The set $A_i$ consists of branch sets wholly
inside $X$ containing a vertex with label $i$; the set $B_i$ consists
of branch sets wholly outside $X$ containing a neighbor of such a
vertex. Every branch set in $A_i$ is adjacent to every branch set
in $B_i$. Since $|A_i|\geq6$ and $J$ contains no
$K_{6,5}$, we have $|B_i|\leq4$, and the branch sets of $B_i$ (green)
cover all edges between $A_i$ and $B_i$.}
\label{fig:clique-width-cut}
\end{figure}
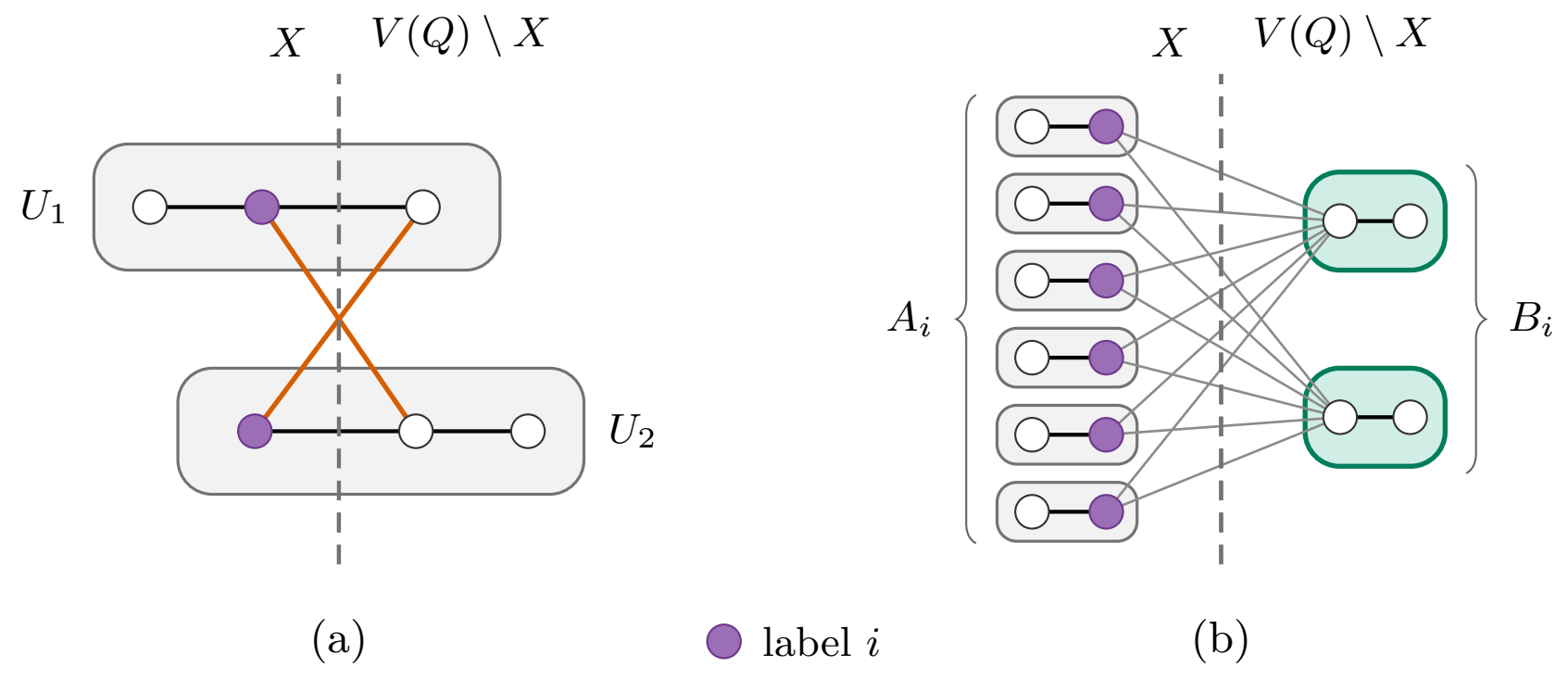

\begin{lemma}\label{lem:clique-induced-minor-treewidth}
Let $Q$ be a nonempty simple graph with clique-width $k$.
Every induced minor $J$ of $Q$ containing neither $K_6$ nor $K_{6,5}$
as a subgraph satisfies $\tw(J)\leq30k-1$.
\end{lemma}
\begin{proof}
If $J$ has no edges, its treewidth is at most zero and the bound holds.
Otherwise, fix a construction
expression for $Q$ using $k$ labels and an induced-minor model of $J$
in $Q$. Thus each vertex of $J$ is represented by a connected branch
set, the branch sets are pairwise disjoint, and two vertices of $J$
are adjacent exactly when an edge of $Q$ joins their branch sets.

Suppress the unary operations in the expression tree, leaving a
binary tree whose leaves are the vertices of $Q$. Choose one
representative vertex in each branch set, take the minimal subtree
containing these representatives, and suppress its degree-two vertices.
Identifying each representative with its vertex of $J$ gives a vertex
branch decomposition of $J$. Every edge cut of this decomposition is
inherited from an edge of the original binary tree. One side $X$ of
that original cut is the vertex set of a subexpression. Vertices
with the same label at this subexpression have identical neighbors in
$V(Q)\setminus X$, since all later operations treat them identically.

Fix such a cut; \Cref{fig:clique-width-cut} illustrates the following
two steps. Call a branch set \emph{mixed} if it meets both $X$
and $V(Q)\setminus X$. By connectedness, each mixed branch set
contains an edge crossing the cut. Choose one such edge and assign
the branch set the label of its endpoint in $X$. Any two mixed
branch sets assigned the same label are adjacent in $J$: the inside
endpoint of either chosen edge is adjacent to the outside endpoint
of the other. Hence each label is assigned at most five mixed branch
sets, since $J$ contains no $K_6$. There are at most $5k$ mixed
branch sets in total.

For each label $i$, let $A_i$ be the vertices of $J$ whose branch
sets lie wholly in $X$ and contain a vertex with label $i$. Let $B_i$
be the vertices of $J$ whose branch sets lie wholly outside $X$ and
contain a neighbor of a vertex with label $i$ in $X$. The uniform
external neighborhoods of label $i$ imply that every vertex of
$A_i$ is adjacent to every vertex of $B_i$. If $|A_i|\leq5$, select
all vertices of $A_i$. Otherwise, $|B_i|\leq4$, since $J$ contains
no $K_{6,5}$, and we select all vertices of $B_i$. This selects at
most five vertices per label. Every edge between branch sets wholly
on opposite sides belongs to some $A_i\times B_i$, so the selected
vertices cover all these edges.

Add all vertices represented by mixed branch sets to the cover. The
resulting set has size at most $10k$ and covers every edge of $J$ crossing the displayed
partition: a branch set wholly on one side has its representative on
that same side. Thus every crossing matching has size at most $10k$.
The branch decomposition has width at most $10k$, and
$\tw(J)+1\leq3\operatorname{mmw}(J)\leq30k$.
\end{proof}

The reduced support is an induced minor of the input support and
excludes both forbidden subgraphs by \Cref{lem:reduced-edges}.
We therefore obtain the following algorithmic consequence.

\begin{corollary}\label{cor:clique-width}
\Gossip can be solved deterministically in
$\Oh(n^2)+2^{\Oh(\cw(G))}\cdot n$ time.
\end{corollary}
\begin{proof}
Apply the \TreeContractionRule exhaustively to obtain $H$.
By \Cref{lem:clique-induced-minor-treewidth},
$\tw(H)\leq30\cw(G)-1$. The claim follows from
\Cref{lem:rule-time,thm:treewidth}. The clique-width expression is
used only to prove the treewidth bound; the algorithm does not need
one as input.
\end{proof}

These applications require more than sparsity alone. Note that every
planar simple graph is reduced, since $|E[X]|\leq\max\{1,3|X|-6\}\leq3|X|-4$
for all $X$ with $|X|\geq2$. Grids are planar, bipartite, and perfect,
and they have unbounded treewidth. Thus vertex deletion distance to planar,
bipartite, or perfect graphs cannot bound the treewidth after
the \TreeContractionRule, even when that distance is zero. Moreover,
\Gossip remains NP-hard on planar simple graphs
(\Cref{thm:deg5-hard}), which are already reduced.
\section{Parameterization by the Excess}
\label{sec:above-bound}

In this section, we study how to find an $M$-witness whose cycle has
length at most a given bound $k$. We give a deterministic XP algorithm
and a randomized FPT algorithm parameterized by $k$. Together with the
polynomial-time recognition of $T$-graphs, these algorithms also yield
algorithms parameterized by the excess $\delta=m-(2n-4)$.

The connection follows from a simple counting bound. The two forests
of an $M$-witness together use exactly $2n-8$ edges, leaving at most
$\delta+4$ edges for its cycle. We can therefore apply the bounded-cycle
algorithms with $k=\delta+4$, as formalized by the following lemma.

\begin{lemma}\label{lem:m-witness-size-bound}
If an $n$-vertex multigraph $G$ admits an $M$-witness, then it is
connected, $n\geq4$, and $m\geq2n-4$. Moreover, if
$m=2n-4+\delta$, every $M$-witness $(C,F_1,F_2)$ satisfies
\[
    |E(C)|\leq \delta+4.
\]
\end{lemma}

\begin{proof}
Every component of $F_1$ meets $C$, so $C\cup F_1$ is connected
and spans $V$. Hence $G$ is connected. Since $C$ has length at least
four, $n\geq4$. Each spanning forest $F_i$ has $n-4$ edges.
Pairwise edge-disjointness therefore gives
\[
    m\geq |E(C)|+|E(F_1)|+|E(F_2)|
      =|E(C)|+2n-8\geq2n-4.
\]
Substituting $m=2n-4+\delta$ yields $|E(C)|\leq \delta+4$.
\end{proof}

\subsection{A Deterministic XP Algorithm}

Our deterministic algorithm enumerates all cycles of length at most $k$
in the support and tests whether any of them extends to an $M$-witness
using \Cref{lem:cycle-extension}. The number of candidates is polynomial
in the input size for fixed $k$, with its dependence on $k$ controlled
by the maximum degree $\Delta$ of the support.

\begin{proposition}\label{prop:above-xp}
Given a multigraph $G$ that is not a $T$-graph and an integer $k\geq4$,
one can decide whether $G$ admits an $M$-witness whose cycle has length
at most $k$ deterministically in
$\Oh(k\cdot m\cdot\Delta^{k-2}\cdot n^2)$ time. Whenever the
algorithm accepts, it returns such a witness.
\end{proposition}
\begin{proof}
Reject if $n<4$, if $G$ is
disconnected, or if $m<2n-4$, by \Cref{lem:m-witness-size-bound}.
We may assume $\Delta\geq2$, since otherwise the support is a forest
and contains no witness cycle. Let $\ell_{\max}=\min\{n,k\}$. By
\Cref{lem:representative-cycle}, it suffices to consider cycles of the
support, using one fixed representative of every parallel class.
We enumerate all cycles of the support with lengths $\ell$ from four to
$\ell_{\max}$ by a depth-first search from each of the at most $2m$ ordered pairs
of adjacent vertices. There are at most $2m\Delta^{\ell-2}$ such vertex
sequences of length $\ell$, and each is generated in $\Oh(\ell)$ time.
For each cycle, we apply \Cref{lem:cycle-extension}, which takes
$\Oh(n^2+\ell^2n)\subseteq\Oh(\ell n^2)$ time since $\ell\leq n$.
This test returns the two forests if the cycle extends to an
$M$-witness. Hence the total time is
\[
\sum_{\ell=4}^{\ell_{\max}}2m\Delta^{\ell-2}\cdot\Oh(\ell n^2)
\subseteq\Oh(\ell_{\max}\cdot m\cdot\Delta^{\ell_{\max}-2}\cdot n^2)
\subseteq\Oh(k\cdot m\cdot\Delta^{k-2}\cdot n^2).
\]
\end{proof}

For the excess parameter, first test for a $T$-graph in $\Oh(n^2)$
time using \Cref{lem:t-graph-test}, handling $n<4$ directly.
Otherwise, by \Cref{thm:characterization,lem:m-witness-size-bound},
the graph is a yes-instance of \Gossip if and only if it has an
$M$-witness with cycle length at most $\delta+4$.
Applying \Cref{prop:above-xp} with $k=\delta+4$ therefore gives a
deterministic algorithm with running time
$\Oh((\delta+4)\cdot m\cdot\Delta^{\delta+2}\cdot n^2)$, and hence
$\Oh((\delta+4)\cdot m\cdot n^{\delta+4})$.

\subsection{A Randomized FPT Algorithm}
\label{sec:excess-fpt}

We next give a randomized FPT algorithm for the same bounded-cycle
problem. As before, the input is assumed not to be a $T$-graph, since
these graphs can be recognized in polynomial time.

\begin{theorem}\label{thm:short-m-witness}
Given a multigraph $G$ that is not a $T$-graph and an integer $k\geq4$,
one can decide whether $G$ admits an $M$-witness whose cycle has length
at most $k$ in randomized $2^{\Oh(k)}m^3$ time. The algorithm never accepts a
no-instance and accepts every yes-instance with probability at least
$1-1/n$. Whenever it accepts, it returns such a witness.
\end{theorem}

For the excess parameter, first test for a $T$-witness in $\Oh(n^2)$
time. Otherwise, apply \Cref{thm:short-m-witness} with $k=\delta+4$.
Since $m=2n-4+\delta\leq(\delta+2)n$, this gives a randomized
$2^{\Oh(\delta)}n^3$-time algorithm for \Gossip with the same one-sided
error guarantee. An accepting run returns a witness, which
\Cref{lem:construct-labeling} converts into a temporally connecting
labeling.

\begin{figure}[t]
\centering
\begin{tikzpicture}[
    every node/.style={font=\footnotesize},
    vertex/.style={circle,draw=black!75,fill=white,inner sep=0pt,minimum size=5.2mm},
    root/.style={vertex,fill=black!18,draw=black,line width=.9pt},
    cycleedge/.style={black,line width=1pt},
    essential/.style={black,line width=2.2pt},
    blueforest/.style={witnessblue,line width=1pt},
    orangeforest/.style={witnessorange,line width=1pt,dash pattern=on 3pt off 2pt},
    removed/.style={black!35,line width=.8pt},
    panel/.style={font=\small}
]
\begin{scope}
  \coordinate (c1) at (-0.95,0.95);
  \coordinate (c2) at (0.95,0.95);
  \coordinate (c3) at (1.45,-0.55);
  \coordinate (c4) at (0,-1.45);
  \coordinate (c5) at (-1.45,-0.55);
  \coordinate (x3) at (0,1.55);
  \coordinate (x1) at (0,2.45);
  \coordinate (x4) at (0,3.35);
  \coordinate (x2) at (0.1,-0.4);
  \fill[black!7,rounded corners=7pt] (-1.3,0.68) rectangle (1.3,1.22);
  \draw[cycleedge] (c1)--(c2);
  \draw[essential] (c2)--(c3) node[midway,right=1pt] {$b_1$};
  \draw[cycleedge] (c3)--(c4) node[midway,below right=-1pt] {$b_2$};
  \draw[essential] (c4)--(c5) node[midway,below left=-1pt] {$b_3$};
  \draw[cycleedge] (c5)--(c1) node[midway,left=1pt] {$b_4$};
  \draw[blueforest] (x3)--(c1) (x3)--(c2) (x4) to[out=190,in=100] (c1) (x4)--(x1) (x2)--(c3);
  \draw[orangeforest] (x3)--(x1) (x1)--(c1) (x1)--(c2) (x4) to[out=-10,in=80] (c2) (x2)--(c5);
  \foreach \v/\l in {c1/c_1,x1/x_1,x3/x_3,x4/x_4,x2/x_2}
    \node[vertex] at (\v) {$\l$};
  \foreach \v/\l in {c2/c_2,c3/c_3,c4/c_4,c5/c_5}
    \node[root] at (\v) {$\l$};
  \node[panel] at (0,-2.2) {(a) $M$-witness and root set $\rho$};
\end{scope}
\begin{scope}[xshift=4.5cm]
  \coordinate (r) at (0.45,-0.55);
  \coordinate (c1) at (-1.05,0.45);
  \coordinate (x1) at (0.45,2.05);
  \coordinate (x3) at (-0.1,0.65);
  \coordinate (x4) at (-1.2,2.25);
  \coordinate (x2) at (1.85,-0.55);
  \draw[removed] (c1) to[bend right=12] (r);
  \draw[removed] (c1) to[bend right=34] (r);
  \path (r) edge[removed,out=-5,in=-50,min distance=12mm,looseness=8] (r);
  \path (r) edge[removed,out=-60,in=-105,min distance=12mm,looseness=8] (r);
  \path (r) edge[removed,out=-115,in=-160,min distance=12mm,looseness=8] (r);
  \draw[blueforest] (x3)--(c1) (x3)--(r) (c1)--(x4) (x4)--(x1) (x2) to[bend right=22] (r);
  \draw[orangeforest] (x3)--(x1) (x1)--(c1) (x1)--(r)
     (x4) to[out=200,in=180,looseness=1.1] ($(c1)+(-0.1,-1.3)$) to[out=0,in=180] (r)
     (x2) to[bend left=22] (r);
  \foreach \v/\l in {c1/c_1,x1/x_1,x3/x_3,x4/x_4,x2/x_2}
    \node[vertex] at (\v) {$\l$};
  \node[root,minimum size=6.5mm] at (r) {$\rho$};
  \node[panel] at (0.3,-2.2) {(b) quotient $G/\rho$};
\end{scope}
\begin{scope}[xshift=7.1cm,yshift=1.9cm]
  \draw[cycleedge] (0,0)--(0.6,0) node[right,text=black] {cycle $C$};
  \draw[essential] (0,-0.5)--(0.6,-0.5) node[right,text=black] {$B_1=\{b_1,b_3\}$};
  \draw[blueforest] (0,-1)--(0.6,-1) node[right,text=black] {forest $F_1$};
  \draw[orangeforest] (0,-1.5)--(0.6,-1.5) node[right,text=black] {forest $F_2$};
  \draw[removed] (0,-2)--(0.6,-2) node[right,text=black] {image of $E(C)$};
  \node[root,minimum size=4mm] at (0.3,-2.55) {};
  \node[right] at (0.6,-2.55) {root in $\rho$};
\end{scope}
\end{tikzpicture}
\caption{The reduction of \Cref{lem:m-witness-to-ric} on the running
example of \Cref{fig:m-graph}(b), with $n=9$ and $m=15$, i.e., with
excess $\delta=1$. (a) An $M$-witness
$(C,F_1,F_2)$. Both forests meet the $5$-cycle $C$ in the intervals
$\{c_1,c_2\}$ (shaded), $\{c_3\}$, $\{c_4\}$, and $\{c_5\}$, which are
separated by the essential edges $b_1,\ldots,b_4$. The endpoints of the
matching $B_1=\{b_1,b_3\}$ form the root set $\rho$, and every component
of each forest contains exactly one root. (b) Identifying $\rho$ turns
$F_1$ and $F_2$ into two edge-disjoint spanning trees of $G/\rho$, that
is, into a basis of $M(G/\rho)\vee M(G/\rho)$. The cycle edges avoid
this basis, so $E(C)$ is independent in the dual matroid, which has rank
$m-(2n-8)=\delta+4=5$. The algorithm therefore searches for a cycle through
the four roots whose edge set is independent in this dual matroid.}
\label{fig:excess-quotient-example}
\end{figure}
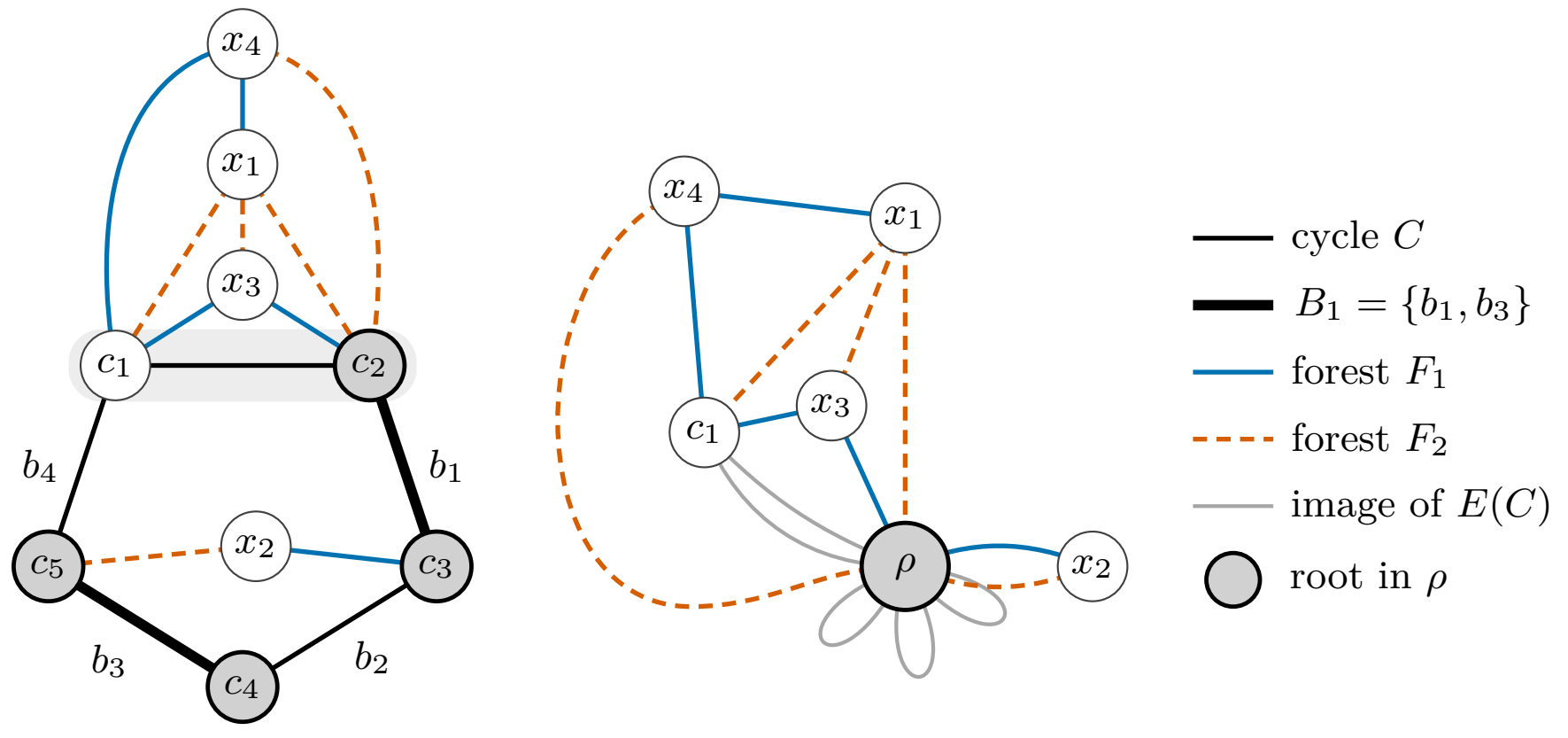

The proof of \Cref{thm:short-m-witness} has three main steps.
\begin{enumerate}
\item \emph{Express the forest requirement as a condition on the cycle.}
By \Cref{lem:four-intervals}, a witness admits four common roots, one
in each component of each forest. These roots are the endpoints of two
vertex-disjoint cycle edges, so we can enumerate their possible
choices. For fixed roots, identifying them turns the desired forests
into two edge-disjoint spanning trees. We encode these trees using
matroid union. Duality then translates the requirement that both trees
remain available after deleting the cycle into independence of the
cycle edges in a dual matroid. Thus, we can search for the cycle
without first choosing the forests.

\item \emph{Reduce the cycle search to a matroid of small rank.}
We construct a randomized matrix representation of the union and
retain a root choice only if the matrix has the rank required to
accommodate both spanning trees. We then represent the dual and
randomly project its columns to at most $k$ coordinates. This bounds
the matroid rank by $k$, and hence also bounds the length of every
independent cycle. The random choices preserve a fixed witness with
high probability, while every cycle accepted by the resulting test
extends to a valid witness. To achieve the claimed running time, we
reuse the main matrix computations across the candidate root sets.

\item \emph{Find an independent cycle through the roots.}
We solve the resulting cycle problem deterministically by growing
paths from one root, recording their endpoint, length, and which roots
they visit. Storing every possible path would be too expensive.
Instead, representative families retain a small collection that
preserves every possible completion to an independent cycle. This
requires preserving both independence of the edges and avoidance of
vertices used by the completion. Once a path visits all four roots,
we test whether an edge closes it into a cycle. The rank bound makes
this search single-exponential in $k$. After finding a cycle, we
recover the two forests by a spanning-tree packing computation.
\end{enumerate}
\Cref{fig:excess-pipeline} summarizes the algorithm.
\Cref{fig:excess-quotient-example} illustrates its matroidal encoding
on our running example: identifying the four roots turns the forests
into two edge-disjoint spanning trees, whose union avoids the cycle.

\begin{figure}[t]
\centering
\begin{tikzpicture}[
    node distance=5mm,
    every node/.style={font=\footnotesize},
    stage/.style={draw=black!70,rounded corners=2pt,fill=black!3,
                  text width=80mm,inner sep=4pt,align=left},
    randstage/.style={stage,draw=auxpurple,fill=auxpurple!6},
    cost/.style={text width=22mm,align=left,font=\footnotesize,anchor=west},
    randtag/.style={fill=auxpurple,text=white,font=\scriptsize\bfseries,
                    rounded corners=1.5pt,inner sep=1.5pt},
    note/.style={font=\scriptsize,text=black!70,align=left},
    arr/.style={-{Latex[length=2mm]},line width=.7pt},
    loopbox/.style={draw=black!60,dashed,rounded corners=5pt,
                    inner xsep=6pt,inner ysep=6pt},
    iv/.style={circle,fill=white,draw=black!70,inner sep=0pt,minimum size=1.6mm},
    ir/.style={circle,fill=black!30,draw=black,inner sep=0pt,minimum size=2mm}
]
  \newcommand{\iconcycle}[1]{%
    \foreach \i in {0,...,5} \coordinate (c\i) at ($(#1)+(60*\i+30:4.2mm)$);
    \draw[black!70,line width=.6pt] (c0)--(c1)--(c2)--(c3)--(c4)--(c5)--cycle;}

  \node (in) {\textbf{Input:} a multigraph $G$ that is not a $T$-graph,
    and an integer $k\geq4$};

  \node[randstage,below=6mm of in] (pre) {\textbf{Matroid union}
     (\Cref{lem:union-representation})\\[1pt]
     The two forests are independent in the union of two graphic
     matroids. A random matrix is used to represent this union.};

  \node[note,below=10mm of pre.south west,anchor=north west,xshift=1mm] (looplabel)
     {for each choice of four roots: the endpoints of two disjoint edges $e,f$};
  \node[stage,below=3mm of looplabel.south west,anchor=north west,xshift=-1mm] (dual)
     {\textbf{Dualize} (\Cref{thm:reform})\\[1pt]
     Keep these roots only if the matrix has rank $2n-8$. The
     cycle must avoid both forests, so by duality its edge set is
     independent in the dual of this union.};
  \node[randstage,below=of dual] (proj) {\textbf{Compress}
     (\Cref{lem:short-cycle-projection})\\[1pt]
     A random projection lowers the rank to at most $k$. A fixed short
     cycle stays independent with high probability.};
  \node[stage,below=of proj] (ric) {\textbf{Find the cycle}
     (\Cref{thm:ric-fpt})\\[1pt]
     Search deterministically for a cycle through the four roots whose
     edge set is independent.};

  \coordinate (I1) at ([xshift=-13mm,yshift=3mm]dual.west|-looplabel);
  \begin{scope}
    \iconcycle{I1}
    \draw[black,line width=1.8pt] (c0)--(c1) (c3)--(c4);
    \node[font=\scriptsize] at ($(c0)!.5!(c1)+(60:2mm)$) {$e$};
    \node[font=\scriptsize] at ($(c3)!.5!(c4)+(240:2mm)$) {$f$};
    \foreach \p in {c2,c5} \node[iv] at (\p) {};
    \foreach \p in {c0,c1,c3,c4} \node[ir] at (\p) {};
    \node[note,below=6mm of I1,align=center] {roots\\ (gray)};
  \end{scope}
  \node[loopbox,fit=(looplabel)(dual)(proj)(ric)] (loop) {};

  \node[stage,text width=52mm,below=11mm of ric.south west,anchor=north west] (rec)
     {\textbf{Complete the witness}\\[1pt]
     After identifying the four roots, two edge-disjoint spanning trees
     of the remaining graph give the two forests.};
  \node[note,anchor=north east,align=center] (rej) at ([yshift=-12mm]ric.south east)
     {no choice of roots\\ succeeds: reject};

  \draw[arr] (in)--(pre);
  \draw[arr] (pre.south)--(pre.south|-loop.north);
  \draw[arr] (dual)--(proj);
  \draw[arr] (proj)--(ric);
  \draw[arr] (ric.south-|rec)--node[right,note]{cycle found}(rec);
  \draw[arr,black!55] (rej.north|-loop.south)--(rej.north);
  \draw[arr,black!55] (ric.west) -- ++(-5mm,0) |-
     node[pos=.25,left,note,align=right]{no cycle:\\next roots} (dual.west);

  \node[randtag,anchor=south east] at ([xshift=-2pt,yshift=1pt]pre.north east) {random};
  \node[randtag,anchor=south east] at ([xshift=-2pt,yshift=1pt]proj.north east) {random};

  \node[cost] at ([xshift=4mm]loop.east|-pre)  {$\Oh(n^2m+km^2)$};
  \node[cost,note] at ([xshift=4mm]loop.east|-looplabel) {per choice of\\ roots ($\leq m^2$):};
  \node[cost] at ([xshift=4mm]loop.east|-dual) {$\Oh(n)$};
  \node[cost] at ([xshift=4mm]loop.east|-proj) {$\Oh(k^2m)$};
  \node[cost] at ([xshift=4mm]loop.east|-ric)  {$2^{\Oh(k)}\cdot m$};
  \node[cost] at ([xshift=4mm]loop.east|-rec)  {$\Oh(n^2)$};
\end{tikzpicture}
\caption{Overview of the algorithm of \Cref{thm:short-m-witness}. The
stages in the dashed box are repeated for at most $m^2$ choices of
roots. Only the purple stages use randomness; they never produce a
wrong witness and miss an existing one with probability below $1/n$. The right column gives the running time of each stage; the total is
$2^{\Oh(k)}\cdot m^3$.}
\label{fig:excess-pipeline}
\end{figure}
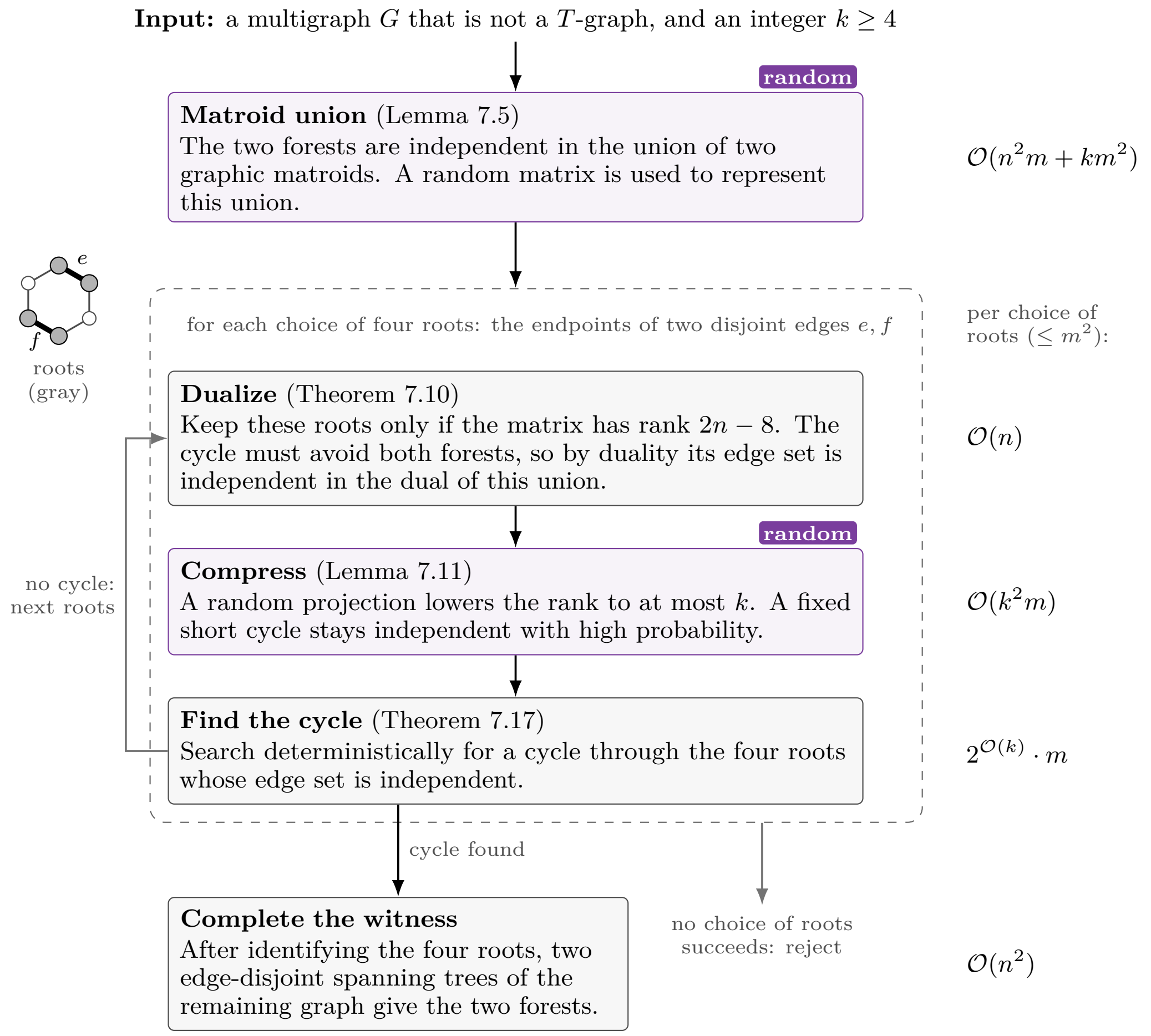

\subsubsection{Representing the Required Matroids}
\label{sec:matroid-preliminaries}

For the randomized reduction, we work over a prime field
$\mathbb{F}$ with $2n^2\leq|\mathbb{F}|\leq4n^2$. By Bertrand's postulate~\cite[Section~5.2]{Shoup08}
such a field exists, and its prime order can be found with the sieve of Eratosthenes in
$\Oh(n^2\log\log n)$ time~\cite[Section~5.4]{Shoup08}. Its elements have $\Oh(\log n)$ bits, so
field operations take constant time in our model, and we assume that
a uniformly random field element can be drawn in constant time.
For a matrix $A$ with column set $E$, we write
$A[X]$ for the submatrix formed by the columns in $X\subseteq E$, and
$A[W,X]$ if additionally only the rows in a set $W$ are kept; we write
$A[W,\cdot]$ and $A[\cdot,X]$ to keep all columns or all rows,
respectively. The kernel (or null space) of $A$ is
$\ker A=\{z\in\mathbb{F}^E:Az=0\}$.
We use the standard incidence-matrix representation of graphic matroids
and four further algebraic tools. For the latter, we include proofs in
\Cref{app:matroid-tools} to make explicit the representation, probability,
and running-time guarantees used by our algorithm.

\begin{lemma}[Representation of graphic matroids
    {\cite{Oxley2011}}]
\label{lem:graphic-representation}
Let $Q$ be a connected multigraph with $n_Q$ vertices and edge set $E$.
Delete one row from the signed vertex-edge incidence matrix of $Q$, with
a zero column for each loop. The resulting $(n_Q-1)\times|E|$ matrix
over $\mathbb{F}$ has full row rank and represents $M(Q)$. It can be
constructed in $\Oh(n_Q|E|)$ time.
\end{lemma}

Instead of representing the union of two matroids exactly, we use a
randomized representation that never creates new independent sets and
preserves any fixed basis with high probability for the field sizes
used here. This suffices for our
one-sided error, and it allows fields of polynomial size. The random
matrix $\Lambda$ depends only on $E$; we use one outcome for all
pairs $A_1,A_2$ that arise.

\begin{lemma}[Randomized representation of matroid unions
    {\cite{Marx09,NarayananSV1994}}]
\label{lem:union-representation}
Let $M_1$ and $M_2$ be matroids on $E$ represented over $\mathbb{F}$
by matrices $A_1$ and $A_2$ with full row ranks $r_1$ and $r_2$. Choose
$x_e\in\mathbb{F}$ independently and uniformly at random for all
$e\in E$, let $\Lambda=\operatorname{diag}(x_e:e\in E)$, and let $\mathcal N$
be the matroid represented by
\[
    A=\begin{pmatrix}A_1\\A_2\Lambda\end{pmatrix}.
\]
Then $\mathcal{I}(\mathcal N)\subseteq\mathcal{I}(M_1\vee M_2)$ for
every outcome. Moreover, for every fixed $B\in\mathcal{I}(M_1\vee M_2)$
with $|B|=r_1+r_2$, we have that $\Pr[B\in\mathcal{I}(\mathcal
N)]\geq1-r_2/|\mathbb{F}|$.
\end{lemma}

We obtain a representation of the dual by taking a basis of the
kernel of the representing matrix.

\begin{lemma}[Representation of dual matroids
    {\cite[Theorem~2.2.8]{Oxley2011}}]
\label{lem:dual-representation}
Let $A$ be a matrix over $\mathbb{F}$ with column set $E$, and let $Z$
be a matrix whose rows form a basis of $\ker A$. Then $Z$ has full row
rank and represents $M[A]^*$.
\end{lemma}

To reuse the linear algebra across root sets, we select bases of the
row and column spaces and invert the nonsingular submatrix at their
intersection.

\begin{lemma}[Nonsingular maximal submatrices {\cite{IbarraMH82}}]
\label{lem:max-submatrix}
Given an $a\times b$ matrix $\Phi$ over $\mathbb{F}$, one can compute
in $\Oh(ab\cdot\min\{a,b\}^{\omega-2})$ time its rank $r$, a set $I$ of
$r$ rows, a set $J$ of $r$ columns such that $\Phi[I,J]$ is
nonsingular, and $\Phi[I,J]^{-1}$. Moreover,
$\Phi=\Phi[\cdot,J]\,\Phi[I,J]^{-1}\,\Phi[I,\cdot]$.
\end{lemma}

The one-sided error of the union construction remains harmless after
dualization, provided that the constructed matroid attains the
required rank. The following lemma is a consequence of standard
matroid duality~\cite{Oxley2011}; we include a direct proof in
\Cref{app:matroid-tools}.

\begin{lemma}[Equal-rank transfer through duality]
\label{lem:equal-rank-dual}
Let $M$ and $N$ be matroids on the same ground set, with
\[
    \mathcal{I}(N)\subseteq\mathcal{I}(M)
    \qquad\text{and}\qquad r(N)=r(M).
\]
Then every basis of $N$ is a basis of $M$, and
\[
    \mathcal{I}(N^*)\subseteq\mathcal{I}(M^*).
\]
\end{lemma}

These representation tools allow us to encode the two forests and the
cycle of an $M$-witness in a single dual matroid.

\subsubsection{A Matroidal Characterization}
\label{sec:matroid-characterization}

By \Cref{lem:m-witness-size-bound}, we may restrict attention to
connected multigraphs with at least four vertices. Let $G=(V,E)$
be such a graph. For a four-element set $\rho\subseteq V$,
let $G/\rho$ be the multigraph obtained by identifying the vertices
of $\rho$ into one vertex while retaining every edge, including
resulting loops, as a distinct ground-set element; for
$X\subseteq E$, we write $(G-X)/\rho$ for $G/\rho$ with the edges
of $X$ deleted.
Recall that $M(Q)$ denotes the graphic matroid of a multigraph $Q$,
whose independent sets are the edge sets of forests. We write $r(M)$
for the rank of a matroid $M$, $M^*$ for its dual, and $M_1\vee M_2$
for the union of two matroids on the same ground set.
The graph $G/\rho$ is connected and has $n-3$ vertices. Every spanning
tree therefore has $n-4$ edges, which is the rank of its graphic matroid:
\begin{equation}\label{eq:rooted-graphic-rank}
    r(M(G/\rho))=n-4.
\end{equation}
Identifying roots converts a spanning forest rooted at those
vertices into a spanning tree, as made precise next.

\begin{lemma}[Rooted forests as bases]\label{lem:pullback}
Let $G=(V,E)$ be a connected multigraph, let $\rho\subseteq V$
have size four, and let $F\subseteq E$. Then $F$ is a basis of
$M(G/\rho)$ if and only if $(V,F)$ is a forest with exactly four
components, each containing exactly one vertex of $\rho$.
\end{lemma}

\begin{proof}
First observe that
\begin{equation}\label{eq:rooted-independence}
    F\in\mathcal{I}(M(G/\rho))
    \quad\Longleftrightarrow\quad
    \begin{array}{l}
    (V,F)\text{ is a forest, and every component}\\
    \text{contains at most one vertex of }\rho.
    \end{array}
\end{equation}
Indeed, the image of a cycle of $G$ contains a cycle in $G/\rho$.
If a component of $(V,F)$ contains two roots, choose a path between
roots with no root as an internal vertex. Its image is a cycle in
$G/\rho$, possibly a loop or a cycle of length two. Either event
contradicts independence.

Conversely, if the image of a forest contains a cycle, then lifting
that cycle yields either a cycle in the original forest or a path
with distinct endpoints in $\rho$. Thus a forest with at most one
root per component remains a forest after identification. This
proves \eqref{eq:rooted-independence}.

If $F$ is a basis, it is such a forest and has $n-4$ edges by
\eqref{eq:rooted-graphic-rank}. Hence it has four components.
The four roots lie in distinct components, so each component
contains exactly one root. Conversely, a forest with four
components and one root per component is independent by
\eqref{eq:rooted-independence} and has size $n-4$, so it is a basis.
\end{proof}

An $M$-witness contains two edge-disjoint forests $F_1,F_2$. We
therefore combine two copies of a graphic matroid by matroid union.
For graphs that are not $T$-graphs, both forests can be rooted at the
same four vertices by \Cref{lem:four-intervals}. For a
four-element set $\rho\subseteq V$, let
$\mathcal{M}_\rho=M(G/\rho)\vee M(G/\rho)$. By
\eqref{eq:rooted-graphic-rank},
\begin{equation}\label{eq:union-rank-upper-bound}
    r(\mathcal{M}_\rho)\leq2\,r(M(G/\rho))=2n-8.
\end{equation}

When $\mathcal M_\rho$ has rank $2n-8$, its bases encode two
edge-disjoint spanning forests, each with exactly one root in every
component. The cycle must avoid both forests. By matroid duality, its
edges are independent in $\mathcal M_\rho^*$ precisely when deleting
them leaves a basis of $\mathcal M_\rho$ available. The following theorem
therefore allows us to search for a cycle through the roots while
leaving the choice of forests implicit. Conversely, every $M$-witness
in a graph that is not a $T$-graph is captured by some root set
consisting of the endpoints of two vertex-disjoint cycle edges.

\begin{theorem}[Matroidal characterization of $M$-witnesses]
\label{thm:reform}
Let $G=(V,E)$ be a connected multigraph with $n\geq4$ vertices.
\begin{enumerate}
    \item\label{it:reform-sound} Let $\rho\subseteq V$ have size four
    with $r(\mathcal{M}_\rho)=2n-8$, and let $C$ be a cycle of $G$ with
    $\rho\subseteq V(C)$ and $E(C)\in\mathcal{I}(\mathcal{M}_\rho^*)$.
    Then $(G-E(C))/\rho$ has two edge-disjoint spanning trees. For
    any two such trees $T_1,T_2$, the graphs $F_i=(V,E(T_i))$ form an
    $M$-witness $(C,F_1,F_2)$ of $G$.
    \item\label{it:reform-complete} Suppose that $G$ is not a
    $T$-graph, and let $(C,F_1,F_2)$ be an $M$-witness of $G$. Then
    there are two vertex-disjoint edges $e,f\in E(C)$ such that
    $\rho=V(\{e,f\})$ satisfies the following: $E(F_1)\mathbin{\dot\cup}E(F_2)$
    is a basis of $\mathcal{M}_\rho$ that is disjoint from $E(C)$. In
    particular, $r(\mathcal{M}_\rho)=2n-8$ and
    $E(C)\in\mathcal{I}(\mathcal{M}_\rho^*)$.
\end{enumerate}
\end{theorem}

\begin{proof}
For~\ref{it:reform-sound}, $E\setminus E(C)$ contains a basis $D$ of
$\mathcal{M}_\rho$ by the definition of dual independence. Write $D=D_1\cup D_2$,
with $D_i$ independent in $M(G/\rho)$. Then
\[
    2n-8=|D_1\cup D_2|
    \leq |D_1|+|D_2|\leq2n-8
\]
by \eqref{eq:rooted-graphic-rank}. Equality holds throughout. Thus
$D_1$ and $D_2$ are disjoint bases of $M(G/\rho)$, that is, edge sets
of two edge-disjoint spanning trees of the connected graph $G/\rho$.
They avoid $E(C)$, so they are spanning trees of $(G-E(C))/\rho$.
Now let $T_1,T_2$ be any two edge-disjoint spanning trees of
$(G-E(C))/\rho$. Each $E(T_i)$ is a basis of $M(G/\rho)$ by
\eqref{eq:rooted-graphic-rank}. By \Cref{lem:pullback}, each $F_i$ is a
spanning forest with four components, each containing exactly one
vertex of $\rho\subseteq V(C)$. The cycle $C$ contains the four
vertices of $\rho$, so it has length at least four. Since
$C,F_1,F_2$ are pairwise edge-disjoint, they form an $M$-witness.

For~\ref{it:reform-complete}, $G$ is an $M$-graph, so
\Cref{lem:four-intervals} provides a matching $B_1=\{e,f\}\subseteq
E(C)$ with exactly one endpoint in each component of each forest.
Let $\rho=V(B_1)$. By \Cref{lem:pullback}, $E(F_i)$ is a basis of
$M(G/\rho)$ for $i\in\{1,2\}$. Hence
$D=E(F_1)\mathbin{\dot\cup}E(F_2)$ is independent in $\mathcal{M}_\rho$
and has size $2n-8$. By \eqref{eq:union-rank-upper-bound}, $D$ is a
basis and $r(\mathcal{M}_\rho)=2n-8$. Since $D\subseteq E\setminus
E(C)$, the definition of dual independence implies
$E(C)\in\mathcal{I}(\mathcal{M}_\rho^*)$.
\end{proof}

\subsubsection{Reduction to a Rooted Independent Cycle}
\label{sec:reduction-to-ric}

By \Cref{thm:reform}, once the four roots are fixed, we need a cycle
through them whose edge set is independent in the dual matroid.
Independence guarantees that the remaining edges contain the two
required forests, so these forests need not be chosen during the cycle
search. This leads to the following auxiliary problem, in which the
matroid is supplied by a matrix representation.

\problemdef{Rooted Independent Cycle (RIC).}{A connected multigraph $G=(V,E)$, a root set
$R\subseteq V$ with $|R|=4$, and a linear matroid
$M=(E,\mathcal{I})$ of rank $r$, given by an $r\times|E|$ matrix over a
finite field $\mathbb{F}$ with $|\mathbb{F}|\geq |V|$.}{Is there a cycle $C$ with
$R\subseteq V(C)$ and $E(C)\in\mathcal{I}$?}

Note that every solution has at most $r$ edges, since its edge set is
independent. We use the rank $r$ as the parameter.

By \Cref{thm:reform}, it suffices to try at most $m^2$ root
sets, namely the endpoint sets of pairs of vertex-disjoint edges.
The equal-rank transfer lemma (\Cref{lem:equal-rank-dual}) ensures soundness for the randomized
union representation. The dual representation may have rank much larger
than $k$, whereas our algorithm for RIC is single-exponential in the
matroid rank. We therefore randomly project its columns into a vector
space of dimension at most $k$. This preserves the independence of a
fixed witness cycle with high probability and cannot turn a dependent
edge set into an independent one.

We use the following elementary projection guarantee. It is needed
only for one fixed solution, rather than simultaneously for all small
independent sets.

\begin{lemma}\label{lem:short-cycle-projection}
Let $B$ be a matrix with $a$ rows over $\mathbb F$, and choose a
uniform random $\ell\times a$ matrix $P$. Every set of columns
independent in $PB$ is independent in $B$. For any fixed set of
$q\leq\ell$ columns independent in $B$, its images are independent
with probability at least $1-q/|\mathbb F|$.
\end{lemma}
\begin{proof}
Multiplication preserves every linear dependence, proving the first
claim. For the second, consider the determinant in the first $q$ rows
of the selected columns of $PB$. This is a polynomial of degree $q$
in the entries of $P$. It is nonzero: these rows of $P$ can select a
nonsingular $q\times q$ row minor of the original columns. The
Schwartz--Zippel bound~\cite[Lemma~10.19]{C+15} proves the claim.
\end{proof}

The following lemma gives the main reduction from finding an
$M$-witness with cycle length at most $k$ to RIC. It constructs at most
$m^2$ instances, corresponding to candidate root sets, each with matroid
rank at most $k$. Every returned cycle extends to a valid $M$-witness.
Conversely, if a short witness exists, then with high probability at
least one instance retains its cycle as a solution. Thus, it remains
to solve RIC efficiently in the matroid rank.

\begin{lemma}\label{lem:m-witness-to-ric}
Let $G$ be a connected non-$T$-graph with $n\geq4$ vertices,
$m\geq2n-4$ edges, and let $k\geq4$. Let
$\delta=m-(2n-4)$ and $\ell=\min\{k,n,\delta+4\}$.
After randomized $\Oh(n^2m+\ell m^2)$ preprocessing, one can compute
at most $m^2$ instances of RIC in $\Oh(n+\ell^2m)$ time each.
Every instance has a representation of full row rank $r\leq\ell$.
For every random outcome, a solution of an instance gives an
$M$-witness with the same cycle, of length at most $k$.
If $G$ has an $M$-witness with cycle length at most $k$, then with
probability at least $1-(n-4+\ell)/|\mathbb F|$ some instance has a solution.
\end{lemma}

The proof has three main ingredients.
\begin{enumerate}
\item \emph{Represent the forest requirement.}
For each candidate root set, we construct a randomized representation
of the union of the two graphic matroids. We discard root sets whose
representation does not have the rank needed for two spanning trees.
For the remaining root sets, dualization turns the requirement that
the cycle leave both trees available into independence of its edges.

\item \emph{Share the computation across root sets.}
The matrices for different root sets are obtained from one common
matrix by deleting a constant number of rows. We exploit this
relationship to compute their dual representations using common
preprocessing and a small additional computation for each root set.

\item \emph{Project and preserve a witness.}
We randomly project the dual representations to rank at most $k$,
obtaining the RIC instances. Every solution of a resulting instance
extends to an $M$-witness. For a fixed short witness, its two forests
survive the randomized union representation and its cycle survives the
projection with high probability.
\end{enumerate}

\begin{proof}[Proof of \Cref{lem:m-witness-to-ric}]
\emph{Setup.} Orient every edge of $G$ arbitrarily, and let
$D\in\mathbb{F}^{V\times E}$ be the signed vertex-edge incidence matrix,
whose column of an edge with tail $u$ and head $v\neq u$ has entry $1$
in row $u$, entry $-1$ in row $v$, and zeros elsewhere; loops give zero
columns. Choose $x_e\in\mathbb{F}$ independently and uniformly at
random for all $e\in E$, once and for all root sets, and let
$\Lambda=\operatorname{diag}(x_e:e\in E)$ and
\[
    \Phi=\begin{pmatrix}D\\D\Lambda\end{pmatrix}\in\mathbb{F}^{(V\times[2])\times E},
\]
where row $(v,1)$ is row $v$ of $D$ and row $(v,2)$ is row $v$ of
$D\Lambda$. Every column of $\Phi$ has at most four nonzero entries.

Let $\rho\subseteq V$ have size four. Deleting the rows of $\rho$ from
$D$ gives the signed incidence matrix of $G/\rho$, with the inherited
orientation, from which the row of the identified vertex has been
deleted. By \Cref{lem:graphic-representation}, this
$(n-4)\times m$ matrix $A_\rho$ has full row rank and represents
$M(G/\rho)$. Apply \Cref{lem:union-representation} with
$A_1=A_2=A_\rho$ and the shared matrix $\Lambda$, and let
$\mathcal{N}_\rho$ denote the matroid represented by
\[
    U_\rho=\begin{pmatrix}A_\rho\\A_\rho\Lambda\end{pmatrix},
\]
that is, by $\Phi$ with the eight rows in $\rho\times[2]$ deleted. For
every outcome,
\[
    \mathcal{I}(\mathcal{N}_\rho)\subseteq\mathcal{I}(\mathcal{M}_\rho),
    \qquad
    r(\mathcal{N}_\rho)\leq r(\mathcal{M}_\rho)\leq2n-8.
\]

The matrix $U_\rho$ is obtained from $\Phi$ by deleting eight rows:
one for each of the four roots in each of the two incidence-matrix
copies. Deleting these equations enlarges the kernel by at most eight
dimensions. Thus, a basis of $\ker U_\rho$ can be obtained by extending
a common basis of $\ker\Phi$ with at most eight additional vectors.
This allows us to compute and project the common basis once and handle
the additional vectors separately for each root set.

\emph{Kernels.} Let $S_\rho\subseteq\mathbb{F}^{V\times[2]}$ be the
eight-dimensional subspace of vectors supported on $\rho\times[2]$.
Then $U_\rho z=0$ if and only if $\Phi z\in S_\rho$. Let $z_1,\ldots,z_t\in
\mathbb{F}^E$ be such that $\Phi z_1,\ldots,\Phi z_t$ form a basis of
$\operatorname{im}\Phi\cap S_\rho$. The linear map
\[
 \ker U_\rho\longrightarrow\operatorname{im}\Phi\cap S_\rho,
 \qquad z\longmapsto\Phi z,
\]
is surjective and has kernel $\ker\Phi$. Thus a basis of $\ker\Phi$
together with arbitrary lifts $z_1,\ldots,z_t$ is a basis of
$\ker U_\rho$. In particular,
\begin{equation}\label{eq:kernel-dimension}
    \dim\ker U_\rho=\dim\ker\Phi+\dim(\operatorname{im}\Phi\cap S_\rho).
\end{equation}

\emph{Preprocessing.} Draw $\Lambda$ and construct $\Phi$ in $\Oh(nm)$
time. By \Cref{lem:max-submatrix}, compute the rank $\eta$ of $\Phi$, row
and column sets $I$ and $J$ with $\Phi[I,J]$ nonsingular, and
$\Phi[I,J]^{-1}$ in $\Oh(n^{\omega-1}m)$ time, since $\Phi$ has $2n$
rows and $m\geq n$ columns. For $u\in\mathbb{F}^{V\times[2]}$, let
$\pi(u)\in\mathbb{F}^E$ agree with $\Phi[I,J]^{-1}u[I]$ on $J$ and
vanish outside $J$. The map $\pi$ is linear, and
$\Phi\pi(u)=\Phi[\cdot,J]\,\Phi[I,J]^{-1}u[I]$. If $u=\Phi z$, then
$u[I]=\Phi[I,\cdot]z$, so the factorization of
\Cref{lem:max-submatrix} gives $\Phi\pi(u)=u$. Hence,
\begin{equation}\label{eq:image-test}
    u\in\operatorname{im}\Phi
    \quad\Longleftrightarrow\quad
    \Phi\pi(u)=u.
\end{equation}
If $m-\eta>\delta+4$, then
$\dim\ker U_\rho>\delta+4$ and thus $r(\mathcal{N}_\rho)=m-\dim\ker U_\rho<2n-8$
for every $\rho$ by \eqref{eq:kernel-dimension}; we output no instance.
Otherwise, we proceed as follows.
By the factorization, every row of $\Phi$ is a linear combination of
the rows in $I$, so $\ker\Phi=\ker\Phi[I,\cdot]$. For every column
$c\notin J$, the vector $y_c=\mathbf{1}_c-\pi(\Phi[\cdot,c])$ satisfies
$\Phi[I,\cdot]y_c=\Phi[I,c]-\Phi[I,J]\Phi[I,J]^{-1}\Phi[I,c]=0$. These
$m-\eta=\dim\ker\Phi$ vectors are linearly independent, since $y_c$ has entry
one at $c$ and zeros at all other columns outside $J$. Hence they form a
basis of $\ker\Phi$. Each $y_c$ takes $\Oh(n)$ time, since $\Phi[I,c]$
has at most four nonzero entries.
Let $Y$ have these $m-\eta$ basis vectors as rows. Conditional on
$\Lambda$, choose $P_0$ uniformly among the $\ell\times(m-\eta)$ matrices
over $\mathbb F$, using fresh randomness, and precompute $Q_0=P_0Y$. Constructing $Y$ densely and multiplying takes
$\Oh(\ell m^2)$ operations. Together with the elimination, the
preprocessing takes $\Oh(n^2m+\ell m^2)$ operations.

\emph{Root sets.} We enumerate the at most $m^2$ ordered pairs $e,f$ of
edges of the support and keep those that are vertex-disjoint. This
generates every set $\rho=V(\{e,f\})$ for vertex-disjoint edges $e,f$,
some of them repeatedly. For a fixed $\rho$, let $b_1,\ldots,b_8$ be the
unit vectors of the rows in $\rho\times[2]$, which form a basis of
$S_\rho$. For each $i$, the vector $\pi(b_i)$ is a column of
$\Phi[I,J]^{-1}$ placed on $J$, or zero, and $d_i=\Phi\pi(b_i)-b_i$ is
computed in $\Oh(n)$ time. By linearity and \eqref{eq:image-test},
$\sum_i\gamma_ib_i\in\operatorname{im}\Phi$ if and only if
$\sum_i\gamma_id_i=0$. Gaussian elimination on the $2n\times8$ matrix with
columns $d_1,\ldots,d_8$ yields, in $\Oh(n)$ time, a basis
$\gamma^{(1)},\ldots,\gamma^{(t)}$ of its null space. Since
$b_1,\ldots,b_8$ are linearly independent, the vectors
$u_j=\sum_i\gamma^{(j)}_ib_i$ form a basis of $\operatorname{im}\Phi\cap
S_\rho$. Let $z_j=\sum_i\gamma^{(j)}_i\pi(b_i)=\pi(u_j)$; then
$\Phi z_j=u_j$ by \eqref{eq:image-test}. By \eqref{eq:kernel-dimension},
$\dim\ker U_\rho=m-\eta+t$, so
$r(\mathcal{N}_\rho)=m-\dim\ker U_\rho=\eta-t$.
Discard $\rho$ unless $r(\mathcal N_\rho)=2n-8$, equivalently
$m-\eta+t=\delta+4$. For a surviving set, the matrix
\[
 B_\rho=\begin{pmatrix}Y\\Z_\rho\end{pmatrix},
 \qquad Z_\rho=\begin{pmatrix}z_1^{\mathsf T}\\\vdots\\z_t^{\mathsf T}\end{pmatrix},
\]
has basis rows for $\ker U_\rho$ and represents
$\mathcal N_\rho^*$ by \Cref{lem:dual-representation}.
Each of the at most eight rows of $Z_\rho$ is supported on $J$.
Choose a fresh uniform $\ell\times t$ matrix $P_\rho$, independently
of all previous choices, and form
\[
 Q_\rho=Q_0+P_\rho Z_\rho
       =\begin{pmatrix}P_0&P_\rho\end{pmatrix}B_\rho.
\]
The correction costs $\Oh(\ell n)$ operations; writing $Q_\rho$
costs $\Oh(\ell m)$. Retain a basis of its rows in
$\Oh(\ell^2m)$ operations. This produces a full-row-rank
representation of a matroid $\widehat{\mathcal N}_\rho$ of rank
$r\leq\ell$. Discard it if $r<4$; otherwise output
$(G,\rho,\widehat{\mathcal N}_\rho)$. Conditional on $\Lambda$ and a
fixed surviving $\rho$, the kernel basis $B_\rho$ is fixed, and
$[P_0\ P_\rho]$ is uniform. Thus
\Cref{lem:short-cycle-projection} applies even though $P_0$ is shared.

\emph{Soundness.} A surviving root set satisfies
$r(\mathcal N_\rho)=r(\mathcal M_\rho)=2n-8$.
By \Cref{lem:short-cycle-projection,lem:equal-rank-dual},
\[
 \mathcal I(\widehat{\mathcal N}_\rho)
 \subseteq\mathcal I(\mathcal N_\rho^*)
 \subseteq\mathcal I(\mathcal M_\rho^*).
\]
A solution cycle therefore extends to an $M$-witness by
\Cref{thm:reform}. Its length is at most $r\leq\ell\leq k$.

\emph{Completeness.} Fix an $M$-witness $(C,F_1,F_2)$ with
$|E(C)|\leq k$ before any random choices. By
\Cref{lem:m-witness-size-bound}, its cycle has at most $\ell$ edges.
By \Cref{thm:reform}, there is an enumerated root set $\rho$ for which
$D'=E(F_1)\mathbin{\dot\cup}E(F_2)$ is a basis of
$\mathcal M_\rho$ avoiding $C$. By \Cref{lem:union-representation},
with probability at least $1-(n-4)/|\mathbb F|$, this remains a basis
of $\mathcal N_\rho$. Then $\rho$ survives and $E(C)$ is independent
in $B_\rho$. Conditional on any such $\Lambda$,
\Cref{lem:short-cycle-projection} preserves these columns with
probability at least $1-|E(C)|/|\mathbb F|$. Combining the two failure
bounds gives at most $(n-4+\ell)/|\mathbb F|$.
Only this fixed witness and root set are needed, so there is no union
bound over root sets.
\end{proof}

We have reduced the bounded-cycle witness problem to at most $m^2$
instances of RIC, each with matroid rank at most $k$. To complete the
FPT algorithm with the desired running time, we solve each instance in
$2^{\Oh(k)}m$ time. Enumerating all candidate cycles would still give
an exponent depending on $k$ in the polynomial factor. We instead use
a dynamic program that grows paths and compresses its collections of
partial solutions using representative families.

\subsubsection{Representative Families}
\label{sec:representative-families}

Our algorithm for RIC grows paths from one root and records their
endpoint, length, and which prescribed roots they visit. Paths with
the same recorded information may still admit different completions:
a path may use a vertex needed by the continuation, or its edges may
become dependent when combined with the remaining cycle edges. We can
neither retain all paths, since there are too many, nor retain just one
arbitrary path for each combination of recorded information, since this
may discard a path needed for a feasible completion. Instead,
representative families let us keep a small collection of paths that
preserves every feasible completion. This extends the approach of
Agrawal et al.~\cite{AgrawalJKS20} for matroid-constrained paths by
tracking the prescribed roots and closing the path into a cycle. To
preserve both independence of the edges and avoidance of vertices used
by a continuation, we include vertices in the matroid ground set,
following Fomin et al.~\cite{FominLPS16}.

A $q$-representative family preserves the possibility of extending a
partial solution by at most $q$ further elements. Whenever an extension
is disjoint from some original partial solution and their union is
independent, it is also disjoint from some retained partial solution
whose union with it is independent.

\begin{definition}[$q$-representative family {\cite[Section~12.3]{C+15}}]
\label{def:representative-family}
Let $M=(E,\mathcal{I})$ be a matroid, let $p,q\geq0$ be integers,
and let $\mathcal{A}\subseteq\mathcal{I}$ consist of sets of
size $p$. A subfamily $\widehat{\mathcal{A}}\subseteq\mathcal{A}$
is \emph{$q$-representative for $\mathcal{A}$ in $M$} if the
following holds for every $Y\subseteq E$ with $|Y|\leq q$:
whenever some $A\in\mathcal{A}$ satisfies
\[
    A\cap Y=\varnothing,\qquad A\cup Y\in\mathcal{I},
\]
some $\widehat A\in\widehat{\mathcal{A}}$ also satisfies
\[
    \widehat A\cap Y=\varnothing,\qquad
    \widehat A\cup Y\in\mathcal{I}.
\]
\end{definition}

This is the definition of Fomin, Lokshtanov, Panolan, and
Saurabh~\cite{FominLPS16}, specialized to families of $p$-element sets.
Their algorithm takes an explicitly listed family of independent
$p$-element sets and retains at most $\binom{p+q}{p}$ members while
preserving all independent extensions by at most $q$ elements. We use
it to compress the collections of partial paths after each extension
step. The bound depends only on the matroid rank $p+q$, rather than on
the number of paths generated.

\begin{theorem}[Computation of representative families
    {\cite[Theorem~1.1]{FominLPS16}}]
\label{thm:representative-families}
Let $M=(E,\mathcal{I})$ be a linear matroid of rank $p+q$, represented
by a $(p+q)\times|E|$ matrix over a field $\mathbb{F}$, and let
$\mathcal{A}\subseteq\mathcal{I}$ be an explicitly listed family of $a$
sets, each of size $p$. One can compute a subfamily
$\widehat{\mathcal{A}}\subseteq\mathcal{A}$ of size at most
$\binom{p+q}{p}$ that is $q$-representative for $\mathcal{A}$ in $M$
using
\[
    \Oh\left(\binom{p+q}{p}a\,p^{\omega}+a\binom{p+q}{q}^{\omega-1}\right)
\]
field operations, where $\omega$ is the matrix multiplication exponent.
\end{theorem}

Our dynamic program extends only retained paths. The following closure
property ensures that these extensions still represent the extensions
of all feasible paths, including those discarded during compression. If
$\widehat{\mathcal A}$ is $(q+d)$-representative for $\mathcal A$ and
$Z$ is a fixed $d$-element set, then
\[
 \{A\cup Z:A\in\widehat{\mathcal A},\ A\cap Z=\varnothing,
                  \ A\cup Z\in\mathcal I\}
\]
is $q$-representative for the analogous extension family of
$\mathcal A$.

To see this, consider a continuation $Y$ of at most $q$ elements that
can extend some $A\cup Z$. Before adding $Z$, the combined extension
$Z\cup Y$ has at most $q+d$ elements. Representativity therefore
provides a retained set $\widehat A$ that also admits this combined
extension. Consequently, $\widehat A\cup Z$ admits the original
continuation $Y$.

We also use two elementary properties when combining transitions and
compressing their results. Taking the union of representative
subfamilies preserves representativity for the union of the original
families. Representativity is also transitive: if $\mathcal A_1$ is
$q$-representative for $\mathcal A_0$, and $\mathcal A_2$ is
$q$-representative for $\mathcal A_1$, then $\mathcal A_2$ is
$q$-representative for $\mathcal A_0$.

\subsubsection{A Dynamic Program over Rooted Paths}
\label{sec:ric-dp}

Let $(G,R,M)$ be an instance of RIC with $n$ vertices and $m$ edges,
where $M=(E,\mathcal{I})$ has rank $r$ and is represented over a field
$\mathbb{F}$ with $|\mathbb{F}|\geq n$. Reject if $r<4$, since every
solution has at least four edges. A solution has as many vertices as
edges, so it has at most $L=\min\{r,n\}$ vertices. Let $\kappa=L-1$
and write $R=\{r_1,\ldots,r_4\}$. Let $r^*=r_1$ and
$V'=V\setminus\{r^*\}$.

\begin{figure}[t]
\centering
\begin{tikzpicture}[
    every node/.style={font=\footnotesize},
    gv/.style={circle,draw=black!75,fill=white,inner sep=0pt,minimum size=4.6mm},
    rv/.style={gv,fill=black!18,draw=black,line width=.9pt},
    pathE/.style={witnessblue,line width=1.4pt},
    oldE/.style={black!35,line width=1pt,dash pattern=on 2pt off 1.5pt},
    contE/.style={witnessorange,line width=1.2pt,dash pattern=on 3pt off 2pt},
    graphE/.style={black!20,line width=.5pt},
    panel/.style={font=\small}
]
\newcommand{\ricgraph}{%
  \coordinate (r1) at (0,0);    \coordinate (p1) at (0.9,0.75);
  \coordinate (r2) at (2.0,0.85); \coordinate (p2) at (3.0,0.35);
  \coordinate (v)  at (3.6,-0.55);
  \coordinate (q1) at (2.9,-1.45); \coordinate (r3) at (1.8,-1.85);
  \coordinate (q2) at (0.8,-1.75); \coordinate (r4) at (0.05,-1.1);
  \coordinate (o1) at (1.05,-0.25); \coordinate (o2) at (2.1,-0.35);
  \draw[graphE] (r1)--(o1)--(o2)--(v) (o1)--(r2) (o2)--(q1) (r2)--(o2) (p1)--(o1) (o2)--(r3) (r1)--(q2);}
\begin{scope}
  \ricgraph
  \draw[pathE] (r1)--(p1)--(r2)--(p2)--(v);
  \draw[contE] (v)--(q1)--(r3)--(q2)--(r4)--(r1);
  \foreach \p in {p1,p2,q1,q2,o1,o2} \node[gv] at (\p) {};
  \node[gv] at (v) {$v$};
  \node[rv] at (r1) {$r^*$};
  \foreach \p/\l in {r2/r_2,r3/r_3,r4/r_4} \node[rv] at (\p) {$\l$};
  \node[panel,align=center] at (1.8,-2.6) {(a) state $(v,\{r^*,r_2\},5)$\\and a continuation path $Q$};
\end{scope}
\begin{scope}[xshift=5.6cm]
  \ricgraph
  \draw[oldE] (r1)--(p1)--(r2)--(p2)--(v);
  \draw[pathE] (r1)--(o1)--(r2)--(o2)--(v);
  \draw[contE] (v)--(q1)--(r3)--(q2)--(r4)--(r1);
  \foreach \p in {p1,p2,q1,q2,o1,o2} \node[gv] at (\p) {};
  \node[gv] at (v) {$v$};
  \node[rv] at (r1) {$r^*$};
  \foreach \p/\l in {r2/r_2,r3/r_3,r4/r_4} \node[rv] at (\p) {$\l$};
  \node[panel,align=center] at (1.8,-2.6) {(b) a retained path $\widehat P$\\with the same continuation path $Q$};
\end{scope}
\begin{scope}[xshift=10.4cm,yshift=0.6cm]
  \draw[pathE] (0,0)--(0.55,0) node[right,text=black,align=left] {path $P$ from $r^*$};
  \draw[contE] (0,-0.5)--(0.55,-0.5) node[right,text=black,align=left] {continuation path $Q$};
  \draw[oldE] (0,-1)--(0.55,-1) node[right,text=black,align=left] {discarded path};
  \node[rv,minimum size=3.5mm] at (0.27,-1.5) {};
  \node[right] at (0.55,-1.5) {root};
\end{scope}
\end{tikzpicture}
\caption{The dynamic program of \Cref{thm:ric-fpt}. (a) A state
$(v,T,s)$ records the endpoint $v$ of a path $P$ grown from the root
$r^*$, the roots $T$ it visits, and its number $s$ of vertices. The path is
encoded as the set $X_P$ of its vertices other than $r^*$ and its edges,
in the direct sum of a uniform matroid on the vertices and the
input edge matroid $M$. A continuation path $Q$ visits the remaining
roots and closes the cycle at $r^*$. (b) Representative families keep
only a few paths per state, but for every continuation path $Q$ of any feasible
path in the state, they retain a path $\widehat P$ that $Q$ also completes:
$\widehat P$ shares only its endpoints $v$ and $r^*$ with $Q$, and the
edges of $\widehat P$ and $Q$ together are independent in $M$.}
\label{fig:ric-dp}
\end{figure}
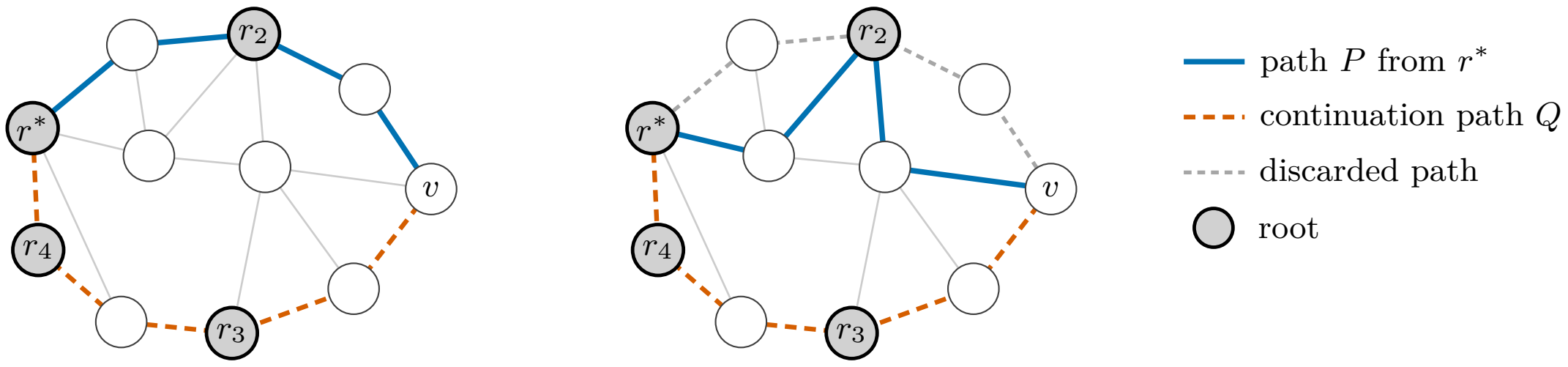

The dynamic program grows simple paths whose edges are independent in
$M$, records which roots they visit, and tests whether they can be
closed into a cycle. Adding vertices to the matroid ground set ensures
that representative-family compression preserves paths avoiding the
vertices of a possible continuation. \Cref{fig:ric-dp} illustrates the
path states and how a retained path preserves a possible completion.

\subparagraph*{Vertices as Matroid Elements.}
Let $U$ be the uniform matroid of rank $\kappa$ on $V'$, whose
independent sets are the subsets of $V'$ of size at most $\kappa$.
Choose pairwise distinct $a_v\in\mathbb{F}$ for $v\in V'$, and let
$A_U$ be the $\kappa\times|V'|$ matrix whose column for $v$ is
$(1,a_v,\ldots,a_v^{\kappa-1})^{\mathsf T}$. This is a Vandermonde
matrix: any $\kappa$ columns form a nonsingular matrix. Hence, $A_U$
has full row rank and represents $U$~\cite[Section~12.1.2]{C+15}.
Let $\widetilde M=U\oplus M$ be the direct sum~\cite[Section~12.1.5]{C+15} on the disjoint
union $V'\mathbin{\dot\cup}E$, whose independent sets are the sets
$X\cup F$ with $X\in\mathcal{I}(U)$ and $F\in\mathcal{I}$. The
block-diagonal matrix formed by the two representations has full row
rank and represents $\widetilde M$, which has rank
\[
    N=\kappa+r\leq2r-1.
\]
For an $r^*$-$v$ path $P$, let
$X_P=(V(P)\setminus\{r^*\})\mathbin{\dot\cup}E(P)$. If $P$ has $s$
vertices, then $|X_P|=2s-2$, and $X_P\in\mathcal{I}(\widetilde M)$ if and
only if $s\leq L$ and $E(P)\in\mathcal{I}$. The set $X_P$ determines $P$,
since parallel edges are distinct ground-set elements.

For $v\in V$, a set $T$ with $r^*\in T\subseteq R$, and $1\leq s\leq L$,
the \emph{state} $(v,T,s)$ has the path family
\[
    \mathcal{P}(v,T,s)=
    \left\{X_P:
    \begin{array}{l}
    P\text{ is an }r^*\text{-}v\text{ path},\ |V(P)|=s,\\
    V(P)\cap R=T,\quad E(P)\in\mathcal{I}
    \end{array}\right\}.
\]
We allow the path consisting only of $r^*$, so
$\mathcal{P}(r^*,\{r^*\},1)=\{\varnothing\}$, and all other states with
$s=1$ have empty families. Every member of $\mathcal{P}(v,T,s)$ is a set
of size $2s-2$ that is independent in $\widetilde M$. Any completion
of a path with $s$ vertices to a solution with at most $L$ vertices
adds at most $L-s$ vertices and $L-s+1$ edges, including the closing
edge. A continuation path $Q$ from $v$ to $r^*$ is encoded by the
extension set
$Y=(V(Q)\setminus\{v,r^*\})\mathbin{\dot\cup}E(Q)$: its endpoints are
already represented by $X_P$ or excluded from the ground set.
We therefore use the parameters
\begin{equation}\label{eq:representative-parameters}
    p=2s-2,\qquad q=N-2s+2,\qquad p+q=N=r(\widetilde M),
\end{equation}
as required by \Cref{thm:representative-families}. Since $\kappa\leq
r-1$, we have $q\geq N-2L+2=r-\kappa\geq1$ for all $s\leq L$, so even a
path on $L$ vertices retains an extension budget for the closing edge.
The uniform part only bounds the number of vertices; vertex-disjointness
is enforced by the condition $\widehat A\cap Y=\varnothing$ in
\Cref{def:representative-family}, applied to the vertices in $Y$.

The dynamic program stores families
$\widehat{\mathcal{P}}(v,T,s)$ and proceeds as follows.
\begin{description}
    \item[Initialization.]
    Set $\widehat{\mathcal{P}}(r^*,\{r^*\},1)=\{\varnothing\}$ and
    initialize every other retained family and every candidate family
    $\mathcal{C}(v,T,s)$ as empty.
    \item[Extension.]
    For $X_P\in\widehat{\mathcal{P}}(v,T,s)$ with $s<L$ and an edge
    $e=vw$ with $w\notin V(P)$, test whether
    $E(P)\cup\{e\}\in\mathcal{I}$. If so, insert $X_P\cup\{w,e\}$ into
    $\mathcal{C}(w,T\cup(\{w\}\cap R),s+1)$ and record a predecessor
    pointer to $X_P$ and $e$. Consider parallel edges separately.
    \item[Reduction.]
    For a state with $s\geq2$, once all incoming candidates have
    been generated, apply \Cref{thm:representative-families}
    in $\widetilde M$ to $\mathcal{C}(v,T,s)$ using
    \eqref{eq:representative-parameters}. Store the resulting family
    as $\widehat{\mathcal{P}}(v,T,s)$. It has size at most
    $\binom{N}{2s-2}$. Keep a predecessor pointer for every retained member.
    \item[Acceptance.]
    For every state $(v,R,s)$, every
    $X_P\in\widehat{\mathcal{P}}(v,R,s)$, and every edge
    $e$ between $v$ and $r^*$, accept if
    $E(P)\cup\{e\}\in\mathcal{I}$. Return the cycle formed by the path
    recovered from the predecessor pointers and $e$.
    Reject if no test succeeds.
\end{description}
Process states by increasing $s$, reducing each noninitial state
before using it for extensions or acceptance. Only retained families
generate candidates; the algorithm never enumerates the full
families $\mathcal{P}(v,T,s)$.

To justify repeated compression, we prove that each retained family
represents the full family of feasible paths in its state.

\begin{lemma}\label{lem:representative-invariant}
At every processed state $(v,T,s)$,
$\widehat{\mathcal{P}}(v,T,s)$ is an
$(N-2s+2)$-representative subfamily of
$\mathcal{P}(v,T,s)$ in $\widetilde M$.
\end{lemma}

\begin{proof}
Induct on $s$. For $s=1$, every retained family equals its path
family, namely $\{\varnothing\}$ for $(r^*,\{r^*\},1)$ and $\varnothing$
otherwise.

For $s\geq2$, let $q=N-2s+2$. We first show
$\mathcal{C}(v,T,s)\subseteq\mathcal{P}(v,T,s)$. A candidate in
$\mathcal{C}(v,T,s)$ has the form $X_{P'}\cup\{v,e\}$, where
$X_{P'}\in\widehat{\mathcal{P}}(u,T',s-1)$, $e=uv$, $v\notin V(P')$,
$E(P')\cup\{e\}\in\mathcal{I}$, and $T=T'\cup(\{v\}\cap R)$. By induction,
$P'$ is an $r^*$-$u$ path with $s-1$ vertices and $V(P')\cap R=T'$.
Appending $e$ gives an $r^*$-$v$ path $P$ with $s$ vertices,
$V(P)\cap R=T$, and $E(P)\in\mathcal{I}$, and the candidate equals
$X_P$. Hence
\[
    \widehat{\mathcal{P}}(v,T,s)
    \subseteq\mathcal{C}(v,T,s)
    \subseteq\mathcal{P}(v,T,s).
\]
For representativity, fix a last edge $e=uv$ and a predecessor state
$(u,T',s-1)$. The induction hypothesis gives a $(q+2)$-representative
family for that state. By the extension closure property established
in \Cref{sec:representative-families}, adding the two elements
$Z=\{v,e\}$ preserves representativity with the extension budget
reduced to $q$. Thus, its valid extensions form a $q$-representative
family for all paths obtained through this transition. Taking the union
over all predecessor states and last edges shows that
$\mathcal C(v,T,s)$ is $q$-representative for
$\mathcal P(v,T,s)$. Finally, the reduction step produces a
$q$-representative family for $\mathcal C(v,T,s)$; transitivity gives
the claim.
\end{proof}

A replacement path may use different non-root vertices, but its
endpoint, its number of vertices, and its set of visited roots are
unchanged. Representativity ensures that the replacement avoids the
vertices of the continuation encoded by the extension set.

\begin{lemma}\label{lem:representative-dp-correctness}
The dynamic program accepts if and only if $(G,R,M)$ is a yes-instance.
Every returned cycle is a solution.
\end{lemma}

\begin{proof}
Suppose that the algorithm accepts at a state $(v,R,s)$ with
$X_P\in\widehat{\mathcal{P}}(v,R,s)$ and an edge $e$ between $v$ and
$r^*$. By \Cref{lem:representative-invariant}, $P$ is an $r^*$-$v$ path
with $V(P)\cap R=R$. Thus $P$ has at least three edges, so its endpoints
are not consecutive on $P$ and $e\notin E(P)$. Hence $P$ and $e$ form a
cycle through $R$, whose edge set $E(P)\cup\{e\}$ is independent by the
acceptance test.

Conversely, let $C$ be a solution with $\ell$ vertices, choose an edge
$e=r^*v$ on $C$, and let $P=C-e$. Then $4\leq\ell\leq L$ and
$X_P\in\mathcal{P}(v,R,\ell)$. The set $Y=\{e\}$ satisfies
$X_P\cap Y=\varnothing$, and $X_P\cup Y=(V(C)\setminus\{r^*\})\cup E(C)$
is independent in $\widetilde M$, since $\ell-1\leq\kappa$ and
$E(C)\in\mathcal{I}$. As $q=N-2\ell+2\geq1$, applying
\Cref{lem:representative-invariant} gives a retained set
$X_{\widehat P}\in\widehat{\mathcal{P}}(v,R,\ell)$ with $e\notin
X_{\widehat P}$ and $X_{\widehat P}\cup\{e\}\in\mathcal{I}(\widetilde M)$,
so $E(\widehat P)\cup\{e\}\in\mathcal{I}$. Hence the corresponding
acceptance test succeeds.
\end{proof}

We now bound the total number of candidates and the cost of reducing
their families.

\begin{theorem}\label{thm:ric-fpt}
Let $(G,R,M)$ be an instance of RIC with $n$ vertices and $m$ edges,
where $M$ has rank $r$ and is represented by an $r\times m$ matrix over a
field $\mathbb{F}$ with $|\mathbb{F}|\geq n$. One can decide $(G,R,M)$
deterministically using
\[
    \Oh\bigl(2^{\omega(2r-1)}\cdot r^{\omega}\cdot m\bigr)
\]
field operations; all other operations take asymptotically at most
the same time. Whenever the algorithm accepts, it returns a solution.
\end{theorem}

\begin{proof}
The initial rejection for $r<4$ is correct, and for $r\geq4$ the
algorithm is correct by \Cref{lem:representative-dp-correctness}.
Constructing the representation of $\widetilde M$ takes $\Oh(rn+rm)$
time, and $n\leq m+1$ since $G$ is connected.

Fix a level $s$ with $2\leq s\leq L$. A retained family at level $s-1$ has
size at most $\binom{N}{2s-4}$. Each of its members generates at most
one candidate per edge incident with its endpoint. Summing over the
endpoints and the at most eight sets $T$ gives at most
\[
    N_s=16m\binom{N}{2s-4}
\]
candidates at level $s$, and hence
$\sum_{s=2}^{L}N_s\leq16m\cdot2^N$. Each candidate requires one
independence test of at most $r$ vectors in $\mathbb{F}^r$, which takes
$\Oh(r^\omega)$ field operations by computing a rank~\cite{IbarraMH82},
and $\Oh(r)$ further time for the test $w\notin V(P)$ and for storing
the set and its pointer. By \Cref{thm:representative-families} with
\eqref{eq:representative-parameters}, reducing a state with $a$
candidates takes $\Oh(a\binom{N}{2s-2}(2r)^{\omega}+a\binom{N}{2s-2}^{\omega-1})
\subseteq\Oh(a\cdot2^{(\omega-1)N}r^{\omega})$ field operations, since
$1\leq\binom{N}{2s-2}\leq2^N$ and $\omega-1\geq1$. There are at most
$2m\sum_{s}\binom{N}{2s-2}\leq2m\cdot2^N$ acceptance tests, each using
$\Oh(r^\omega)$ field operations. In total, the algorithm uses
\[
    \Oh\bigl(m\cdot2^N\cdot2^{(\omega-1)N}r^{\omega}\bigr)
    =\Oh\bigl(2^{\omega N}r^{\omega}m\bigr)
    \subseteq\Oh\bigl(2^{\omega(2r-1)}r^{\omega}m\bigr)
\]
field operations, and all other operations are within the same bound.
\end{proof}

\subsubsection{Completing the Randomized FPT Algorithm}
\label{sec:complete-main-proof}

We now combine the randomized reduction of \Cref{lem:m-witness-to-ric}
with the deterministic RIC algorithm of \Cref{thm:ric-fpt}. The reduction
produces at most $m^2$ instances of rank at most $k$, each solvable in
$2^{\Oh(k)}m$ time, giving the desired $2^{\Oh(k)}m^3$ total running
time. All randomness lies in the reduction, and every returned cycle
can be completed to an $M$-witness.

\begin{proof}[Proof of \Cref{thm:short-m-witness}]
Handle $n<4$, disconnected graphs, and $m<2n-4$ by rejection, as
justified by \Cref{lem:m-witness-size-bound}. We may now assume $m\geq n$.
Let $\ell=\min\{k,n,\delta+4\}$.
Use the sieve of Eratosthenes to list all primes up to $4n^2$, and
choose a prime $p$ between $2n^2$ and $4n^2$, whose existence follows
from Bertrand's postulate. Set $\mathbb F=\mathbb Z/p\mathbb Z$.
This takes $\Oh(n^2\log\log n)$ time~\cite[Section~5.4]{Shoup08}. Apply
\Cref{lem:m-witness-to-ric}, and solve every produced instance with
\Cref{thm:ric-fpt}. Each has rank $r\leq\ell\leq k$, so a call costs
$2^{\Oh(k)}m$ operations. Together with preprocessing and at most
$m^2$ calls, the total is
\[
 \Oh(n^2\log\log n+n^2m+\ell m^2
       +m^2(n+\ell^2m+2^{\Oh(k)}m))
 =2^{\Oh(k)}m^3.
\]
We use the same unit-cost field-arithmetic convention as in the RIC
algorithm; the field elements have $\Oh(\log n)$ bits.
Soundness follows from \Cref{lem:m-witness-to-ric}, for every outcome.
For a yes-instance its failure probability is at most
\[
 \frac{n-4+\ell}{|\mathbb F|}
 \leq\frac{2n-4}{2n^2}<\frac1n.
\]
An accepting call returns a cycle $C$. By \Cref{thm:reform}, the graph
$(G-E(C))/\rho$ has two edge-disjoint spanning trees. Delete its loops
and compute these trees in $\Oh(n^2)$ time with \Cref{thm:pebble} for
$c=2$, and lift them to the
two forests of an $M$-witness as in \Cref{thm:reform}. This preserves
$C$ and hence its length bound. Reconstruction is performed once.
\end{proof}

\section{A Sharp Maximum-Degree Threshold and Few Edges of Multiplicity One}
\label{sec:degree-complexity}

We establish a sharp boundary, unless P${}={}$NP, between
polynomial-time solvability and NP-completeness at maximum degrees four
and five, where degrees are counted with multiplicity. The same
reductions also yield lower bounds under ETH. Finally, we show that
\Gossip remains NP-complete on multigraphs with at most four edges of
multiplicity one.

We begin with the polynomial-time side of the degree threshold, which
follows from the XP algorithm for the excess.

\begin{corollary}\label{cor:degree-four}
\Gossip can be solved in $\Oh(n^3)$ time on multigraphs of maximum
degree at most four, where degrees count parallel edges with
multiplicity.
\end{corollary}
\begin{proof}
Handle $n<4$ directly. For $n\geq4$,
reject when $m<2n-4$: by \Cref{thm:characterization}, a yes-instance
has a $T$-witness, which uses at least $2n-3$ edges, or an $M$-witness,
which uses at least $2n-4$ edges by \Cref{lem:m-witness-size-bound}.
Otherwise, $m\leq2n$ implies $0\leq\delta=m-(2n-4)\leq4$.
Test whether $G$ is a $T$-graph in $\Oh(n^2)$ time using
\Cref{lem:t-graph-test}, and accept if so. Otherwise, by
\Cref{thm:characterization,lem:m-witness-size-bound}, $G$ is a
yes-instance if and only if it admits an $M$-witness with cycle length
at most $k=\delta+4\leq8$. Since the support has maximum degree
$\Delta\leq4$, \Cref{prop:above-xp} gives running time
$\Oh(8\cdot2n\cdot4^6\cdot n^2)=\Oh(n^3)$.
\end{proof}

\subsection{Planar Graphs of Maximum Degree Five}

For the hardness direction, we use the intermediate problem in which
one path and two spanning trees must be pairwise edge-disjoint. We first
recall its connection to \Gossip and then strengthen its hardness to
planar simple graphs of maximum degree five.

\problemdef{PathTreeTree (PTT)}{An undirected multigraph $G$ and distinct
vertices $s,t$.}{Are there an $s$-$t$ path $P$ and two spanning trees
$T_1,T_2$ that are pairwise edge-disjoint?}

G\"obel, Cerdeira, and Veldman established hardness of PTT in both
simple and restricted multigraphs.
\begin{theorem}[{\cite[Theorem~16 and its proof]{gobel1991label}}]\label{thm:ptt-hard-goebel}
PTT is NP-hard on simple graphs and on multigraphs in which every
parallel class has exactly two edges.
\end{theorem}
We strengthen the simple-graph result by imposing planarity and
maximum degree five. The multigraph result will also give hardness
when only a few edges have multiplicity one.

The following classical reduction transfers a PTT instance to
\Gossip by adding only a constant-size path.
\begin{figure}[t]
\centering
\begin{tikzpicture}[every node/.style={font=\footnotesize},
  pathP/.style={black,line width=1.6pt},
  newv/.style={vertex,fill=black!10}]
\begin{scope}
  \node[draw,rounded corners,minimum width=38mm,minimum height=32mm,
        fill=black!3] (G1) at (0,0) {};
  \node at (-1.6,1.25) {$G$};
  \coordinate (s) at (-1.9,0);
  \coordinate (t) at (1.9,0);
  \coordinate (u1) at (-0.6,-0.65);
  \coordinate (u2) at (0.6,-0.65);
  \coordinate (w1) at (-0.9,0.75);
  \coordinate (w2) at (0.9,0.75);
  \draw[blueforest] (s)--(w1)--(w2)--(t) (w1)--(u1) (w2)--(u2);
  \draw[orangeforest] (s)--(w2) (w2)--(u1) (s) to[out=70,in=110] (t) (t)--(w1) (w1)--(u2);
  \draw[pathP] (s) to[bend right=25] (u1) to[bend right=10] (u2) to[bend right=25] (t);
  \foreach \v in {u1,u2,w1,w2} \node[smallv] at (\v) {};
  \node[vertex] at (s) {$s$};
  \node[vertex] at (t) {$t$};
  \node[panel,text width=45mm,align=center] at (0,-2.1)
     {(a) PTT solution $(P,T_1,T_2)$ in $G$};
\end{scope}
\draw[-{Latex[length=3mm]},thick] (2.45,0)--(3.3,0);
\begin{scope}[xshift=5.8cm]
  \node[draw,rounded corners,minimum width=38mm,minimum height=32mm,
        fill=black!3] (G2) at (0,0) {};
  \node at (-1.6,1.25) {$G$};
  \coordinate (s) at (-1.9,0);
  \coordinate (t) at (1.9,0);
  \coordinate (u1) at (-0.6,-0.65);
  \coordinate (u2) at (0.6,-0.65);
  \coordinate (w1) at (-0.9,0.75);
  \coordinate (w2) at (0.9,0.75);
  \coordinate (z1) at (-1.35,-1.75);
  \coordinate (z2) at (0,-2.05);
  \coordinate (z3) at (1.35,-1.75);
  \draw[blueforest] (s)--(w1)--(w2)--(t) (w1)--(u1) (w2)--(u2);
  \draw[orangeforest] (s)--(w2) (w2)--(u1) (s) to[out=70,in=110] (t) (t)--(w1) (w1)--(u2);
  \draw[pathP] (s) to[bend right=25] (u1) to[bend right=10] (u2) to[bend right=25] (t);
  \draw[pathP] (s)--(z1)--(z2)--(z3)--(t);
  \foreach \v in {u1,u2,w1,w2} \node[smallv] at (\v) {};
  \foreach \v/\l in {z1/z_1,z2/z_2,z3/z_3} \node[newv] at (\v) {$\l$};
  \node[vertex] at (s) {$s$};
  \node[vertex] at (t) {$t$};
  \node[panel,text width=55mm,align=center] at (0,-2.75)
     {(b) $M$-witness $(P\cup Q,\,T_1,\,T_2)$ in $G'$};
\end{scope}
\end{tikzpicture}
\caption{The classical reduction from PTT to \Gossip
(\Cref{thm:red-ptt-to-gossip}). The graph $G'$ is obtained from $G$ by
adding an $s$-$t$ path $Q$ of length four with three new vertices
$z_1,z_2,z_3$. (a) A solution of the PTT instance: an $s$-$t$ path $P$
(bold) and two spanning trees $T_1$ (solid) and $T_2$ (dashed), pairwise
edge-disjoint, shown for a small graph~$G$. (b) The cycle $P\cup Q$
together with $T_1$ and $T_2$ is an $M$-witness of $G'$: each $T_i$ is a
spanning forest of $G'$ whose four components are $V(G)$ and the three
singletons $\{z_1\},\{z_2\},\{z_3\}$, all of which meet the cycle.}
\label{fig:ptt-to-gossip}
\end{figure}
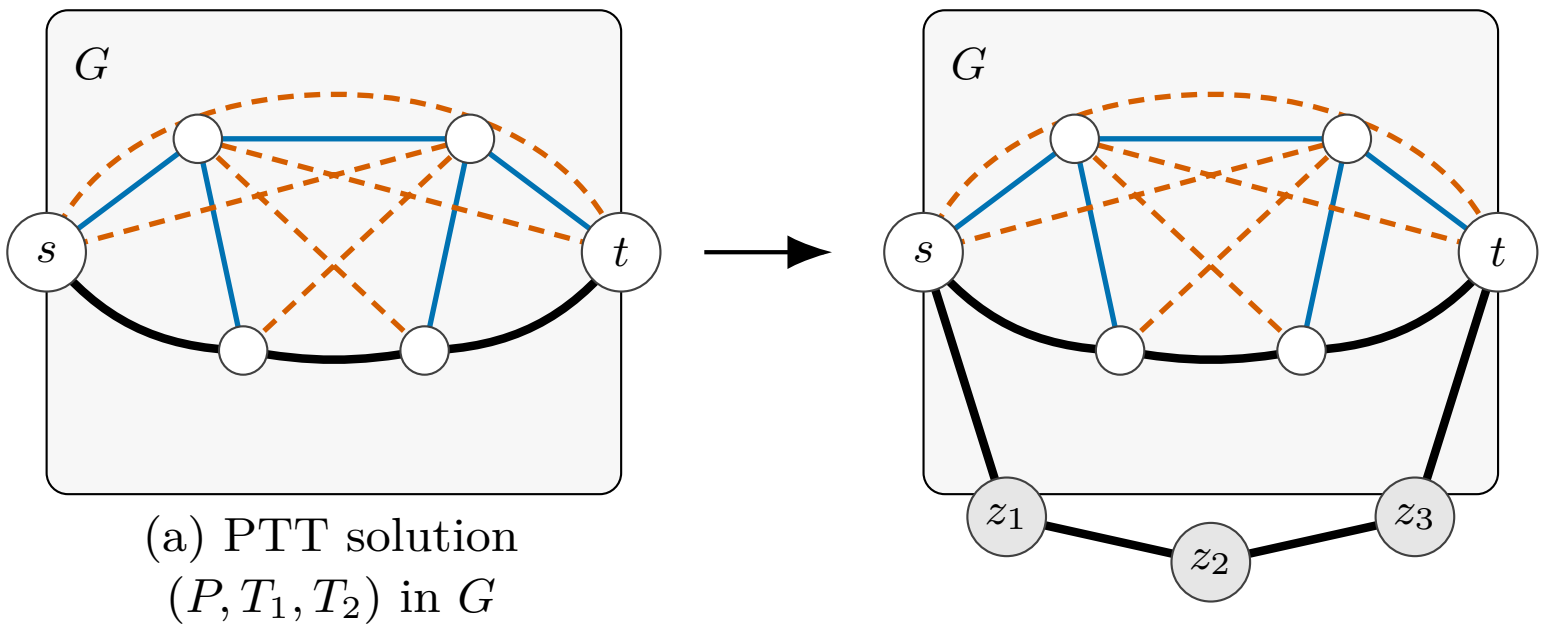

\begin{theorem}[{\cite[proof of Theorem~17]{gobel1991label}}]\label{thm:red-ptt-to-gossip}
Given a PTT instance $(G,s,t)$, add an $s$-$t$ path of length four with
three new internal vertices. The resulting graph is a yes-instance
of \Gossip if and only if $(G,s,t)$ is a yes-instance of PTT.
\end{theorem}
\Cref{fig:ptt-to-gossip} illustrates how a PTT solution extends to an
$M$-witness after adding the path. We include the proof of the converse
as well.
\begin{proof}
Let $Q=(s,z_1,z_2,z_3,t)$ be the added path and $G'$ the resulting
graph. Each $z_j$ has degree two in $G'$, and both of its edges lie
on $Q$.
We first note that every subgraph $F$ of $G'$ in which each $z_j$ is
connected to $s$ or $t$ contains at least three edges of $Q$.
Otherwise, $Q$ loses two edges, and the vertices of $Q$ strictly
between them form a nonempty set of vertices $z_j$ without any other
incident edges in $F$.

Suppose that $(P,T_1,T_2)$ solves $(G,s,t)$. Then $C=P\cup Q$ is a
cycle of length at least five. In $G'$, each $T_i$ is a spanning forest
whose four components are $V(G)$ and $\{z_1\},\{z_2\},\{z_3\}$, and
each component meets $C$. The three subgraphs are pairwise
edge-disjoint, so $(C,T_1,T_2)$ is an $M$-witness of $G'$. By
\Cref{thm:characterization}, $G'$ is a yes-instance.

Conversely, suppose that $G'$ is a yes-instance. By
\Cref{thm:characterization}, it has a $T$-witness or an $M$-witness.
By the first paragraph, every spanning tree of $G'$ contains at least
three of the four edges of $Q$, so any two share at least two edges.
Hence $G'$ has an $M$-witness $(C,F_1,F_2)$. If $C$ contains
some $z_j$, then it uses both edges of $z_j$ and, repeating this
argument, contains $Q$. Otherwise $V(C)\subseteq V(G)$, so every
$z_j$ is connected in $F_i$ to a vertex of $C$, and thus to $s$ or $t$.
Then $F_1$ and $F_2$ each contain three edges of $Q$, which is
impossible for edge-disjoint subgraphs. Hence $Q\subseteq C$, and
$P=C-\{z_1,z_2,z_3\}$ is an $s$-$t$ path in $G$. Since $F_1,F_2$
avoid $E(Q)$, each $z_j$ is an isolated vertex of each $F_i$. These are
three of the four components of $F_i$, so $T_i=F_i-\{z_1,z_2,z_3\}$ is a
spanning tree of $G$. The subgraphs $P,T_1,T_2$ are pairwise
edge-disjoint, as required.
\end{proof}

To obtain hardness under a degree bound, we need a planar attachment
gadget that supplies two spanning trees. The next lemma provides
arbitrarily many attachment vertices while keeping the degree at most
five.
\begin{lemma}\label{lem:docking}
For each positive integer $h$, one can construct in $\Oh(h)$ time a
simple plane graph with $\Oh(h)$ vertices, maximum degree five, two
edge-disjoint spanning trees, and at least $h$ specified degree-two
vertices incident with the outer face. Each specified vertex appears
exactly once along the outer facial walk.
\end{lemma}
\begin{proof}
Call the specified vertices \emph{ports}. Start with two disjoint
copies of $K_4$, with vertices $u_i,v_i,w_i,z_i$ for $i=1,2$, drawn
with outer triangle $u_iv_iw_i$ and $z_i$ inside.
Add edges $u_1u_2$ and $v_1v_2$, drawing the two triangles side by side.
For each $i$, add a new vertex adjacent to $u_i,w_i$ and another new
vertex adjacent to $v_i,w_i$, placing both outside the triangles.
The four new vertices are ports. The outer facial walk is the
ten-vertex cycle through $u_1$, the port at $u_1,w_1$, $w_1$, the
port at $v_1,w_1$, $v_1$, $v_2$, the port at $v_2,w_2$, $w_2$, the
port at $u_2,w_2$, and $u_2$; it contains each port once.
The resulting graph has twelve vertices. Each $u_i,v_i,w_i$ has
degree five, each $z_i$ has degree three, and each port has degree two.
In each $K_4$, the paths with edge sets
$\{u_iv_i,v_iw_i,w_iz_i\}$ and
$\{u_iw_i,u_iz_i,v_iz_i\}$ are edge-disjoint spanning trees.
Use one joining edge in each tree, and attach each port by a different
one of its incident edges to each tree. This certifies two
edge-disjoint spanning trees of the whole base graph.

To combine two gadgets with port sets $\Pi_1,\Pi_2$, choose consecutive
ports $o_1,o_2$ along the outer facial walk of the first and a port
$o_3$ of the second. Add $o_1o_3,o_2o_3$. Embed the second gadget in the
outer-face sector adjacent to the boundary arc from $o_1$ to $o_2$ that
has no other port. The new outer facial walk follows the walk of the
first gadget from $o_2$ to $o_1$ through its other ports, the edge
$o_1o_3$, the complete outer facial walk of the second gadget from
$o_3$ back to $o_3$, and the edge $o_3o_2$. It is not a cycle,
since $o_3$ is a cut vertex, but it contains every remaining port
exactly once. The new port set is
$(\Pi_1\cup\Pi_2)\setminus\{o_1,o_2,o_3\}$. Degrees remain at most five.
Join the first trees by $o_1o_3$ and the second trees by $o_2o_3$.
This gives edge-disjoint spanning trees and $|\Pi_1|+|\Pi_2|-3$ ports.
Combining $g$ base gadgets therefore gives $g+3$ ports on $12g$
vertices. Take $g=\max\{1,h-3\}$. The construction and its plane
embedding take linear time.
\end{proof}

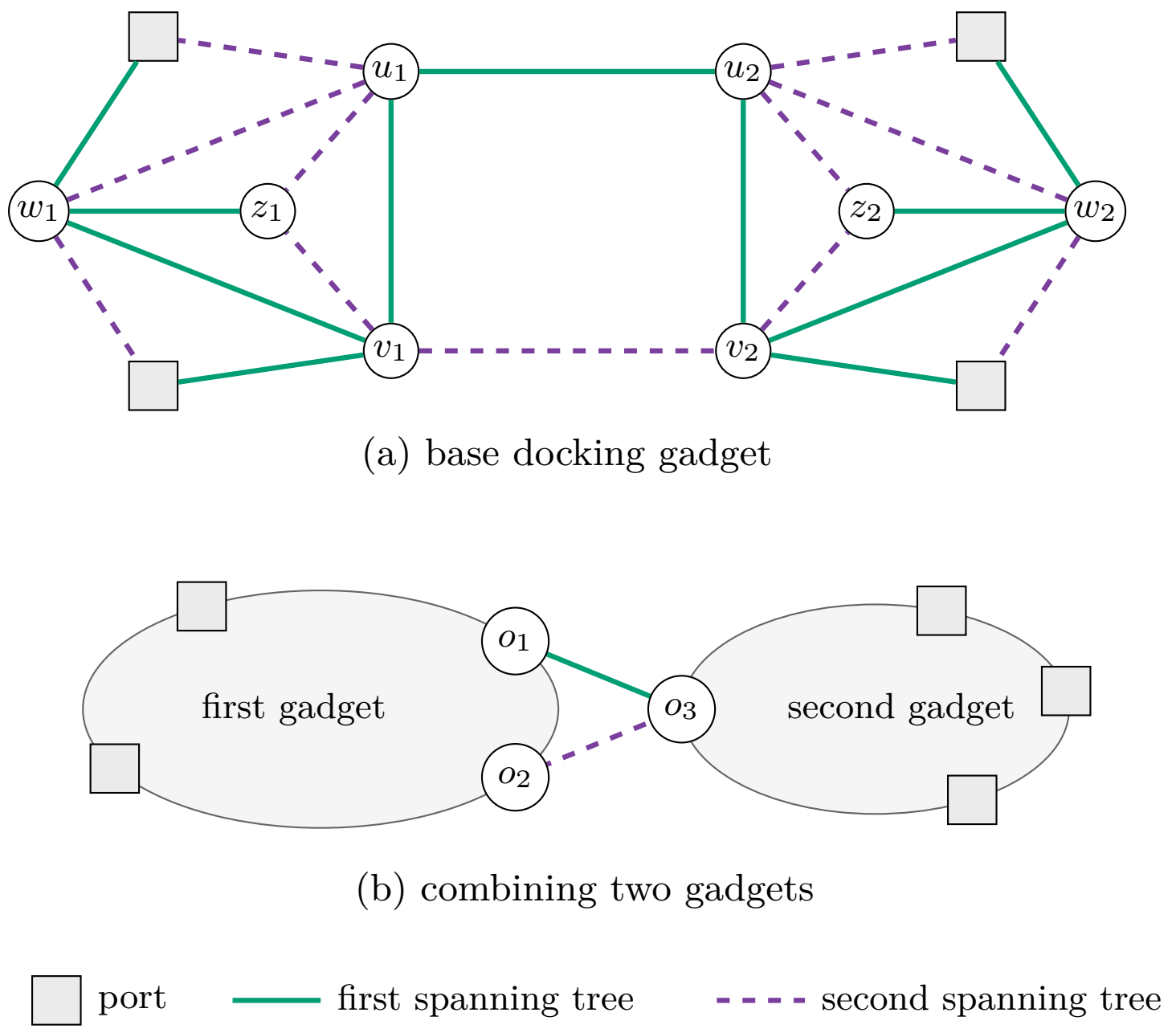
\begin{figure}[t]
\centering
\begin{tikzpicture}[
  x=1.45cm,y=1.15cm,
  every node/.style={font=\footnotesize},
  v/.style={circle,draw,fill=white,inner sep=1pt,minimum size=4.5mm},
  port/.style={rectangle,draw,fill=black!8,inner sep=1.5pt,minimum size=4mm},
  first/.style={auxgreen,very thick},
  second/.style={auxpurple,very thick,dashed}
]
  \foreach \i/\x in {1/-1,2/1}{
    \node[v] (A\i) at (\x,1) {$u_\i$};
    \node[v] (B\i) at (\x,-1) {$v_\i$};
    \node[v] (C\i) at ({3*\x},0) {$w_\i$};
    \node[v] (D\i) at ({1.7*\x},0) {$z_\i$};
    \node[port] (P\i) at ({2.35*\x},1.25) {};
    \node[port] (Q\i) at ({2.35*\x},-1.25) {};
    \draw[first] (A\i)--(B\i)--(C\i)--(D\i);
    \draw[second] (A\i)--(C\i) (A\i)--(D\i) (B\i)--(D\i);
    \draw[first] (C\i)--(P\i) (B\i)--(Q\i);
    \draw[second] (A\i)--(P\i) (C\i)--(Q\i);
  }
  \draw[first] (A1)--(A2);
  \draw[second] (B1)--(B2);
  \node[font=\small] at (0,-1.75) {(a) base docking gadget};

  \begin{scope}[yshift=-4.1cm]
    \draw[fill=black!4,draw=black!60] (-1.4,0) ellipse (1.35 and 0.85);
    \draw[fill=black!4,draw=black!60] (1.75,0) ellipse (1.1 and 0.75);
    \node at (-1.55,0) {first gadget};
    \node at (1.9,0) {second gadget};
    \coordinate (o1) at ($(-1.4,0)+(35:1.35 and 0.85)$);
    \coordinate (o2) at ($(-1.4,0)+(-35:1.35 and 0.85)$);
    \coordinate (p1) at ($(-1.4,0)+(120:1.35 and 0.85)$);
    \coordinate (p2) at ($(-1.4,0)+(210:1.35 and 0.85)$);
    \coordinate (o3) at ($(1.75,0)+(180:1.1 and 0.75)$);
    \coordinate (p3) at ($(1.75,0)+(70:1.1 and 0.75)$);
    \coordinate (p4) at ($(1.75,0)+(-60:1.1 and 0.75)$);
    \coordinate (p5) at ($(1.75,0)+(10:1.1 and 0.75)$);
    \draw[first] (o1)--(o3);
    \draw[second] (o2)--(o3);
    \foreach \p in {p1,p2,p3,p4,p5} \node[port] at (\p) {};
    \node[v,minimum size=5.5mm] at (o1) {$o_1$};
    \node[v,minimum size=5.5mm] at (o2) {$o_2$};
    \node[v,minimum size=5.5mm] at (o3) {$o_3$};
    \node[font=\small] at (0.1,-1.3) {(b) combining two gadgets};
  \end{scope}

  \node[port] (legport) at (-2.9,-5.65) {};
  \node[anchor=west] at (-2.75,-5.65) {port};
  \draw[first] (-1.9,-5.65)--(-1.4,-5.65)
    node[right,text=black] {first spanning tree};
  \draw[second] (0.85,-5.65)--(1.35,-5.65)
    node[right,text=black] {second spanning tree};
\end{tikzpicture}
\caption{The docking gadgets of \Cref{lem:docking}. (a) The
twelve-vertex base gadget. Solid green and dashed purple edges
partition its edges into two spanning trees; the four square vertices
are degree-two ports on the outer face. (b) Combining two gadgets: the
consecutive ports $o_1,o_2$ of the first gadget are joined to a port
$o_3$ of the second gadget. The edge $o_1o_3$ joins the first spanning
trees and $o_2o_3$ joins the second ones. The vertices $o_1,o_2,o_3$
are no longer ports, so the combined gadget has $4+4-3=5$ ports, all on
the outer face.}
\label{fig:docking-gadget}
\end{figure}

The base case of the construction and its two spanning trees are shown
in \Cref{fig:docking-gadget}(a); larger gadgets are obtained by
repeatedly joining copies at three ports, as shown in
\Cref{fig:docking-gadget}(b).

The docking gadget allows us to preserve two spanning trees while
providing sufficiently many attachment vertices of bounded degree.
We now use it to prove hardness of PTT under the required restrictions.
\begin{theorem}\label{thm:ptt-hard-deg5}
PTT is NP-hard on planar simple graphs of maximum degree five, even
when $s,t$ have degree three and lie on the outer face of a supplied
plane embedding.
\end{theorem}
\begin{proof}
The reduction encodes each variable by a cycle with two sign branches.
A solution path selects the branches that represent an assignment,
while the two spanning trees enforce that every clause has a satisfied
literal. Docking gadgets supply the required trees without allowing the
solution path to enter them. We first normalize the source instance,
then describe the construction and verify its degree and planarity
bounds before proving correctness.

\subparagraph*{Source Problem and Normalization.}
We reduce from \emph{Sided Var-Linked Planar Monotone 3SAT} with at most
two occurrences of each literal. This problem is NP-hard by
Brunner et al.~\cite{BrunnerEtAlSoloChess}; see Lemma~B.1 in Appendix~B
of the full version~\cite{BrunnerEtAlSoloChessFull}. We use exactly the
following properties of its instances. Each clause consists of at most
three literals, which are either all positive or all negative. Each
literal occurs in at most two clauses. A rectilinear drawing is
supplied with the instance: the variables are disjoint horizontal
segments on a common horizontal line, the positive clauses are
horizontal segments above this line, the negative clauses are
horizontal segments below it, and each occurrence is a vertical
connection between the segments of its variable and its clause.
No two of these segments and connections cross.
The left-to-right order of the variables on the line supplies the
backbone order required below; we use it to preserve planarity.

Apply pure-literal elimination until each remaining variable has at
least one occurrence of each sign. This preserves satisfiability and
the restricted drawing. Trivial resulting instances can be mapped
to fixed yes- or no-instances. The elimination is needed because the
construction below gives each sign of each variable two occurrence
vertices. An occurrence vertex without a clause gadget would have
degree two, and it would be isolated in $G-E(P)$ whenever the path $P$
uses its branch.
Duplicate any literal with a single
occurrence within its clause. Each literal now has exactly two
occurrences, clauses have size at most six, and the duplicate
incidence connections can be drawn next to the originals.
Occurrences, including repeated literals in one clause, are treated
as distinct objects.

\subparagraph*{Construction.}
Let the variables be $x_1,\ldots,x_p$ in backbone order.
For each $x_i$ create the cycle
\[
 C_i=(s_i^+,s_i,s_i^-,x_i^{-,1},x_i^-,x_i^{-,2},
       t_i^-,t_i,t_i^+,x_i^{+,2},x_i^+,x_i^{+,1}).
\]
\begin{figure}[t]
\centering
\begin{tikzpicture}[
  x=1.0cm,y=0.95cm,
  every node/.style={font=\footnotesize},
  v/.style={circle,draw,fill=white,inner sep=0pt,minimum size=6.4mm,font=\scriptsize},
  occ/.style={v,fill=black!15,minimum size=7.2mm},
  port/.style={rectangle,draw,fill=black!8,inner sep=0pt,minimum size=1.8mm},
  attach/.style={black!55,line width=.6pt},
  backbone/.style={line width=2pt},
  clause/.style={black!60,line width=.6pt}
]
  \coordinate (sp) at (0,2);   \coordinate (s) at (0,0);   \coordinate (sm) at (0,-2);
  \coordinate (xp1) at (2,2);  \coordinate (xp) at (4,2);  \coordinate (xp2) at (6,2);
  \coordinate (xm1) at (2,-2); \coordinate (xm) at (4,-2); \coordinate (xm2) at (6,-2);
  \coordinate (tp) at (8,2);   \coordinate (t) at (8,0);   \coordinate (tm) at (8,-2);
  \draw[rounded corners=4pt,fill=black!4,draw=black!60] (1.5,-1.15) rectangle (6.5,1.15);
  \node[font=\footnotesize,text=black!70] at (4,0) {docking gadget $H_i$};
  \foreach \u/\pa/\pb in {
      sp/{1.85,1.15}/{1.5,0.8},  s/{1.5,0.25}/{1.5,-0.25},  sm/{1.5,-0.8}/{1.85,-1.15},
      xp/{3.7,1.15}/{4.3,1.15},  xm/{3.7,-1.15}/{4.3,-1.15},
      tp/{6.15,1.15}/{6.5,0.8},  t/{6.5,0.25}/{6.5,-0.25},  tm/{6.5,-0.8}/{6.15,-1.15}}{
    \draw[attach] (\u)--(\pa) (\u)--(\pb);
    \node[port] at (\pa) {}; \node[port] at (\pb) {};
  }
  \draw[line width=.8pt] (sp)--(s)--(sm)--(xm1)--(xm)--(xm2)--(tm)--(t)--(tp)
        --(xp2)--(xp)--(xp1)--(sp);
  \draw[backbone] (-1,2)--(sp) (tp)--(9,2);
  \draw[backbone] (-1,0)--(s) (t)--(9,0);
  \draw[backbone] (-1,-2)--(sm) (tm)--(9,-2);
  \foreach \u in {xp1,xp2} \draw[clause] ($(\u)+(-0.1,0)$)--++(-0.12,0.85) ($(\u)+(0.1,0)$)--++(0.12,0.85);
  \foreach \u in {xm1,xm2} \draw[clause] ($(\u)+(-0.1,0)$)--++(-0.12,-0.85) ($(\u)+(0.1,0)$)--++(0.12,-0.85);
  \foreach \u/\l in {sp/s_i^+,s/s_i,sm/s_i^-,xp/x_i^+,xm/x_i^-,tp/t_i^+,t/t_i,tm/t_i^-}
    \node[v] at (\u) {$\l$};
  \foreach \u/\l in {xp1/x_i^{+,1},xp2/x_i^{+,2},xm1/x_i^{-,1},xm2/x_i^{-,2}}
    \node[occ] at (\u) {$\l$};
  \draw[decorate,decoration={brace,amplitude=5pt}] (-0.2,3.05)--(8.2,3.05)
    node[midway,above=5pt] {positive sign branch $B_i^+$ (to positive clauses)};
  \draw[decorate,decoration={brace,amplitude=5pt,mirror}] (-0.2,-3.05)--(8.2,-3.05)
    node[midway,below=5pt] {negative sign branch $B_i^-$ (to negative clauses)};
  \node[align=center,font=\scriptsize] at (-1.55,0) {from\\$x_{i-1}$};
  \node[align=center,font=\scriptsize] at (9.55,0) {to\\$x_{i+1}$};
\end{tikzpicture}
\caption{The variable gadget for $x_i$ in the reduction of
\Cref{thm:ptt-hard-deg5}. The twelve displayed vertices form the cycle
$C_i$. Each of the eight white cycle vertices has two edges to distinct
ports (small squares) of the internal docking gadget $H_i$, in cyclic
order. Each gray occurrence vertex has two edges to ports of its clause
gadget. The three backbone tracks (thick) connect consecutive variable
gadgets. A solution path crosses the gadget along exactly one sign
branch.}
\label{fig:variable-gadget}
\end{figure}
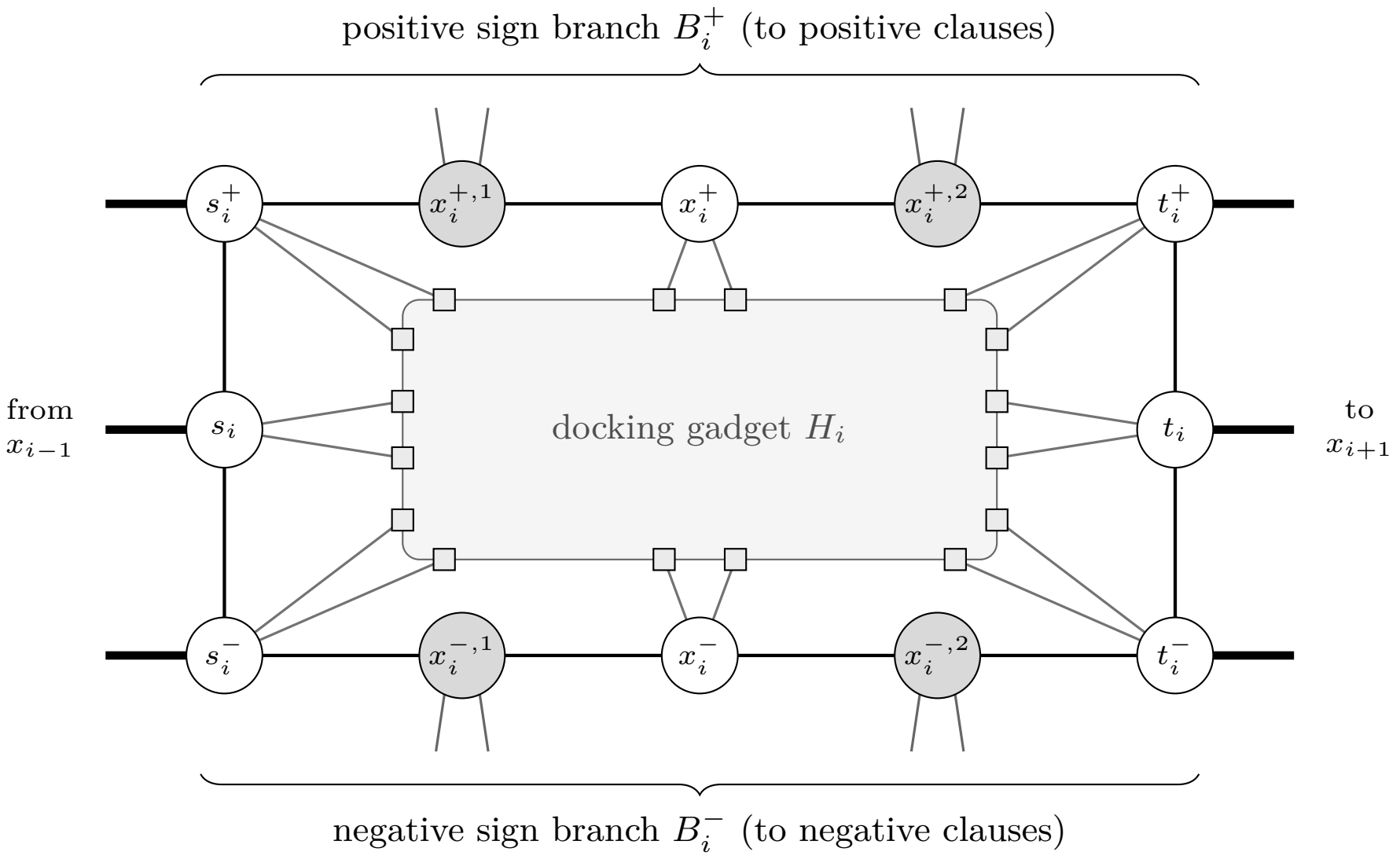

\begin{figure}[t]
\centering
\begin{tikzpicture}[
  x=0.62cm,y=0.62cm,
  every node/.style={font=\scriptsize},
  d/.style={circle,draw=black!80,fill=white,inner sep=0pt,minimum size=2.2mm},
  od/.style={d,fill=black!35},
  term/.style={circle,draw,fill=white,inner sep=0pt,minimum size=5mm,font=\footnotesize},
  attach/.style={black!45,line width=.5pt},
  backbone/.style={line width=1.4pt},
  hl/.style={auxgreen!45,line width=5pt,line cap=round,line join=round},
  gadget/.style={rounded corners=3pt,fill=black!5,draw=black!55}
]
  \foreach \i/\o in {1/0,2/6}{
    \coordinate (sp\i) at (\o+0,1.5);  \coordinate (s\i) at (\o+0,0);  \coordinate (sm\i) at (\o+0,-1.5);
    \coordinate (a\i)  at (\o+1,1.5);  \coordinate (xp\i) at (\o+2,1.5); \coordinate (b\i) at (\o+3,1.5);
    \coordinate (c\i)  at (\o+1,-1.5); \coordinate (xm\i) at (\o+2,-1.5); \coordinate (e\i) at (\o+3,-1.5);
    \coordinate (tp\i) at (\o+4,1.5);  \coordinate (t\i) at (\o+4,0);  \coordinate (tm\i) at (\o+4,-1.5);
  }
  \coordinate (sstar) at (-2,0);
  \coordinate (tstar) at (12,0);
  \draw[hl] (sstar)--(s1)--(sm1)--(c1)--(xm1)--(e1)--(tm1)--(t1)--(s2)--(sp2)--(a2)--(xp2)--(b2)--(tp2)--(t2)--(tstar);
  \draw[gadget] (0.3,3.2) rectangle (9.7,4.4);
  \node[font=\footnotesize] at (5,3.8) {clause gadget $H_{c^+}$ for $c^+=(x_1\lor x_2)$};
  \draw[gadget] (0.3,-3.2) rectangle (9.7,-4.4);
  \node[font=\footnotesize] at (5,-3.8) {clause gadget $H_{c^-}$ for $c^-=(\lnot x_1\lor\lnot x_2)$};
  \foreach \i in {1,2}{
    \draw[gadget] ($(s\i)+(0.55,-0.75)$) rectangle ($(t\i)+(-0.55,0.75)$);
    \node at ($(s\i)!0.5!(t\i)$) {$H_{\i}$};
    \foreach \u/\dx/\dy in {sp\i/0.45/-0.6,s\i/0.55/0,sm\i/0.45/0.6,xp\i/0/-0.75,xm\i/0/0.75,tp\i/-0.45/-0.6,t\i/-0.55/0,tm\i/-0.45/0.6}
      \draw[attach] ($(\u)$)--($(\u)+(\dx,\dy)+(0.08,0.08)$) ($(\u)$)--($(\u)+(\dx,\dy)+(-0.08,-0.08)$);
    \foreach \u in {a\i,b\i}
      \draw[attach] (\u)--($(\u)+(-0.15,1.7)$) (\u)--($(\u)+(0.15,1.7)$);
    \foreach \u in {c\i,e\i}
      \draw[attach] (\u)--($(\u)+(-0.15,-1.7)$) (\u)--($(\u)+(0.15,-1.7)$);
    \draw[line width=.7pt] (sp\i)--(s\i)--(sm\i)--(c\i)--(xm\i)--(e\i)--(tm\i)--(t\i)--(tp\i)--(b\i)--(xp\i)--(a\i)--(sp\i);
  }
  \draw[backbone] (tp1)--(sp2) (t1)--(s2) (tm1)--(sm2);
  \draw[backbone] (sstar)--(sp1) (sstar)--(s1) (sstar)--(sm1);
  \draw[backbone] (tstar)--(tp2) (tstar)--(t2) (tstar)--(tm2);
  \foreach \i in {1,2}{
    \foreach \u in {sp\i,s\i,sm\i,xp\i,xm\i,tp\i,t\i,tm\i} \node[d] at (\u) {};
    \foreach \u in {a\i,b\i,c\i,e\i} \node[od] at (\u) {};
    \node[font=\footnotesize] at ($(s\i)+(2,2.3)$) {$x_\i$};
  }
  \node[term] at (sstar) {$s^*$};
  \node[term] at (tstar) {$t^*$};
\end{tikzpicture}
\caption{The reduction of \Cref{thm:ptt-hard-deg5} for the formula
$(x_1\lor x_2)\land(\lnot x_1\lor\lnot x_2)$. After normalization, each
literal occurs twice (the single occurrences are duplicated within
their clauses), so each clause gadget is attached to four occurrence
vertices (dark gray). Short double strokes indicate the two edges from
each cycle vertex to the ports of the docking gadgets. The highlighted
$s^*$-$t^*$ path corresponds to the satisfying assignment $x_1=1$,
$x_2=0$: in each variable gadget, it traverses the sign branch of the
false literal. The occurrence vertices of the true literals keep both
cycle edges, which attach each clause gadget to the two spanning
trees.}
\label{fig:construction-example}
\end{figure}
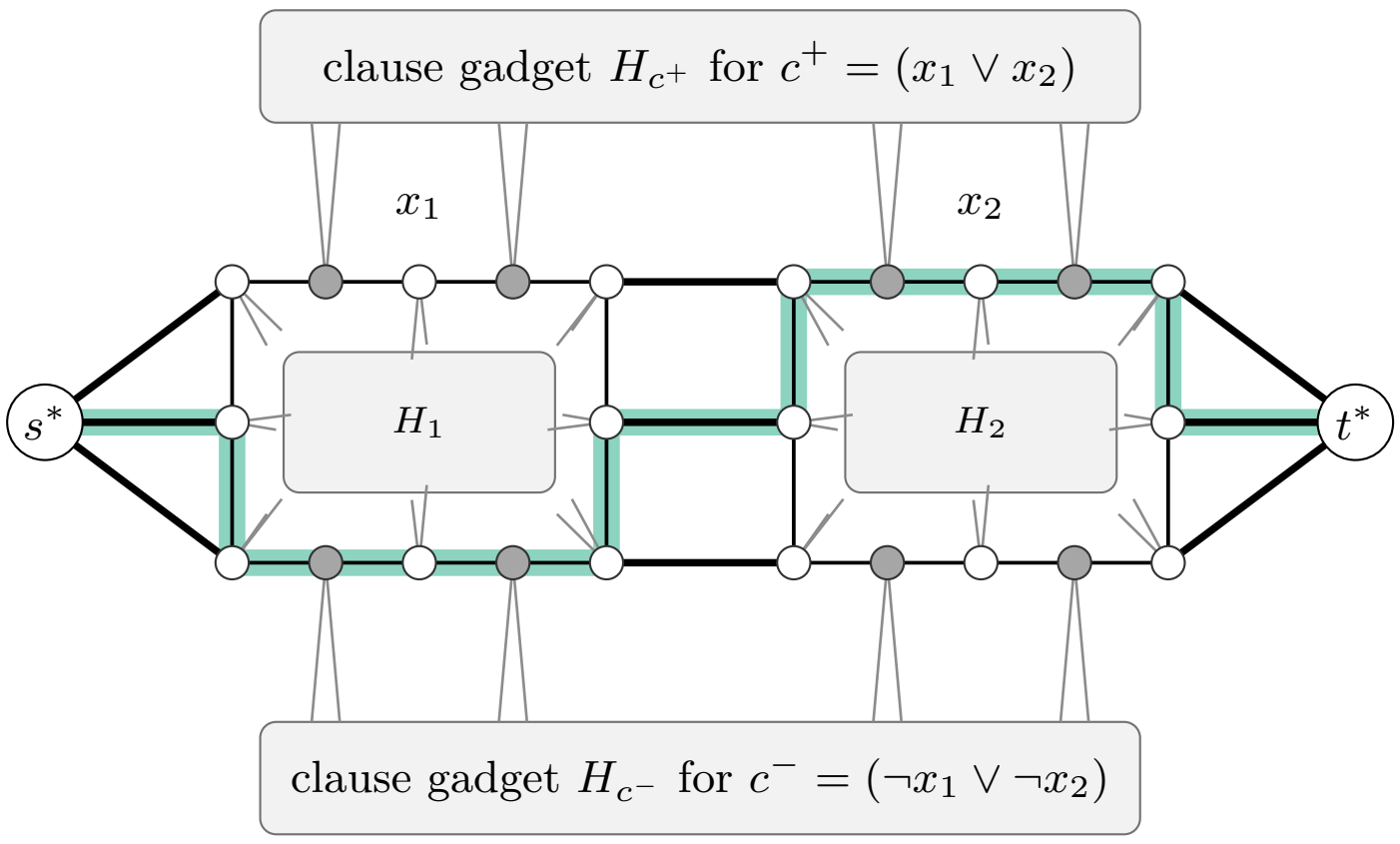

The layout in \Cref{fig:variable-gadget} makes the three backbone
tracks and the two sign branches explicit, and
\Cref{fig:construction-example} shows the complete construction for a
small formula.
The four vertices $x_i^{\circ,j}$ with $\circ\in\{+,-\}$ and
$j\in\{1,2\}$ are the occurrence vertices. For each sign, associate
its two occurrences bijectively with these two vertices in the
order of the supplied drawing. This also handles duplicate literals.
Call the other eight vertices $V_i$.
Place inside $C_i$ a docking gadget $H_i$ from \Cref{lem:docking} with
at least sixteen ports. Attach each vertex of $V_i$ to two distinct
ports, using sixteen different ports altogether in cyclic order.
The attachment edges can be drawn without crossings in the annulus
between $C_i$ and $H_i$.

For every $i<p$, add the three edges
$t_i^+s_{i+1}^+$, $t_is_{i+1}$, and $t_i^-s_{i+1}^-$.
Add terminals $s^*,t^*$ adjacent, respectively, to
$s_1^+,s_1,s_1^-$ and to $t_p^+,t_p,t_p^-$.

For each clause $c$, let $U_c$ be its set of occurrence vertices.
Take a fresh docking gadget $H_c$ with at least $2|U_c|$ ports.
Attach each vertex of $U_c$ to two distinct ports, using different
ports for different occurrences, in the cyclic order of the
incidences around $c$.
Every occurrence vertex belongs to exactly one such clause set.

\subparagraph*{Degree Bound, Planarity, and Size.}
The graph is simple. Used ports have degree three, and unused ports
retain degree two. Each cycle vertex has two cycle edges and two edges
to a docking gadget: occurrence vertices attach to their clause gadget,
while the remaining vertices attach to the variable gadget. Together
with at most one backbone or terminal edge, this gives degree at most
five. The terminals have degree three.
For planarity, replace each variable segment by a thin disk containing
its cycle and internal docking gadget. Draw the three backbone edges
in a narrow corridor along the variable line. Place each clause
gadget in a disk around its clause segment, and route each pair of
attachment edges alongside the corresponding incidence connection.
The docking edges of $V_i$ run inside $C_i$ to the outer facial walk
of $H_i$. The clause edges leave $C_i$ outward, upward from $x_i^{+,1},x_i^{+,2}$
and downward from $x_i^{-,1},x_i^{-,2}$. Along the outer facial walk of $H_i$ and
of each $H_c$, the ports used by consecutive attached vertices are
consecutive, in the order in which these vertices appear around $C_i$
or from left to right along the clause segment, respectively. Since
each port appears exactly once along the walk by \Cref{lem:docking},
the attachments inside the disks do not cross. All disks and corridors may be chosen
disjoint apart from their designated attachments.
Place $s^*,t^*$ beyond the leftmost and rightmost variable disks,
respectively. Both lie on the outer face.
The graph has size linear in the number of literal occurrences.

\subparagraph*{Correctness.}
\begin{claim}
Every solution $(P,T_1,T_2)$ determines a satisfying assignment: the
path $P$ avoids all docking gadgets and traverses exactly one sign
branch of every variable gadget.
\end{claim}
\begin{claimproof}
Every internal vertex of $P$ has degree at least four in $G$: two
edges of $P$, and at least one edge of each of the two spanning trees.
Only attachment edges leave a docking gadget, and their endpoints in
the gadget are used ports of degree three. Since $s^*,t^*$ lie outside
all docking gadgets, the first gadget vertex on $P$ would be such a
port and an internal vertex of $P$. Hence, $P$ visits no docking
gadget.
Consequently $P$ lies in the subgraph formed by the terminals, the
backbone edges, and the cycles $C_i$.
In this subgraph, the occurrence vertices and the vertices
$x_i^\pm$ have degree two.
For each variable, it traverses exactly one of the two sign
branches
\[
 B_i^\circ=(s_i^\circ,x_i^{\circ,1},x_i^\circ,
                       x_i^{\circ,2},t_i^\circ).
\]
Indeed, in this subgraph, the two branches are the only connections
between the part containing $s^*$ and the part containing $t^*$ that
are separated by variable $i$. All internal vertices of a branch have
degree two in this subgraph, so entering a branch forces traversing
it entirely. Since $P$ starts and
ends on different sides, it crosses this separation an odd number of
times, and each branch can be crossed at most once. Hence, $P$
traverses exactly one branch.

Set $x_i$ to true precisely when $P$ traverses $B_i^-$.
If $x_i$ is false then $B_i^+$ is used; if $x_i$ is true then $B_i^-$
is used. Thus every occurrence of a false literal has both incident
cycle edges in $P$.
If a clause were false, deleting $E(P)$ would separate
$V(H_c)\cup U_c$ from the rest of the graph. This contradicts the
existence of a spanning tree avoiding $P$. The assignment satisfies
the formula.
\end{claimproof}

\begin{claim}
Every satisfying assignment yields an $s^*$-$t^*$ path and two
edge-disjoint spanning trees that avoid it.
\end{claim}
\begin{claimproof}
For each variable choose
the sign $\circ$ of its false literal and use the path
\[
 (s_i,s_i^\circ,x_i^{\circ,1},x_i^\circ,
                         x_i^{\circ,2},t_i^\circ,t_i).
\]
Concatenate these paths with the middle backbone edges and with
$s^*s_1,t_pt^*$ to obtain $P$.

We construct two edge-disjoint spanning trees in $G-E(P)$.
Inside every $H_i$ choose its two trees, and attach each vertex of
$V_i$ to them using its two docking edges, one per tree.
Join these local trees using the upper backbone edges for the first
tree and the lower backbone edges for the second. Attach $s^*,t^*$
by their upper edges to the first tree and lower edges to the second.
This produces two edge-disjoint trees on all nonoccurrence cycle vertices,
variable docking gadgets, and terminals.
For each clause, its docking gadget and all its occurrence vertices
similarly have two edge-disjoint local spanning trees.
Choose an occurrence of a satisfied literal in that clause.
Its two cycle edges avoid $P$ and lead to vertices of the already
constructed global trees. Use one edge to join each local tree to
its corresponding global tree.
The clause vertex sets are disjoint and each is attached once per
tree. Repeating this step yields two edge-disjoint spanning trees avoiding
$P$, as required.
\end{claimproof}
\end{proof}

The constant-size reduction to \Gossip preserves planarity and the
maximum-degree bound on these instances.
\begin{theorem}\label{thm:deg5-hard}
\Gossip is NP-complete on planar simple graphs of maximum degree five.
\end{theorem}
\begin{proof}
Apply \Cref{thm:red-ptt-to-gossip} to the instances of
\Cref{thm:ptt-hard-deg5}. Since $s^*,t^*$ lie on the outer face,
the added path can be drawn in the outer face. Its internal vertices
are new, so it creates no parallel edges and the graph stays simple.
The terminal degrees increase from three to four, and the new
vertices have degree two. Membership in NP follows by guessing an
ordering of the edges and checking strict temporal reachability
from every vertex.
\end{proof}
Hence, unless P${}={}$NP, for every integer $d$, \Gossip on
multigraphs of maximum degree at most $d$, counting multiplicities, is
polynomial-time solvable if $d\leq4$ (\Cref{cor:degree-four}) and
NP-complete if $d\geq5$; the hard instances are even planar and
simple.

The same construction, without the planarity requirement, preserves a
linear relation between the source and target sizes. This yields the
first ETH lower bound.
\begin{corollary}\label{cor:eth}
Unless ETH fails, neither PTT nor \Gossip has a
$2^{o(n+m)}$-time algorithm, even on simple graphs of maximum degree five.
\end{corollary}
\begin{proof}
The sparsification lemma~\cite{ImpagliazzoPaturiZane2001} implies that
sparse 3SAT cannot be solved in subexponential time under ETH.
Replace each variable with one copy per occurrence and impose a
cyclic chain of implications between its copies. For at least two
copies these are clauses $(\neg y_i\vee y_{i+1})$, with cyclic indices;
they force all copies to have the same truth value.
Each copy then occurs at most twice with each sign. A variable with
one occurrence needs no implication, and may be removed by
pure-literal elimination. Eliminate any remaining pure literals,
and duplicate single occurrences within their clause as above.
The size increases only linearly, and each literal occurs exactly twice.

Apply the preceding construction without requiring a plane embedding.
For mixed clauses attach their respective occurrence vertices
regardless of sign. Neither direction of the correctness proof uses
monotonicity or planarity. The output still has maximum degree five
and linear size. The constant-size transformation in
\Cref{thm:red-ptt-to-gossip} gives the same bound for \Gossip. The
added path has new internal vertices, so it creates no parallel edges
and the graph stays simple; the terminal degrees become four.
\end{proof}

Starting instead from the corresponding planar SAT lower bound gives
the planar, square-root-exponent version.
\begin{corollary}\label{cor:planar-eth}
Unless ETH fails, neither PTT nor \Gossip admits a
$2^{o(\sqrt n)}n^{\Oh(1)}$-time algorithm on planar simple graphs
of maximum degree five.
\end{corollary}
\begin{proof}
Bentert, Fomin, and Golovach~\cite[Proposition~1]{BentertFominGolovach2026}
prove that Rectilinear Monotone 3SAT with $p$ variables and $q$
clauses cannot be solved in $2^{o(\sqrt{p+q})}$ time under ETH,
even if each variable appears in at most eight clauses.
In the supplied rectilinear embedding, the variables lie along a
horizontal line, positive clauses lie above it, and negative clauses
lie below it; consecutive variables can be linked along that
line~\cite[definition preceding Proposition~1]{BentertFominGolovach2026}.
The reduction of Brunner et al.~\cite[Appendix~B]{BrunnerEtAlSoloChessFull}
makes each literal occur at most twice and preserves the sided
variable-linked embedding. Its size is linear: a variable with $d$
occurrences is replaced by $2d-1$ variables and $2d-2$ additional
clauses. Thus the total number of variables and clauses remains
$\Oh(p+q)$. Pure-literal elimination and duplication of single
occurrences do not increase this bound.
The construction in \Cref{thm:ptt-hard-deg5} produces
$n=\Oh(p+q)$ vertices, and \Cref{thm:red-ptt-to-gossip} adds only three
vertices and four edges.
As in the proof of \Cref{thm:deg5-hard}, the added path lies in the
outer face and the resulting graph is planar, simple, and of maximum
degree five.
A $2^{o(\sqrt n)}n^{\Oh(1)}$-time algorithm on the resulting graphs would
therefore contradict the cited lower bound.
\end{proof}

Planar $n$-vertex graphs have treewidth $\Oh(\sqrt n)$.
Consequently, \Cref{cor:planar-eth} rules out a
$2^{o(\tw)}n^{\Oh(1)}$-time algorithm under ETH, even on planar
simple graphs of maximum degree five. The single-exponential
treewidth dependence in \Cref{thm:treewidth} is therefore optimal up to
the constant in the exponent.

\subsection{Few Edges of Multiplicity One}

Another natural parameter counts the \emph{edges of multiplicity one},
that is, edges belonging to parallel classes of size one. A connected
graph without such edges contains two edge-disjoint spanning trees:
choose a spanning tree of its support and use two copies of each edge.
Such a graph is thus a yes-instance of \Gossip by
\Cref{thm:characterization}.
Nevertheless, four edges of multiplicity one already suffice for
NP-completeness.

\begin{lemma}\label{lem:multiplicity-one}
\Gossip is NP-complete on multigraphs with at most four edges of
multiplicity one.
\end{lemma}
\begin{proof}
Membership in NP follows as in \Cref{thm:deg5-hard}.
By \Cref{thm:ptt-hard-goebel}, PTT is NP-hard when every parallel class
has size two. Applying \Cref{thm:red-ptt-to-gossip} gives an equivalent
\Gossip instance whose only edges of multiplicity one are the four
added edges.
\end{proof}
Thus the problem is trivial without edges of multiplicity one, since
then exactly the connected instances are yes-instances, and it is
NP-complete with at most four. The obvious attempt to reduce the number further
fails. Let $G$ be connected with every parallel class of size two,
let $Q=(s,z_1,z_2,z_3,t)$ be the added path, and add a second copy of one
edge $f$ of $Q$. Then $G$ has two edge-disjoint spanning trees
$A_1,A_2$. Choose distinct edges $e_1,e_2\neq f$ of $Q$, and let
$A_i'$ be $A_i$ together with the three edges of $Q-e_i$, using
different copies of $f$ for $i=1,2$. Each $A_i'$ is a spanning tree,
since $Q-e_i$ consists of two paths starting at $s$ and at $t$. The
trees $A_1',A_2'$ share only the edge of $Q$ other than $e_1,e_2,f$, so the result is
always a $T$-graph.

\section{Conclusion}
\label{sec:conclusion}
We studied the recognition of label-connected graphs through their
spanning witnesses. Our results combine structural characterizations,
algorithms, and matching complexity lower bounds:
\begin{itemize}
\item \emph{Witness structure and preprocessing.} We give a
self-contained proof of the classical characterization and a
linear-time construction of a labeling from either witness. Our
induced-cycle result yields an $\Oh(3^{n/3}n^3)$-time exact algorithm,
while the \TreeContractionRule reduces dense regions without changing
the answer.
\item \emph{Structural parameters.} We obtain a deterministic
$2^{\Oh(\tw)}\cdot n$-time algorithm. Applying the contraction rule
extends the single-exponential dependence to tree-independence number,
chordal vertex deletion number, and clique-width, and gives a linear vertex kernel for
independence number.
\item \emph{Excess and short witness cycles.} We give a deterministic
XP algorithm and a randomized $2^{\Oh(\delta)}n^3$-time algorithm for
excess $\delta=m-(2n-4)$, also allowing a potentially smaller bound on
witness-cycle length. Randomness is confined to the matroid
representation and rank reduction; the \textsc{Rooted Independent
Cycle} subroutine is deterministic.
\item \emph{Hardness and tractability boundaries.} Recognition is
polynomial-time solvable at maximum degree four but NP-complete on
planar simple graphs of maximum degree five. 
Recognition also remains NP-complete with at most four edges of
multiplicity one.
\end{itemize}

These results leave the following questions open.
\begin{enumerate}
\item Can the algorithm for the excess be derandomized while keeping a
running time of $2^{\Oh(\delta)}n^{\Oh(1)}$? For this parameterization,
we can omit the random projection and work directly with the dual
matroid of rank $\delta+4$. It therefore suffices to derandomize the
representation of $M(G/\rho)\vee M(G/\rho)$, the $(2,2)$-sparsity
matroid of $G/\rho$. To the best of our knowledge, no deterministic
polynomial-time representation of these matroids is known.

\item Does \Gossip admit a polynomial kernel parameterized by the
excess? Our contraction rule guarantees local sparsity but does not
bound the instance size in terms of~$\delta$.

\item For instances not rejected by the edge lower bound, maximum degree
four forces the excess $\delta$ to be at most four. More generally,
the total degree surplus above four,
$\sigma=\sum_{v\in V(G)}\max\{0,\deg(v)-4\}$, satisfies
$\delta\leq4+\sigma/2$. The excess algorithm derived from
\Cref{thm:short-m-witness} therefore also gives an FPT algorithm
parameterized by $\sigma$. Is \Gossip in XP, or even
fixed-parameter tractable, parameterized by the number of vertices of
degree at least five?

\end{enumerate}

\bibliographystyle{plainurl}
\bibliography{bibliography}

\newpage
\appendix
\section{Proof Details for the Classical Characterization}
\label{app:characterization}

We give a self-contained proof of \Cref{thm:characterization}, which
states the following: a multigraph admits a temporally connecting
labeling if and only if it has at most one vertex, has two spanning
trees sharing at most one edge, or admits an $M$-witness---a cycle of
length at least four and two spanning forests, pairwise edge-disjoint,
each forest having four components that all meet the cycle.
First, we construct a complete call order from either type of witness.
For the converse, we contract maximal vertex sets supporting two
edge-disjoint spanning trees and analyze the resulting quotient, in
which each contracted set is represented by a single vertex.
Quotients with fewer than four vertices yield a $T$-witness directly.
For larger quotients, the number of edges leaves two cases: one yields
a $T$-witness; in the other, we construct an $M$-witness. The main difficulty is
choosing the cycle so that the quotient witness can be lifted to the
original graph.

The contraction, exchange, and lifting framework follows Burosch and
Laborde~\cite{BuroschL92}; the closed-tree counting argument follows
Bumby~\cite{Bumby81} and Kleitman and Shearer~\cite{KleitmanS80}.
We give these arguments in full and use information paths in the
original graph to ensure that the lifted cycle and forests are
edge-disjoint.

\subsection{Sufficiency: Constructing a Temporal Labeling}
\label{app:construct-labeling}

We prove \Cref{lem:four-intervals,lem:construct-labeling}. The first
states that, in an $M$-graph, the components of both witness forests
meet the cycle in the same four consecutive intervals, whose boundary
edges form two alternating matchings. The second gives a linear-time
construction of a temporally connecting labeling from either type of
witness. We follow Burosch and Laborde~\cite[Lemmas~7--8]{BuroschL92}
and Krumme~\cite[Theorem~3]{Krumme96}.
For each forest $F_j$ of an $M$-witness with cycle $C$, contract each
component of $F_j$ in $C\cup F_j$, retaining all cycle edges, including
any resulting loops. This produces a connected multigraph on four
vertices. Choosing three cycle edges that form a spanning tree of this
quotient completes $F_j$ to a spanning tree of the original graph.
The corresponding graphic matroids therefore have
rank three. The next lemma helps us coordinate these choices for the
two forests, including when the graph also admits a $T$-witness.

A \emph{coloop} of a matroid is an element belonging to every basis.
In a graphic matroid, coloops correspond to bridges: edges that every
spanning tree must use. Every cut of either quotient corresponds to a
cut of the cycle and therefore has even size. Hence neither quotient
has a bridge. Thus no cycle edge is
forced into every completion of either forest, which is why the next
lemma assumes that neither matroid has a coloop.

\begin{lemma}\label{lem:rank-three}
Let $M_1,M_2$ be rank-three matroids on the same ground set $E$, neither
having a coloop. Given independence oracles for $M_1$ and $M_2$, in
$\Oh(|E|)$ oracle calls one can either find bases $J_i$ of $M_i$ with
$|J_1\cap J_2|\leq1$, or find a four-element set $W\subseteq E$ such
that, in each matroid, the bases are exactly the three-element subsets
of $W$.
\end{lemma}
\begin{proof}
All bases below are found greedily. Whenever we construct a basis in
one matroid, we compare it with every basis already constructed in the
other matroid and return if some intersection has size at most one.

Fix a basis $B=\{y_1,y_2,y_3\}$ of $M_2$. Since $M_1$ has no coloop,
for each $i$ it has a basis $A_i$ avoiding $y_i$. Unless we have already
returned,
\[
  A_i=(B\setminus\{y_i\})\cup\{x_i\}
  \qquad\text{for some }x_i\notin B.
\]
For each $h$, choose a basis $B_h$ of $M_2$ avoiding $y_h$. For distinct
$i,j,h$, the basis $B_h$ must contain $\{y_j,x_i\}$ to meet $A_i$ twice,
and $\{y_i,x_j\}$ to meet $A_j$ twice. Since $|B_h|=3$, this implies
$x_i=x_j$. Thus all $x_i$ are the same element $x$, and
$B_h=(B\setminus\{y_h\})\cup\{x\}$. Set $W:=B\cup\{x\}$.

Next choose a basis $A_0$ of $M_1$ avoiding $x$. If no comparison
returns a pair of bases, then $A_0$ must contain $B\setminus\{y_h\}$
for every $h$, and hence $A_0=B$. Consequently, both matroids contain
all four three-element subsets of $W$ as bases.

It remains to inspect $E\setminus W$. If some $e\in E\setminus W$ is
a non-loop of either matroid, extend $e$ to a basis $A$ of that matroid.
Then $|A\cap W|\leq2$. Choosing $w\in A\cap W$ when this intersection
is nonempty, and any $w\in W$ otherwise, gives
$|A\cap(W\setminus\{w\})|\leq1$; the set $W\setminus\{w\}$ is a basis
of the other matroid, so we return this pair. Otherwise, every element
outside $W$ is a loop in both matroids, and their bases are exactly the
three-element subsets of $W$.

We perform only constantly many greedy basis computations and one scan
of $E\setminus W$. Each greedy computation uses $\Oh(|E|)$ oracle calls,
which proves the claimed bound.
\end{proof}

The rank-three lemma gives a linear-time procedure for finding either
two almost edge-disjoint spanning trees or the common interval
partition needed by the cycle schedule.

\begin{lemma}\label{lem:constructive-cross-bases}
Given an $M$-witness $(C,F_1,F_2)$, in $\Oh(n)$ time one can either
find two spanning trees sharing at most one edge, or find four
edges of $C$ giving the common interval partition of
\Cref{lem:four-intervals}.
\end{lemma}
\begin{proof}
For each $j$, identify vertices of $C$ belonging to the same component
of $F_j$, retaining the individual cycle edges and any resulting loops.
The resulting quotient graph $Q_j$ is connected on four vertices.
Every cut has even size, since it comes from a cut of a cycle, so
$Q_j$ has no bridge. Its graphic matroid $M_j$ has rank three and
no coloops, regardless of whether $G$ is a $T$-graph.
A basis of $M_j$ completes $F_j$ to a spanning tree, and two such
trees intersect exactly in their chosen cycle edges.
Apply \Cref{lem:rank-three} to $M_1$ and $M_2$. If it returns two bases
whose intersection has size at most one, these bases complete
$F_1,F_2$ to the required spanning trees. Otherwise it returns a
four-element set $W$ such that every edge outside $W$ is a loop in both
quotients.

Deleting the four edges of $W$ from $C$ leaves four nonempty paths,
allowing single-vertex paths. Each lies in one component of each
forest. Since all four components of each forest meet $C$, the paths
belong to distinct components in both forests. Thus their vertex sets
give the common four-interval partition.

The quotient has only four vertices, so each independence test takes
constant time. Hence \Cref{lem:rank-three} takes $\Oh(|E(C)|)$ time.
Computing the forest components takes $\Oh(n)$ time.
This proves the bound.
\end{proof}

\begin{proof}[Proof of \Cref{lem:four-intervals}]
Apply \Cref{lem:constructive-cross-bases} to the given witness. Since $G$
is an $M$-graph, it is not a $T$-graph. The procedure therefore cannot
return two spanning trees sharing at most one edge and must return the
four intervals of the first statement.
Since the intervals are pairwise disjoint, $b_1,b_3$ are
vertex-disjoint, and so are $b_2,b_4$. Each interval $X_i$ contains
exactly one endpoint of $b_1$ or $b_3$ and exactly one endpoint of $b_2$
or $b_4$. The second statement follows, since the components of each
forest meet $V(C)$ exactly in $X_1,\ldots,X_4$.
\end{proof}

We now prove \Cref{lem:construct-labeling}: given either type of
witness, we construct a temporally connecting labeling in $\Oh(n+m)$
time. In the call interpretation, we first gather secrets at roots,
exchange them between roots when necessary, and finally distribute
them to all vertices.
\begin{proof}[Proof of \Cref{lem:construct-labeling}]
For two edge-disjoint spanning trees, choose an arbitrary common root.
Gather all secrets along the first tree, calling each nonroot vertex's
edge to its parent after all edges from that vertex to its children.
Then distribute the secrets along the second tree, calling each
nonroot vertex's edge to its parent before all edges from that vertex
to its children. If the trees share one edge $uv$,
delete that edge from both trees. Gather along the two components of
the first tree toward $u$ and $v$, make the call $uv$, and distribute
along the two components of the second tree rooted at $u$ and $v$.
In either case every selected edge is used once.

Now suppose an $M$-witness is given. Apply
\Cref{lem:constructive-cross-bases}. If it returns two spanning trees,
use the preceding construction. Otherwise it returns the common
four-interval partition directly, even if $G$ happens to be
a $T$-graph for another reason.
Index the four intervals $X_1,\ldots,X_4$ in cyclic order, and let
$b_1,\ldots,b_4$ be the essential edges as in \Cref{sec:witnesses}, that
is, $b_i$ joins $X_i$ to $X_{i+1}$, with indices modulo four.
In each interval $X_i$, let $p_i$ be the endpoint incident with
$\{b_1,b_3\}$ and $q_i$ the endpoint incident with $\{b_2,b_4\}$.
They coincide if the interval has just one vertex.
First gather along the blue forest $F_1$, rooting its component meeting
$X_i$ at $p_i$.
Call $b_1,b_3$. Next, in each interval, call the edges of its path in
order from $p_i$ to $q_i$. At this point $q_1,q_2$ know the secrets
from the first two components of $F_1$, and $q_3,q_4$ know those from
the other two. Call $b_2,b_4$; all four vertices $q_i$ now know every
secret. Finally distribute along the orange forest $F_2$, rooting its
component meeting $X_i$ at $q_i$.
\Cref{fig:labeling-example} shows the resulting schedule on the
running example.

The selected edges are pairwise distinct. Label them by their positions
in the gossip scheme, and label any remaining edges later in arbitrary
order. Additional calls cannot destroy temporal connectivity.
All forest traversals and cycle traversals take $\Oh(n)$ time.
Together with \Cref{lem:constructive-cross-bases} and the $\Oh(m)$ time
to label unused edges, this gives the claimed bound.
\end{proof}

In particular, \Cref{lem:construct-labeling} proves the sufficiency direction
of \Cref{thm:characterization}. The algorithm accepts an arbitrary
$M$-witness and does not require a separate test for whether the graph
is a $T$-graph.

\subsection{Necessity}
\label{sec:necessity-self-contained}
\begin{figure}[p]
\centering
\begin{tikzpicture}[
    node distance=4mm,
    every node/.style={font=\footnotesize,execute at begin node={\hyphenpenalty=10000\exhyphenpenalty=10000}},
    stage/.style={draw=black!70,rounded corners=2pt,fill=black!3,
                  text width=68mm,inner sep=4pt,align=left},
    tool/.style={stage,text width=29mm,draw=auxgreen!80!black,fill=auxgreen!7},
    exitT/.style={draw=witnessblue,rounded corners=2pt,fill=witnessblue!7,
                  text width=25mm,inner sep=4pt,align=left},
    outbox/.style={stage,text width=40mm,draw=black,fill=black!10},
    note/.style={font=\scriptsize,text=black!70,align=left},
    figtag/.style={fill=black!55,text=white,font=\scriptsize\bfseries,
                   rounded corners=1.5pt,inner sep=1.5pt},
    arr/.style={-{Latex[length=2mm]},line width=.7pt},
    usearr/.style={-{Latex[length=2mm]},line width=.7pt,auxgreen!80!black,
                   dash pattern=on 3pt off 2pt},
    band/.style={draw=black!45,dashed,rounded corners=5pt,
                 inner xsep=5pt,inner ysep=4pt},
    steplabel/.style={anchor=north east,text width=21mm,align=right,inner sep=0pt}
]
  \node (in) {\textbf{Input:} a multigraph $G$ on at least two vertices
    with a complete call order};

  \node[stage,below=11mm of in] (quot) {\textbf{Contract blocks}
     (\Cref{lem:characterization-union,lem:characterization-quotient})\\[1pt]
     Contract the maximal vertex sets that support two edge-disjoint
     spanning trees. The quotient $Q$ on $r$ vertices is simple.};
  \node[stage,below=7mm of quot] (exch) {\textbf{Spanning-tree exchange}
     (\Cref{lem:characterization-exchange,cor:characterization-edges})\\[1pt]
     If $r\geq4$, then $Q$ splits into a spanning tree and a spanning
     forest, so $|E(Q)|\leq2r-3$.};
  \coordinate (b1top) at ([yshift=3.5mm]quot.north);
  \node[band,fit=(quot)(exch)(b1top)] (b1) {};
  \node[steplabel] (sl1) at ([xshift=-4mm,yshift=-1mm]b1.north west) {\textbf{Step 1}\\ construct the quotient and bound its size};

  \node[stage,below=13mm of exch] (lower) {\textbf{Call lower bound}
     (\Cref{lem:characterization-poset-transfer,lem:characterization-closed-tree-reduction})\\[1pt]
     The call order carries over to $Q$ and forces $|E(Q)|\geq2r-4$.
     Hence $|E(Q)|=2r-4$ in the remaining case.};
  \node[stage,below=of lower] (canon) {\textbf{Canonical form}
     (\Cref{lem:characterization-canonical-ideal,cor:characterization-early-components})\\[1pt]
     The early calls form two trees $X,Y$ and unicyclic components. The
     future calls form two trees, rooted at the endpoints $x_0,x_1$ of
     the last early call $e$ of $X$.};
  \coordinate (b3top) at ([yshift=3.5mm]lower.north);
  \node[band,fit=(lower)(canon)(b3top)] (b3) {};
  \node[steplabel] (sl3) at ([xshift=-4mm,yshift=-1mm]b3.north west) {\textbf{Step 3}\\ use the call order to obtain canonical structure};

  \node[stage,below=13mm of canon] (filter) {\textbf{Filter routes}
     (\Cref{lem:characterization-root-good-routes,lem:characterization-actual-closed-set})\\[1pt]
     Keep the information routes that enter the root blocks suitably,
     and choose an inclusion-minimal nonempty closed set $S$ such that
     $Y[S]$ is connected.};
  \node[stage,below=of filter] (select) {\textbf{Select a four-cycle}
     (\Cref{lem:characterization-same-index-rerouting,lem:characterization-compatible-actual-cycle})\\[1pt]
     Two adjacent vertices $U_0,U_1$ of $S$ with future calls to
     different roots give the cycle $x_0,U_0,U_1,x_1$.};
  \node[stage,below=of select] (extend) {\textbf{Extend and lift}
     (\Cref{lem:characterization-protected-extension,cor:characterization-optimal-quotient-lift})\\[1pt]
     Extend the four-cycle to an $M$-witness of $Q$ and lift it to $G$.};
  \coordinate (b4top) at ([yshift=3.5mm]filter.north);
  \node[band,fit=(filter)(select)(extend)(b4top)] (b4) {};
  \node[steplabel] (sl4) at ([xshift=-4mm,yshift=-1mm]b4.north west) {\textbf{Step 4}\\ select a compatible cycle and lift the witness};

  \node[tool,anchor=north west] (lift) at ([xshift=6mm]b4.east|-filter.south)
     {\textbf{Lifting condition}\\
     (\Cref{lem:characterization-m-lift,lem:characterization-threshold})\\[1pt]
     Each block on the cycle needs two spanning trees and a path between
     its ports, pairwise edge-disjoint.};
  \coordinate (b2top) at ([yshift=3.5mm]lift.north);
  \node[band,fit=(lift)(b2top)] (b2) {};
  \node[font=\footnotesize,anchor=south west,inner sep=1pt,align=left] (sl2) at ([yshift=1mm]b2.north west) {\textbf{Step 2}: establish the\\ lifting condition};

  \node[exitT,anchor=west] (t1) at ([xshift=6mm]b1.east|-quot) {$r\leq3$:\\ $T$-witness
     (\Cref{lem:characterization-t-lift})};
  \node[exitT,anchor=west] (t2) at ([xshift=6mm]b1.east|-exch) {$|E(Q)|=2r-3$:\\ $T$-witness
     (\Cref{cor:characterization-t-case})};
  \node[outbox,below=8mm of b4.south] (outM) {\textbf{Output:} an $M$-witness of $G$};

  \draw[arr] (in)--(quot);
  \draw[arr] (quot)--node[right,note]{$r\geq4$}(exch);
  \draw[arr] (quot.east)--(t1.west);
  \draw[arr] (exch.east)--(t2.west);
  \draw[arr] (exch.south)--node[right,note]{$|E(Q)|\leq2r-4$}(lower.north);
  \draw[arr] (lower)--(canon);
  \draw[arr] (canon.south)--(filter.north);
  \draw[arr] (filter)--(select);
  \draw[arr] (select)--(extend);
  \draw[arr] (extend.south)--(outM.north);
  \draw[usearr] (b2.west|-select)--(select.east);
  \draw[usearr] (b2.west|-extend)--(extend.east);

  \node[figtag,anchor=south east] at ([xshift=-2pt,yshift=1pt]exch.north east) {\Cref{fig:exchange}};
  \node[figtag,anchor=south east] at ([xshift=-2pt,yshift=1pt]lower.north east) {\Cref{fig:closed-tree}};
  \node[figtag,anchor=south east] at ([xshift=-2pt,yshift=1pt]canon.north east) {\Cref{fig:necessity-canonical}(a)};
  \node[figtag,anchor=south east] at ([xshift=-2pt,yshift=1pt]filter.north east) {\Cref{fig:filtered-routes}};
  \node[figtag,anchor=south east] at ([xshift=-2pt,yshift=1pt]select.north east) {\Cref{fig:static-selection,fig:port-selection}};
  \node[figtag,anchor=south east] at ([xshift=-2pt,yshift=1pt]extend.north east) {\Cref{fig:necessity-canonical}(b,c)};
  \node[figtag,anchor=south east] at ([xshift=-2pt,yshift=1pt]lift.north east) {\Cref{fig:nonliftable}};

  \foreach \b/\s in {1/{\S\ref{sec:necessity-blocks}--\ref{sec:necessity-exchange}},
                     3/{\S\ref{sec:necessity-canonical}},
                     4/{\S\ref{sec:necessity-path-selection}--\ref{sec:necessity-ports}}}
    \node[note,anchor=north east,align=right] at ([yshift=-2pt]sl\b.south east) {\s};
  \node[note,anchor=north west] at ([yshift=-1pt]b2.south west) {\S\ref{sec:necessity-lifting}};
\end{tikzpicture}
\caption{Roadmap of the necessity proof of \Cref{thm:characterization}.
The dashed boxes are the four steps listed at the beginning of
\Cref{sec:necessity-self-contained}. Small quotients and the upper edge
count give a $T$-witness (blue). Otherwise, the call order fixes the
edge count and the canonical structure, which guides the choice of a
four-cycle that satisfies the lifting condition of Step~2 (green, dashed
arrows). The gray tags name the figures illustrating each part.}
\label{fig:necessity-roadmap}
\end{figure}
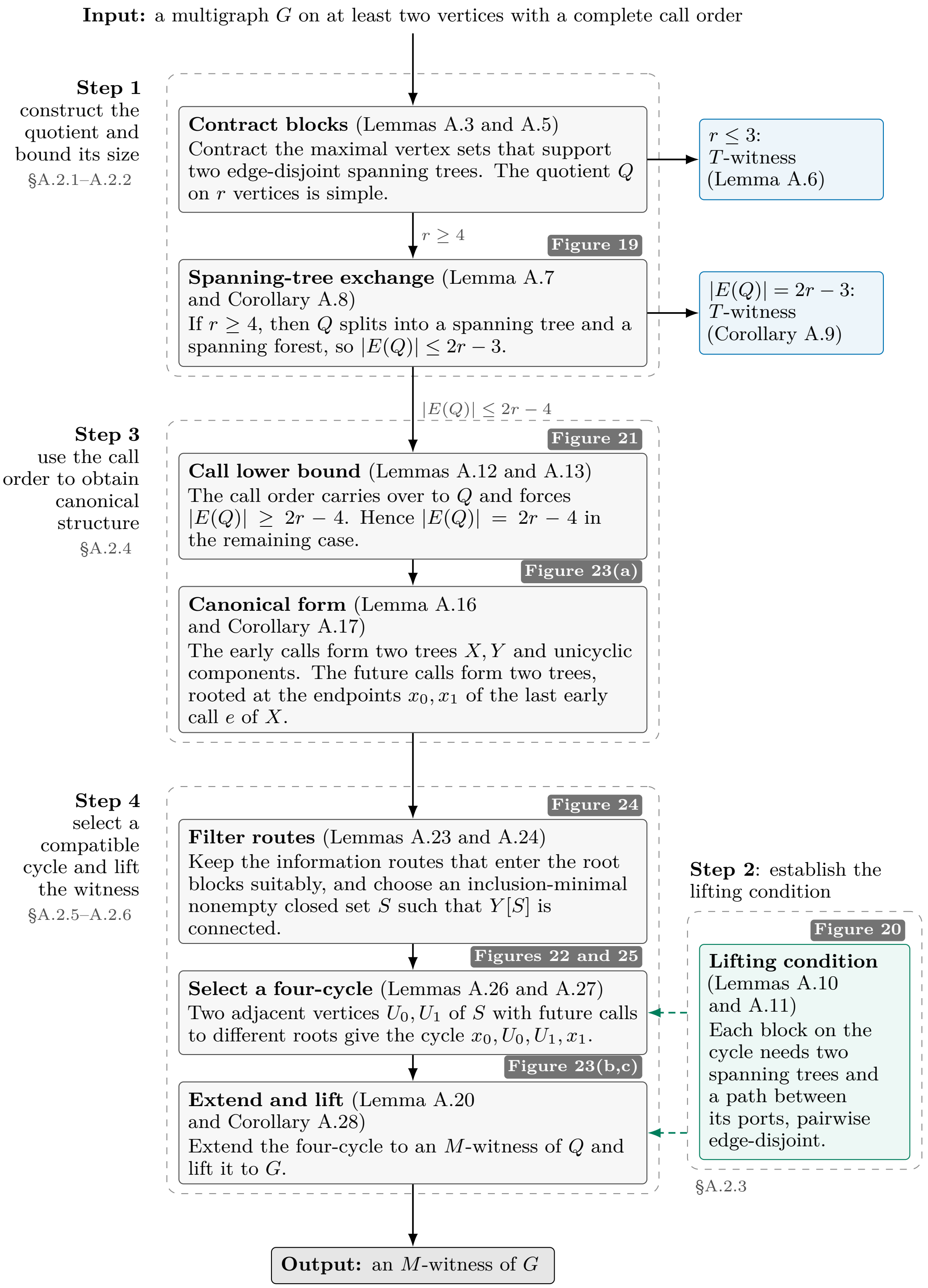

We prove the necessity direction of \Cref{thm:characterization}.
Graphs with at most one vertex satisfy the characterization directly.
For a graph with at least two vertices, we construct a $T$-witness or
an $M$-witness from a complete call order. The proof has four stages.
\begin{enumerate}
\item \emph{Construct the quotient and bound its size.}
Contract maximal sets supporting two edge-disjoint spanning trees.
An exchange argument bounds the number of edges in the quotient and
gives a $T$-witness when the upper bound is attained
(\Cref{sec:necessity-blocks,sec:necessity-exchange}).
\item \emph{Establish the lifting condition.}
To lift a quotient $M$-witness, inside each block on its cycle we need
two spanning trees and a path between its attachment vertices, all
edge-disjoint. We establish a sufficient condition for finding these
subgraphs (\Cref{sec:necessity-lifting}).
\item \emph{Use the call order to obtain canonical structure.}
The call order gives a lower edge bound. At equality, it yields a
canonical division into early and future calls
(\Cref{sec:necessity-canonical}).
\item \emph{Select a compatible cycle and lift the witness.}
Use the canonical structure to select a four-cycle satisfying the
lifting condition, extend it to a quotient $M$-witness, and lift it to
the original graph
(\Cref{sec:necessity-path-selection,sec:necessity-ports}).
\end{enumerate}
\Cref{fig:necessity-roadmap} summarizes these steps and the case distinctions between them.

\subsubsection{Two-Tree Blocks and Their Quotient}
\label{sec:necessity-blocks}
We first construct the quotient and show that its $T$-witnesses lift
to the original graph. Fix a multigraph $G$ with a complete call
order. Within this argument graphs are loopless: deleting loops
preserves completeness, since they transmit no new secret. Paths may
be trivial, and spanning subgraphs retain isolated vertices. For an
edge set $A$, the notation $H[A]$ denotes the spanning subgraph with
edge set $A$; for a vertex set it denotes the induced subgraph.
Deleting a path means deleting its edges. A path is \emph{nonseparating}
if this deletion leaves its ambient graph connected.

In a graph $H$, a nonempty vertex set $X\subseteq V(H)$ is a
\emph{two-tree set} if $H[X]$ contains two edge-disjoint spanning trees.
It is \emph{nontrivial} if $|X|\geq2$. A singleton is a two-tree set,
with two empty trees. These are the vertex sets of the $T^*$-subgraphs of
Burosch and Laborde~\cite{BuroschL92}.

\begin{lemma}\label{lem:characterization-union}
If $X$ and $Y$ are two-tree sets and $X\cap Y\neq\varnothing$, then
$X\cup Y$ is a two-tree set.
\end{lemma}
\begin{proof}
Choose edge-disjoint spanning trees $A_X,B_X$ of $G[X]$ and
$A_Y,B_Y$ of $G[Y]$. The graph with edge set $E(A_X)\cup E(A_Y)$
is connected on $X\cup Y$. Extend $A_X$ to a spanning tree $A$ of
this graph. Every edge of $A$ outside $A_X$ belongs to $A_Y$ and has
at least one endpoint outside $X$: otherwise it would form a cycle
with the spanning tree $A_X$ on $X$.

Consider the graph on $X\cup Y$ with edge set
\[
 E(B_X)\cup\bigl(E(B_Y)\setminus E(A_X)\bigr).
\]
It is edge-disjoint from $A$. Indeed, $B_X$ avoids $A_X$, and its
edges have both endpoints in $X$, so it also avoids $A-A_X$.
The retained edges of $B_Y$ avoid $A_X$ by definition and avoid
$A-A_X$ because the latter edges belong to $A_Y$.
The graph is connected: every deleted edge of $B_Y$ belongs to $A_X$
and has both endpoints in $X\cap Y$, so its endpoints can instead
be joined through $B_X$. Thus it contains a spanning tree $B$
edge-disjoint from $A$.
\end{proof}

We use this union property to decompose the graph into maximal
two-tree sets.

\begin{corollary}\label{cor:characterization-partition}
The inclusion-maximal two-tree sets form a partition $\mathcal B$ of
$V(G)$.
\end{corollary}
\begin{proof}
Every vertex belongs to a two-tree set and hence, by finiteness, to a
maximal one. Two distinct maximal sets cannot intersect by
\Cref{lem:characterization-union}.
\end{proof}

We call the sets in $\mathcal B$ \emph{blocks}. Let $Q$ be obtained by
contracting each block to one vertex. Delete edges internal to a block, and retain each external edge
as a distinct edge of $Q$, together with its original endpoints.

\begin{lemma}\label{lem:characterization-quotient}
The quotient $Q$ is simple and contains no nontrivial two-tree set. If $G$ is connected, then $Q$ is connected.
\end{lemma}
\begin{proof}
The quotient $Q$ has no loops, since all edges internal to a block
were deleted. Suppose a set of at least two vertices of $Q$ supports
two edge-disjoint spanning trees. Lift their edges to $G$. In each
represented block, add one internal spanning tree to the first lift
and the other to the second. Each lift is connected and has exactly
one less edge than vertices, so the two lifts are edge-disjoint
spanning trees on the union of the blocks. This contradicts their
maximality. Two parallel edges of $Q$ would themselves be two
edge-disjoint spanning trees on their two endpoints, so no parallel
edges occur. Connectivity is preserved by contracting connected sets.
\end{proof}

A $T$-witness of the quotient lifts to the original graph by using
the two internal spanning trees in each block.

\begin{lemma}\label{lem:characterization-t-lift}
If $Q$ has a $T$-witness, then $G$ has a $T$-witness. If $Q$ has one
vertex, then $G$ has two edge-disjoint spanning trees.
\end{lemma}
\begin{proof}
Lift each quotient tree and add one of the two internal spanning trees
in every block. The two resulting graphs are spanning trees, and their
intersection consists precisely of the lifts of shared quotient edges.
The one-vertex case is the definition of a two-tree block.
\end{proof}

\subsubsection{Spanning-Tree Exchange}
\label{sec:necessity-exchange}

The following lemma lets us replace a chosen spanning tree so that
the remaining edges form a forest, without losing connections already
present in the complement. Components of the complement may merge,
but none split. This lets us bound the quotient's edge count and later
modify trees while preserving specified complementary components.
The lemma is adapted from Burosch and
Laborde~\cite[Lemma~3]{BuroschL92}. A partition \emph{coarsens}
another if every part of the latter is contained in a part of the former.

\begin{figure}[t]
\centering
\begin{tikzpicture}[
  every node/.style={font=\footnotesize},
  v/.style={circle,draw=black!80,fill=white,inner sep=0pt,minimum size=2.4mm},
  sE/.style={black,line width=1pt},
  kE/.style={witnessblue,line width=1pt,dash pattern=on 3pt off 1.8pt},
  hlE/.style={witnessorange,line width=2.2pt},
  newE/.style={auxgreen,line width=1.4pt,dash pattern=on 3pt off 1.8pt},
  regionR/.style={fill=witnessblue!7,draw=witnessblue!45,rounded corners=10pt},
  regionO/.style={fill=black!4,draw=black!30,rounded corners=8pt},
  piece/.style={draw=black!50,rounded corners=6pt,fill=white,fill opacity=.6},
  lbl/.style={font=\footnotesize,text=black!70},
]
\begin{scope}
  \draw[regionR] (-0.85,-0.4) rectangle (4.2,1.8);
  \node[lbl,anchor=north west] at (-0.85,1.8) {$R$};
  \draw[regionO] (0.35,-1.95) rectangle (3.25,-1.05);
  \foreach \i/\x in {1/0,2/1.8,3/3.6} {
    \draw[piece] (\x-0.4,-0.25) rectangle (\x+0.4,1.15);
    \node[lbl,anchor=south] at (\x,1.15) {$X_\i$};}
  \coordinate (u1) at (0,0.9);   \coordinate (u2) at (0,0.1);
  \coordinate (v1) at (1.8,0.9); \coordinate (v2) at (1.8,0.1);
  \coordinate (w1) at (3.6,0.9); \coordinate (w2) at (3.6,0.1);
  \coordinate (o1) at (0.9,-1.5); \coordinate (o2) at (2.7,-1.5);
  \draw[kE] (o1)--(o2);
  \draw[kE] (v1)--(w1)--(v2)--(u1) (u2)--(v2)--(w2);
  \draw[hlE] (u1)--(v1);
  \draw[kE] (u1)--(v1) node[pos=.35,above=1pt,text=black] {$e$};
  \draw[sE] (u1)--(u2) (v1)--(v2) (w1)--(w2) (o1)--(v2)--(o2)--(w2);
  \draw[hlE] (u2)--(o1);
  \draw[sE] (u2)--(o1) node[pos=.65,right=2pt,text=black] {$f$};
  \foreach \p in {u1,u2,v1,v2,w1,w2,o1,o2} \node[v] at (\p) {};
  \node[panel] at (1.8,-2.45) {(a) a nonbridge $e$ of $K[R]$ between pieces};
\end{scope}
\begin{scope}[xshift=7.4cm]
  \draw[regionR] (-0.75,-1.95) rectangle (3.65,1.75);
  \node[lbl,anchor=north west] at (-0.75,1.75) {$R$};
  \draw[piece] (-0.5,-1.3) rectangle (1.4,1.35);  \node[lbl,anchor=north] at (0.45,-1.32) {$X$};
  \draw[piece] (2.35,-1.3) rectangle (3.35,1.35); \node[lbl,anchor=north] at (2.85,-1.32) {$X_2$};
  \coordinate (a1) at (0,1.0); \coordinate (a2) at (1.0,1.0); \coordinate (a3) at (0.5,0.35);
  \coordinate (b1) at (0.5,-0.35); \coordinate (b2) at (1.0,-0.95);
  \coordinate (y1) at (2.85,1.0); \coordinate (y2) at (2.85,-0.95);
  \draw[kE] (a2)--(a3)--(a1);
  \draw[kE] (b1)--(b2);
  \draw[kE] (y1)--(y2);
  \draw[hlE] (a2)--(y1);
  \draw[kE] (a2)--(y1) node[midway,above=1pt,text=black] {$e'$};
  \draw[kE] (b2)--(y2);
  \draw[newE] (a3)--(b1) node[midway,right=1pt,font=\scriptsize,text=auxgreen!60!black] {new};
  \foreach \p in {a1,a2,a3,b1,b2,y1,y2} \node[v] at (\p) {};
  \node[font=\scriptsize,text=witnessblue,anchor=east] at (-0.05,0.55) {$K_1$};
  \node[font=\scriptsize,text=witnessblue,anchor=east] at (0.42,-0.65) {$K_2$};
  \node[panel] at (1.45,-2.45) {(b) after replacing $S[X]$};
\end{scope}
\end{tikzpicture}
\caption{The two steps in the proof of \Cref{lem:characterization-exchange}.
Black edges belong to the spanning tree $S$, and blue dashed edges to its
complement $K$. For a component $R$ of $K$ (blue region), the boxes are
the vertex sets $X_i$ of the components of $S[R]$.
(a) If a complementary edge $e$ joining two pieces were not a bridge of
$K[R]$, the $S$-path between its endpoints would leave $R$ through an
edge $f$. Exchanging $e$ and $f$ keeps $K[R]-e$ connected and joins $R$ to
another component of the complement, reducing the number of components.
(b) If $K[R]$ contains a cycle, it lies in one piece $X$, here in the
component $K_1$ of $K[X]$. Replacing $S[X]$ by induction makes the
complement inside $X$ a forest that keeps every component of $K[X]$
connected and joins two of them. Here, the replacement removes a cycle
edge inside $K_1$ and adds the green connection between $K_1$ and $K_2$.
The old path between them leaves $X$ through a bridge $e'$ between
pieces, which now lies on a cycle of the complement. The same argument is used in
\Cref{lem:characterization-protected-extension}.}
\label{fig:exchange}
\end{figure}
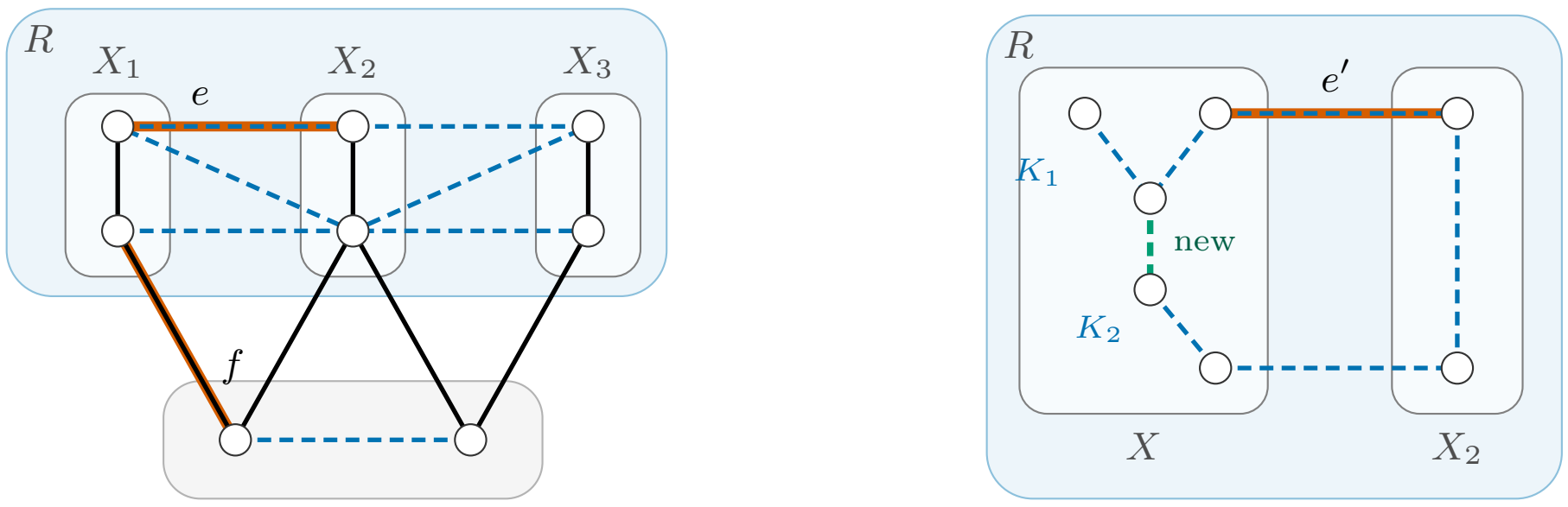

\begin{lemma}\label{lem:characterization-exchange}
Let $H$ be connected and loopless, with no nontrivial two-tree set.
For every spanning tree $T$ of $H$, there is a spanning tree $S$ such
that $H-E(S)$ is a forest and its component partition coarsens that
of $H-E(T)$.
\end{lemma}

We choose a spanning tree whose complement has as few components as
possible, subject to the coarsening condition. If the complement
contains a cycle, we modify the tree inside that component using
induction. The modified tree admits an exchange that reduces the
number of complementary components, contradicting minimality.
\Cref{fig:exchange} illustrates both steps.

\begin{proof}[Proof of \Cref{lem:characterization-exchange}]
Induct on $|V(H)|$. The assertion is immediate for one vertex.
Among the spanning trees satisfying the required coarsening condition,
choose $S$ so that $H-E(S)$ has as few components as possible. Such a
tree exists because $T$ satisfies the condition. Write $K=H-E(S)$.

Fix a component of $K$ with vertex set $R$, and let $X_1,\ldots,X_s$
be the vertex sets of the components of $S[R]$.
Every edge $e$ of $K[R]$ joining different sets $X_i$ is a bridge
of $K[R]$. Otherwise, the unique $S$-path between its endpoints must
leave $R$, since its endpoints are in different components of $S[R]$.
Choose an edge $f$ of this path with one endpoint in $R$ and the other
outside $R$. Replacing $f$ by $e$ in $S$ gives another spanning tree.
Deleting $e$ from $K[R]$ does not disconnect it, whereas adding $f$
joins it to another component of $K$. The new complementary partition
therefore coarsens that of $K$ and has fewer components, a contradiction.
This bridge property holds for every minimizing tree $S$.

Suppose now that $K[R]$ contains a cycle. Since $H$ is loopless,
$|R|\geq2$. None of its edges is a
bridge, so the cycle lies entirely inside one of the sets $X_i$;
call that set $X$. If $s=1$, then $S[R]$ and a spanning
tree of $K[R]$ would be two edge-disjoint spanning trees on $R$, contrary
to the hypothesis. Hence $s\geq2$, and in particular
$|X|<|V(H)|$.

Apply the inductive hypothesis to the connected induced graph $H[X]$
and its spanning tree $S[X]$. Obtain a spanning tree $S_X$ whose
complement is a forest and whose complementary partition coarsens
that of $K[X]$. Replace $S[X]$ in $S$ by $S_X$, obtaining a spanning
tree $\widetilde S$ of $H$. Its complementary partition is exactly
that of $K$. Indeed, each old component of $K[X]$ remains connected,
all edges outside $H[X]$ are retained, and all newly added complementary
edges have endpoints in $R$; thus no old component splits or merges
with a different old component. Moreover, the component vertex sets
of $\widetilde S[R]$ are still $X_1,\ldots,X_s$.

Let $q$ be the number of components of $K[X]$. Since it contains a
cycle, it has at least $|X|-q+1$ edges. The new complement in $H[X]$
has the same number of edges, because both replaced trees have
$|X|-1$ edges. Being a forest, it therefore has at most $q-1$
components. Consequently, it connects two distinct old components
of $K[X]$.

Choose vertices in these two old components and a simple path between
them in $K[R]$. This path must leave $X$. It thus contains an edge $e$
joining different sets $X_i$, which is a bridge of $K[R]$ by the
bridge property. Removing $e$ separates the two selected vertices.
Every component of $K[X]$ lies on one side of this cut, and its
vertices remain connected after the replacement. The edges outside
$H[X]$, other than $e$, are unchanged. Hence the old connections on
each side of the cut remain, while the new connection inside $X$
joins the two sides without using $e$. The edge $e$ is consequently
not a bridge of $H-E(\widetilde S)$.
But $\widetilde S$ is also a minimizing tree, and $e$ still joins
different components of $\widetilde S[R]$. This contradicts the
bridge property. Thus $K$ is a forest.
\end{proof}

The partition into a spanning tree and a complementary forest bounds
the number of edges.

\begin{corollary}\label{cor:characterization-edges}
If $H$ is connected and loopless, has $r\geq2$ vertices, and contains
no nontrivial two-tree set, then $|E(H)|\leq2r-3$.
\end{corollary}
\begin{proof}
By \Cref{lem:characterization-exchange}, the edges partition into a spanning
tree and a spanning forest. The forest has at least two components,
since otherwise $H$ itself would be a two-tree set. The two parts
have at most $r-1$ and $r-2$ edges, respectively.
\end{proof}

When the quotient attains this bound, the complementary forest has
exactly two components, which yields a $T$-witness.

\begin{corollary}\label{cor:characterization-t-case}
If the connected quotient $Q$ has $r\geq2$ vertices and $2r-3$ edges,
then $G$ has a $T$-witness.
\end{corollary}
\begin{proof}
The partition supplied by \Cref{lem:characterization-exchange} consists of a
spanning tree $A$ and a spanning forest $B$ with exactly two components.
Choose an edge $e$ of $A$ joining these components. The trees $A$
and $B+e$ share exactly $e$. Apply \Cref{lem:characterization-t-lift}.
\end{proof}

\subsubsection{Lifting Witnesses}
\label{sec:necessity-lifting}

Two internal spanning trees suffice to lift a quotient $T$-witness.
For a quotient cycle, those same two trees may leave no internal
path between its attachment vertices. The next lemma isolates the
additional condition needed for lifting.

\begin{lemma}\label{lem:characterization-m-lift}
Suppose $Q$ has an $M$-witness $(D,L_1,L_2)$. For each block $X$
represented on $D$, let $p_X,q_X$ be the original endpoints in $X$
of its two incident cycle edges. Suppose $G[X]$ has two spanning
trees $A_X,B_X$ and a $p_X$-$q_X$ path $P_X$ whose three edge sets
are pairwise disjoint. Then $G$ has an $M$-witness.
\end{lemma}
\begin{proof}
In blocks off $D$, choose any two edge-disjoint internal spanning
trees and call them $A_X,B_X$ as well. Lift $L_1,L_2$ and add all
trees $A_X,B_X$, respectively. Replacing each quotient vertex by an
internal tree preserves the component partition. A cycle in the lift
could not lie inside one block, whose internal edges form a tree.
Its external edges would therefore project to a nonempty closed trail
in the quotient forest, which is impossible. Thus each lift is a
forest with exactly four components.
The two lifted forests are edge-disjoint: their external edges come
from the edge-disjoint quotient forests $L_1,L_2$, and their internal edges
come from the edge-disjoint trees $A_X,B_X$ in each block.
Replace each vertex $X$ on $D$ by $P_X$. Paths in different blocks
are vertex-disjoint, and each block occurs only once on $D$, so this
gives a cycle of length at least four. A path may be trivial if
$p_X=q_X$. The lifted cycle avoids both forests by the local
edge-disjointness assumptions and the edge-disjointness of the quotient
witness. Every forest component still meets the lifted cycle.
\end{proof}

To obtain the local condition in \Cref{lem:characterization-m-lift}, we will find
a nonseparating path on one side of a call threshold and a spanning
tree on the other side. The two threshold graphs have disjoint edges.

\begin{lemma}\label{lem:characterization-threshold}
Let $J$ be a graph and let $E^-,E^+\subseteq E(J)$ be disjoint.
Suppose $(V(J),E^+)$ is connected and there is a $p$-$q$ path
$P$ in $(V(J),E^-)$ such that $(V(J),E^-\setminus E(P))$ is
connected. Then $J$ contains two spanning trees and a $p$-$q$
path whose edge sets are pairwise disjoint.
\end{lemma}
\begin{proof}
Choose one spanning tree from $E^-\setminus E(P)$ and another
from $E^+$. Both avoid $P$, and they avoid each other.
\end{proof}

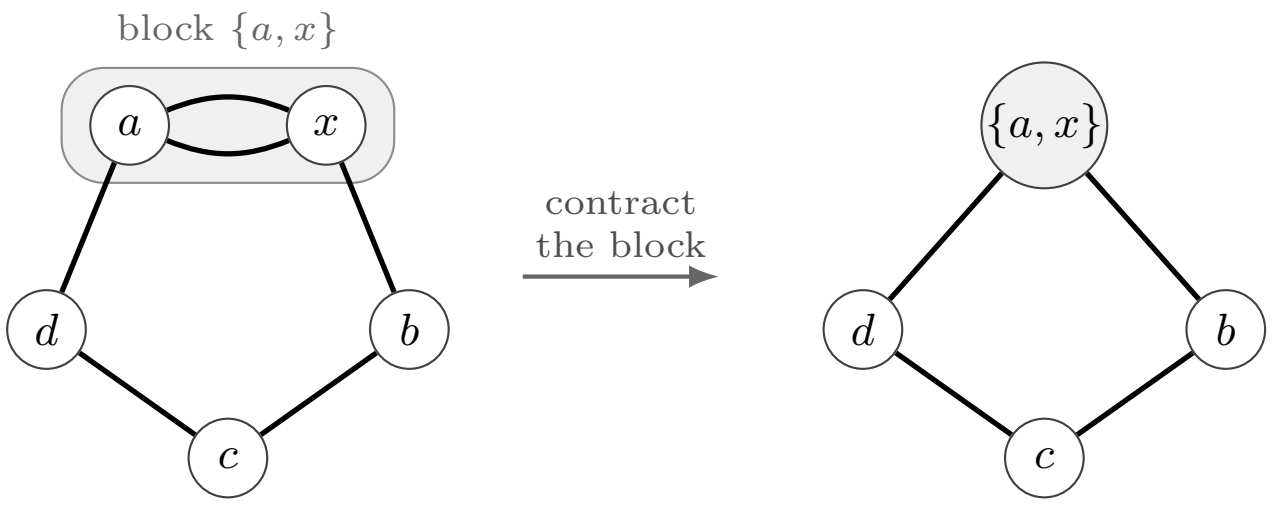
\begin{figure}[t]
\centering
\begin{tikzpicture}[
  every node/.style={font=\footnotesize},
  blk/.style={rounded corners=8pt,draw=black!45,fill=black!6},
  arr/.style={-{Latex[length=2mm]},line width=.8pt,black!60},
]
\begin{scope}
  \draw[blk] (-0.45,0.62) rectangle (1.75,1.38);
  \node[vertex] (a) at (0,1) {$a$};
  \node[vertex] (x) at (1.3,1) {$x$};
  \node[vertex] (b) at (1.85,-0.35) {$b$};
  \node[vertex] (c) at (0.65,-1.2) {$c$};
  \node[vertex] (d) at (-0.55,-0.35) {$d$};
  \draw[cycleedge] (a) to[bend left=22] (x);
  \draw[cycleedge] (a) to[bend right=22] (x);
  \draw[cycleedge] (x)--(b)--(c)--(d)--(a);
  \node[font=\scriptsize,text=black!60,anchor=south] at (0.65,1.38) {block $\{a,x\}$};
  \node[panel] at (0.65,-1.95) {(a) the graph $G$};
\end{scope}
\draw[arr] (2.6,0)--node[above,font=\scriptsize,text=black!70,align=center]{contract\\the block}(3.9,0);
\begin{scope}[xshift=5.4cm]
  \node[vertex,fill=black!6,minimum size=6mm] (ax) at (0.65,1) {$\{a,x\}$};
  \node[vertex] (b) at (1.85,-0.35) {$b$};
  \node[vertex] (c) at (0.65,-1.2) {$c$};
  \node[vertex] (d) at (-0.55,-0.35) {$d$};
  \draw[cycleedge] (ax)--(b)--(c)--(d)--(ax);
  \node[panel] at (0.65,-1.95) {(b) the quotient $Q$};
\end{scope}
\end{tikzpicture}
\caption{A quotient $M$-witness that does not lift. (a) The graph $G$ has
six edges on five vertices, so it has no $T$-witness. Its only cycles of
length at least four have length five, leaving one edge for two forests,
so it has no $M$-witness either. (b) The quotient $Q$ is a four-cycle,
which is an $M$-witness with empty forests. The two internal edges of the
block cannot supply both two spanning trees and an $a$-$x$ path. Compare
the \TreeContractionRule (\Cref{fig:contraction-rule}), where a third
internal tree supplies this path.}
\label{fig:nonliftable}
\end{figure}

Finding an $M$-witness in the quotient is not sufficient: its cycle
may fail to lift while remaining edge-disjoint from the two forests.
The following example explains why we must also control the cycle's
attachment vertices inside each block. Consider the
five-cycle $(a,x,b,c,d,a)$ and add a second copy of the edge $ax$.
The set $\{a,x\}$ is a two-tree block, and contracting it produces a
four-cycle. The original graph has no $T$-witness: two spanning trees
sharing at most one edge require at least seven edges, but it has six.
It also has no $M$-witness: its only cycles of length at least four
have length five, leaving just one edge for two forests that would
each need one edge. Thus an $M$-witness of the quotient need not lift (\Cref{fig:nonliftable}).
The two internal edges of the block cannot supply both two spanning
trees and a path from $a$ to $x$.

\subsubsection{Causal Orders and Canonical Form}
\label{sec:necessity-canonical}
We now use the complete call order to establish a lower bound on the
quotient's number of edges and describe its structure when this bound
is attained. Rather than fix one chronological order, we retain the
comparisons between calls that preserve information propagation.
This allows us to rearrange calls while keeping the scheme complete.
We first define this partial order on the original calls and then
restrict it to the quotient calls, so that the later lifting argument
retains access to paths inside the original blocks.

For a partial order, an \emph{ideal} is a set closed under
predecessors. A comparison $a<b$ is a \emph{cover} if no element lies
strictly between $a,b$. A \emph{linear extension} is a total order
respecting every comparison. We write $a<S$ when $a<s$ for every
$s\in S$.

Fix a complete call order of $G$. Its \emph{causal order} $P$ is the
transitive closure of the relations $a<b$ between chronologically
ordered calls sharing an original endpoint. A linear extension of $P$
preserves every information path and hence remains complete. Let $P_Q$
be the restriction to the external calls, identified with $E(Q)$.

\begin{lemma}\label{lem:characterization-poset-transfer}
Every linear extension of $P_Q$ is complete on $Q$.
Every cover of $P_Q$ relates incident edges of $Q$.
\end{lemma}
\begin{proof}
For any ordered pair of quotient vertices, choose an original vertex
in each corresponding block and an increasing information path
between them. Project this path to $Q$, omitting internal calls.
The retained external calls increase in $P_Q$.
Removing closed subwalks preserves this property. Thus $P_Q$ is
admissible: it provides an increasing path between every ordered pair
of quotient vertices. Every linear extension preserves these paths.

For a cover $a<_{P_Q}b$, take an original causal chain from $a$ to
$b$. It has no intermediate external call, since that would lie
strictly between $a$ and $b$ in $P_Q$. Its intermediate internal calls
all belong to one block: successive calls sharing an original endpoint
cannot move between blocks without an external call. Therefore $a,b$
share that block in $Q$. The same conclusion holds for a chain with
no intermediate call.
\end{proof}

This order must not be replaced by the usual causal order obtained by
treating every quotient block as one vertex. Such a replacement would
add comparisons between calls using different original attachment
vertices, without providing internal paths between those vertices.

\paragraph{Closed-Tree Normalization and Counting}

By \Cref{cor:characterization-edges}, a quotient on $r\geq4$ vertices
has at most $2r-3$ edges. We now show that completeness requires at
least $2r-4$ edges. The case of $2r-3$ edges already yields a
$T$-witness by \Cref{cor:characterization-t-case}, so only the case of
$2r-4$ edges remains for the subsequent $M$-witness construction.
We rearrange the early calls
to obtain a closed tree component. If this component is nontrivial,
each of its maximal calls precedes every future call. Placing one
such call last among the early calls lets us count distinct first
future calls at all vertices except its two endpoints. A singleton
component gives a separate, stronger bound. The closedness condition
below enables the exchanges needed to obtain this division.

Let $H$ be a finite simple graph on $r\ge4$ vertices, with a partial
order $P$ whose covers relate incident edges and whose linear
extensions are complete. The ideal-exchange argument follows
Bumby~\cite{Bumby81} and Kleitman and Shearer~\cite{KleitmanS80}.
Relative to an ideal $J$, we call its edges early calls and the
remaining edges future calls. A linear extension placing $J$ first
realizes this division chronologically.

A tree component $A$ of an ideal $J$ is
\emph{closed} if there is no ideal $J'\supsetneq J$ such that
$V(A)$ is the vertex set of a component of $H[J']$ and
$(J'\setminus J)\cap E(H[V(A)])\neq\varnothing$. Thus no larger
ideal can add an internal edge while retaining this component's vertex set.

The following elementary observations will be used throughout.
\begin{enumerate}
\item A tree component is closed precisely when no minimal edge outside
$J$ is internal to it. Indeed, an internal minimal edge can be added;
conversely, a saturated predecessor chain among the edges added by
any larger ideal cannot move from an internal edge to an entirely
outside edge, since covers are adjacent.
\item An edge maximal in a
component of $H[J]$ is maximal in $J$, since a saturated chain in $J$
cannot leave that component.
\item For a maximal edge $a\in J$ and a
minimal edge $b\notin J$, a comparison $a<b$ is a cover; if they
are incomparable, $(J\setminus\{a\})\cup\{b\}$ is an ideal.
\item Closure persists
under enlargement of the ideal while the component is retained.
\end{enumerate}

If an ideal has two tree components, exchanges preserving its size
produce an ideal with a closed tree component. To see this, take two
tree components $A,B$. If neither is closed, choose a minimal
remaining internal edge $b$ of $B$ and a maximal edge $a$ of $A$.
The two edges are incomparable: the reverse comparison contradicts
ideality, whereas $a<b$ would be an adjacent cover between disjoint
components. Swapping $a,b$ splits $A$ into two trees and makes $B$
cyclic. Repeat with the two pieces of $A$. Their total vertex count
is strictly smaller than $|V(A)|+|V(B)|$, so induction terminates. A
singleton component is already closed.

\begin{figure}[t]
\centering
\begin{tikzpicture}[
  every node/.style={font=\footnotesize},
  v/.style={circle,draw=black!80,fill=white,inner sep=0pt,minimum size=4.4mm},
  rv/.style={circle,draw=black,fill=black!18,line width=.9pt,inner sep=0pt,minimum size=4.6mm},
  early/.style={black,line width=.9pt},
  lastcall/.style={black,line width=2.4pt},
  futE/.style={witnessorange,line width=1pt,dash pattern=on 3pt off 1.8pt},
  firstE/.style={witnessorange,line width=1.1pt,dash pattern=on 3pt off 1.8pt,
                 -{Latex[length=2mm,width=1.6mm]}},
  comp/.style={rounded corners=8pt,fill=black!5,draw=black!25},
  newB/.style={rounded corners=7pt,fill=auxgreen!12,draw=auxgreen!80!black,line width=.8pt},
  arr/.style={-{Latex[length=2mm]},line width=.8pt,black!60},
  lbl/.style={font=\scriptsize,text=black!60},
]
\begin{scope}
  \draw[comp] (-0.4,-1.65) rectangle (1.0,1.65);
  \node[lbl] at (0.3,1.85) {$X$};
  \node[rv] (x0) at (0.3,0.45) {};  \node[rv] (x1) at (0.3,-0.45) {};
  \node[v] (t1) at (0.3,1.2) {};    \node[v] (t2) at (0.3,-1.2) {};
  \node[v] (o) at (1.85,1.2) {};
  \draw[early] (t1)--(x0) (x1)--(t2);
  \draw[lastcall] (x0)--(x1) node[midway,left=1pt] {$e$};
  \draw[futE] (t1)--(o) node[pos=.65,above] {$f$};
  \node[panel] at (0.75,-2.15) {before};
\end{scope}
\draw[arr] (2.4,-0.6)--node[above,font=\scriptsize,text=black!70,align=center]{swap $e,f$\\if incomparable}(4.0,-0.6);
\begin{scope}[xshift=4.75cm]
  \draw[comp] (-0.4,0.25) rectangle (2.25,1.65);
  \draw[newB] (-0.25,-1.75) rectangle (0.85,-0.3);
  \node[font=\scriptsize,text=auxgreen!60!black,anchor=west,align=left] at (0.95,-1.0) {$B$: smaller closed\\ tree component};
  \node[rv] (x0) at (0.3,0.6) {};  \node[rv] (x1) at (0.3,-0.65) {};
  \node[v] (t1) at (0.3,1.3) {};    \node[v] (t2) at (0.3,-1.35) {};
  \node[v] (o) at (1.85,1.3) {};
  \draw[early] (t1)--(x0) (x1)--(t2) (t1)--(o) node[midway,above] {$f$};
  \draw[futE,line width=1.3pt] (x0)--(x1);
  \node[anchor=east,inner sep=1.5pt,fill=white,text=witnessorange!90!black] at (0.12,0.0) {$e$};
  \node[panel] at (0.9,-2.15) {after};
\end{scope}
\node[panel] at (3.2,-2.75) {(a) every maximal call of $X$ precedes all future calls};
\begin{scope}[xshift=10.3cm]
  \draw[comp] (-0.4,-1.65) rectangle (1.0,1.65);
  \node[lbl] at (0.3,1.85) {$X$};
  \node[rv] (x0) at (0.3,0.45) {$x_0$};  \node[rv] (x1) at (0.3,-0.45) {$x_1$};
  \node[v] (t1) at (0.3,1.2) {};    \node[v] (t2) at (0.3,-1.2) {};
  \draw[early] (t1)--(x0) (x1)--(t2);
  \draw[lastcall] (x0)--(x1);
  \node[v] (o1) at (2.0,1.2) {};
  \node[v] (o2) at (2.0,0.0) {};
  \node[v] (o3) at (2.0,-1.2) {};
  \draw[early] (o2)--(o3);
  \draw[firstE] (x0)--(o1);
  \draw[firstE] (o1)--(t1);
  \draw[firstE] (x1)--(o2);
  \draw[firstE] (x1)--(o3);
  \draw[firstE] (o3)--(t2);
  \node[panel,align=center] at (0.85,-2.45) {(b) distinct first\\ future calls};
\end{scope}
\end{tikzpicture}
\caption{The closed-tree reduction
(\Cref{lem:characterization-closed-tree-reduction}). Black edges are early
calls in the ideal; orange dashed edges are future calls.
(a) Let $e$ be a maximal call of a minimum closed tree component $X$, and
let $f$ be a minimal future call. If $e$ and $f$ were incomparable,
exchanging them would leave the part $B$ of $X-e$ not touched by $f$ as a
smaller closed tree component. Hence $e$ precedes every future call.
(b) Placing the maximal call $e=x_0x_1$ last among the early calls, every
other vertex still needs a future call. The first such calls (arrows,
pointing to the vertex for which they are first) are pairwise distinct,
which gives at least $r-2$ future calls.}
\label{fig:closed-tree}
\end{figure}
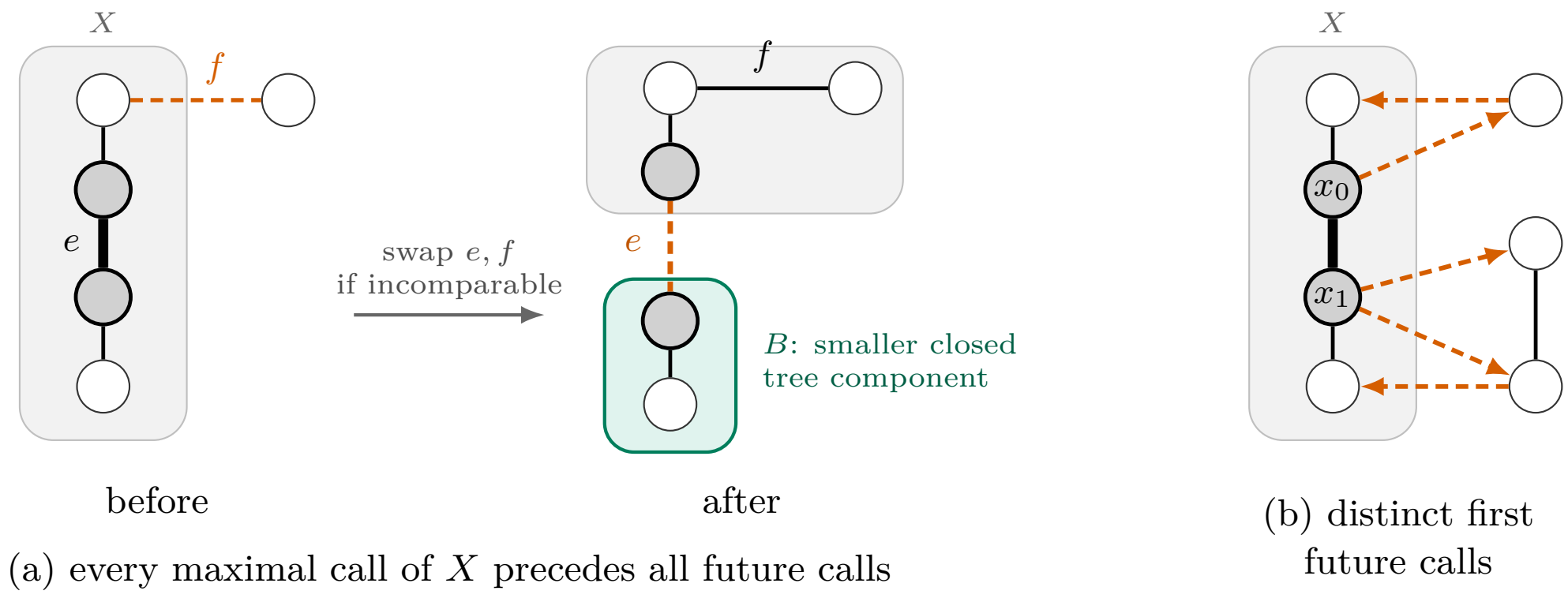

\begin{lemma}[Closed-tree reduction]\label{lem:characterization-closed-tree-reduction}
Let $J$ be an ideal with a closed tree component $A$. There are an
ideal $I$ with $|I|\ge|J|$ and a closed tree component $X$ with
$V(X)\subseteq V(A)$ such that every minimal edge of
$F=E(H)\setminus I$ crosses the boundary of $X$.
If $X$ is a singleton, then $|F|\ge r-1$. Otherwise every maximal
edge of $X$ precedes every edge of $F$. Moreover, for any maximal
edge $e=x_0x_1$ of $X$ and any linear extension placing $I$ first
and $e$ last in $I$, the first calls in $F$ of the $r-2$ vertices
other than $x_0,x_1$ exist and are pairwise distinct. In particular,
$|F|\ge r-2$.
\end{lemma}

The proof has three steps.
\begin{enumerate}
\item Choose a smallest closed tree component contained in $V(A)$
among ideals of size at least $|J|$, then enlarge the ideal while retaining that component.
Every minimal remaining call must cross its boundary.
\item If the component is nontrivial, exchanging a maximal internal
call with an incomparable minimal remaining call would produce a
smaller closed tree component. Thus every maximal internal call
precedes all remaining calls.
\item In the nontrivial case, place one such maximal call $e=x_0x_1$
last in the early ideal.
Every vertex other than its endpoints still needs a future call.
The order constraints prevent one future call from being the first
for two such vertices, giving the bound $|F|\ge r-2$.
\end{enumerate}
\Cref{fig:closed-tree} illustrates the second and third steps.

If the component is a singleton, spreading its secret to all other
vertices directly gives the stronger bound $|F|\ge r-1$.

\begin{proof}[Proof of \Cref{lem:characterization-closed-tree-reduction}]
Among ideals of size at least $|J|$ with a closed tree component
contained in $V(A)$, choose one with a minimum-size such component
$X$. This family is nonempty because it contains $(J,A)$. Enlarge
the chosen ideal maximally while retaining $X$ and its edge set as
a component; call the result $I$. Closure persists. A minimal edge
outside $I$ is not internal to $X$, by closedness, and is not
entirely outside $X$, by maximality of $I$. Thus it crosses $X$.

If $X$ is a singleton, its secret has reached nobody else at the end
of $I$ in an extension placing $I$ first. Each future call informs
at most one new vertex about that secret, so $|F|\ge r-1$.

Suppose $|V(X)|\ge2$ and let $e$ be any maximal edge of $X$.
For a minimal edge $f\in F$, the relation $f<e$ is impossible by
ideality. If $e,f$ were incomparable, interchange them. The edge
$f$ touches only one part of $X-e$; the untouched part $B$ remains
a tree component of $(I\setminus\{e\})\cup\{f\}$. It is closed. Indeed, a minimal edge
of the new complement is either $e$, an old minimal edge of the complement,
or a cover-successor of $f$: for a newly minimal old complementary
edge, all its former complementary predecessors must have been $f$.
Any element strictly between $f$ and that edge would also have been
outside $I$ and would still remain in the complement, a contradiction.
The first crosses $B$, the second
crosses $X$, and the third meets an endpoint of $f$, neither of
which belongs to $B$. None is internal to $B$. This contradicts
minimality of $X$, since $V(B)\subseteq V(A)$ and the ideal size
is unchanged. Consequently $e<f$ for every minimal $f\in F$, and
hence $e<F$.

Take any extension placing $I$ first and $e=x_0x_1$ last in $I$.
Every vertex other than $x_0,x_1$ needs a future call. A vertex outside $X$ lacks the
secrets of $X$; a vertex of $X-\{x_0,x_1\}$ lacks secrets from
the opposite component of $X-e$, since only the endpoints of the
last early call $e$ receive those secrets at that call.

If a future call $f$ were first at two vertices outside $\{x_0,x_1\}$,
it would be minimal
in $F$: otherwise a saturated chain from a predecessor in $F$ to
$f$ would end with an earlier incident future call, contrary to
first participation at both endpoints. But $e<f$ would then be a
cover, forcing $f$ to touch an endpoint of $e$. This is a
contradiction. The first future calls are therefore pairwise distinct.
\end{proof}

Starting with $r-2$ early calls, the reduction leaves at least $r-2$
future calls. Together, these give the classical lower bound.

\begin{corollary}[Call lower bound]\label{cor:characterization-call-lower-bound}
The graph $H$ has at least $2r-4$ edges.
\end{corollary}
\begin{proof}
Completeness implies connectivity, so an ideal of size $r-2$ exists.
Such an ideal has at least two tree components: if there are $t$
tree components, its edge count is at least $r-t$.
The preceding exchanges give a closed tree component without changing
its size. Apply \Cref{lem:characterization-closed-tree-reduction}. The resulting
ideal has at least $r-2$ edges and its complement has at least $r-2$
edges.
\end{proof}

If the total number of calls is exactly $2r-4$, an ideal containing
at least $r-2$ calls and a closed tree component cannot be enlarged
while retaining that component. Moreover, every minimal remaining
call must cross its boundary.

\begin{corollary}[Saturation of a closed early tree]
\label{lem:characterization-closed-tree-saturated}
Suppose $|E(H)|=2r-4$. If an ideal $J$ has $|J|\ge r-2$ and a
closed tree component $A$, then $|J|=r-2$, and every minimal edge
outside $J$ crosses the boundary of $A$.
\end{corollary}
\begin{proof}
\Cref{lem:characterization-closed-tree-reduction} gives
$|J|\le|I|\le(2r-4)-(r-2)=r-2$, so equality holds.
A minimal edge outside $J$ cannot be internal to $A$, by closedness.
If it were entirely outside $A$, adding it would retain $A$ as a
closed tree component in an ideal of size $r-1$, contradicting the
just proved bound. Thus it crosses $A$.
\end{proof}

\paragraph{Canonical Form at Equality}
It remains to handle a quotient on $r\geq4$ vertices with exactly
$2r-4$ edges. At this lower bound, every inequality in
\Cref{lem:characterization-closed-tree-reduction} is tight. We use this
equality to divide the calls into early and future calls, with the
future calls forming two trees ordered outward from their roots.
This structure will guide the choice of a quotient cycle and the
paths needed to lift it through the original blocks.

\begin{figure}[t]
\centering
\begin{tikzpicture}[every node/.style={font=\footnotesize},
  early/.style={black,line width=.9pt},
  lastcall/.style={black,line width=2.4pt},
  fut0/.style={auxgreen,line width=1.1pt,-{Latex[length=2.1mm,width=1.7mm]}},
  fut1/.style={auxpurple,line width=1.1pt,dash pattern=on 3.5pt off 1.5pt,
               -{Latex[length=2.1mm,width=1.7mm]}},
  comp/.style={rounded corners=8pt,fill=black!6,draw=black!22},
  witcycle/.style={black,line width=1.5pt,preaction={draw,black!15,line width=5.5pt}},
  blk/.style={rounded corners=6pt,draw=black!45,fill=black!4},
  port/.style={circle,fill=black,inner sep=0pt,minimum size=1.8mm},
  inner/.style={circle,fill=black!35,inner sep=0pt,minimum size=1.2mm},
  ipath/.style={line width=1.1pt,decorate,decoration={snake,amplitude=.55mm,
               segment length=3.2mm,pre length=1mm,post length=1mm},
               preaction={draw,black!15,line width=5.5pt}},
  calledge/.style={witcycle},
  note/.style={font=\scriptsize,text=black!70}]
\def\placeQ{%
  \node[root] (x0) at (0,0.9) {$x_0$};
  \node[root] (x1) at (0,-0.9) {$x_1$};
  \node[vertex] (w) at (-1.15,0) {$w$};
  \node[vertex] (U0) at (3,0.9) {$U_0$};
  \node[vertex] (U1) at (3,-0.9) {$U_1$};
  \node[vertex] (y) at (4.15,0) {$y$};
  \node[vertex] (d1) at (1,0.42) {$d_1$};
  \node[vertex] (d2) at (1,-0.42) {$d_2$};
  \node[vertex] (d3) at (2,-0.42) {$d_3$};
  \node[vertex] (d4) at (2,0.42) {$d_4$};}
\begin{scope}
  \fill[comp] (-1.55,-1.3) rectangle (0.38,1.3);
  \fill[comp] (0.62,-0.8) rectangle (2.38,0.8);
  \fill[comp] (2.62,-1.3) rectangle (4.55,1.3);
  \node[text=black!60] at (-1.3,1.05) {$X$};
  \node[text=black!60] at (4.3,1.05) {$Y$};
  \node[text=black!60] at (1.5,0) {$Z$};
  \placeQ
  \draw[early] (w)--(x0) (y)--(U1) (d1)--(d2)--(d3)--(d4)--(d1);
  \draw[early] (U0)--(U1) node[midway,right=-1pt] {$g$};
  \draw[lastcall] (x0)--(x1) node[midway,left=1pt] {$e$};
  \draw[fut0] (x0)--(U0) node[midway,above=-1pt] {$f_0$};
  \draw[fut0] (x0)--(d1);
  \draw[fut0] (U0)--(d4);
  \draw[fut0] (U0)--(y);
  \draw[fut1] (x1)--(U1) node[midway,below=-1pt] {$f_1$};
  \draw[fut1] (x1)--(d2);
  \draw[fut1] (U1)--(d3);
  \draw[fut1] (x1)--(w);
  \node[panel] at (1.5,-1.75) {(a) canonical form of a complete order};
\end{scope}
\begin{scope}[xshift=6.95cm]
  \placeQ
  \draw[witcycle] (x0)--(U0)--(U1)--(x1)--(x0);
  \draw[blueforest] (d1)--(d2)--(d3)--(d4) (w)--(x0) (y)--(U1) (x0)--(d1);
  \draw[orangeforest] (d4)--(d1) (x1)--(d2) (U0)--(d4) (U1)--(d3)
                      (x1)--(w) (U0)--(y);
  \node[panel] at (1.5,-1.75) {(b) $M$-witness of $Q$};
\end{scope}
\begin{scope}[yshift=-4.5cm,xshift=-0.15cm]
  \foreach \cx/\cy/\hx/\vy/\nm/\nt in {0/1.2/1/-1/{$B_{x_0}$}/{after $e$},
      5.2/1.2/-1/-1/{$B_{U_0}$}/{before $f_0$},
      5.2/-1.2/-1/1/{$B_{U_1}$}/{before $f_1$},
      0/-1.2/1/1/{$B_{x_1}$}/{after $e$}}{
    \draw[blk] (\cx-1.2,\cy-0.65) rectangle (\cx+1.2,\cy+0.65);
    \node[note] at (\cx-\hx*0.6,\cy-\vy*0.9) {\nm};
    \node[note] at (\cx-\hx*0.42,\cy+\vy*0.05) {\nt};
    \foreach \dx/\dy in {-0.88/0.42,0.02/-0.45,0.25/0.38}
      \node[inner] at (\cx+\hx*\dx,\cy+\vy*\dy) {};
    \coordinate (v) at (\cx+\hx*0.6,\cy+\vy*0.65);
    \coordinate (h) at (\cx+\hx*1.2,\cy-\vy*0.2);
    \pgfmathsetmacro{\bend}{-25*\hx*\vy}
    \draw[ipath] (v) to[bend right=\bend] (h);
  }
  \draw[calledge] (0.6,0.55)--(0.6,-0.55) node[midway,right=3pt] {$e$};
  \draw[calledge] (1.2,1.4)--(4.0,1.4) node[midway,above=1pt] {$f_0$};
  \draw[calledge] (4.6,0.55)--(4.6,-0.55) node[midway,left=3pt] {$g$};
  \draw[calledge] (4.0,-1.4)--(1.2,-1.4) node[midway,below=1pt] {$f_1$};
  \foreach \p in {(0.6,0.55),(1.2,1.4),(4.0,1.4),(4.6,0.55),
                  (4.6,-0.55),(4.0,-1.4),(1.2,-1.4),(0.6,-0.55)}
    \node[port] at \p {};
  \node[anchor=east,inner sep=2.5pt] at (0.6,0.55) {$a_0$};
  \node[anchor=east,inner sep=2.5pt] at (0.6,-0.55) {$a_1$};
  \node[panel] at (2.6,-2.45) {(c) lifting the cycle into the blocks};
\end{scope}
\begin{scope}[xshift=7.75cm,yshift=-2.85cm]
  \node[anchor=west,font=\scriptsize\bfseries] at (-0.05,0.38) {panel (a)};
  \draw[early] (0,0)--(0.55,0) node[right,text=black] {early call (in $I$)};
  \draw[lastcall] (0,-0.38)--(0.55,-0.38) node[right,text=black] {last early call $e$ of $X$};
  \draw[fut0] (0,-0.76)--(0.55,-0.76) node[right,text=black] {future call, tree of $x_0$};
  \draw[fut1] (0,-1.14)--(0.55,-1.14) node[right,text=black] {future call, tree of $x_1$};
  \node[anchor=west,font=\scriptsize\bfseries] at (-0.05,-1.62) {panels (b) and (c)};
  \draw[witcycle] (0,-2.0)--(0.55,-2.0) node[right,text=black] {witness cycle};
  \draw[blueforest] (0,-2.38)--(0.55,-2.38) node[right,text=black] {forest $L_1$};
  \draw[orangeforest] (0,-2.76)--(0.55,-2.76) node[right,text=black] {forest $L_2$};
  \draw[ipath] (0,-3.14)--(0.65,-3.14) node[right,text=black] {internal path $P_X$};
\end{scope}
\end{tikzpicture}
\caption{The necessity proof of \Cref{thm:characterization} on a quotient
$Q$ with $r=10$ vertices and $2r-4=16$ edges that contains no nontrivial
two-tree set. A complete call order is $d_1d_2$, $d_3d_4$, $d_2d_3$,
$d_4d_1$, $wx_0$, $yU_1$, $U_0U_1$, $x_0x_1$, followed by $x_0U_0$,
$x_1U_1$, $x_0d_1$, $x_1d_2$, $U_0d_4$, $U_1d_3$, $x_1w$, $U_0y$; every
linear extension of its causal order is complete.
(a) The canonical form of
\Cref{lem:characterization-canonical-ideal,cor:characterization-early-components}.
The early calls form the tree components $X$ and $Y$ and the unicyclic
component $Z$; the call $e=x_0x_1$ is the unique maximal call of $X$
and precedes every future call. The future calls form two trees rooted at
$x_0$ and $x_1$ and are ordered outward from the roots; the minimal
future calls $f_0,f_1$ join $X$ to $Y$. Here
$f_i=f_{U_i}=x_iU_i$ for $i\in\{0,1\}$.
(b) The $M$-witness of $Q$
obtained from the four-cycle $(x_0,U_0,U_1,x_1)$ by
\Cref{lem:characterization-protected-extension}: the exchange moves the
early call $d_4d_1$ into the spanning tree $Q[F]+e$ and the future call
$x_0d_1$ out of it, which removes the unicyclic component.
(c) Lifting to $G$ (\Cref{lem:characterization-m-lift}), shown
schematically. Each quotient vertex represents an original block.
To lift the witness, each block on the cycle needs two spanning trees
and a path between its two ports---the original endpoints of its
incident cycle edges---with all three edge sets pairwise disjoint.
In $B_{x_i}$, the path uses calls after $e$, and deleting it leaves
those calls connected, providing one spanning tree; the calls before
$e$ provide the other. In $B_{U_i}$, the path and the first spanning
tree use calls before $f_i$, while the second spanning tree uses
calls after $f_i$
(\Cref{lem:characterization-threshold,lem:characterization-compatible-actual-cycle}).}
\label{fig:necessity-canonical}
\end{figure}
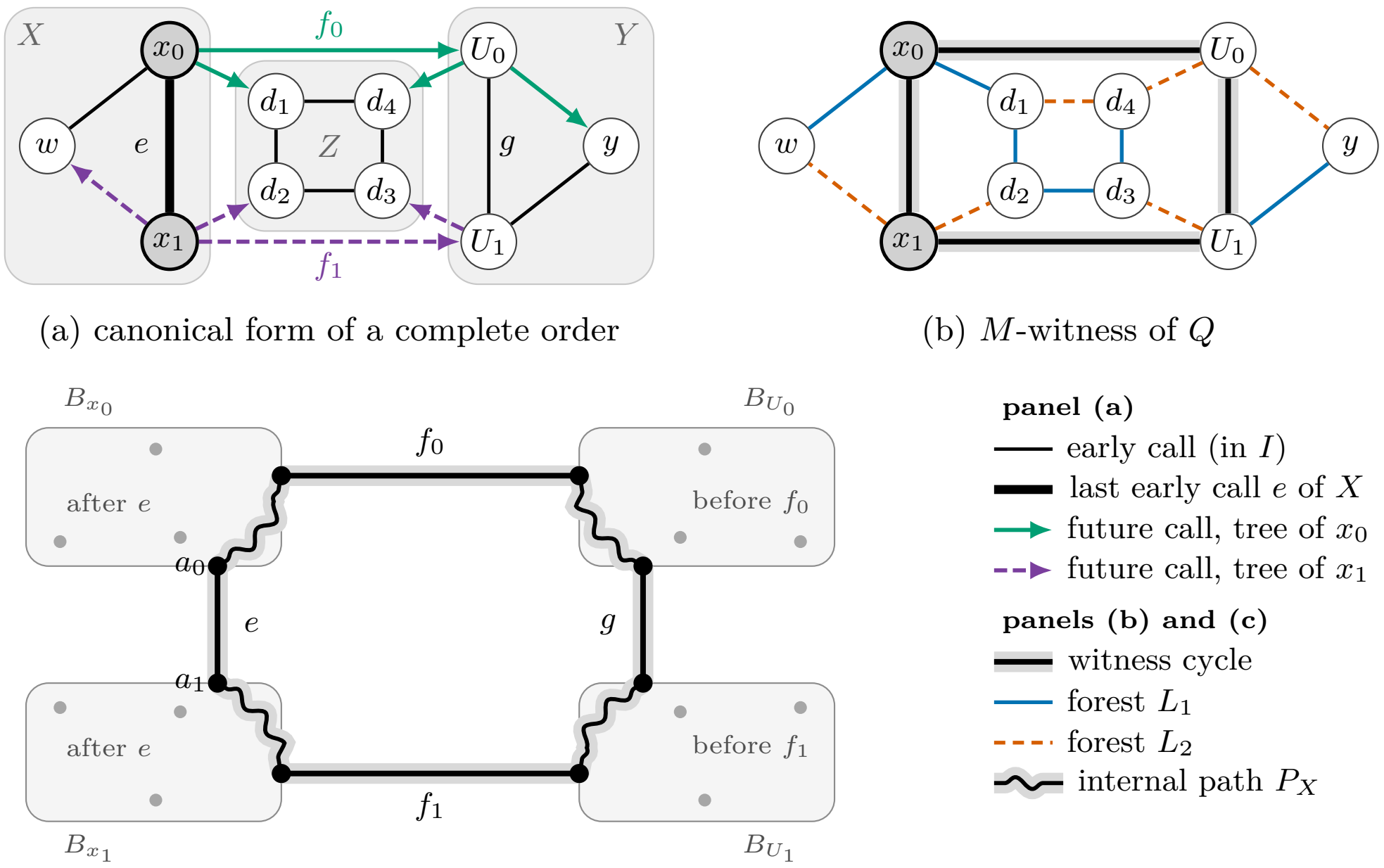

\Cref{fig:necessity-canonical} illustrates this canonical form and the
subsequent construction and lifting of a quotient $M$-witness.

\begin{lemma}[Canonical ideal normalization]\label{lem:characterization-canonical-ideal}
Let $H$ be a finite simple graph on $r\ge4$ vertices with $2r-4$
edges. Let $P$ be a partial order on its edges such that every linear
extension is a complete call order, and every cover relation of $P$
relates incident edges. Then there is an ideal $I$ with complement $F$,
a tree component $X$ of $H[I]$, and an edge $e=x_0x_1$ of $X$ such that:
\begin{enumerate}
\item $|I|=|F|=r-2$ and $|V(X)|\ge2$;
\item $e$ is the unique maximal edge of $X$, and every other edge of
 $X$ precedes $e$ in $P$;
\item $e<_P f$ for every $f\in F$;
\item $H[F]$ is a spanning forest with two components rooted at
 $x_0,x_1$. Its edges are ordered outward: on each root-to-vertex path,
 successive edges increase in $P$;
\item every $P$-increasing trail ending at $x_0$ or $x_1$ contains at
 most one edge of $F\cup\{e\}$, and that edge is last.
\end{enumerate}
Moreover, every $P$-minimal edge of $F$ crosses the boundary of $X$.
Here an increasing trail has distinct edges, may repeat vertices, and
its successive edges are strictly increasing in $P$.
\end{lemma}
\begin{proof}
An ideal of size $r-2$ has at least two tree components. The preceding
exchanges produce an ideal $J$ of that size with a closed tree
component. Apply \Cref{lem:characterization-closed-tree-reduction}, obtaining
$I,X,F$. The singleton case would give at least $2r-3$ edges, so
$X$ is nontrivial. The first-call injection and the total $2r-4$
force $|I|=|F|=r-2$. Fix a maximal edge $e=x_0x_1$ of $X$.
In every linear extension placing $I$ first and $e$ last in $I$,
each future edge is the first future call of exactly one nonroot.
Choose one such extension.

Direct each future edge toward its uniquely assigned nonroot. In the
chosen extension, its other endpoint is a root or has participated in
an earlier future call. Starting with the two roots, the future calls
therefore add one new vertex at a time. They form a spanning forest
with exactly two rooted components, ordered outward in that extension.
This outward order holds in $P$ itself. Indeed, the first-call
injection and equality argument applies to \emph{every} linear
extension putting $I$ first and $e$ last in $I$. In every such
extension every prefix of the future calls is a forest each of whose
nontrivial components contains a root. For the fixed graph $H[F]$, this forces
every ancestor edge to occur before every descendant edge. All linear
extensions of $P$ restricted to $F$ occur in such extensions of $P$:
place $I$ first in any extension ending at its maximal edge $e$.
Therefore every ancestor edge precedes its descendants in $P$.
Indeed, if such a pair were incomparable, adjoining its reverse
comparison would remain acyclic and admit a linear extension,
contrary to the order forced in every extension.

Both roots must participate in $F$, since $X$ is a proper component
of $H[I]$ and neither root yet knows secrets outside $X$. Let $f_i$
be the first future edge at $x_i$. It is also first at its other
endpoint, because every future edge is assigned to a nonroot and
its root endpoint cannot be that assigned vertex. Hence $f_i$ is
minimal in $F$. It crosses the boundary of $X$.
If $s$ is any maximal edge of $X$,
\Cref{lem:characterization-closed-tree-reduction} gives $s<_P F$.
Thus $s<_P f_0$ and $s<_P f_1$
are covers. Since $s$ is internal to $X$, adjacency forces its
endpoints to be $x_0,x_1$. Simplicity gives $s=e$. Finiteness then
implies that every other edge of $X$ precedes $e$.

Consider now an increasing trail ending at a root. Once it enters
$F$, all its subsequent edges belong to $F$, because $I$ is an ideal.
A trail in a forest is a path. A path of at least two future edges
ending at a root runs inward and therefore has decreasing consecutive
edges, which is impossible. If the trail uses $e$, any subsequent
early edge would have to be internal to $X$ and follow $e$, also
impossible. Its remaining suffix therefore lies in the tree $H[F]+e$.
From an endpoint of $e$, future edges run outward, so a nonempty
such suffix cannot end at either root. Consequently $e$, when used,
is last and no future edge precedes it. This proves the final claim.
\end{proof}

\begin{corollary}[The other early components are unicyclic]
\label{cor:characterization-early-components}
For the ideal $I$ in \Cref{lem:characterization-canonical-ideal}, the graph
$H[I]$ has exactly two tree components, one of which is $X$.
All its other components are unicyclic. In particular, the second
tree component $Y$ is closed, and every minimal edge of $F$
has one endpoint in $X$ and the other in $Y$.
\end{corollary}
\begin{proof}
There are at least two tree components because $|I|=r-2$.
Every tree component other than $X$ is closed: every minimal
edge of $F$ crosses the boundary of $X$ and consequently cannot
be internal to a different component. The closure criterion
therefore applies. By \Cref{lem:characterization-closed-tree-saturated}, every
minimal edge of $F$ crosses the boundary of each such tree
component as well as that of $X$. Three tree components would
require a minimal edge of the nonempty set $F$ to meet three
disjoint vertex sets, which is impossible. There are exactly two
tree components, and every minimal edge of $F$ joins them.

For each connected component $D$ of $H[I]$, set
$\beta(D)=|E(D)|-|V(D)|+1$. This is zero for a tree, one for a
unicyclic component, and at least two for any other component.
The equality $|I|=r-2$ gives
\[
 2=\sum_D(1-\beta(D))
   =2-\sum_{D:\,\beta(D)\ge1}(\beta(D)-1).
\]
Each summand on the right is nonnegative, so every nontree
component has $\beta(D)=1$, as claimed.
\end{proof}

\subsubsection{Selecting Paths and Extending Cycles}
\label{sec:necessity-path-selection}
To construct an $M$-witness that lifts to the original graph, we need
two kinds of tools. Inside each block on the cycle, we must select a
path between its attachment vertices while leaving enough edges for
the two internal spanning trees. At the quotient level, we must
extend the selected four-cycle to an $M$-witness without changing
that cycle. We establish these tools here;
\Cref{sec:necessity-ports} uses them to select a four-cycle with
compatible attachment paths.

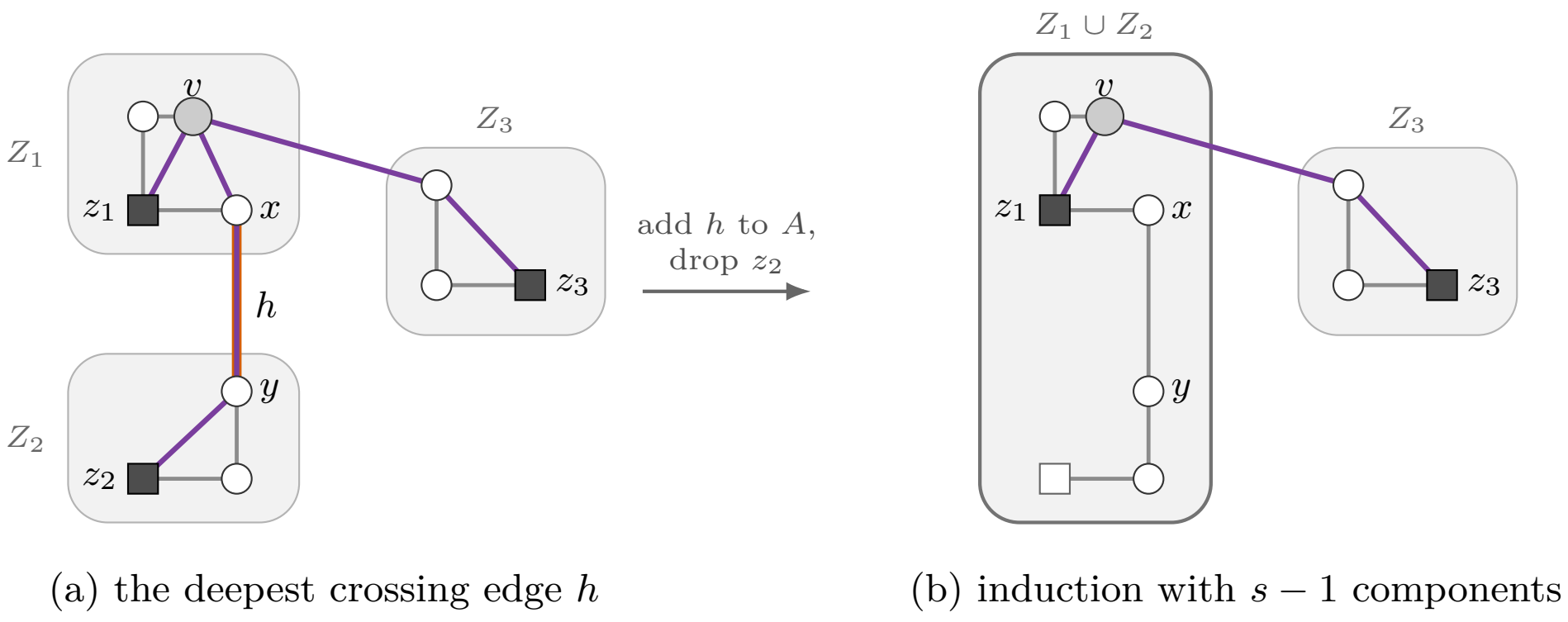
\begin{figure}[t]
\centering
\begin{tikzpicture}[
  every node/.style={font=\footnotesize},
  v/.style={circle,draw=black!80,fill=white,inner sep=0pt,minimum size=2.4mm},
  sp/.style={rectangle,draw=black,fill=black!70,inner sep=0pt,minimum size=2.4mm},
  spoff/.style={rectangle,draw=black!60,fill=white,inner sep=0pt,minimum size=2.4mm},
  aE/.style={black!45,line width=.9pt},
  tE/.style={auxpurple,line width=1.2pt},
  hE/.style={witnessorange,line width=2.2pt},
  zc/.style={rounded corners=9pt,fill=black!5,draw=black!30},
  zm/.style={rounded corners=9pt,fill=black!5,draw=black!55,line width=.8pt},
  arr/.style={-{Latex[length=2mm]},line width=.8pt,black!60},
  lbl/.style={font=\scriptsize,text=black!60},
]
\newcommand{\staticbase}{%
  \coordinate (v) at (1.0,1.7); \coordinate (a1) at (0.6,1.7); \coordinate (z1) at (0.6,0.95);
  \coordinate (p) at (1.35,0.95);
  \coordinate (y) at (1.35,-0.5); \coordinate (a2) at (1.35,-1.2); \coordinate (z2) at (0.6,-1.2);
  \coordinate (q) at (2.95,1.15); \coordinate (a3) at (2.95,0.35); \coordinate (z3) at (3.7,0.35);}
\newcommand{\staticlabels}{%
  \node[anchor=south,inner sep=4.5pt] at (v) {$v$};
  \node[anchor=west,inner sep=5pt] at (p) {$x$};
  \node[anchor=west,inner sep=5pt] at (y) {$y$};}
\begin{scope}
  \staticbase
  \draw[zc] (0,0.6) rectangle (1.85,2.2);  \node[lbl,anchor=east] at (-0.05,1.4) {$Z_1$};
  \draw[zc] (0,-1.55) rectangle (1.85,-0.2); \node[lbl,anchor=east] at (-0.05,-0.88) {$Z_2$};
  \draw[zc] (2.55,-0.05) rectangle (4.3,1.45); \node[lbl,anchor=south] at (3.42,1.45) {$Z_3$};
  \draw[aE] (v)--(a1)--(z1)--(p) (y)--(a2)--(z2) (q)--(a3)--(z3);
  \draw[tE] (v)--(z1) (v)--(p) (y)--(z2) (v)--(q)--(z3);
  \draw[hE] (p)--(y);
  \draw[tE] (p)--(y);
  \node[anchor=west,inner sep=4pt] at (1.35,0.2) {$h$};
  \foreach \u in {p,y,a1,a2,a3,q} \node[v] at (\u) {};
  \node[v,fill=black!20,minimum size=3mm] at (v) {};
  \foreach \u in {z1,z2,z3} \node[sp] at (\u) {};
  \node[anchor=east,inner sep=5.5pt] at (z1) {$z_1$};
  \node[anchor=east,inner sep=5.5pt] at (z2) {$z_2$};
  \node[anchor=west,inner sep=5.5pt] at (z3) {$z_3$};
  \staticlabels
  \node[panel] at (2.05,-2.1) {(a) the deepest crossing edge $h$};
\end{scope}
\draw[arr] (4.6,0.3)--node[above,font=\scriptsize,text=black!70,align=center]{add $h$ to $A$,\\drop $z_2$}(5.95,0.3);
\begin{scope}[xshift=7.3cm]
  \staticbase
  \draw[zm] (0,-1.55) rectangle (1.85,2.2); \node[lbl,anchor=south] at (0.92,2.2) {$Z_1\cup Z_2$};
  \draw[zc] (2.55,-0.05) rectangle (4.3,1.45); \node[lbl,anchor=south] at (3.42,1.45) {$Z_3$};
  \draw[aE] (v)--(a1)--(z1)--(p) (y)--(a2)--(z2) (q)--(a3)--(z3) (p)--(y);
  \draw[tE] (v)--(z1) (v)--(q)--(z3);
  \foreach \u in {p,y,a1,a2,a3,q} \node[v] at (\u) {};
  \node[v,fill=black!20,minimum size=3mm] at (v) {};
  \foreach \u in {z1,z3} \node[sp] at (\u) {};
  \node[spoff] at (z2) {};
  \node[anchor=east,inner sep=5.5pt] at (z1) {$z_1$};
  \node[anchor=west,inner sep=5.5pt] at (z3) {$z_3$};
  \staticlabels
  \node[panel] at (2.05,-2.1) {(b) induction with $s-1$ components};
\end{scope}
\end{tikzpicture}
\caption{The induction step of \Cref{lem:characterization-static-selection}.
Gray edges belong to $A$ and form the components $Z_i$, each with a
specified vertex $z_i$ (squares); purple edges form a minimal tree
$T$ with $E(T)\subseteq D$ containing $v$ and all $z_i$, rooted at $v$. (a) The
crossing edge $h=xy$ of $T$ with $x$ deepest has its whole subtree below
$y$ inside one component, here $Z_2$, and only the $v$-$z_2$ path uses it.
(b) Adding $h$ to $A$ merges $Z_2$ with the component of $x$, and $z_2$ is
no longer specified. The path selected by induction avoids $h$ and is
also nonseparating in the original graph.}
\label{fig:static-selection}
\end{figure}

The first lemma selects a path whose deletion preserves connectivity.
Its ordered version will let us apply this selection inside a block
using calls before or after a specified call.
\Cref{fig:static-selection} illustrates the induction step in the proof
of the static path-selection lemma.

\begin{lemma}\label{lem:characterization-static-selection}
Let $J$ be a graph with disjoint edge sets $A,D$. The graph
$(V(J),A)$ has components $Z_1,\ldots,Z_s$, and each $Z_i$ contains
a specified vertex $z_i$. Suppose $D$ connects a vertex $v$ to every
$z_i$. Then some $v$-$z_i$ path in $D$ can be deleted from $J$
while leaving $J$ connected.
\end{lemma}
\begin{proof}
We induct on $s$. It suffices to retain $A$ and a minimal tree $T$
containing $v,z_1,\ldots,z_s$, with $E(T)\subseteq D$.
If $v$ is one of the specified vertices, choose
the trivial path. If $s=1$, deleting any $v$-$z_1$ path of $T$
leaves the connected spanning graph $(V(J),A)$.

For $s\ge2$, root $T$ at $v$. Choose an edge $h=xy$ joining
distinct components of $(V(J),A)$, with $x$ the parent of $y$ and
$x$ as deep as possible. Such an edge exists because $T$ connects
specified vertices in different components. By the choice of $h$,
all vertices in the subtree rooted at $y$ lie in the same component
$Z_i$. Minimality of $T$ ensures that this subtree contains a specified
vertex, which must be $z_i$. Thus $h$ is used by the $v$-$z_i$ path
and by none of the other specified root paths. Let $Z_j$ be the
component containing $x$, where $j\ne i$. Add $h$ to $A$, merging
$Z_i,Z_j$.
Discard $z_i$ from the specified list and let $D'$ be the union of
the edge sets of the remaining root paths, keeping the full vertex set and
$A\cup\{h\}$. The merged component retains $z_j$, and every other
component retains its specified vertex. The remaining root paths
avoid $h$, so induction applies with $s-1$ components and the
disjoint connecting edge set $D'$. Its selected path also leaves the original
graph $J$ connected when deleted.
\end{proof}

In the ordered setting, gathering paths reach the endpoints of
external calls before those calls occur, and connecting paths leave
those endpoints afterward. Restricting to maximal external calls
makes the two collections of internal edges disjoint. Paths between
comparable external calls ensure that the retained endpoints still
suffice for gathering, allowing us to apply the preceding lemma.

\begin{lemma}\label{lem:characterization-antichain-selection}
Let $J$ be a subgraph of a graph whose calls carry a partial order
$P$. Let $t_1,\ldots,t_k$ be external calls with endpoints $z_i$ in
$J$, and fix $v\in V(J)$. All comparisons and increasing paths below
use the restriction of $P$ to $E(J)\cup\{t_1,\ldots,t_k\}$. Suppose:
\begin{enumerate}
\item every vertex of $J$ has an internal increasing path to some
 $z_i$, all of whose calls precede $t_i$;
\item every $z_i$ has an internal increasing path to $v$, all of
 whose calls follow $t_i$;
\item if $t_i<t_j$, there is an internal increasing path from $z_i$
 to $z_j$ whose calls lie strictly between $t_i$ and $t_j$.
\end{enumerate}
Then some $z_i$-$v$ path can be deleted while retaining connectivity
of $J$. The selected path uses only calls from the paths in (2).
\end{lemma}
\begin{proof}
Retain only maximal calls among $t_1,\ldots,t_k$. Condition (3)
extends every gathering path in (1) to the port of a maximal call,
so these ports still suffice. Let $A$ be the union of the edge sets of
their gathering paths and $D$ the union of the edge sets of their
paths in (2). These edge sets are
disjoint: a common call would satisfy $t_i<a<t_j$ for two maximal
calls, a contradiction. Every component of $(V(J),A)$ contains a
port. Choose one port in each component and apply
\Cref{lem:characterization-static-selection}.
\end{proof}

The next lemma completes the quotient forest construction once a
compatible four-cycle has been selected. It does not select the
original attachment paths.

\begin{lemma}\label{lem:characterization-protected-extension}
Suppose $Q$ contains no nontrivial two-tree set, has $2r-4$ edges,
and has the canonical partition $I,F$ of
\Cref{lem:characterization-canonical-ideal}. Let $Y$ be its other early tree.
Suppose $g=u_0u_1$ is an edge of $Y$ and
$f_i=x_iu_i\in F$ for $i=0,1$. Then the four-cycle with edges
$e,g,f_0,f_1$ extends to an $M$-witness of $Q$ whose two forests
each have exactly one of $x_0,x_1,u_0,u_1$ in every component.
\end{lemma}
\begin{proof}
Start with the spanning tree $B=Q[F]+e$. Its complement
$K=Q[I\setminus\{e\}]$
has three distinguished tree components: the two parts $A_0,A_1$
of $X-e$ and the tree $Y$. Every other component is unicyclic by
\Cref{cor:characterization-early-components}. Among spanning trees $B$
containing $e,f_0,f_1$ but not $g$, whose complements have three
distinct tree components containing $x_0,x_1,u_0$ and all other
components unicyclic, minimize the number of complementary components.
The initial tree $Q[F]+e$ satisfies these conditions. Write
$K=Q-E(B)$ for the complement of the chosen tree. Since $g=u_0u_1$
remains in the complement, $u_1$ belongs to the component of $u_0$.

For a remaining unicyclic component $Z$, every edge of $K[V(Z)]$
joining different components of $B[V(Z)]$ is a bridge of $K[V(Z)]$.
Otherwise add such a nonbridge edge $h$ to $B$ and remove a tree
edge $t$ on its fundamental cycle crossing the boundary of $Z$.
The edge $t$ is none of $e,f_0,f_1$: all their endpoints lie in the
three distinguished components, whereas $t$ crosses the boundary of
$Z$. The edge $g$ remains in the complement. The new complement consists of $Z-h$ joined to
another component by $t$. If that component is a tree, the merge
is a tree; if it is unicyclic, the merge is unicyclic. None of the
distinguished components merge with each other. Thus the three specified
vertices remain in distinct tree components, and all other components
remain unicyclic. This reduces the component count, a
contradiction.

Since $Z$ is unicyclic, the bridge property puts its unique cycle
inside the vertex set $W$ of one component of $B[V(Z)]$.
The set $W$ is a proper subset of $V(Z)$: otherwise
$B[V(Z)]$ and a spanning tree of $K[V(Z)]$ would be two edge-disjoint spanning
trees on $V(Z)$. Replace the tree $B[W]$ using
\Cref{lem:characterization-exchange} on $Q[W]$. The new internal complement
is a forest whose partition coarsens the old one. If the old internal
complement has $q$ components, it has $|W|-q+1$ edges, since exactly
one component contains the unique cycle of $Z$. The new complement has
the same number of edges and is a forest, so it has $q-1$ components.
Since its partition coarsens the old one, two old internal complementary
components merge.

Globally the complementary vertex components remain unchanged, and
the component $Z$ stays connected. Its number of edges is unchanged,
so it remains unicyclic. The vertex components of $B[V(Z)]$ also remain
unchanged. Since $W\subseteq V(Z)$, the edges $e,f_0,f_1,g$ are untouched,
and the new tree satisfies the same minimizing conditions.
But an old complementary path
between the newly merged pieces leaves $W$ and crosses an unchanged
bridge joining distinct components of $B[V(Z)]$. The new internal
connections preserve every old internal complementary component
without using the bridge, since the bridge leaves $W$. Thus the
old connections on either side of its cut remain, and the newly
joined pieces connect the sides without the bridge. This makes
that bridge a nonbridge, contradicting the bridge
property for the equally minimizing new tree. Thus no unicyclic
component remains.

The resulting complement $K'=Q-E(B)$ is a forest with $r-3$ edges and three
components containing $x_0,x_1,u_0$ separately. Set
\[
 L_1=K'-g,\qquad L_2=B-\{e,f_0,f_1\}.
\]
Since $g$ lies in the tree component containing $u_0,u_1$, deleting it
separates these two vertices. The first forest therefore has four
components separating the four specified vertices.
The second does too, because its three removed edges form the path
$u_0,x_0,x_1,u_1$ in a spanning tree. Both avoid the four-cycle,
and they are edge-disjoint from each other. Each of their components
contains exactly one cycle vertex.
\end{proof}

\subsubsection{Selecting Compatible Ports Using Original Routes}
\label{sec:necessity-ports}

We now carry out the final stage of the necessity proof: select a
four-cycle in the quotient whose witness can be lifted to the original
graph. Finding the cycle alone is not enough. In each original block
on the cycle, its two attachment vertices---called \emph{ports}---must
be joined by a path such that this path and two spanning trees of the
block are pairwise edge-disjoint. We use information routes in the
original graph to select the quotient edges and guarantee these
internal paths. We then extend the four-cycle to an $M$-witness using
\Cref{lem:characterization-protected-extension} and lift that witness
using \Cref{lem:characterization-m-lift}.

We keep the original causal order $P$ and its restriction $P_Q$ to
external calls. Throughout this section, $Q$ has $r\ge4$ vertices and
$2r-4$ edges, and $I,F,X,Y,e=x_0x_1$ are as in
\Cref{lem:characterization-canonical-ideal} and
\Cref{cor:characterization-early-components}. Let $B_U$ denote the original
block represented by a quotient vertex $U$. For an external call $a$
incident with $U$, write $p_U(a)$ for its original endpoint in $B_U$.
In particular, let $a_i=p_{x_i}(e)$ for $i\in\{0,1\}$.

An \emph{original route} is a path in $G$ whose successive calls
increase in $P$. Completeness supplies such a route between every
ordered pair of original vertices. Its external projection is a
$P_Q$-increasing trail: its edges are distinct, although its quotient
vertices may repeat. For an original block $B$ and a call $t$, define
\[
 J_B^-(t)=\bigl(B,\{a\in E(G[B]):a<_P t\}\bigr),\qquad
 J_B^+(t)=\bigl(B,\{a\in E(G[B]):t<_P a\}\bigr).
\]
These graphs retain every vertex of $B$. In the following, calls before
or after $t$ mean predecessors or successors of $t$ in the causal
order $P$. All paths below may be
trivial. Deleting the edges of a trivial path leaves the graph
unchanged.

The selection has three steps. First, retain routes whose final
attachment to $B_{x_i}$ admits an internal path whose deletion
preserves the required connectivity. Second, use these routes to find
a connected set of blocks in which internal return paths can be
forced. Third, select two adjacent blocks with routes to different
roots and compatible ports at both ends. We begin with the confinement and connectivity facts
needed in the first two steps.

\begin{lemma}[Confinement to a quotient forest]
\label{lem:characterization-forest-confinement}
Let $a<_{P_Q}b$ be external calls sharing a quotient vertex $U$.
Suppose a quotient forest $R$ contains $a,b$ and every external call
$c$ with $a<_{P_Q}c<_{P_Q}b$. Then there is an internal original
route from $p_U(a)$ to $p_U(b)$ whose calls lie strictly between
$a$ and $b$ in $P$.
\end{lemma}
\begin{proof}
Place a marker at both endpoints immediately after call $a$.
A causal chain from $a$ to $b$ implies that at least one endpoint of
$b$ receives the marker before $b$. Use only the intermediate calls
of this chain to transmit the marker. Tracing these transmissions gives
an increasing original path, using calls strictly between $a,b$ in $P$,
from an endpoint of $a$ to an endpoint of $b$.

Every external call of this path belongs to $R-\{a,b\}$. Deleting
$a,b$ from the forest separates their two nonshared endpoints from
each other and from their common endpoint $U$. Thus the only possible
match between an endpoint component of $a$ and an endpoint component
of $b$ is the component containing $U$. The path therefore starts at
$p_U(a)$ and ends at $p_U(b)$. Its external projection is a closed
trail from $U$ to $U$ in a forest, so it is empty. The path is internal
to $B_U$.
\end{proof}

We will apply this lemma in three settings. Comparisons between early
calls in the same early tree are confined to that tree. Indeed, $I$
is an ideal, and a causal chain between its external calls cannot
leave an early component: consecutive external calls of a causal
chain share a quotient vertex. Comparisons between future calls are
confined to $F$, since $I$ is an ideal. Finally, comparisons from $e$
to future calls are confined to the tree $Q[F]+e$. To see this, $e$ is
maximal in $I$: a chain of incident early calls cannot leave $X$,
and $e$ is maximal in $X$. Hence no other early call lies above $e$.

\begin{lemma}[Internal connectivity at the relevant thresholds]
\label{lem:characterization-internal-threshold-connectivity}
For each $i\in\{0,1\}$, the graphs
$J_{B_{x_i}}^-(e)$ and $J_{B_{x_i}}^+(e)$ are connected.
Moreover, if $U\in V(Y)$ and $f=Ux_i\in F$, then $J_{B_U}^+(f)$
is connected.
\end{lemma}
\begin{proof}
Fix an original vertex $z\in B_{x_i}$ and an original route from
$z$ to $a_{1-i}$. Its first external call lies in $I$. Otherwise
all its external calls would lie in $F$, and a trail in $F$ could
not connect its different root components. Denote this first call
by $a$. The call $a$ belongs to $X$ and is incident with $x_i$.
If $a=e$, the initial internal segment gathers $z$ to $a_i$ before
$e$. Otherwise $a<_{P_Q}e$, and
\Cref{lem:characterization-forest-confinement}, applied in $X$, gives an internal
route from $p_{x_i}(a)$ to $a_i$ between $a,e$. Concatenate it with
the initial internal segment. Erasing closed subwalks gives an
internal route from $z$ to $a_i$ before $e$. These routes prove that
$J_{B_{x_i}}^-(e)$ is connected.

Next take an original route from $a_{1-i}$ to $z$.
\Cref{lem:characterization-canonical-ideal} applies to its external trail ending
at $x_i$. Its last external call is either $e$ or a future call $t$
incident with $x_i$. Indeed, if there were no future call, the unique
early-tree path from $x_{1-i}$ to $x_i$ would use $e$.
If the last call is $e$, its internal suffix starts at $a_i$ and
lies after $e$. Otherwise $e<_{P_Q}t$, and confinement in $Q[F]+e$
gives an internal route from $a_i$ to $p_{x_i}(t)$ between $e,t$.
Append the internal suffix following $t$. This proves that
$J_{B_{x_i}}^+(e)$ is connected.

Finally, fix $U\in V(Y)$ with $f=Ux_i\in F$ and an original
vertex $z\in B_U$. Take an original route from $a_i$ to $z$.
Some external call of this route lies in $F$, since its endpoints
belong to different early components. All external calls after the
first future call remain in $F$. Let $t$ be its last external call,
which is incident with $U$. If $t=f$, its internal suffix gathers
$p_U(f)$ to $z$ after $f$. Otherwise $t$ is an edge from $U$ to one
of its future children. The outward order gives $f<_{P_Q}t$.
Confinement in $F$ gives an internal route from $p_U(f)$ to $p_U(t)$
between $f,t$; append the suffix following $t$. Thus every original
vertex of $B_U$ is internally reachable from $p_U(f)$ after $f$.
\end{proof}

\paragraph{Filtering Routes}

We first ensure that each root block $B_{x_i}$, for $i\in\{0,1\}$,
contains two spanning trees and a path between its two cycle ports,
with all three edge sets pairwise disjoint, as required by
\Cref{lem:characterization-m-lift}. A route entering $B_{x_i}$ through
a future call $t$ determines the port $p_{x_i}(t)$. We retain only
routes for which this port can be joined to the port $a_i$ of $e$ by
a path in $J_{B_{x_i}}^+(e)$ whose deletion leaves that graph connected.
The remaining edges of this graph supply one spanning tree, and
$J_{B_{x_i}}^-(e)$ supplies the other. We show that every original source in a
block represented in $Y$ still has a retained route to each root block.

For $i\in\{0,1\}$, call a future edge $t=Vx_i$ with $V\in V(Y)$
\emph{good at $x_i$} if there is an $a_i$-$p_{x_i}(t)$ path in $J_{B_{x_i}}^+(e)$ whose
deletion leaves $J_{B_{x_i}}^+(e)$ connected. This property concerns
only the edge $t$ and its endpoint $p_{x_i}(t)$ in $B_{x_i}$.

\begin{lemma}\label{lem:characterization-root-good-routes}
For every original vertex $u$ in a block represented in $Y$ and each
$i\in\{0,1\}$, there is an original route from $u$ to some original
vertex of $B_{x_i}$ whose last external call is good at $x_i$.
\end{lemma}
\begin{proof}
For every original vertex $v\in B_{x_i}$, choose an original
$u$-$v$ route. Its external projection ends at $x_i$.
The last assertion of \Cref{lem:characterization-canonical-ideal} implies that
this projection is an early path in $Y$ followed by a single future
edge $t_v=V_vx_i$. In particular, it cannot use $e$: an early prefix
starting in $Y$ cannot reach $X$.

The internal suffix from $p_{x_i}(t_v)$ to $v$ lies after $t_v$.
Confinement in $Q[F]+e$ gives an internal route from $a_i$ to
$p_{x_i}(t_v)$ between $e,t_v$. Whenever two selected future calls
are comparable, confinement in $F$ gives an internal route between
their ports strictly between those calls.

Apply \Cref{lem:characterization-antichain-selection} with the call order reversed to the
internal graph $J_{B_{x_i}}^+(e)$, with distinguished vertex $a_i$
and external calls $t_v$. The reversed suffixes supply its gathering
paths, the reversed routes from $a_i$ to $p_{x_i}(t_v)$ supply its connecting paths,
and the confined between-call routes supply its third condition.
Repeated calls $t_v$ are identified; all occurrences have the same
endpoint in $B_{x_i}$, and their suffixes still cover all targets.
Every used internal call lies after $e$. The lemma selects one
$t_v$ good at $x_i$, and the associated chosen original route proves
the assertion. Its target may be any vertex of $B_{x_i}$; no claim
that it ends specifically at $a_i$ is needed.
\end{proof}

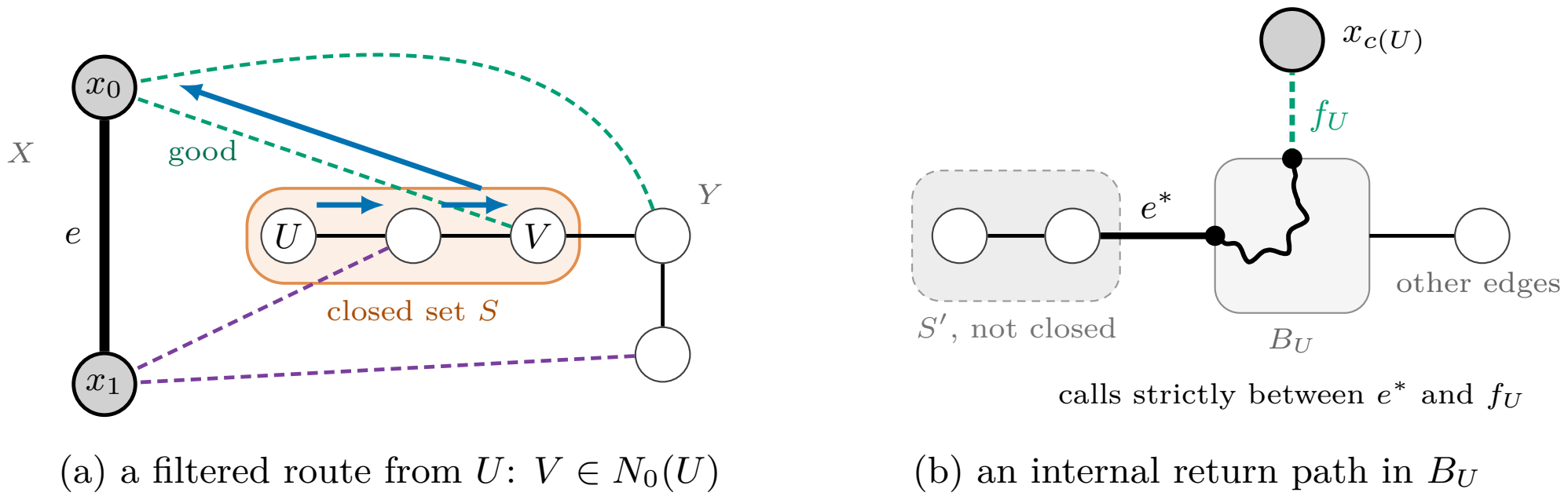
\begin{figure}[t]
\centering
\begin{tikzpicture}[
  every node/.style={font=\footnotesize},
  qv/.style={circle,draw=black!75,fill=white,inner sep=0pt,minimum size=4.6mm},
  rootv/.style={circle,draw=black,fill=black!18,line width=.9pt,inner sep=0pt,minimum size=5.2mm},
  treeE/.style={black,line width=.9pt},
  lastE/.style={black,line width=2.4pt},
  fut0/.style={auxgreen,line width=.9pt,dash pattern=on 2.5pt off 1.5pt},
  fut1/.style={auxpurple,line width=.9pt,dash pattern=on 2.5pt off 1.5pt},
  route/.style={-{Latex[length=2mm,width=1.6mm]},line width=1.3pt,witnessblue},
  setS/.style={fill=witnessorange!10,draw=witnessorange!70,rounded corners=9pt,line width=.7pt},
  setSp/.style={fill=black!7,draw=black!40,rounded corners=7pt,dash pattern=on 2pt off 1.5pt},
  blk/.style={rounded corners=6pt,draw=black!45,fill=black!4},
  port/.style={circle,fill=black,inner sep=0pt,minimum size=1.7mm},
  ipath/.style={line width=1.1pt,decorate,decoration={snake,amplitude=.5mm,segment length=3mm,pre length=1mm,post length=1mm}},
  good/.style={font=\scriptsize,text=auxgreen!70!black},
]
\begin{scope}
  \node[rootv] (x0) at (0,1.25) {$x_0$};
  \node[rootv] (x1) at (0,-1.25) {$x_1$};
  \draw[lastE] (x0)--(x1) node[midway,left=1pt] {$e$};
  \node[font=\scriptsize,text=black!60] at (-0.7,0.7) {$X$};
  \draw[setS] (1.2,-0.4) rectangle (4.0,0.4);
  \node[qv] (y1) at (1.55,0) {$U$};
  \node[qv] (y2) at (2.6,0) {};
  \node[qv] (y3) at (3.65,0) {$V$};
  \node[qv] (y4) at (4.7,0) {};
  \node[qv] (y5) at (4.7,-1.0) {};
  \draw[treeE] (y1)--(y2)--(y3)--(y4)--(y5);
  \node[font=\scriptsize,text=black!60] at (5.1,0.35) {$Y$};
  \node[font=\scriptsize,text=witnessorange!80!black,anchor=north] at (2.6,-0.42) {closed set $S$};
  \draw[fut0] (y3)--(x0);
  \node[good] at ($(y3)!0.8!(x0)+(0.1,-0.3)$) {good};
  \draw[fut1] (y2)--(x1);
  \draw[fut1] (y5)--(x1);
  \draw[fut0] (y4) to[out=110,in=10] (x0);
  \draw[route] ([yshift=2.6mm]y1.east)--([yshift=2.6mm]y2.west);
  \draw[route] ([yshift=2.6mm]y2.east)--([yshift=2.6mm]y3.west);
  \draw[route] ($(y3)!0.15!(x0)+(0.07,0.21)$)--($(y3)!0.85!(x0)+(0.07,0.21)$);
  \node[panel,align=center] at (2.4,-2.0) {(a) a filtered route from $U$: $V\in N_0(U)$};
\end{scope}
\begin{scope}[xshift=7.4cm]
  \node[rootv] (r) at (2.6,1.65) {}; \node[anchor=west,inner sep=4pt] at (r.east) {$x_{c(U)}$};
  \draw[setSp] (-0.6,-0.55) rectangle (1.15,0.55);
  \node[font=\scriptsize,text=black!55] at (0.28,-0.78) {$S'$, not closed};
  \node[qv] (s1) at (-0.2,0) {};
  \node[qv] (s2) at (0.75,0) {};
  \draw[treeE] (s1)--(s2);
  \draw[blk] (1.95,-0.65) rectangle (3.25,0.65);
  \node[font=\scriptsize,text=black!60] at (2.6,-0.88) {$B_U$};
  \coordinate (pe) at (1.95,0); \coordinate (pf) at (2.6,0.65);
  \draw[treeE,line width=1.6pt] (s2)--(pe) node[midway,above] {$e^*$};
  \draw[fut0,line width=1.3pt] (pf)--(r) node[midway,right] {$f_U$};
  \draw[ipath] (pe) .. controls (2.35,-0.35) and (2.75,0.0) .. (pf);
  \node[qv] (s3) at (4.2,0) {};
  \draw[treeE] (3.25,0)--(s3);
  \node[font=\scriptsize,text=black!60,anchor=north west] at (3.35,-0.18) {other edges};
  \foreach \p in {pe,pf} \node[port] at (\p) {};
  \node[font=\scriptsize,anchor=north] at (2.6,-1.1) {calls strictly between $e^*$ and $f_U$};
  \node[panel,align=center] at (1.8,-2.0) {(b) an internal return path in $B_U$};
\end{scope}
\end{tikzpicture}
\caption{Filtered routes and minimal closed sets
(\Cref{lem:characterization-root-good-routes,lem:characterization-actual-closed-set}).
(a) A filtered route projects to an early path in $Y$ followed by one
future call to a root block, here from $V$ to $x_0$; its last call is good
at that root. The vertex $V$ is then in $N_0(U)$. A set $S$ of vertices of
$Y$ is closed if every filtered route starting in a block represented
in $S$ makes its final departure toward a root from a block represented
in $S$. (b) Let $S$ be inclusion-minimal among nonempty closed sets
inducing a connected subgraph of $Y$, and let $U\in S$. Let $e^*$ be a
maximal edge of $Y[S]$ at $U$. The vertex set $S'$ of the
component of $Y[S]-e^*$ not containing $U$ is not closed, so some filtered route
starting in $S'$ leaves it. That route crosses $e^*$ into the block $B_U$
and leaves it through the future call $f_U$ to a root. Its segment inside
$B_U$ (wavy) uses only calls between $e^*$ and $f_U$.}
\label{fig:filtered-routes}
\end{figure}

A \emph{filtered route with root index $i$} starts at an original vertex of
a block in $Y$, ends at any original vertex of $B_{x_i}$, and has a
last external call good at $x_i$. Every such route projects to an
early path in $Y$ followed by that last future call. For $U\in V(Y)$,
let $N_i(U)$ be the set of all vertices $V$ that occur immediately
before $x_i$ in these projections, ranging over \emph{all} filtered
routes starting at \emph{every} original vertex of $B_U$.
A nonempty set $S\subseteq V(Y)$ is \emph{closed} if
\[
 N_0(U)\cup N_1(U)\subseteq S\qquad\text{for every }U\in S.
\]
Thus every filtered route starting in a block represented in $S$
makes its final departure toward a root from a block still
represented in $S$.
The whole set $V(Y)$ is closed and induces a connected graph. Choose
$S$ inclusion-minimal among nonempty closed sets for which $Y[S]$ is
connected. \Cref{fig:filtered-routes} illustrates filtered routes and
closed sets, together with the internal return paths obtained from
the minimality of $S$.

\paragraph{Obtaining Internal Return Paths}

Filtering provides suitable attachment paths inside the root blocks.
We now prepare the corresponding paths inside blocks represented by
the other early tree $Y$. Recall that $S$ is an inclusion-minimal
nonempty closed set for which $Y[S]$ is connected. We show that each
block represented in $S$ has a future call to a root and that every
incident edge of $Y[S]$ can be followed by an internal route to the
port of this future call. These routes will supply the connecting
paths needed by the ordered path-selection lemma,
\Cref{lem:characterization-antichain-selection}.

\begin{lemma}[Minimal closed sets supply internal returns]
\label{lem:characterization-actual-closed-set}
The graph $Y[S]$ is a tree with at least two vertices. Every $U\in S$
has a unique future edge $f_U=Ux_{c(U)}$ incident with a root, for a
root index $c(U)\in\{0,1\}$, and this edge is good at $x_{c(U)}$.
For every edge $h$ of $Y[S]$ incident with $U$, we have
$h<_{P_Q}f_U$, and there is an internal
original route from $p_U(h)$ to $p_U(f_U)$ with all calls strictly
between $h$ and $f_U$.
\end{lemma}
\begin{proof}
By its choice, $Y[S]$ is a connected subgraph of the tree $Y$, hence a tree.
A singleton closed set would require future edges from its sole
vertex to both roots, since every original source has a filtered
route of each root index by \Cref{lem:characterization-root-good-routes}. Those edges
would connect the two root components of $F$, which is impossible.
Consequently $|S|\ge2$.

Fix $U\in S$ and an edge $e^*$ maximal among the edges of $Y[S]$
incident with $U$. Let $S'$ be the vertex set of the component of
$Y[S]-e^*$ not containing $U$. Then $S'$ is a nonempty proper subset
of $S$, and $Y[S']$ is connected. By minimality of $S$, the set $S'$
is not closed. A witnessing filtered
route has its starting block in $S'$ and its exit in $S\setminus S'$.
It crosses $e^*$ into $U$ and cannot use another early edge, by
maximality of $e^*$. Its next external call is therefore a future edge
$f_U$ from $U$ to a root. This call is good at that root, and the
actual route gives $e^*<_{P_Q}f_U$, even if its internal segment is
trivial. That segment is the required return path.
There cannot be future edges from $U$ to both roots because the roots
lie in different components of $F$; simplicity gives uniqueness for
the chosen root. This proves the assertion for every maximal $e^*$.

For an arbitrary incident edge $h$, choose a maximal incident edge
$e^*$ with $h\le_{P_Q}e^*$. If $h\ne e^*$, confinement in the early tree
$Y$ supplies an internal original route from $p_U(h)$ to $p_U(e^*)$
between $h,e^*$. Append the return path just obtained between $e^*,f_U$.
In particular $h<_{P_Q}f_U$. The ordered call intervals give an
increasing internal walk; erase
closed subwalks. The resulting path lies strictly between $h,f_U$.
\end{proof}

\paragraph{Selecting Neighbors with Different Root Indices}
Each vertex $U\in S$ now has a future edge $f_U$ to the root
$x_{c(U)}$. We seek adjacent vertices $U_0,U_1$ with different root
indices, so that their future edges, their common early edge, and
$e=x_0x_1$ form a four-cycle. At both ends of the common early edge,
we also need an internal path to the future port whose deletion
leaves the corresponding graph $J_{B_U}^-(f_U)$ connected. We first find
admissible neighbor choices at every vertex, then use a rerouting
argument to obtain a pair that chooses each other and has different
root indices.

\begin{lemma}\label{lem:characterization-safe-first-edges}
Fix $U\in S$. For every original vertex $u\in B_U$, choose a
filtered route with root index $1-c(U)$. Among their first external edges
there is an edge $h$ such that $J_{B_U}^-(f_U)$ contains a
$p_U(h)$-$p_U(f_U)$ path whose deletion leaves this graph connected.
\end{lemma}
\begin{proof}
The early projection of every chosen route stays in $Y[S]$ because
its starting block and its exit belong to $S$. Its first external
edge $h_u$ is early: the only root exit from $U$ has root index $c(U)$,
whereas the chosen route has the other root index. Its initial internal
segment gathers $u$ to $p_U(h_u)$ before $h_u$.
\Cref{lem:characterization-actual-closed-set} supplies an internal route from
$p_U(h_u)$ to $p_U(f_U)$ between $h_u,f_U$. All these gathering and
connecting paths lie in $J_{B_U}^-(f_U)$.

For any comparable chosen first edges $h_u<_{P_Q}h_v$, confinement
in $Y$ supplies the internal between-port route required by the
third condition of \Cref{lem:characterization-antichain-selection}. Its first
condition is supplied by the gathering prefixes, and its second by
the return paths to $p_U(f_U)$. That lemma gives the claimed edge
and nonseparating path. Repeated occurrences of the same first call
may be identified; they have the same original port.
\end{proof}

To find an early edge with compatible internal paths at both ends,
we encode the admissible choices at each block in a directed graph.

Define a directed graph $D$ on $S$, using only edges of $Y[S]$.
Include an arc $U\to V$ if there is a filtered route starting in
$B_U$ with root index $1-c(U)$ whose first external edge is $UV$, and
there is a
$p_U(UV)$-$p_U(f_U)$ path in $J_{B_U}^-(f_U)$ whose deletion
leaves that graph connected. The route and the nonseparating path
need not use the same internal edges.
\Cref{lem:characterization-safe-first-edges} shows that every vertex of $D$ has
positive outdegree. Its formulation also allows us to prescribe a
family of candidate routes before choosing the outgoing arc.

\begin{lemma}\label{lem:characterization-same-index-rerouting}
If $U\to V$ is an arc of $D$ and $c(U)=c(V)$, then $V$ has an
outgoing arc to a neighbor other than $U$.
\end{lemma}
\begin{proof}
Take a filtered route witnessing $U\to V$. Its terminal exit has
root index $1-c(U)$, so its early projection cannot end at $V$. Let $VW$
be its next external edge after $UV$. Between these calls the route
has an actual internal segment in $B_V$ from $p_V(UV)$ to $p_V(VW)$,
with all calls strictly between $UV,VW$.

For every original vertex $v\in B_V$, choose a filtered route with
root index $1-c(V)$. If its first external edge differs from $VU$, keep
it. Otherwise keep its internal prefix before $VU$, then append the
actual internal $V$-segment of the $U$-route and the suffix of that
route beginning with $VW$. The prefix ends at $p_V(VU)$ before
$VU$; the replacement segment starts at that same port after $UV$.
Hence this concatenation is an increasing original walk. Its first
external call is $VW$, and its last external call is still good at
the opposite root. The suffix after $VW$ never returns to $V$,
since its early projection is a path in $Y$. Erasing closed
subwalks inside $B_V$ therefore gives a filtered original route
whose first external edge differs from $VU$.

We have obtained such a candidate route for every original vertex
of $B_V$. Apply \Cref{lem:characterization-safe-first-edges} to these candidates.
The resulting outgoing arc of $V$ avoids $U$.
\end{proof}

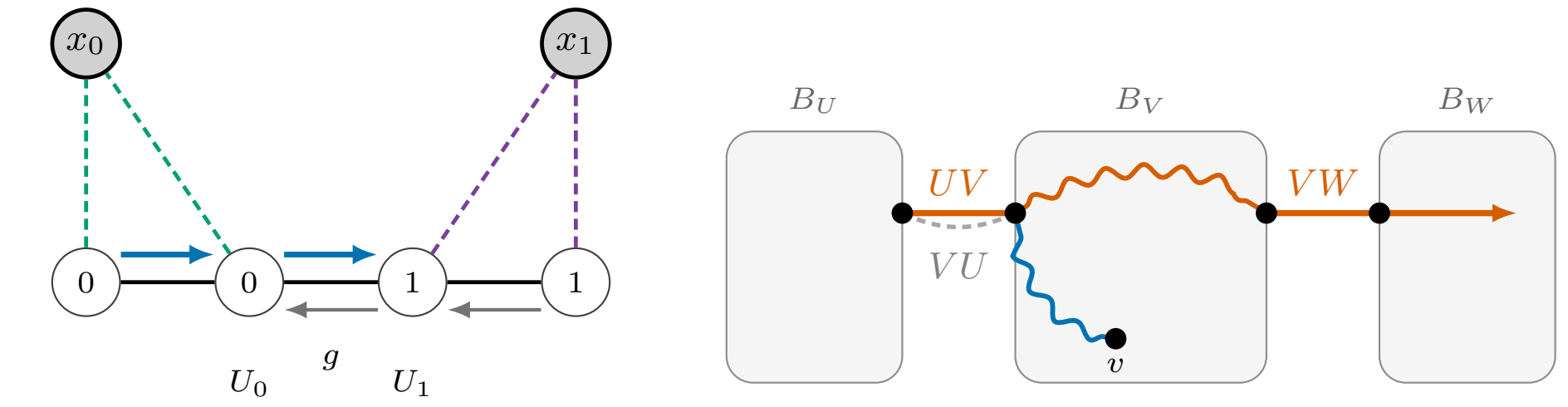
\begin{figure}[t]
\centering
\begin{tikzpicture}[
    every node/.style={font=\footnotesize},
    qv/.style={circle,draw=black!75,fill=white,inner sep=0pt,minimum size=5.4mm},
    rootv/.style={circle,draw=black,fill=black!18,line width=.9pt,inner sep=0pt,minimum size=5.4mm},
    treeE/.style={black,line width=.9pt},
    fut0/.style={auxgreen,line width=.9pt,dash pattern=on 2.5pt off 1.5pt},
    fut1/.style={auxpurple,line width=.9pt,dash pattern=on 2.5pt off 1.5pt},
    arc/.style={-{Latex[length=1.8mm,width=1.4mm]},line width=.8pt,black!55},
    patharc/.style={-{Latex[length=2mm,width=1.6mm]},line width=1.3pt,witnessblue},
    gE/.style={black,line width=2.4pt},
    blk/.style={rounded corners=6pt,draw=black!45,fill=black!4},
    port/.style={circle,fill=black,inner sep=0pt,minimum size=1.7mm},
    ipath/.style={line width=1.1pt,decorate,decoration={snake,amplitude=.5mm,segment length=3mm,pre length=1mm,post length=1mm}},
    panel/.style={font=\small}
]
\begin{scope}
  \node[rootv] (x0) at (0.0,1.9) {$x_0$};
  \node[rootv] (x1) at (3.9,1.9) {$x_1$};
  \node[qv] (A) at (0,0) {};
  \node[qv] (B) at (1.3,0) {};
  \node[qv] (C) at (2.6,0) {};
  \node[qv] (E) at (3.9,0) {};
  \draw[treeE] (A)--(B)--(C)--(E);
  \draw[fut0] (A)--(x0) (B)--(x0);
  \draw[fut1] (C)--(x1) (E)--(x1);
  \foreach \v/\c in {A/0,B/0,C/1,E/1} \node[font=\scriptsize] at (\v.center) {$\c$};
  \draw[patharc] ([yshift=2.2mm]A.east)--([yshift=2.2mm]B.west);
  \draw[patharc] ([yshift=2.2mm]B.east)--([yshift=2.2mm]C.west);
  \draw[arc] ([yshift=-2.2mm]C.west)--([yshift=-2.2mm]B.east);
  \draw[arc] ([yshift=-2.2mm]E.west)--([yshift=-2.2mm]C.east);
  \node[font=\scriptsize] at (B.south) [below=3mm] {$U_0$};
  \node[font=\scriptsize] at (C.south) [below=3mm] {$U_1$};
  \node[font=\scriptsize] at (1.95,-0.62) {$g$};
  \node[panel,align=center] at (1.95,-1.65) {(a) the digraph $D$ on the tree $Y[S]$};
\end{scope}
\begin{scope}[xshift=6.0cm,yshift=0.2cm]
  \draw[blk] (-0.9,-1.0) rectangle (0.5,1.0);  \node[font=\scriptsize,text=black!60] at (-0.2,1.25) {$B_U$};
  \draw[blk] (1.4,-1.0) rectangle (3.4,1.0);   \node[font=\scriptsize,text=black!60] at (2.4,1.25) {$B_V$};
  \draw[blk] (4.3,-1.0) rectangle (5.7,1.0);   \node[font=\scriptsize,text=black!60] at (5.0,1.25) {$B_W$};
  \coordinate (pu) at (0.5,0.35); \coordinate (pv) at (1.4,0.35);
  \coordinate (qv) at (3.4,0.35); \coordinate (qw) at (4.3,0.35);
  \coordinate (v) at (2.2,-0.65);
  \draw[witnessorange,line width=1.4pt] (pu)--(pv) node[midway,above] {$UV$};
  \draw[witnessorange,ipath] (pv) .. controls (2.0,0.75) and (2.8,0.75) .. (qv);
  \draw[witnessorange,line width=1.4pt] (qv)--(qw) node[midway,above] {$VW$};
  \draw[witnessorange,line width=1.4pt,-{Latex[length=2mm]}] (qw)--(5.4,0.35);
  \draw[witnessblue,ipath] (v) .. controls (1.7,-0.6) and (1.5,-0.1) .. (pv);
  \draw[black!35,line width=1pt,dash pattern=on 2pt off 1.5pt] (pv) to[bend left=25] node[midway,below=2pt,text=black!50] {$VU$} (pu);
  \foreach \p in {pu,pv,qv,qw} \node[port] at (\p) {};
  \node[port] at (v) {}; \node[below=1pt,font=\scriptsize] at (v) {$v$};
  \node[panel,align=center] at (2.4,-1.85) {(b) rerouting when $c(U)=c(V)$};
\end{scope}
\end{tikzpicture}
\caption{Selecting adjacent blocks with different root indices
(\Cref{lem:characterization-same-index-rerouting,lem:characterization-compatible-actual-cycle}).
(a) The vertices of the tree $Y[S]$ carry their root indices $c(U)$,
and dashed edges show the future edges $f_U$ to the roots $x_0,x_1$.
Gray and blue arrows are arcs of $D$. A directed path that cannot be
extended (blue) ends with an arc $U_0\to U_1$; because $Y[S]$ is a tree,
its last vertex can only point back to its predecessor, so
$U_1\to U_0$ is also an arc. If the two root indices were equal,
rerouting would give another outgoing arc, so they differ, and
$g=U_0U_1$ is the required edge. (b) The rerouting step: a route from
$v\in B_V$ that would leave through $VU$ retains its internal prefix
(blue), then follows the internal segment and the suffix of a route
witnessing $U\to V$ (orange), and
leaves $B_V$ through $VW$.}
\label{fig:port-selection}
\end{figure}

\Cref{fig:port-selection} illustrates this rerouting step and the
selection of adjacent blocks with different root indices in the next
lemma.

\begin{lemma}[Compatible four-cycle]
\label{lem:characterization-compatible-actual-cycle}
There is an edge $g=U_0U_1$ of $Y$ with $f_{U_i}=U_ix_i$ for
$i\in\{0,1\}$ such that the four-cycle
\[
 x_0,U_0,U_1,x_1,x_0
\]
has the local lifting property of \Cref{lem:characterization-m-lift} in each
of its four original blocks.
\end{lemma}
\begin{proof}
Choose a directed path of $D$ that cannot be extended at its last
vertex without repeating a vertex. Since every vertex has positive
outdegree, this path has a last arc $U\to V$. Every outgoing neighbor
of $V$ lies on the path. A neighbor other than its predecessor $U$
would form a cycle with a segment of that path in the tree $Y[S]$.
Thus $U$ is its only possible outgoing neighbor, and $V\to U$ is an arc.
If $c(U)=c(V)$, \Cref{lem:characterization-same-index-rerouting} gives
an outgoing neighbor of $V$ other than $U$, a contradiction.
Relabel the endpoints as $U_0,U_1$ so that $c(U_i)=i$, and let
$g=U_0U_1$. Both orientations of $g$ belong to $D$.

For each $i$, the arc $U_i\to U_{1-i}$ supplies a path between
the $U_i$-ports of $g$ and $f_{U_i}$ whose deletion leaves
$J_{B_{U_i}}^-(f_{U_i})$ connected.
\Cref{lem:characterization-internal-threshold-connectivity} gives a connected
internal graph strictly after $f_{U_i}$. Apply
\Cref{lem:characterization-threshold} with these two disjoint edge sets. The
block $B_{U_i}$ therefore contains two spanning trees and a path
between its cycle ports, with all three edge sets pairwise disjoint.

By \Cref{lem:characterization-actual-closed-set}, the edge $f_{U_i}$ is good at
$x_i$. Hence the ports of $e$ and $f_{U_i}$ in $B_{x_i}$ are joined
by a path whose deletion leaves $J_{B_{x_i}}^+(e)$ connected.
The graph $J_{B_{x_i}}^-(e)$ is connected by
\Cref{lem:characterization-internal-threshold-connectivity}. Another application
of \Cref{lem:characterization-threshold}, with the two sides of $e$ interchanged,
gives the same local lifting property in $B_{x_i}$.
The four quotient vertices are distinct because $U_0,U_1$ lie in
the other early component $Y$.
\end{proof}

The selected four-cycle has compatible internal paths in all four
blocks. We can therefore extend it to a quotient $M$-witness and lift
that witness to $G$.

\begin{corollary}\label{cor:characterization-optimal-quotient-lift}
Under the standing hypotheses, $G$ has an $M$-witness.
\end{corollary}
\begin{proof}
Apply \Cref{lem:characterization-protected-extension} to the edge $U_0U_1$ and
the two future edges from \Cref{lem:characterization-compatible-actual-cycle}.
It extends their four-cycle to a quotient $M$-witness. The local
lifting property just proved satisfies all hypotheses of
\Cref{lem:characterization-m-lift}, which gives an $M$-witness of $G$.
\end{proof}

We now complete the necessity proof by distinguishing the small
quotients and the two possible edge counts for larger quotients.

\begin{proof}[Proof of necessity in \Cref{thm:characterization}]
Discard loops and assume that $G$ has at least two vertices.
Completeness implies that $G$ is connected. Form its maximal
two-tree blocks and their quotient $Q$. By
\Cref{lem:characterization-quotient}, the graph $Q$ is connected and
simple and contains no nontrivial two-tree set. By
\Cref{lem:characterization-poset-transfer}, every linear extension of
the induced order $P_Q$ is complete on $Q$. If $Q$ has one vertex,
\Cref{lem:characterization-t-lift} gives two edge-disjoint spanning trees.
If $Q$ has two vertices, it is a single edge. If it has three,
completeness forces a triangle: otherwise it is a two-edge path,
and a secret sent on its second call cannot reach the endpoint
whose only call was first. These quotients have two
spanning trees sharing one edge, which lift by
\Cref{lem:characterization-t-lift}.

Suppose $Q$ has $r\ge4$ vertices. By
\Cref{lem:characterization-poset-transfer,cor:characterization-call-lower-bound,cor:characterization-edges},
we have $2r-4\le |E(Q)|\le2r-3$. The upper case gives a $T$-witness of $G$ by
\Cref{cor:characterization-t-case}. In the lower case,
\Cref{lem:characterization-canonical-ideal} and
\Cref{cor:characterization-early-components} give the standing hypotheses
of this section. \Cref{cor:characterization-optimal-quotient-lift} then
gives an $M$-witness of $G$.
\end{proof}

This completes the characterization. For sufficiency, either witness
yields a complete call order by \Cref{lem:construct-labeling}. For
necessity, contracting the two-tree blocks reduces the problem to a
simple quotient. Small quotients and the upper edge-count case yield
a $T$-witness. In the remaining case, the call order supplies the
canonical structure used to select a four-cycle with compatible
internal paths. Extending this cycle to a quotient $M$-witness and
lifting it through the blocks yields an $M$-witness of the original
graph. Thus every graph admitting a temporally connecting labeling
has one of the structures in \Cref{thm:characterization}.

\section{Matroid Representation Tools}\label{app:matroid-tools}

We prove the standard representation lemmas of
\Cref{sec:matroid-preliminaries}, using the notation introduced there.

\begin{proof}[Proof of \Cref{lem:union-representation}]
Let $X\in\mathcal{I}(\mathcal N)$. Then some set $W$ of $|X|$ rows
satisfies $\det A[W,X]\neq0$. Write $W=W_1\cup W_2$, where $W_1$ and
$W_2$ are the rows coming from $A_1$ and from $A_2\Lambda$, respectively. By
the generalized Laplace expansion along the rows $W_1$,
\[
    \det A[W,X]=\sum_{X_1\,\dot\cup\,X_2=X,\ |X_1|=|W_1|}
    \pm\det A_1[W_1,X_1]\cdot\det A_2[W_2,X_2]\cdot
    \prod_{e\in X_2}x_e.
\]
Some summand is nonzero. Hence, the columns of $X_1$ are independent in
$A_1$ and those of $X_2$ are independent in $A_2$, so
$X=X_1\cup X_2\in\mathcal{I}(M_1\vee M_2)$.

Now let $B\in\mathcal{I}(M_1\vee M_2)$ with $|B|=r_1+r_2$. Then $B$ is
the disjoint union of a basis $D_1$ of $M_1$ and a basis $D_2$ of $M_2$.
Consider $\det A[B]$ as a polynomial in the variables $x_e$, using the
same expansion with all rows. Distinct sets $X_2$ yield distinct
monomials $\prod_{e\in X_2}x_e$, so no cancellation occurs, and the
coefficient of $\prod_{e\in D_2}x_e$ is nonzero. Thus $\det A[B]$ is a
nonzero polynomial of total degree $r_2$. By the Schwartz--Zippel
lemma~\cite{Schwartz1980}, a uniformly random evaluation is zero with
probability at most $r_2/|\mathbb{F}|$.
\end{proof}

The same expansion shows that $A$ represents
$M_1\vee M_2$ over the field of rational functions in the variables
$x_e$; this is the classical construction used by
Marx~\cite{Marx09} and, in a parallel setting, by Narayanan, Saran, and
Vazirani~\cite{NarayananSV1994}. We only need the two properties stated
in \Cref{lem:union-representation}.

\Cref{lem:union-representation} is applied to one outcome of $\Lambda$
for several pairs $A_1,A_2$ on the same ground set in
\Cref{lem:m-witness-to-ric}. This is allowed: the first statement holds
for every outcome and hence for all pairs simultaneously, and the
second statement concerns a single fixed pair and a single fixed set $B$.

\begin{proof}[Proof of \Cref{lem:dual-representation}]
Deleting rows of $A$ that are linear combinations of other rows changes
neither $M[A]$ nor $\ker A$, so we may assume that $A$ has full row rank
$r$. Choose a set $B$ of $r$ linearly independent columns and reorder
the columns as $A=(A[B]\ A[\bar B])$ with $\bar B=E\setminus B$. The
matrix $A'=A[B]^{-1}A$ equals $(I_r\ A'')$ for some $A''$, and
$M[A']=M[A]$ and $\ker A'=\ker A$. The matrix
$Z_0=(-A''^{\mathsf T}\ I_{|E|-r})$ represents
$M[A]^*$~\cite[Theorem~2.2.8]{Oxley2011}. Since
$A'Z_0^{\mathsf T}=-A''+A''=0$, its $|E|-r=\dim\ker A$ rows lie in
$\ker A$, and they are linearly independent, so they form a basis of
$\ker A$. Hence $Z=TZ_0$ for an invertible matrix $T$. Multiplying from
the left by an invertible matrix preserves linear dependencies among the
columns, so $M[Z]=M[Z_0]=M[A]^*$, and $Z$ has full row rank.
\end{proof}

We next give the linear-algebra computation used to share the
preprocessing across root sets.

\begin{proof}[Proof of \Cref{lem:max-submatrix}]
By transposing $\Phi$ and exchanging the roles of $I$ and $J$, we may
assume $a\leq b$. The LSP decomposition of Ibarra, Moran, and
Hui~\cite{IbarraMH82} computes, in $\Oh(a^{\omega-1}b)$ time, a
factorization $\Phi=LSP$, where $L$ is a unit lower triangular $a\times
a$ matrix, $P$ is a permutation matrix, and $S$ is an $a\times b$ matrix
whose nonzero rows, $r=\operatorname{rk}\Phi$ many, form an upper
triangular matrix with nonzero diagonal entries after deleting the
zero rows. For every $i$, the first
$i$ rows of $\Phi$ are obtained from the first $i$ rows of $SP$ by
multiplying with the invertible leading $i\times i$ block of $L$, so both
span the same space. Hence the rank of the first $i$ rows of $\Phi$
increases exactly at the indices of the nonzero rows of $S$, and these
indices form a set $I$ of $r$ linearly independent rows of $\Phi$.
Applying the same decomposition to the $r\times b$ matrix $\Phi[I,\cdot]$
of full row rank yields, in $\Oh(r^{\omega-1}b)$ time, a set $J$ of $r$
linearly independent columns of $\Phi[I,\cdot]$, namely the columns that
$P$ maps to the positions of the diagonal of the triangular part. Then
$\Phi[I,J]$ is nonsingular, and we invert it in $\Oh(r^\omega)$ time. The
total time is $\Oh(a^{\omega-1}b)=\Oh(ab\cdot\min\{a,b\}^{\omega-2})$.

The rows in $I$ are $r=\operatorname{rk}\Phi$ linearly independent rows,
so they span the row space of $\Phi$. Thus $\Phi=X\,\Phi[I,\cdot]$ for
some matrix $X$. Restricting to the columns in $J$ gives
$\Phi[\cdot,J]=X\,\Phi[I,J]$, so $X=\Phi[\cdot,J]\,\Phi[I,J]^{-1}$. This
proves the factorization.
\end{proof}

\Cref{lem:equal-rank-dual} is a special case of the standard fact that
rank-preserving weak maps are preserved by duality (see, e.g.,
Oxley~\cite{Oxley2011}); we include the short proof.

\begin{proof}[Proof of \Cref{lem:equal-rank-dual}]
A basis of $N$ is independent in $M$ and has size $r(N)=r(M)$, so it is
a basis of $M$. If $X\in\mathcal{I}(N^*)$, then the definition of
dual independence implies that $E\setminus X$ contains a basis of $N$.
This is also a basis of $M$, so $X\in\mathcal{I}(M^*)$ by the same definition.
\end{proof}

\end{document}